\documentclass[onecolumn,showpacs,preprintnumbers,amsmath,amssymb,floatfix]{revtex4}
\usepackage{url}
\usepackage{bm}		
\usepackage{graphicx}	
\usepackage{dcolumn}	
\usepackage{fancybox}	

\begin{document}

\title[Perturbation theory of self-similar black holes]{Perturbation theory of spherically symmetric self-similar black holes}

\author{Andrew J S Hamilton}
\affiliation{JILA and
Dept.\ Astrophysical \& Planetary Sciences,
Box 440, U. Colorado, Boulder, CO 80309, USA}
\email{Andrew.Hamilton@colorado.edu}
\homepage{http://casa.colorado.edu/~ajsh/}

\newcommand{\dd}{d}
\newcommand{\im}{i} 
\newcommand{\ee}{e} 
\newcommand{\perpperp}{\perp\!\!\perp}

\newcommand{\proper}[1]{\tilde{#1}}

\newcommand{\Msun}{\textrm{M}_\odot}

\newcommand{\vel}{\mbox{\sl v\/}}

\newcommand{\uv}{{\overset{\scriptstyle u}{\scriptstyle v}}}
\newcommand{\vu}{{\overset{\scriptstyle v}{\scriptstyle u}}}

\newcommand{\Dr}[2]{{\partial^{#1} {#2} \over \partial r^{\ast {#1}}}}
\newcommand{\cz}{{\tilde c}}
\newcommand{\Cz}{{\tilde C}}
\newcommand{\fz}{{\tilde f}}
\newcommand{\Fz}{{\tilde F}}
\newcommand{\Jz}{{\tilde J}}
\newcommand{\Kz}{{\tilde K}}
\newcommand{\Nz}{{\tilde N}}
\newcommand{\Rz}{{\tilde R}}
\newcommand{\Sz}{{\tilde S}}

\newcommand{\plus}{{\scriptscriptstyle +}}
\newcommand{\minus}{{\scriptscriptstyle -}}
\newcommand{\plusminus}{{\scriptscriptstyle \pm}}
\newcommand{\minusplus}{{\scriptscriptstyle \mp}}
\newcommand{\zero}{{\scriptstyle 0}}
\newcommand{\one}{{\scriptstyle 1}}
\newcommand{\two}{{\scriptstyle 2}}
\newcommand{\three}{{\scriptstyle 3}}
\newcommand{\comma}{{\scriptscriptstyle ,}}
\newcommand{\smallzero}{{\scriptscriptstyle 0}}
\newcommand{\smallone}{{\scriptscriptstyle 1}}

\newcommand{\gvec}{\bm{g}}
\newcommand{\pvec}{\bm{p}}
\newcommand{\xvec}{\bm{x}}
\newcommand{\gammavec}{\bm{\gamma}}
\newcommand{\partialvec}{\bm{\partial}}

\newcommand{\gravM}{\calM}
\newcommand{\gravN}{\calN}
\newcommand{\emQ}{\calQ}
\newcommand{\emS}{\calS}

\newcommand{\frakC}{\mathfrak{C}}
\newcommand{\calD}{\mathcal{D}}
\newcommand{\frakD}{\mathfrak{D}}
\newcommand{\calE}{\mathcal{E}}
\newcommand{\frakE}{\mathfrak{E}}
\newcommand{\calM}{\mathcal{M}}
\newcommand{\calN}{\mathcal{N}}
\newcommand{\calQ}{\mathcal{Q}}
\newcommand{\calS}{\mathcal{S}}
\newcommand{\frakS}{\mathfrak{S}}
\newcommand{\ncz}[1]{{}_{#1\,} \! \cz}
\newcommand{\nd}[1]{{}_{#1\,} \! d}
\newcommand{\nfrakC}[1]{{}_{#1\,} \! \frakC}
\newcommand{\ncalD}[1]{{}_{#1\,} \! \calD}
\newcommand{\nfrakD}[1]{{}_{#1\,} \! \frakD}
\newcommand{\nfrakDzero}[1]{{}_{#1\,} \! \overset{\smallzero}{\frakD}}
\newcommand{\nfz}[1]{{}_{#1\,} \! \fz}
\newcommand{\nDelta}[1]{{}_{#1\,} \! \Delta}
\newcommand{\nDeltazero}[1]{{}_{#1\,} \! \overset{\smallzero}{\Delta}}
\newcommand{\ndelta}[1]{{}_{#1\,} \! \delta}
\newcommand{\ndeltazero}[1]{{}_{#1\,} \! \overset{\smallzero}{\delta}}
\newcommand{\ncalQ}[1]{{}_{#1\,} \! \calQ}
\newcommand{\nJ}[1]{{}_{#1\,} \! J}
\newcommand{\nL}[1]{{}_{#1\,} \! L}
\newcommand{\nY}[1]{{}_{#1\,} \! Y}
\newcommand{\nz}[1]{{}_{#1\,} \! z}
\newcommand{\nZ}[1]{{}_{#1\,} \! Z}
\newcommand{\nsquare}[1]{{}_{#1\,} \! \square}
\newcommand{\npsi}[1]{{}_{#1\,} \! \psi}

\hyphenpenalty=3000

\newcommand{\penrosernfig}{
    \begin{figure}
    \includegraphics[scale=1]{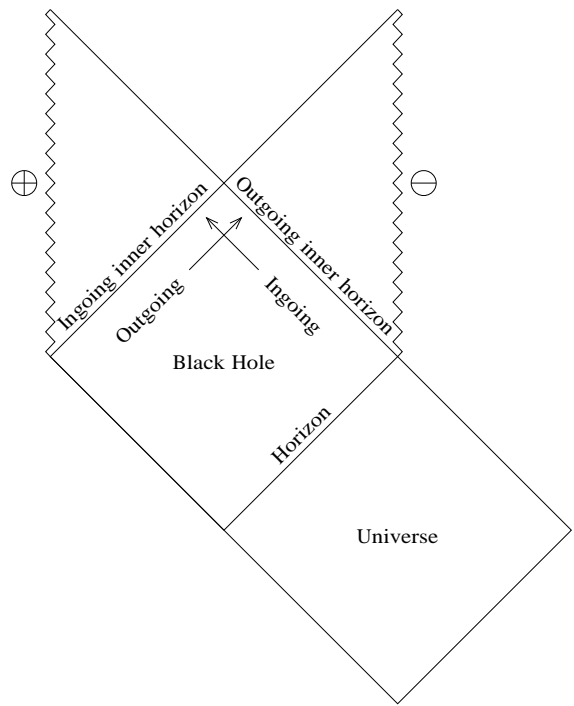}
    \caption[1]{
    \label{penrosern}
Partial Penrose diagram of the Reissner-Nordstr\"om geometry
for a stationary charged black hole,
illustrating that in order to fall through the inner horizon,
ingoing (positive energy) and outgoing (negative energy) streams
must fall through separate ingoing and outgoing inner horizons
into causally separate regions of spacetime.
    }
    \end{figure}
}

\begin{abstract}
The theory of perturbations of spherically symmetric self-similar black holes
is presented, in the Newman-Penrose formalism.
It is shown that
the wave equations for gravitational, electromagnetic, and scalar waves
are separable, though not decoupled.
A generalization of the Teukolsky equation is given.
Monopole and dipole modes are treated.
The Newman-Penrose wave equations
governing polar and axial spin-$0$ perturbations
are explored.
\end{abstract}


\date{\today}

\maketitle

\section{Introduction}

It has been known for decades that something strange and dramatic must happen
near the inner horizon of a realistic black hole.
Penrose (1968) \cite{Penrose68}
first pointed out that a person passing through the Cauchy horizon
(the outgoing inner horizon)
of an empty charged
(Reissner-Nordstr\"om)
or rotating
(Kerr-Newman)
black hole
will see the outside universe infinitely blueshifted.
Empty here means empty of charge, matter, or radiation
(but not of a static electric field)
except at the singularity.
Subsequent perturbation theory investigations,
starting with Simpson \& Penrose (1973)
\cite{SP73}
and culminating with Chandrasekhar \& Hartle (1982)
\cite{CH82},
confirmed that an observer crossing the Cauchy horizon
of the Reissner-Nordstr\"om geometry
will experience an infinite flux of radiation
from the outside world.
The prediction that
an infinite energy density will collect at the Cauchy horizon
is at odds with the assumption that the black hole is empty,
except in
the exceptional case of an uncharged, non-rotating
(Schwarzschild) black hole,
whose inner horizon coincides with the singularity at zero radius.

The full nonlinear nature of the instability near the inner horizon
was clarified in a seminal paper by Poisson \& Israel (1990) \cite{PI90}.
Poisson \& Israel argued that,
if ingoing and outgoing streams are simultaneously present
just above the inner horizon of a spherical charged black hole,
then relativistic counter-streaming between the ingoing and outgoing fluids
will lead to an exponentially growing instability which they dubbed
``mass inflation''.
During mass inflation,
the interior, or Misner-Sharpe \cite{MS64}, mass,
a gauge-invariant scalar quantity,
exponentiates to huge values.
Other gauge-invariant measures, such as
the proper density, the proper pressure, and the Weyl scalar,
exponentiate along with the interior mass.
The phenomenon of mass inflation has been confirmed
analytically and numerically
in many papers
\cite{Ori91,BDIM94,BS95,BO98,Burko97,Burko03,Dafermos04b,HKN05,HP05b}.

Real black holes probably rotate (and rotate rapidly),
but probably have very little charge,
thanks to the huge charge-to-mass ratio of individual protons and electrons,
$e / m_\textrm{p} \approx 10^{18}$,
where $e$ is the dimensionless charge of the proton or electron,
the square root of the fine-structure constant,
and $m_\textrm{p}$ is the proton mass in units of the Planck mass.
The literature on mass inflation in rotating black holes is
exceedingly modest
\cite{Ori92,Ori99}
(see also the review \cite{Brady99}),
and there is much room for further investigation
of this hard problem.
In the meantime the present paper follows the bulk of the literature
in assuming that charge is a satisfactory surrogate for angular momentum
(e.g.~\cite{Dafermos04b}).
Physically,
the negative pressure, or tension, of the radial electric field
of a charged black hole
produces a gravitational repulsion analogous to,
and perhaps an adequate model for,
that produced by the centrifugal force in a rotating black hole.
It is this gravitational repulsion
that is responsible for the presence of inner horizons
inside empty black holes,
within which the flow of space becomes subluminal.

Physically, what causes mass inflation?
It is clear that relativistic counter-streaming
between ingoing and outgoing streams is a central ingredient,
as originally suggested by Poisson \& Israel \cite{PI90}.
Hansen et al{.} \cite{HKN05}
have suggested that adiabatic compression may be a contributing factor.
In our own previous investigations
we considered feeding a spherical black hole
in a self-similar fashion
both with a single fluid of charged baryons
\cite{HP05a},
and with two
counter-streaming fluids, charged baryons and neutral dark matter
\cite{HP05b}.
We found that mass inflation did not occur in the
case of a single fluid,
but did occur with two fluids.
For the baryons we assumed a relativistic equation of state
$p = \frac{1}{3} \rho$,
so that the sound speed, at $\sqrt{1/3}$ of the speed of light,
was less than the speed of light.
Thus in the case of the single fluid
there was no possibility of relativistic counter-streaming,
and indeed no mass inflation occurred.
In the case of two fluids,
outgoing charged baryons could stream relativistically
through ingoing neutral dark matter,
and mass inflation duly occurred.

\penrosernfig

I offer the following conceptual picture of the physical mechanism
that causes mass inflation inside black holes.
First,
in order to drop from the superluminally infalling region
of the black hole
through an inner horizon into a subluminal region,
it is necessary that ingoing and outgoing fluids drop
through separate ingoing and outgoing inner horizons
into causally separated pieces of spacetime,
as seen most clearly in a Penrose diagram,
such as the Penrose diagram of the Reissner-Nordstr\"om geometry
shown in Figure~\ref{penrosern}.
If the ingoing and outgoing subluminal spacetimes were causally connected,
then time would go in one direction for the ingoing fluid,
and in the opposite direction for the outgoing fluid,
which cannot be.
This is most evident in stationary or self-similar black hole spacetimes,
where there is a Killing vector associated with (conformal)
time-translation symmetry,
and that Killing vector must point in one direction
for the ingoing fluid
(positive energy geodesics),
and in the opposite direction for the outgoing fluid
(negative energy geodesics).
The only way for the ingoing and outgoing fluids
to pass into causally separated subluminal spacetimes
is to move faster than the speed of light through each other,
which is impossible.
In their attempt to exceed the speed of light,
the ingoing and outgoing fluids counter-stream ever faster,
and in so doing they create a growing positive radial pressure.
Mass inflation begins when the gravitational force
produced by the radial pressure of the counter-streaming fluids
exceeds the gravitational force produced by the mass of the black hole.
The gravitational force produced by the radial pressure is inwards,
but the inward direction is in opposite directions
for the ingoing and outgoing fluids,
being towards the black hole for the ingoing fluid,
and away from the black hole for the outgoing fluid.
The result is an exponential feedback instability,
in which the increasing counter-streaming velocity
increases the pressure, which increases the gravity,
which accelerates the fluids ever faster through each other.
Mass inflation thus appears to depend
on an intriguing alternation between gravitational attraction and repulsion:
first, there must be sufficient gravitational attraction
to produce gravitational collapse and an event horizon;
next, there must be sufficient gravitational repulsion
to bring the black hole to the brink of having an inner horizon;
and finally,
mass inflation produces an exponentially growing inward
gravitational attraction sourced by the pressure
of counter-streaming ingoing and outgoing streams.

Most published analytic and numerical investigations
of mass inflation
have considered feeding a spherical charged black hole
with a massless scalar field,
usually uncharged
\cite{Christodoulou86,GP87,GG93,BS95,Brady95,BO98,Burko97,Burko99,HO01,Burko02a,Burko03,MG03,Dafermos04b,HKN05},
but sometimes charged
\cite{HP96,HP98b,HP99,SP01,OP03,Dafermos04a,DR05}.
These investigations have confirmed that mass inflation occurs generically
when a spherical charged black hole is fed with a massless scalar field.
Since a massless scalar field supports waves moving at the speed of light,
it is consistent to hypothesize that mass inflation in the scalar field
arises physically from relativistic counter-streaming
between ingoing and outgoing waves.
This idea provides one of the principal motivations behind the present paper:
can one see explicitly from a perturbation analysis
how counter-streaming between waves moving at the speed of light
near an inner horizon might account for mass inflation?
The present paper does not answer this question,
but it lays a foundation for being able to address the question
in future work.

Gravitational waves are of particular relevance
to the issue of mass inflation,
because even if there were no other way in which relativistic
counter-streaming could occur
(perhaps there is no massless scalar field,
and perhaps the fluid is so opaque that electromagnetic waves cannot propagate),
certainly gravitational waves will allow it.
However,
spherically symmetric gravitational waves do not exist
-- they must be of at least quadrupole ($l = 2$) order --
so it is necessary to go beyond spherical symmetry
to admit gravitational waves.
Of course,
as remarked for example by \cite{Nolan06},
perturbations of a massless scalar field can be considered as a model
of gravitational perturbations.
However, this is something that,
like the assumption that charge models rotation,
should be checked.

The purpose of this paper is to present the foundations of the perturbation
theory of spherically symmetric, self-similar black holes.
The work presented here is a natural generalization
of the extensive literature on perturbations
of spherically symmetric stationary black holes
initiated by Regge \& Wheeler (1957)
\cite{RW57};
see especially the monograph by Chandrasekhar (1983)
\cite{Chandrasekhar}
and the reviews by
\cite{KS99,NR05,WP05} and references therein;
and more recently
\cite{MP05,FGR06,FS06,DGP07,NZFR07,Petrov07,SAM07}.
General relativistic spherically symmetric self-similar solutions
were first applied to the problem
of gravitational collapse by
Ori and Piran (1990) \cite{OP90},
following earlier work
on general relativistic self-similar solutions in cosmology.
Since then there have been many investigations;
see especially the reviews
\cite{CC99,MHIO02,CG03,Harada03} and references therein;
and more recently
\cite{HM03,MH04,PW04,WWW04,NW05,BCSVF06}.

There are two reasons why it is useful to consider self-similar solutions
and their perturbations
as a way to understand mass inflation better.
The first reason is that,
although the geometry in the vicinity of an accreting black hole may
be well approximated by an empty stationary solution
(the Reissner-Nordstr\"om geometry, in the case of charged spherical black holes),
the geometry inevitably departs hugely from the empty stationary geometry
in the region near the inner horizon where mass inflation takes place.
The simplest way to model this departure from the stationary case
in a fully non-linear fashion is with self-similar solutions.
The second reason is that numerical gravitational simulations are notoriously plagued
with numerical instabilities
\cite{Lehner01,CPST02,Font03,LNRT04}.
The mass inflation instability is a real instability,
not a numerical one,
and it is essential that numerics be under tight control
to explore mass inflation reliably.
Self-similar solutions
offer a numerically robust and fast approach to modeling mass inflation
with some degree of flexibility.
Although  much of the literature on self-similar black holes
is restricted to the case of a single, ideal fluid
\cite{CC00,Harada01,MHIO02,CG03},
the scope of self-similar solutions is actually much broader than this.
It is possible to consider non-ideal fluids,
for example with a finite conductivity
\cite{HP05a},
and multiple, possibly interacting, fluids
\cite{HP05b}.
Thus self-similar solutions provide a broad stage
upon which to carry out investigations of mass inflation.

Several papers have considered
spherical perturbations to spherical self-similar spacetimes;
see the review by \cite{Harada03},
and more recently
\cite{MTH05,MT06}.
By comparison, there have been few investigations of non-spherical
perturbations of spherical self-similar spacetimes.
The investigations that have been done have been based on the
Gerlach and Sengupta (1979)
\cite{GS79}
framework,
which treats
perturbations of arbitrary spherically symmetric spacetimes,
not necessarily stationary or self-similar
(see also the independent work by
\cite{Karlovini02} on arbitrary spherical spacetimes).
Gundlach and Mart\'in-Garcia
\cite{MG99}
review the Gerlach-Sengupta formalism,
and use it to analyze perturbations of the Choptuik
\cite{Choptuik93,MG03}
critical solution that describes the discretely self-similar collapse
of a scalar field at the threshold of gravitational collapse
(see \cite{Gundlach03} for a review of critical gravitational collapse).
Gundlach and Mart\'in-Garcia
\cite{GM00,MG01}
extend the Gerlach-Sengupta formalism
to develop equations
that could be used to investigate the
spherical collapse of a star numerically.
Frolov \cite{Frolov99}
used the Gerlach-Sengupta formalism
to investigate perturbations to the
Roberts \cite{Roberts89}
self-similar solution,
an analytic solution describing the gravitational collapse
of an uncharged scalar field.
Nolan et al{.} \cite{NW05}
used the Gerlach-Sengupta formalism
in their study of perturbations of self-similar Vaidya spacetimes
admitting naked singularities.
The approach of the present paper is more restricted than
that of Gerlach-Sengupta,
but has the advantage that it yields a complete separation
of variables of the wave equations.

Nolan et al{.} \cite{NW02,NW05,Nolan06}
consider spherically symmetric self-similar collapse models
that produce naked singularities.
In these models,
gravitational collapse leads to the formation of a naked singularity,
which subsequently becomes covered by an event horizon,
as illustrated by the Penrose diagram in Fig.~2 of \cite{NW02}.
The naked singularity is a source of unpredictability,
and the Cauchy horizon,
defined in the usual way to be the boundary of predictability,
is thus a null cone that extends outwards
from the formation event of the naked singularity.
\cite{NW02,Nolan06}
show that the flux of (arbitrary, non-spherical)
scalar radiation impinging on this Cauchy horizon is finite,
and \cite{NW05}
show that,
if the background self-similar solution is that of null dust
(Vaidya geometry),
then the flux of gravitational and matter perturbations
remains similarly finite on the Cauchy horizon.
Nolan et al.\ conclude that
the finite flux points to the stability of
the Cauchy horizon in this situation.
This conclusion is consistent with the conceptual picture described above.
Mass inflation, as envisaged by Poisson \& Israel,
must occur within a superluminally infalling region
beneath an event horizon,
for only then can the inward direction (towards smaller radius)
point in opposite directions for ingoing and outgoing streams.
Thus the Cauchy horizon considered by \cite{Nolan06,NW05,NW02},
which lies outside the event horizon,
is not expected to be subject to the mass inflation instability.
Curiously,
\cite{NW05} go on to find that the flux
diverges on the event horizon that eventually covers the naked singularity.
The physical origin of this divergence is not clear.

The most fundamental result of this paper is a generalization
of the Teukolsky spin-$s$ wave equation
\cite{BP73,T72,T73}
to the case of
spherically symmetric, self-similar black holes.
There is no additional restriction on the form of the
unperturbed energy-momentum tensor
beyond spherical symmetry and self-similarity.
As in stationary black holes,
eigenfunctions of the wave equations
for the propagating components of spin-$s$ waves
are separable,
taking the form
$\left[ {(1 + V) / (1 - V)} \right]^{\mp s/2} A^{- (s + 1)} \ee^{- \im \omega t} \npsi{s}_{\omega l}(r) \, \nY{s}_{lm}(\theta,\phi)$
in conformal polar coordinates
$\{ t , r , \theta , \phi \}$,
where
$\left[ {(1 + V) / (1 - V)} \right]^{\mp s/2}$
is a radial Lorentz boost factor
(which equals unity in a frame at rest in the conformal frame),
$A$ is a conformal factor,
$\omega$ is a conformal frequency,
$\nY{s}_{lm}$ are spin-$s$ spherical harmonics,
and $\npsi{s}_{\omega l}(r)$ are radial eigenfunctions
of the following generalization of the Teukolsky equation
\begin{equation}
\label{teukolsky}
  \left\{
  - \,
  {\dd^2 \over \dd r^{\ast 2}}
  +
  ( \im \omega \pm |s| \kappa )^2
  +
  \bigl[
    l ( l + 1 )
    - s^2
    - ( 1 + s^2 ) ( R - P )
    - 2 |s| P_\perp
    + 2 ( 1 + 2 s^2 ) M
  \bigr]
  \Delta
  \right\}
  \npsi{s}_{\omega l}
  =
  \mbox{sources}
\end{equation}
where the $\pm$ sign in front of $\kappa$ is $+$ for ingoing, $-$ for outgoing waves.
Here
$r^\ast$ is a generalization of the Regge-Wheeler \cite{RW57} radial coordinate;
$\Delta$ is the dimensionless horizon function;
$\kappa \equiv {\textstyle \frac{1}{2}} \dd \ln \Delta / \dd r^\ast$;
$R \equiv 4 \pi A^2 \rho$,
$P \equiv 4 \pi A^2 p$,
and
$P_\perp \equiv 4 \pi A^2 p_\perp$
are the dimensionless energy density, radial pressure, and transverse pressure;
and
$M \equiv m / A$
is the dimensionless interior mass,
with $m$ the interior, or Misner-Sharpe \cite{MS64}, mass.
The present paper considers only integral spin waves --
gravitational waves (spin 2),
electromagnetic waves (spin 1),
and scalar waves (spin 0)
--
so it is not guaranteed that equation~(\ref{teukolsky})
holds for half-integral spin, although it may well do so.

It should be emphasized that
in the general self-similar case of non-vanishing background energy-momentum,
the system of wave equations is inextricably coupled,
so that waves of one kind excite perturbations
which in turn provide a source for waves of another kind.
Thus as long as waves are present,
the source terms on the right hand side of equation~(\ref{teukolsky})
cannot in general be made to vanish,
which is unlike the case of the Schwarzschild or Kerr geometries.

The structure of this paper is as follows.
Section~\ref{unperturbed}
describes the unperturbed self-similar background,
\S\ref{npformalism} sets up the Newman-Penrose formalism
\cite{NP62,GHP73},
and \S\ref{perturbations}
uses the Newman-Penrose formalism to
develop the equations governing
perturbations of the self-similar background.
Sections~\ref{gravitationalwaves},
\ref{electromagneticwaves},
and
\ref{scalarwaves}
proceed to derive wave equations
for gravitational, electromagnetic, and scalar waves.
Section~\ref{summary}
summarizes the results.
Because of the length and technicality of the paper,
application to problems of interest,
notably mass inflation near the inner horizon,
is deferred to subsequent work.

The conventions of this paper are for the most part those of
Misner, Thorne \& Wheeler \cite{MTW}
(with the exception that
the indices on the electromagnetic field tensor $F_{mn}$ are swapped compared to MTW).
All complicated equations have been calculated and checked with
the algebraic manipulation program Mathematica.
It is to be hoped that if there are errors in the equations,
then they are errors of transcription, and they do not propagate.

In many places in this paper,
starting from
\S\ref{commutingoperators},
symmetrically related pairs of equations
are given as single equations with two sets of alternative indices.
The notation is used partly for compactness,
and partly to manifest the
$u \leftrightarrow v$, $+ \leftrightarrow -$
symmetry of the equations
in the Newman-Penrose formalism
\cite{GHP73}.
To reduce the possibility of error,
the algebra for each of a pair of symmetrically related equations
was calculated separately,
the pair were then merged,
and finally the indices were checked for the expected symmetry.

\section{Unperturbed background}
\label{unperturbed}

This section presents the equations governing
the spherically symmetric self-similar black hole solutions
that provide the unperturbed background to the subsequent perturbation theory.
In basic respects the development is similar to that of \cite{HP05a,HP05b},
but the gauge choices that permit separation of variables in the wave equations
prove to be different from the gauge choices of \cite{HP05a,HP05b},
so it is necessary to re-develop the formalism ab initio.
The most fundamental difference from \cite{HP05a,HP05b}
is to make the conformal character
of the self-similar solutions explicit from the outset.

\subsection{Self-similarity and the choice of coordinates and tetrad}

The assumption of self-similarity
is the assumption that the system possesses a
conformal time-translation symmetry.
This implies
that there exists a conformal time coordinate $t$
such that the geometry at any one time is conformally related
to the geometry at any other time
\begin{equation}
\label{metricconformal}
  \dd s^2
  =
  a(t)^2 g^{(c)}_{\mu\nu} \dd x^\mu \dd x^\nu
\end{equation}
in which the conformal factor $a(t)$ depends only the conformal time $t$,
and the conformal metric coefficients $g^{(c)}_{\mu\nu}$
are all independent of the conformal time $t$.
The condition~(\ref{metricconformal}) on the metric
is a necessary but not sufficient condition for self-similarity:
self-similarity requires that the entire system,
including
notably
the energy-momentum tensor,
possesses conformal time-translation symmetry.
For example,
the Friedmann-Robertson-Walker metric of cosmology satisfies
equation~(\ref{metricconformal}),
but in general does not evolve in a self-similar fashion.

The assumption of spherical symmetry
implies that the metric can be reduced further to the form,
in conformal polar coordinates
$x^\mu \equiv \{ t , r , \theta , \phi \}$,
\begin{equation}
\label{metricspherical}
  \dd s^2
  =
  a(t)^2 \left[
  g^{(c)}_{tt}(r) \dd t^2 + 2 \, g^{(c)}_{tr}(r) \dd t \dd r + g^{(c)}_{rr}(r) \dd r^2
  + r^2 \dd o^2 \right]
\end{equation}
in which
$\dd o^2 \equiv \dd \theta ^2 + \sin^2\!\theta \, \dd \phi^2$
is the metric of the surface of a unit sphere,
and the conformal metric coefficients
$g^{(c)}_{tt}$, $g^{(c)}_{tr}$, and $g^{(c)}_{rr}$
are functions only of the conformal radial coordinate $r$,
independent of either the conformal time $t$
or the angular coordinates,
the polar angle $\theta$ and the azimuthal angle $\phi$.

In place of the conformal factor $a$
it is convenient to work with the conformal factor $A$
\begin{equation}
\label{A}
  A \equiv a(t) r
\end{equation}
since it proves to be $A$, not $a$,
that appears ubiquitously in formulae.
The fact that the angular part
of the metric~(\ref{metricspherical})
is $a^2 r^2 \dd o^2 = A^2 \dd o^2$
implies that the conformal factor $A$
is equal to the proper circumferential radius,
with the defining property that the proper circumference 
measured by observers at fixed conformal time $t$
is $2 \pi A$.
Being a proper quantity,
$A$ is coordinate (and tetrad) gauge-invariant,
whereas neither the conformal time $t$
nor the conformal radius $r$
are themselves coordinate gauge-invariant.
The conformal factor $A$ is a dimensional quantity,
with units of length.
In self-similar solutions,
all quantities are proportional to some power of $A$,
and that power can be determined by dimensional analysis.


In this paper I choose to work in the tetrad formalism.
Orthonormal tetrads are considered in this section, \S\ref{unperturbed},
and associated Newman-Penrose \cite{NP62,GHP73} null tetrads
will be considered starting from \S\ref{npformalism}.

Let $\gvec_\mu$ denote the basis of tangent vectors of the coordinates.
By definition,
the scalar products of the tangent vectors constitute the metric $g_{\mu\nu}$
\begin{equation}
\label{gg}
  \gvec_\mu \cdot \gvec_\nu = g_{\mu\nu}
  \ .
\end{equation}
In the tetrad formalism,
a set of frames is erected at each point of the spacetime,
with axes
$\gammavec_m$, the tetrad,
whose dot products form the tetrad metric
$\gamma_{mn}$
\begin{equation}
  \gammavec_m \cdot \gammavec_n = \gamma_{mn}
  \ .
\end{equation}
In an orthonormal tetrad,
as considered throughout this section,
the axes are chosen to be locally inertial,
so that the tetrad metric $\gamma_{mn}$ is the Minkowski metric $\eta_{mn}$
\begin{equation}
  \gamma_{mn}
  =
  \eta_{mn}
  \ .
\end{equation}
In a Newman-Penrose tetrad,
considered in \S\ref{npformalism} and thereafter,
the tetrad metric is not Minkowski,
but the tetrad metric components $\gamma_{mn}$,
equation~(\ref{npmetric}), are still constants,
so that for example equation~(\ref{Gammaantisym}) below,
and thus also equation~(\ref{gammaGamma}), remains valid.
In this paper
dummy latin indices signify tetrad frames,
while dummy greek indices signify coordinate frames.
The axes $\gammavec_m$ of the locally inertial frames are related to
the tangent vectors $\gvec_\mu$
by the vierbein
${e_m}^\mu$
and its inverse
${e^m}_\mu$
\begin{equation}
  \gammavec_m = {e_m}^\mu \gvec_\mu
  \ , \quad
  \gvec_\mu = {e^m}_\mu \gammavec_m
  \ .
\end{equation}
The vierbein provide the means of transforming between
the tetrad components
and coordinate components of any 4-vector or tensor object.
For example,
the tetrad components $p^m$
and the coordinate components $p^\mu$ of the 4-vector
$\pvec \equiv p^\mu \gvec_\mu = p^m \gammavec_m$
are related by
$p^m = {e^m}_\mu p^\mu$.
Tetrad components are raised, lowered, and contracted
with the tetrad metric $\gamma_{mn}$,
while coordinate components are raised, lowered, and contracted
with the coordinate metric $g_{\mu\nu}$.

The most general vierbein that is manifestly spherically symmetric contains
6 free parameters \cite{Robertson32},
of which 2 correspond to physically distinct spacetimes,
2 are associated with coordinate gauge freedoms of time and radius,
and 2 are associate with the tetrad gauge freedoms
of a Lorentz boost in the radial direction,
and a spatial rotation about the radial direction.
I choose to eliminate one of the coordinate gauge freedoms
by choosing the conformal factor $A$, as above, equation~(\ref{A}),
to be the proper circumferential radius,
and I choose to eliminate the tetrad freedom
associated with spatial rotations about the radial direction
by aligning the tetrad axes with a non-spinning locally inertial frame
in radial free-fall.
Specifically, I choose the vierbein to take the following form,
whose slightly complicated form yields a somewhat simpler inverse vierbein,
equation~(\ref{inversevierbein}),
\begin{equation}
\label{vierbein}
  {e_m}^\mu
  =
  {1 \over A}
  \left(
  \begin{array}{cccc}
    {\displaystyle - {\lambda^t \over \xi {\cdot} \lambda}} &  {\displaystyle {r \xi^r \over \xi {\cdot} \lambda}} & 0 & 0 \\
    {\displaystyle {\lambda^r \over \xi {\cdot} \lambda}} & {\displaystyle - {r \xi^t \over \xi {\cdot} \lambda}} & 0 & 0 \\
    0 & 0 & 1 & 0 \\
    0 & 0 & 0 & {\displaystyle {1 / \sin\theta}}
  \end{array}
  \right)
  \ .
\end{equation}
The choice of vierbein in equation~(\ref{vierbein}) is more general than necessary,
but it is convenient to defer the choice of gauge of the conformal radius $r$
(which will in \S\ref{radialgauge} be dictated by the requirement that eigensolutions
of wave equations be separable in the coordinate $r$),
and to retain the freedom of an arbitrary Lorentz boost
in the radial direction
(so that, for example,
one could choose the tetrad to be comoving with the
center-of-mass frame of infalling matter).
The vierbein~(\ref{vierbein}) contain 4 free parameters,
which in self-similar solutions turn out to constitute
two dimensionless tetrad-frame 4-vectors
$\xi^m \equiv \{ \xi^t , \xi^r , 0 , 0 \}$
and $\lambda^m \equiv \{ \lambda^t , \lambda^r , 0 , 0 \}$.
Dimensionless means a function only of the conformal radius $r$,
independent of the conformal factor $A$.
The conformal factor $A$ itself is not a free parameter,
being fixed equal to the circumferential radius,
equation~(\ref{A}).
The quantity
$\xi \cdot \lambda = - \, \xi^t \lambda^t + \xi^r \lambda^r$
in equation~(\ref{vierbein})
is the scalar product of the dimensionless 4-vectors $\xi$ and $\lambda$.
The inverse of the vierbein~(\ref{vierbein}) is
\begin{equation}
\label{inversevierbein}
  {e^m}_\mu
  =
  A
  \left(
  \begin{array}{cccc}
    \xi^t & \lambda^r / r & 0 & 0 \\
    \xi^r & \lambda^t / r & 0 & 0 \\
    0 & 0 & 1 & 0 \\
    0 & 0 & 0 & \sin\theta
  \end{array}
  \right)
  \ .
\end{equation}
The metric $\dd s^2 = \gamma_{mn} {e^m}_\mu {e^n}_\nu \dd x^\mu \dd x^\nu$
corresponding to the inverse vierbein
${e^m}_\mu$ of equation~(\ref{inversevierbein}) is
\begin{equation}
\label{metric}
  \dd s^2 = A^2
  \biggl[
    - \, \Bigl( \xi^t \dd t + {\lambda^r \over r} \dd r \Bigr)^2
    + \Bigl( {\lambda^t \over r} \dd r + \xi^r \dd t \Bigr)^2
    + \dd o^2
  \biggr]
\end{equation}
and it apparent that the assumption of self-similarity,
equation~(\ref{metricconformal}),
is equivalent to the hypothesis that,
as claimed above,
$\xi^m$ and $\lambda^m$ are dimensionless,
independent of the conformal time $t$.
The determinant of the metric is equal to the square $(1/e)^2$
of the determinant $1/e$  of the inverse vierbein, which is
\begin{equation}
  {1 \over e}
  =
  A^4 {- \xi {\cdot} \lambda \over r} \sin\theta
  \ .
\end{equation}

Directed derivatives $\partial_m$ are defined
to be the spacetime derivatives along the axes $\gammavec_m$
of the tetrad frame:
\begin{equation}
\label{directedderivative}
  \partial_m
  \equiv
  \gammavec_m \cdot \partialvec
  =
  {e_m}^\mu {\partial \over \partial x^\mu}
\end{equation}
where
$\partialvec = \gvec^\mu \partial / \partial x^\mu = g^{\mu\nu} \gvec_\nu \partial / \partial x^\mu$
is the invariant spacetime vector derivative.
The directed derivatives $\partial_m$
depend only on the choice of tetrad frame,
and are independent of the choice of coordinate system.
Unlike the coordinate partial derivatives $\partial / \partial x^\mu$,
the directed derivatives $\partial_m$ do not commute,
equation~(\ref{partialcommute}) below.

The coordinate 4-velocity
$\upsilon^\mu$
of an object at rest in the tetrad frame is
\begin{equation}
  \upsilon^\mu = \partial_t x^\mu = {e_t}^\mu
  =
  {1 \over A \, \xi \cdot \lambda}
  \{ - \lambda^t , r \xi^r , 0 , 0 \}
  \ .
\end{equation}
If the tetrad frame were chosen to be at rest in the similarity frame,
so that $\partial_t r = 0$,
then this would correspond to the gauge choice $\xi^r = 0$.
However,
I choose to retain the freedom of an arbitrary radial Lorentz boost,
that is, I allow $\xi^r$ to be arbitrary.

\subsection{Gauge-invariant scalars}

The fact that quantities in self-similar solutions are,
modulo powers of the conformal factor $A$,
independent of conformal time $t$
(with vanishing conformal time derivative
$\partial / \partial t$)
means that the tangent vector $\gvec_t$
is a conformal Killing vector:
\begin{equation}
  \gvec_t \cdot \partialvec
  =
  {\partial \over \partial t}
  \ .
\end{equation}
The tetrad-frame components of the Killing vector are
$A \xi^m$, because
\begin{equation}
  \gammavec^m \cdot \gvec_t
  =
  {e^m}_t
  =
  A \xi^m
\end{equation}
or equivalently
\begin{equation}
  {\partial \over \partial t}
  =
  A \xi^m \partial_m
  \ .
\end{equation}
This explains the choice of the quantities
$\xi^m$ in the definition~(\ref{vierbein}) of the vierbein,
and also demonstrates that $\xi^m$
is a tetrad-frame 4-vector, as claimed.

Minus the square of the dimensionless Killing vector $\xi^m$
defines the horizon function $\Delta$,
a coordinate and tetrad gauge-invariant dimensionless scalar:
\begin{equation}
\label{Delta}
  \Delta
  \equiv
  -
  \xi_m \xi^m
  =
  - A^{-2} g_{tt}
  \ .
\end{equation}
According to the metric~(\ref{metric}),
the proper time $\tau$ experienced by observers at fixed conformal position
in the black hole,
$\dd r = \dd \theta = \dd \phi = 0$,
is related to the conformal time $t$ by
$\dd \tau = A \Delta^{1/2} \dd t$,
the vanishing of which defines the positions of horizons.
The horizon function $\Delta$
is positive outside the horizon,
zero at the horizon,
and negative inside the horizon,
reflecting the fact that the Killing vector $\gvec_t$
is timelike outside the horizon, null at the horizon,
and spacelike inside the horizon.

Besides the dimensionless Killing vector $\xi^m$,
there is another coordinate gauge-invariant dimensionless tetrad-frame 4-vector
that arises naturally in self-similar solutions,
the radial 4-gradient $\beta_m = \{ \beta_t , \beta_r , 0 , 0 \}$
defined by
\begin{equation}
\label{beta}
  \beta_m
  \equiv
  \partial_m A
  \ .
\end{equation}
Physically,
$\beta_t$ and $\beta_r$ are the proper time and radial derivatives
of the circumferential radius $A$ as measured by observers
at rest in the tetrad frame.
That $\beta_m$ is coordinate gauge-invariant
follows from the fact that it is a tetrad-frame derivative
of the coordinate gauge-invariant quantity $A$.
The fact that $\beta_m$ is dimensionless
follows from the fact that,
if $f(r)$ is some arbitrary dimensionless function,
then
\begin{equation}
\label{dimensionlessbeta}
  \partial_m A f(r)
  =
  A \partial_m f(r)
  +
  \beta_m f(r)
\end{equation}
is a sum of two terms,
the first of which is manifestly dimensionless because
$A \partial_m f(r) = A {e_m}^r \partial f / \partial r$
and ${e_m}^\mu \propto 1/A$, equation~(\ref{vierbein}),
from which it follows that the second term
$\beta_m f(r)$ must also be dimensionless,
so that $\beta_m$ itself must be dimensionless as asserted.
Expressions involving directed derivatives of powers of the conformal factor $A$
times dimensionless quantities,
such as on the left hand side of equation~(\ref{dimensionlessbeta}),
appear ubiquitously in the theory of self-similar black holes,
for example in the definitions of the tetrad-frame connections and Riemann tensor.

As an aside,
it is worth mentioning
that the argument from equation~(\ref{dimensionlessbeta})
that $\beta_m$ is dimensionless
fails for the Friedmann-Robertson-Walker metric,
which possesses spatial translation symmetry,
so that physical quantities depend only on conformal time $t$,
not on position,
and consequently $f(r)$ in equation~(\ref{dimensionlessbeta})
cannot be taken to be arbitrary.
Thus in the Friedmann-Robertson-Walker metric
a more general class of self-similar solutions exists.
However,
the present paper concerns itself with self-similar solutions of black holes,
not of cosmology,
so for this paper $\beta_m$ is indeed dimensionless.

The scalar product
$\xi^m \beta_m$
of the Killing vector $\xi^m$
with the radial 4-gradient $\beta_m$
defines a second coordinate and tetrad gauge-invariant dimensionless scalar $\vel$
\begin{equation}
\label{vel}
  \vel
  \equiv
  \xi^m \beta_m
  =
  {1 \over A} {\partial A \over \partial t}
  =
  {1 \over a} {\dd a \over \dd t}
  \ .
\end{equation}
Since $a(t)$ is a function only of conformal time $t$,
equation~(\ref{metricconformal}),
the last expression of equation~(\ref{vel}),
$\vel = ( \dd a / \dd t ) / a$,
shows that
$\vel$
is a function only of conformal time $t$.
But the fact that $\vel$ is dimensionless
(a function only of $r$)
in self-similar solutions
implies that $\vel$ must also be independent of conformal time $t$.
It then follows that $\vel$ must be a constant,
independent of either $t$ or $r$.
In \cite{HP05a,HP05b}
$\vel$ was set to unity,
which corresponds to a certain gauge choice of the scaling
of conformal time $t$.
Here instead $\vel$ is allowed to be an arbitrary constant,
which has the virtue of making
it transparent how to take the limit of a stationary black hole,
which corresponds to $\vel \rightarrow 0$.
Physically, the constant $\vel$ is proportional to the
expansion velocity of the black hole
at some fixed (but arbitrary) conformal radius $r$,
and thus a measure of the rate at which the black hole is accreting and growing.
Specifically,
the velocity $\vel$ is related to the rate
$\dd A / \dd \tau$
at which an observer at rest in the conformal frame measures the
circumferential radius $A$ to vary with proper time
$\dd\tau = A \Delta^{1/2} \dd t$
by
\begin{equation}
  \vel
  =
  \Delta^{-1/2}
  {\dd A \over \dd \tau}
  \ .
\end{equation}

In terms of the expansion velocity $\vel$ and the 4-vectors $\xi^m$ and $\lambda^m$
in the vierbein~(\ref{vierbein}),
the components
$\beta_m = {e_m}^\mu \partial A / \partial x^\mu$
of the radial 4-gradient are
\begin{equation}
  \beta_t
  =
  {\xi^r - \vel \lambda^t \over \xi \cdot \lambda}
  \ ,
  \quad
  \beta_r
  =
  {- \, \xi^t + \vel \lambda^r \over \xi \cdot \lambda}
  \ .
\end{equation}

A third coordinate and tetrad gauge-invariant dimensionless scalar,
besides the horizon function $\Delta$ and the expansion velocity $\vel$,
is given by the square $\beta^m \beta_m$ of the radial 4-gradient $\beta_m$.
In terms of this a gauge-invariant dimensionless scalar,
the dimensionless interior mass $M$
can be defined by
\begin{equation}
\label{M}
  M
  \equiv
  {\textstyle \frac{1}{2}} \left( 1 - \beta_m \beta^m \right)
  \ .
\end{equation}
The dimensionless interior mass $M$ is related to the
the interior, or Misner-Sharpe \cite{MS64}, mass $m$ by
$M = m / A$
(this notation differs from that of
\cite{HP05a,HP05b},
who defined $M$ to be the interior mass,
as opposed to a dimensionless version thereof).
The physical interpretation of $m = A M$ as an interior mass
emerges from the spherically symmetric Einstein equations.

\subsection{Tetrad-frame connections}

In the tetrad formalism,
the covariant derivatives $D_n$
of the covariant,
$p_k \equiv \gammavec_k \cdot \pvec$,
and contravariant,
$p^k \equiv \gammavec^k \cdot \pvec$,
tetrad-frame components of a 4-vector
$\pvec \equiv p^\mu \gvec_\mu = p^m \gammavec_m$
are
\begin{eqnarray}
  D_n p_k &=& \partial_n p_k - \Gamma^m_{kn} p_m
\nonumber
\\
  D_n p^k &=& \partial_n p^k + \Gamma^k_{mn} p^m
\end{eqnarray}
where the tetrad-frame connection coefficients $\Gamma^k_{mn}$,
also known as the Ricci rotation coefficients,
are defined by the directed derivatives $\partial_n$ of the tetrad axes $\gammavec_m$
\begin{equation}
\label{gammaGammadef}
  \partial_n \gammavec_m
  \equiv
  \Gamma^k_{mn} \gammavec_k
\end{equation}
in much the same way that
the usual coordinate-frame connection coefficients, the Christoffel symbols
$\Gamma^\kappa_{\mu\nu}$,
are defined by the coordinate derivatives $\partial / \partial_\nu$
of the coordinate tangent vectors $\gvec_\mu$
\begin{equation}
\label{Gammadef}
  {\partial \gvec_\mu \over \partial x^\nu}
  \equiv
  \Gamma^\kappa_{\mu\nu} \gvec_\kappa
  \ .
\end{equation}
Let
$d^k_{mn}$
denote the vierbein derivatives
\begin{equation}
\label{gammad}
  d^k_{mn}
  \equiv
  {e^k}_\kappa \, \partial_n {e_m}^\kappa
  =
  {e^k}_\kappa \, {e_n}^\nu \, {\partial {e_m}^\kappa \over \partial x^\nu}
  \ .
\end{equation}
The fact that
$\gammavec_m = {e_m}^\mu \gvec_\mu$
implies that
the tetrad-frame connection coefficients
$\Gamma^k_{mn}$
defined by equation~(\ref{gammaGammadef})
are related to
the coordinate-frame connection coefficients
$\Gamma^\kappa_{\mu\nu}$
defined by equation~(\ref{Gammadef})
by
\begin{equation}
\label{GammaGamma}
  \Gamma^k_{mn}
  =
  d^k_{mn}
  +
  {e^k}_\kappa {e_m}^\mu {e_n}^\nu
  \Gamma^\kappa_{\mu\nu}
  \ .
\end{equation}
The definition~(\ref{gammaGammadef})
and the fact that the tetrad metric coefficients are constant,
$\partial_n ( \gammavec_k \cdot \gammavec_m ) = \partial_n \gamma_{km} = 0$,
implies that the tetrad-frame connection coefficients with all indices lowered
$\Gamma_{kmn} \equiv \gamma_{kl} \Gamma^l_{mn}$
are antisymmetric in their first two indices,
\begin{equation}
\label{Gammaantisym}
  \Gamma_{kmn} = - \Gamma_{mkn}
\end{equation}
which expresses mathematically
the fact that $\Gamma_{kmn}$
for each given final index $n$
is the generator of a Lorentz transformation between tetrad frames
parallel transported along the axis $\gammavec_n$.
The tangent vectors
$\gvec_\mu$
can be regarded as coordinate derivatives of the invariant 4-vector interval
$d \xvec \equiv \gvec_\mu d x^\mu$,
that is,
$\gvec_\mu = \partial \xvec / \partial x^\mu$,
and the commutativity of partial derivatives,
$\partial \gvec_\mu / \partial x^\nu
=
\partial^2 \xvec / \partial x^\nu \partial x^\mu
=
\partial^2 \xvec / \partial x^\mu \partial x^\nu
=
\partial \gvec_\nu / \partial x^\mu$,
implies the usual no-torsion condition of general relativity that the
coordinate-frame connections are symmetric in their last two indices
\begin{equation}
\label{notorsion}
  \Gamma^\kappa_{\mu\nu}
  =
  \Gamma^\kappa_{\nu\mu}
  \ .
\end{equation}
Unlike coordinate derivatives,
the directed derivatives do not commute.
The commutator of directed derivatives is
\begin{equation}
\label{partialcommute}
  \left[ \partial_n , \partial_m \right]
  =
  \bigl( \Gamma^k_{mn} - \Gamma^k_{nm} \bigr) \partial_k
  =
  \bigl( d^k_{mn} - d^k_{nm} \bigr) \partial_k
\end{equation}
the middle expression of which follows from
equation~(\ref{gammaGammadef}),
and the last expression from
equation~(\ref{directedderivative}).
From
the relation~(\ref{GammaGamma}),
the Lorentz antisymmetry~(\ref{Gammaantisym}),
and
the no-torsion condition~(\ref{notorsion}),
it can be deduced that
the tetrad-frame connection coefficients
$\Gamma_{kmn}$
can be expressed in terms of the vierbein derivatives
$d_{kmn} \equiv \gamma_{kl} d^l_{mn}$,
equation~(\ref{gammad}),
as
\begin{eqnarray}
\label{gammaGamma}
  \Gamma_{kmn}
  =
  {\textstyle \frac{1}{2}}
  \left(
  d_{kmn} - d_{mkn} + d_{nmk} - d_{nkm} + d_{mnk} - d_{knm}
  \right)
\end{eqnarray}
which provides an explicit way to compute $\Gamma_{kmn}$
from the vierbein ${e_m}^\mu$.

For the vierbein ${e_m}^\mu$ given by equation~(\ref{vierbein}),
the non-vanishing tetrad-frame connection coefficients
$\Gamma_{kmn}$
are
\begin{subequations}
\begin{eqnarray}
  \Gamma_{trr}
  &=&
  {h_t / A}
  \ ,
\\
  \Gamma_{trt}
  &=&
  {h_r / A}
  \ ,
\\
  \Gamma_{\phi t\phi}
  =
  \Gamma_{\theta t\theta}
  &=&
  {\beta_t / A}
  \ ,
\\
  \Gamma_{\phi r\phi}
  =
  \Gamma_{\theta r\theta}
  &=&
  {\beta_r / A}
  \ ,
\\
  \Gamma_{\phi\theta\phi}
  &=&
  {\cot\theta / A}
  \ ,
\end{eqnarray}
\end{subequations}
where
\begin{equation}
  h_t
  \equiv
  {1 \over \xi \cdot \lambda}
  \left(
  {\partial \lambda^t \over \partial t}
  +
  \vel \lambda^t
  -
  {\partial r \xi^t \over \partial r}
  \right)
  \ ,
  \quad
  h_r
  \equiv
  - \,
  {1 \over \xi \cdot \lambda}
  \left(
  {\partial \lambda^r \over \partial t}
  +
  \vel \lambda^r
  -
  {\partial r \xi^r \over \partial r}
  \right)
  \ .
\end{equation}
Physically,
the connection coefficient
$h_t / A \equiv \Gamma_{trr}$
is minus the local ``Hubble parameter'' of the radial flow,
that is,
minus the proper radial gradient,
$- D_r u^r$,
of the proper radial velocity
between objects each of which is comoving with the tetrad frame at its position;
while
$h_r / A \equiv \Gamma_{trt}$
is the proper gravitational force,
or equivalently minus the proper acceleration,
$- D_t u^r$,
experienced by
an observer at rest in the tetrad frame.
These assertions follow from the fact that
observers at rest in the tetrad frame
have 4-velocity $u^m = \{1,0,0,0\}$,
so that
\begin{subequations}
\begin{eqnarray}
  D_r u^r
  &=&
  \partial_r u^r + \Gamma^r_{tr} u^t
  =
  \Gamma^r_{tr}
  =
  - h_t / A
  \ ,
\\
  D_t u^r
  &=&
  \partial_t u^r + \Gamma^r_{tt} u^t
  =
  \Gamma^r_{tt}
  =
  - h_r / A
  \ .
\end{eqnarray}
\end{subequations}

\subsection{Riemann, Ricci, Einstein, and Weyl tensors}
\label{riemann}

The Riemann tensor
$R_{klmn}$
in the tetrad frame
is defined in the usual way
by the commutator of the covariant derivative,
$R_{klmn} p^n \equiv \left[ D_k , D_l \right] p_m$,
and is given in terms of the tetrad-frame connection coefficients by
\begin{equation}
\label{gammaRiemann}
  R_{klmn}
  =
  \partial_l \Gamma_{nmk} - \partial_k \Gamma_{nml}
  + \Gamma^j_{ml} \Gamma_{jnk} - \Gamma^j_{mk} \Gamma_{jnl}
  + ( \Gamma^j_{lk} - \Gamma^j_{kl} ) \Gamma_{nmj}
\end{equation}
which has two extra terms (the last two) compared to
the usual coordinate expression for the Riemann tensor
in terms of Christoffel symbols.
The Ricci tensor and scalar
are given by the usual contractions
of the Riemann tensor,
$R_{km} = \gamma^{ln} R_{klmn}$
and
$R = \gamma^{km} R_{km}$,
and the Einstein tensor is given
by the usual expression in terms of the
Ricci tensor and scalar,
$G_{km} = R_{km} - \frac{1}{2} R \, \gamma_{km}$.
The Weyl tensor $C_{klmn}$ is defined in the usual way to be
the totally trace-free,
or tidal,
part of the Riemann tensor
$R_{klmn}$,
\begin{equation}
\label{Weyl}
  C_{klmn}
  \equiv
  R_{klmn}
  - {\textstyle \frac{1}{2}} \left( \gamma_{km} R_{ln} - \gamma_{kn} R_{lm} + \gamma_{ln} R_{km} - \gamma_{lm} R_{kn} \right)
  + {\textstyle \frac{1}{6}} \left( \gamma_{km} \gamma_{ln} - \gamma_{kn} \gamma_{lm} \right)
  \ .
\end{equation}

In the present case,
define the symbols $F$, $R$ (not the Ricci scalar!), $P$, and $P_\perp$ by
\begin{subequations}
\label{FRPPA}
\begin{eqnarray}
\label{FA}
  F
  &\equiv&
  A \partial_t \beta_r + \beta_t h_r
  =
  A \partial_r \beta_t + \beta_r h_t
  \ ,
\\
  R
  &\equiv&
  - A \partial_r \beta_r - \beta_t h_t + M
  \  = \ 
  {1 \over \beta_r} \left[
    \partial_r ( A M )
    - \beta_t F
  \right]
  \ ,
\\
  P
  &\equiv&
  - A \partial_t \beta_t - \beta_r h_r - M
  \  = \ 
  - {1 \over \beta_t} \left[
    \partial_t ( A M )
    + \beta_r F
  \right]
  \ ,
\\
  P_\perp
  &\equiv&
  {1 \over 2 \beta_r} \Bigr[
    A ( \partial_r P + \partial_t F )
    - h_r (R + P)
    - 2 h_t F
  \Bigr]
  \  = \ 
  - {1 \over 2 \beta_t} \Bigr[
    A ( \partial_t R + \partial_r F )
    - h_t (R + P)
    - 2 h_r F
  \Bigr]
  \ .
\end{eqnarray}
\end{subequations}
The definitions~(\ref{FRPPA}) of $F$, $R$, $P$, and $P_\perp$
differ from the corresponding definitions in \cite{HP05a,HP05b}
in that appropriate factors of the conformal factor $A$
have been included so as to make $F$, $R$, $P$, and $P_\perp$
dimensionless,
that is,
functions only of the conformal radial coordinate $r$
in self-similar solutions.
In terms of these quantities,
the non-vanishing components of the Einstein tensor $G_{mn}$ are
\begin{subequations}
\label{einstein}
\begin{eqnarray}
\label{einsteintt}
  G_{tt}
  &=&
  2 R / A^2
  \ ,
  \qquad
\\
\label{einsteintr}
  G_{tr}
  &=&
  - 2 F / A^2
  \ ,
\\
\label{einsteinrr}
  G_{rr}
  &=&
  2 P / A^2
  \ ,
\\
\label{einsteinpp}
  G_{\phi\phi}
  =
  G_{\theta\theta}
  &=&
  2 P_\perp / A^2
  \ .
\end{eqnarray}
\end{subequations}
The Einstein equations
\begin{equation}
  G_{mn} = 8 \pi T_{mn}
\end{equation}
imply that the dimensionless quantities
$R$, $F$, $P$, and $P_\perp$
are related to the tetrad-frame
energy density $\rho \equiv T^{tt}$,
radial energy flux $f \equiv T^{tr}$,
radial pressure
$p \equiv T^{rr}$,
and transverse pressure
$p_\perp \equiv T^{\theta\theta} = T^{\phi\phi}$,
by
\begin{equation}
\label{einsteinRFPP}
  R = 4 \pi A^2 \rho
  \ , \quad
  F = 4 \pi A^2 f
  \ , \quad
  P = 4 \pi A^2 p
  \ , \quad
  P_\perp = 4 \pi A^2 p_\perp
  \ .
\end{equation}
The non-vanishing components of the Weyl tensor $C_{klmn}$ are
\begin{eqnarray}
  C_{trtr}
  &\!\!=\!\!&
  -
  C_{\theta\phi\theta\phi}
  =
  - 2
  C_{t \theta t \theta}
  =
  - 2
  C_{t \phi t \phi}
  =
  2
  C_{r \theta r \theta}
  =
  2
  C_{r \phi r \phi}
  =
  C / A^2
\end{eqnarray}
where $C$ is the dimensionless Weyl scalar
\begin{equation}
\label{C}
  C
  \equiv
  {\textstyle {1 \over 3}}
  ( R - P + P_\perp - 3 M )
  \ .
\end{equation}


\section{Newman-Penrose formalism}
\label{npformalism}

The perturbation theory presented in \S\ref{perturbations}
follows the Newman-Penrose formalism
\cite{NP62,Penrose65b,GHP73,BP73},
and this section sets up the necessary apparatus.
Subsection~\ref{radialgauge}
defines the background radial gauge
with respect to which wave equations prove to be separable.
Subsection~\ref{newmanpenrose}
introduces the Newman-Penrose tetrad.
Subsections~\ref{commutingoperators}--\ref{spinradialoperators}
define angular and radial raising and lowering operators,
in terms of which wave operators turn out to be most elegantly expressed.
Finally,
\S\ref{asymptotic}
discusses the asymptotic behavior of wave solutions
in a hypothetical asymptotically flat spacetime
at large distance from the black hole,
a question that can be addressed independently of the black hole itself.

\subsection{Radial gauge and the Regge-Wheeler coordinate}
\label{radialgauge}

Up to this point the gauge of the conformal radial coordinate $r$
in the unperturbed background has been left unspecified.
It turns out that the requirement that eigenfunctions of wave equations be
separable functions of the conformal coordinates $t$, $r$, $\theta$, and $\phi$
forces the radial coordinate $r$ into a certain gauge.
It should be commented that in much of the rest of this paper
the background radial gauge is not actually fixed until the last moment.
That is,
the definitions~(\ref{calDvu}) and (\ref{Dvu})
of the radial operators
$\calD_\vu$
and
$\ncalD{s}_\vu$
are independent of the background radial gauge:
only when these operators are converted to coordinates,
equation~(\ref{gaugecalDvu}),
is the radial gauge actually fixed.

Recall that the coordinate time derivative is the
directed derivative along the Killing vector
$A \, \xi^m = A \, \{ \xi^t , \xi^r , 0 , 0 \}$
\begin{equation}
\label{partialt}
  {\partial \over \partial t}
  =
  A \, \xi^m \partial_m
  \ .
\end{equation}
Let $\chi^m \equiv \{ \xi^r , \xi^t , 0 , 0 \}$
denote the 4-vector in the $r$-$t$ plane
orthogonal to the Killing vector,
with the same magnitude as $\xi^m$.
The directed derivative along this orthogonal direction is
\begin{equation}
\label{partialr}
  A \, \chi^m \partial_m
  =
  - {r \Delta \over \xi {\cdot} \lambda}
  {\partial \over \partial r}
  +
  {\xi^t \lambda_r + \xi^r \lambda_t \over \xi {\cdot} \lambda}
  {\partial \over \partial t}
  \ .
\end{equation}
Separability of the wave equations
turns out to require that
the directed derivative along the orthogonal direction $\chi^m$,
equation~(\ref{partialr}),
be a purely radial coordinate derivative.
This imposes the gauge condition that
$\xi^t \lambda_r + \xi^r \lambda_t = 0$.
The corresponding values of $\lambda^m$ are
\begin{equation}
  \lambda^m
  =
  {\xi^m
  \over
  \left[ (1 - 2 M) \Delta + \vel^2 \right]^{1/2}}
  \ .
\end{equation}
In this gauge, the directed derivative
in the direction $\chi^m$ orthogonal to the Killing vector $\xi^m$ is
a purely radial derivative, as desired,
\begin{equation}
\label{partialra}
  A \, \chi^m \partial_m
  =
  \left[ (1 - 2 M) \Delta + \vel^2 \right]^{1/2}
  r
  {\partial \over \partial r}
  \ .
\end{equation}
The coordinate metric in this gauge reduces to the diagonal form
\begin{equation}
  \dd s^2 = A^2
  \biggl[
  - \, \Delta \, \dd t^2
  + {\Delta \over \left[ (1 - 2 M) \Delta + \vel^2 \right] r^2} \, \dd r^2
  + \dd o^2
  \biggr]
  \ .
\end{equation}
Note that even though the coordinate metric
$g_{\mu\nu}$
is diagonal in this gauge,
the vierbein
${e_m}^\mu$
is not diagonal,
thanks to the (desirable)
freedom of an arbitrary
Lorentz boost in the radial direction.

In self-similar evolution,
the quantities $\Delta$, $\vel$, and $M$,
equations~(\ref{Delta}), (\ref{vel}), and (\ref{M}),
are all dimensionless,
which is to say functions only of the conformal radius $r$,
and a Regge-Wheeler \cite{RW57} (tortoise)
coordinate $r^\ast$ can be defined such that
\begin{equation}
\label{rast}
  A \, \chi^m \partial_m
  =
  {\partial \over \partial r^\ast}
  \ .
\end{equation}
In terms of the Regge-Wheeler coordinate $r^\ast$,
the metric then simplifies further to
\begin{equation}
  \dd s^2 = A^2
  \left[
  \Delta
  \left(
  - \, \dd t^2 + \dd r^{\ast 2}
  \right)
  + \dd o^2
  \right]
  \ .
\end{equation}

Stationary solutions
are obtained in the limit of zero expansion velocity,
$\vel \rightarrow 0$.
In this case
the time-dependent part $a(t)$ of the conformal factor $A \equiv a(t) r$,
equation~(\ref{A}),
can be set to unity without loss of generality, $a(t) = 1$,
so that it is consistent to take $A = r$ in stationary solutions.

The Reissner-Nordstr\"om geometry
is not only stationary but also empty
aside from a central singularity and a static electric field.
To convert to
the Reissner-Nordstr\"om geometry,
set
$\Delta = ( 1 - 2 M ) / r^2$,
the dimensionless interior mass $M$
being related to the black hole's
constant mass $m_\bullet$ and charge $q_\bullet$ by
$M = m_\bullet / r - \frac{1}{2} q_\bullet^2 / r^2$.
The conformal time $t$
and conformal radius $r$
then coincide with the usual
Reissner-Nordstr\"om time $t$
and circumferential radius $r$.

\subsection{Newman-Penrose tetrad}
\label{newmanpenrose}

Newman-Penrose null tetrads
\cite{NP62,GHP73}
are particularly well adapted to
exploring fields that propagate at the speed of light
\cite{Penrose65b}.
Define the Newman-Penrose null tetrad
$\{ \gamma_v , \gamma_u , \gamma_\plus , \gamma_\minus \}$
in terms of the orthonormal tetrad
$\{ \gamma_t , \gamma_r , \gamma_\theta , \gamma_\phi \}$
by
\begin{subequations}
\label{nptetrad}
\begin{eqnarray}
  \gamma_v
  &\equiv&
  \frac{1}{\sqrt{2}} ( \gamma_t + \gamma_r )
  \ ,
\\
  \gamma_u
  &\equiv&
  \frac{1}{\sqrt{2}} ( \gamma_t - \gamma_r )
  \ ,
\\
  \gamma_\plus
  &\equiv&
  \frac{1}{\sqrt{2}} ( \gamma_\theta + \im \gamma_\phi )
  \ ,
\\
  \gamma_\minus
  &\equiv&
  \frac{1}{\sqrt{2}} ( \gamma_\theta - \im \gamma_\phi )
  \ .
\end{eqnarray}
\end{subequations}
The Newman-Penrose
tetrad metric is
\begin{equation}
\label{npmetric}
  \gamma_{mn}
  \equiv
  \gamma_m \cdot \gamma_n
  =
  \left(
  \begin{array}{cccc}
  0 & -1 & 0 & 0 \\
  -1 & 0 & 0 & 0 \\
  0 & 0 & 0 & 1 \\
  0 & 0 & 1 & 0
  \end{array}
  \right)
\end{equation}
with indices
$m$, $n$ running over $v, u, +, -$.

It is useful to define the operation of angular conjugation,
designated $^\star$
(to be distinguished from complex conjugation~$^\ast$),
as flipping the azimuthal tetrad axis:
\begin{equation}
\label{angularconjugate}
  \gamma^\star_\phi
  \equiv
  - \gamma_\phi
  \ .
\end{equation}
In a Newman-Penrose tetrad,
angular conjugation corresponds to swapping the angular axes, $+ \leftrightarrow -$,
\begin{equation}
\label{npangularconjugate}
  \gamma^\star_v
  =
  \gamma_v
  \ ,
  \quad
  \gamma^\star_u
  =
  \gamma_u
  \ ,
  \quad
  \gamma^\star_\plus
  =
  \gamma_\minus
  \ ,
  \quad
  \gamma^\star_\minus
  =
  \gamma_\plus
  \ .
\end{equation}
If a quantity is real in an orthonormal tetrad,
then in a Newman-Penrose tetrad
its angular conjugate is the same as its complex conjugate,
but with waves it is convenient to allow 
a complex time dependence $\sim \ee^{- \im \omega t}$
(and an associated complex radial dependence),
in which case angular conjugation is not the same as complex conjugation.
The effect of angular conjugation
is to swap covariant angular indices $+ \leftrightarrow -$
on tensors (such as the Riemann tensor) and other natively real objects
in a Newman-Penrose tetrad (such as the tetrad connections).
In general,
an object can be resolved into polar $(p)$ and axial $(a)$ parts
defined by the property that the polar and axial parts respectively
do not and do change sign under angular conjugation.
For objects that are real in an orthonormal tetrad,
the polar and axial parts correspond to the real and imaginary
parts of the object in a Newman-Penrose tetrad.

\subsection{Commuting radial and angular operators}
\label{commutingoperators}

The radial derivative operators
$\calD_v$
and
$\calD_u$
and angular operators
$\calD_\plus$
and
$\calD_\minus$
defined as follows,
equations~(\ref{calD}),
play a fundamental role
because they provide the basis for the definitions of
radial and angular raising and lowering operators
[equations~(\ref{Dvu}) and (\ref{Dpm}) below]
which appear ubiquitously
in the equations describing perturbations
of spherically symmetric self-similar black holes:
\begin{subequations}
\label{calD}
\begin{eqnarray}
\label{calDvu}
  \calD_\vu
  &\!\equiv\!&
  {1 \over A} \partial_\vu A^2
  \ ,
\\
\label{calDpm}
  \calD_\plusminus
  &\!\equiv\!&
  A \partial_\plusminus
  \ .
\end{eqnarray}
\end{subequations}
In the radial gauge specified in \S\ref{radialgauge},
equation~(\ref{rast}),
the operators
$\calD_m$ become
\begin{subequations}
\label{gaugecalD}
\begin{eqnarray}
\label{gaugecalDvu}
  \calD_\vu
  &=&
  {1 \over 2 A^2 \xi^\vu} \left( {\partial \over \partial t} \pm \Dr{}{} \right) A^2
  \ ,
\\
\label{gaugecalDpm}
  \calD_\plusminus
  &=&
  {1 \over \sqrt{2}} \left(
  {\partial \over \partial\theta} \pm {\im \over \sin\theta} {\partial \over \partial\phi}
  \right)
  \ .
\end{eqnarray}
\end{subequations}
The radial operators
$\calD_v$
and
$\calD_u$
commute with the angular operators
$\calD_\plus$
and
$\calD_\minus$
(whereas the radial directed derivatives
$\partial_v$
and
$\partial_u$
do not commute with
the angular directed derivatives $\partial_\plus$ and $\partial_\minus$).
The placement of factors of $A$ in equation~(\ref{calDvu}) for
$\calD_\vu$
is such as to eliminate factors of $A$ in the relation~(\ref{Dvuh}) below.

The eigenfunctions of wave equations are mutually orthogonal.
The measure with respect to which inner products of eigenfunctions
are defined is the invariant 4-volume element
$\dd^4 x = e^{-1} \dd t \dd \phi \dd \theta \dd r$,
where $e$ is the determinant of the vierbein ${e_m}^\mu$.
In the present case the invariant 4-volume
$\dd^4 x$ is a product
of radial and angular parts $\dd {\diamond}$ and $\dd o$
\begin{equation}
\label{measure}
  \dd^4 x
  =
  \dd {\diamond}
  \,
  \dd o
\end{equation}
given by
\begin{subequations}
\label{measures}
\begin{eqnarray}
\label{radialmeasure}
  \dd {\diamond}
  &\equiv&
  A^4 \Delta \, \dd t \dd r^\ast
  \ ,
\\
\label{angularmeasure}
  \dd o
  &\equiv&
  \sin\theta \, \dd \theta \dd \phi
  \ .
\end{eqnarray}
\end{subequations}

The contravariant components of
the operators given by equations~(\ref{calD}) are
$\calD^v = - \calD_u$,
$\calD^u = - \calD_v$,
and
$\calD^\plus = \calD_\minus$,
$\calD^\minus = \calD_\plus$,
in accordance with the Newman-Penrose metric~(\ref{npmetric}),
and the Hermitian conjugates of these
with respect to the radial and angular measures
given by equations~(\ref{measures}) are
\begin{subequations}
\begin{eqnarray}
  \calD^{\dagger\vu}
  &=&
  A^2
  \partial^{\dagger\vu} {1 \over A}
  =
  {1 \over \xi^\vu}
  \calD_\uv
  \xi^\vu
  =
  {1 \over A^2 \Delta}
  \left( {\partial \over \partial t} \mp \Dr{}{} \right)
  A^2 \xi^\vu
  \ ,
\\
  \calD^{\dagger\plusminus}
  &=&
  \partial^{\dagger\plusminus} A
  =
  - \, {1 \over \sin\theta}
  \calD_\minusplus
  \sin\theta
  \ .
\end{eqnarray}
\end{subequations}

\subsection{Spin raising and lowering angular operators, and spin-weighted spherical harmonics}
\label{spinangularoperators}

This subsection summarizes the familiar properties
\cite{Goldberg67}
of the spin raising and lower angular operators
and the spin-weighted spherical harmonics
introduced by Newman and Penrose
\cite{NP62}.
The aim is not to introduce any new ideas,
but rather to establish notation,
and because the pattern provides a template
for the spin radial operators
presented in \S\ref{spinradialoperators} below.

Define spin raising and lowering angular operators
$\ncalD{s}_\plus$
and
$\ncalD{s}_\minus$
(these are,
modulo a factor of $-1/\sqrt{2}$,
the same as Newman and Penrose's $\eth$ and $\bar{\eth}$
\cite{NP66,Goldberg67})
by
\begin{subequations}
\label{Dpm}
\begin{eqnarray}
  \ncalD{s}_\plus
  &\equiv&
  ( \sin\theta )^s \, \calD_\plus \, ( \sin\theta )^{-s}
  \ ,
\\
  \ncalD{s}_-
  &\equiv&
  ( \sin\theta )^{-s} \, \calD_\minus \, ( \sin\theta )^s
  \ .
\end{eqnarray}
\end{subequations}
The Hermitian conjugates of the contravariant components of these operators are
\begin{subequations}
\label{Dpmh}
\begin{eqnarray}
  \ncalD{s}^{\dagger \plus}
  &=&
  - \,
  \ncalD{s\plus \one}_\minus
  \ ,
\\
  \ncalD{s}^{\dagger \minus}
  &=&
  - \,
  \ncalD{s\minus \one}_\plus
  \ .
\end{eqnarray}
\end{subequations}
The products of the
raising and lowering operators
$\ncalD{s}_\plusminus$
with their contravariant Hermitian conjugates
$\ncalD{s}^{\dagger \plusminus}$
are the Hermitian operators
\begin{subequations}
\label{DpmhDpm}
\begin{eqnarray}
\label{DphDp}
  2 \,
  \ncalD{s}^{\dagger \plus}
  \,
  \ncalD{s}_\plus
  &=&
  \nL{s}^2
  - s ( s {+} 1 )
\\
\label{DmhDm}
  2 \,
  \ncalD{s}^{\dagger \minus}
  \,
  \ncalD{s}_\minus
  &=&
  \nL{s}^2
  - s ( s {-} 1 )
\end{eqnarray}
\end{subequations}
where
$\nL{s}^2$
is the squared total angular momentum operator
\begin{equation}
  \nL{s}^2
  \equiv
  - \,
  {1 \over \sin\theta} {\partial \over \partial \theta} \sin\theta {\partial \over \partial \theta}
  + {1 \over \sin^2 \theta}
  \left(
  m^2
  {+} s^2
  {+} 2 m s \cos\theta
  \right)
\end{equation}
with $\partial / \partial \phi \equiv \im m$.

An important feature of the operators
$\ncalD{s}^{\dagger \plus} \, \ncalD{s}_\plus$
and
$\ncalD{s}^{\dagger \minus} \, \ncalD{s}_\minus$
is that they differ by a constant,
equations~(\ref{DpmhDpm}),
\begin{equation}
\label{diffDDpm}
  \ncalD{s}^{\dagger \plus} \, \ncalD{s}_\plus
  -
  \ncalD{s}^{\dagger \minus} \, \ncalD{s}_\minus
  =
  - s
  \ ,
\end{equation}
since it is this property that ensures that
the ladder of raising and lowering angular operators
$\ncalD{s}_\pm$
yields a mutually consistent set of eigenfunctions.
The operator
$\ncalD{s}^{\dagger \plus} \, \ncalD{s}_\plus$
represents the operation of raising followed by lowering,
and thus connects spin-$s$
with spin-$(s{+}1)$ functions,
while
$\ncalD{s}^{\dagger \minus} \, \ncalD{s}_\minus$
represents the operation of lowering followed by raising,
and thus connects functions on the adjacent level,
spin-$s$ with spin-$(s{-}1)$.
The fact that the two operators differ by a constant,
equation~(\ref{diffDDpm}),
implies that the spin-$s$ eigenfunctions
are the same for both levels.
This is true for any spin $s$,
and so it follows that, as asserted,
the ladder of raising and lowering angular operators
yields a mutually consistent set of eigenfunctions.

The eigenfunctions of the squared total angular momentum operator
$\nL{s}^2$
are the spin-weighted spherical harmonics $\nY{s}_{lm}(\theta, \phi)$,
with
$l \, {\ge} \, |s|, |m|$,
satisfying
\begin{equation}
\label{Ls}
  \nL{s}^2 \,
  \nY{s}_{lm}(\theta, \phi)
  =
  l ( l + 1) \,
  \nY{s}_{lm}(\theta, \phi)
  \ .
\end{equation}
The spin $s$, harmonic number $l$, and azimuthal number $m$
must be either all integral, or all half-integral,
although only integral spin is considered in this paper.
The spin raising and lowering angular operators
raise and lower by one unit the spin $s$ of the spin-weighted spherical harmonics
\begin{equation}
\label{angularraiselower}
  \ncalD{s}_\plusminus \,
  \nY{s}_{lm}(\theta, \phi)
  =
  \mp
  \left[ {(l \mp s) ( l \pm s + 1) \over 2} \right]^{1/2}
  \!
  \nY{s \plusminus \one}_{lm}(\theta, \phi)
  \ .
\end{equation}
The normalization factor in equation~(\ref{angularraiselower})
follows, modulo a conventional choice of sign
(per \cite{NP66,Goldberg67}),
from the fact that the raising and lowering operators
$\ncalD{s}_\plus$ and $- \ncalD{s \plus \one}_\minus$
are Hermitian conjugates of each other,
equation~(\ref{Dpmh}),
and that their product
$- \ncalD{s \plus \one}_\minus \, \ncalD{s}_\plus$
has eigenvalue
$(l {-} s) ( l {+} s {+} 1) / 2$,
in accordance with equation~(\ref{DphDp}).
Spin-harmonics of opposite spins $s$ are related by
\begin{equation}
\label{Yconjugate}
  \nY{s}_{lm}
  =
  (-)^{m+s}
  \nY{\minus s}^\ast_{l \comma \minus m}
  \ .
\end{equation}

\subsection{Spin raising and lowering radial operators}
\label{spinradialoperators}

As will be seen in \S\ref{perturbations},
radial operators
$\ncalD{s}_v$
and
$\ncalD{s}_u$
analogous to the angular operators
$\ncalD{s}_\plus$
and
$\ncalD{s}_\minus$
of the previous subsection,
\S\ref{spinangularoperators},
appear ubiquitously in the perturbation theory
of spherically self-similar black holes.
The structure of the radial operators resembles that of
the angular operators,
and it is convenient to refer to the radial operators
as spin raising and lowering operators.
However,
whereas the angular operators
yield a consistent ladder of angular eigenfunctions,
the radial operators do not,
because whereas in the angular case
the Hermitian operators connecting
adjacent levels of spin differ by a constant,
equation~(\ref{diffDDpm}),
in the radial case they do not,
equation~(\ref{diffDDvu})
(except in the case that the Weyl scalar $C$ is constant,
which occurs in flat space, where $C = 0$,
as considered in the next subsection, \S\ref{asymptotic}).
Thus the designation of
$\ncalD{s}_v$
and
$\ncalD{s}_u$
as spin raising and lowering radial operators
has a limited meaning.
It is true that for each spin $s$,
the Hermitian operator
$\ncalD{s}^{\dagger v} \, \ncalD{s}_v$
has two eigenfunctions, and that
$\ncalD{s}_v$ ``raises''
the ``spin-$s$'' eigenfunction to the ``spin-$(s{+}1)$'' eigenfunction,
while
$\ncalD{s}^{\dagger v} = \ncalD{s+1}_u$ ``lowers''
the ``spin-$(s{+}1)$'' eigenfunction to the ``spin-$s$'' eigenfunction.
But the eigenfunctions of operators
$\ncalD{s}^{\dagger v} \, \ncalD{s}_v$
with different spins $s$ are different,
so there is no consistent ladder of spin radial eigenfunctions
(except in the case that $C$ is constant,
considered in the next subsection, \S\ref{asymptotic}).

Define the spin raising and lowering radial operators
$\ncalD{s}_v$
and
$\ncalD{s}_u$
by
\begin{subequations}
\label{Dvu}
\begin{eqnarray}
  \ncalD{s}_v
  &\equiv&
  (\xi^u)^{s} \, \calD_v \, (\xi^u)^{-s}
  \ ,
\\
  \ncalD{s}_u
  &\equiv&
  (\xi^v)^{-s} \, \calD_u \, (\xi^v)^{s}
  \ .
\end{eqnarray}
\end{subequations}
The Hermitian conjugates of the contravariant components of these operators are
\begin{subequations}
\label{Dvuh}
\begin{eqnarray}
  \ncalD{s}^{\dagger v}
  &=&
  \ncalD{s\plus \one}_u
\\
  \ncalD{s}^{\dagger u}
  &=&
  \ncalD{s\minus \one}_v
  \ .
\end{eqnarray}
\end{subequations}

%
The products of the
raising and lowering operators
$\ncalD{s}_\vu$
with their contravariant Hermitian conjugates
$\ncalD{s}^{\dagger \vu}$
are the Hermitian operators
\begin{subequations}
\label{DvuhDvu}
\begin{eqnarray}
  2 \, \ncalD{s}^{\dagger v} \, \ncalD{s}_v
  &=&
  {1 \over A^2 \Delta}
  \left(
  {\partial \over \partial t} - \Dr{}{}
  - s \Dr{}{\ln\xi^v}
  \right)
  \left(
  {\partial \over \partial t} + \Dr{}{}
  - s \Dr{}{\ln\xi^u}
  \right)
  A^2
  \ ,
\\
  2 \, \ncalD{s}^{\dagger u} \, \ncalD{s}_u
  &=&
  {1 \over A^2 \Delta}
  \left(
  {\partial \over \partial t} + \Dr{}{}
  - s \Dr{}{\ln\xi^u}
  \right)
  \left(
  {\partial \over \partial t} - \Dr{}{}
  - s \Dr{}{\ln\xi^v}
  \right)
  A^2
  \ .
\end{eqnarray}
\end{subequations}
These Hermitian operators can be re-expressed as
\begin{subequations}
\label{DDradial}
\begin{eqnarray}
  2 \, \ncalD{s}^{\dagger v} \, \ncalD{s}_v
  &=&
  {1 \over A^2 \Delta}
  \biggl( {\xi^v \over \xi^u} \biggr)^{-s/2}
  \left[
  - \,
  \Dr{2}{}
  +
  ( \im \omega + s \kappa )^2
  +
  s ( 1 + 6 C ) \Delta
  \right]
  \biggl( {\xi^v \over \xi^u} \biggr)^{s/2}
  A^2
  \ ,
\\
  2 \, \ncalD{s}^{\dagger u} \, \ncalD{s}_u
  &=&
  {1 \over A^2 \Delta}
  \biggl( {\xi^v \over \xi^u} \biggr)^{-s/2}
  \left[
  - \,
  \Dr{2}{}
  +
  ( \im \omega + s \kappa )^2
  -
  s ( 1 + 6 C ) \Delta
  \right]
  \biggl( {\xi^v \over \xi^u} \biggr)^{s/2}
  A^2
  \ ,
\end{eqnarray}
\end{subequations}
where $\partial / \partial t \equiv - \im \omega$,
the dimensionless scalar $\kappa$ is
\begin{equation}
\label{kappa}
  \kappa
  \equiv
  {1 \over 2 \Delta}
  \Dr{}{\Delta}
  =
  \xi^u h_u - \xi^v h_v - \Dr{}{\ln A}
  \ ,
\end{equation}
and $C$ is the dimensionless unperturbed Weyl scalar, equation~(\ref{C}).
The factor $1 + 6 C$ in the potential $s ( 1 + 6 C ) \Delta$
in equations~(\ref{DDradial})
comes from
\begin{equation}
\label{Dkappa}
  {1 \over \Delta} \Dr{}{\kappa}
  =
  1 + 6 C
  \ .
\end{equation}

The Hermitian operators
$\ncalD{s}^{\dagger v} \, \ncalD{s}_v$
and
$\ncalD{s}^{\dagger u} \, \ncalD{s}_u$
differ by a function,
according to equations~(\ref{DDradial}),
\begin{equation}
\label{diffDDvu}
  \ncalD{s}^{\dagger v} \, \ncalD{s}_v
  -
  \ncalD{s}^{\dagger u} \, \ncalD{s}_u
  =
  s \kappa
  =
  s ( 1 + 6 C )
  \ .
\end{equation}
This difference, equation~(\ref{diffDDvu}), is generally not constant
because the Weyl scalar $C$ is generally not constant,
and it follows that,
in contrast to the angular raising and lowering operators $\ncalD{s}_{\pm}$,
the ladder of radial raising and lowering operators
$\ncalD{s}_\vu$
does not yield a mutually consistent set of radial eigenfunctions
(except in the case that $C$ is constant,
such as the case $C = 0$
considered in the next subsection, \S\ref{asymptotic}).

In practice,
the radial operators 
$\ncalD{s}^{\dagger v} \, \ncalD{s}_v$
and
$\ncalD{s}^{\dagger u} \, \ncalD{s}_u$
occur in the
gravitational, electromagnetic, and scalar
wave equations~(\ref{waveCz}),
(\ref{waveCza}),
(\ref{waveFz}),
(\ref{waveFza}),
and
(\ref{wavepsi})
in combination with angular parts
$\ncalD{s}^{\dagger \plus} \, \ncalD{s}_\plus$
and
$\ncalD{s}^{\dagger \minus} \, \ncalD{s}_\minus$
as follows,
with positive spin $s$:
\begin{equation}
\label{wave}
  \ncalD{\plusminus s}^{\dagger \uv} \, \ncalD{\plusminus s}_\uv
  +
  \ncalD{\plusminus s}^{\dagger \minusplus} \, \ncalD{\plusminus s}_\minusplus
  =
  {1 \over 2 A^2 \Delta}
  \biggl( {\xi^v \over \xi^u} \biggr)^{\mp s/2}
  \left\{
  - \,
  \Dr{2}{}
  +
  ( \im \omega \pm s \kappa )^2
  +
  \bigl[
    l ( l + 1 )
    - s^2
    - 6 s C
  \bigr]
  \Delta
  \right\}
  \biggl( {\xi^v \over \xi^u} \biggr)^{\pm s/2}
  A^2
\end{equation}
in which the total angular momentum operator
$\nL{s}^2$
in the angular operators
$\ncalD{s}^{\dagger \plus} \, \ncalD{s}_\plus$
and
$\ncalD{s}^{\dagger \minus} \, \ncalD{s}_\minus$,
equations~(\ref{DpmhDpm}),
has been replaced by its eigenvalue $l(l+1)$,
equation~(\ref{Ls}).
The wave operators in the wave equations for
gravitational, electromagnetic, and scalar waves
prove to be the same as that given by equation~(\ref{wave}),
apart from the addition of further dimensionless terms,
functions only of conformal radius $r$,
in the factor inside square brackets
multiplying the dimensionless horizon function $\Delta$
within the curly braces on the right hand side.

The operator given by equation~(\ref{wave})
has homogeneous solutions $\Psi_{\plusminus s}$
\begin{equation}
\label{DDPsi}
  \Bigl(
  \ncalD{\plusminus s}^{\dagger \uv} \, \ncalD{\plusminus s}_\uv
  +
  \ncalD{\plusminus s}^{\dagger \minusplus} \, \ncalD{\plusminus s}_\minusplus
  \Bigr)
  \Psi_{\plusminus s}
  =
  0
\end{equation}
of definite conformal frequency $\omega$
and definite spherical harmonic numbers $lm$,
separating as
\begin{equation}
\label{Psi}
  \Psi_{\plusminus s}
  =
  ( {\xi^v / \xi^u} )^{\mp s/2}
  A^{-2} \,
  \ee^{- \im \omega t} \,
  \npsi{\plusminus s}_{\omega l}(r) \,
  \nY{\plusminus s}_{lm}(\theta,\phi)
\end{equation}
where
$\npsi{\plusminus s}_{\omega l}(r)$
depends only on conformal radius $r$,
not on conformal time $t$.
The factors of $( {\xi^v / \xi^u} )^{\mp s/2}$
in the solutions $\Psi_{\plusminus s}$
embody their dependence on radial Lorentz boosts.
In a tetrad frame at rest in the similarity frame,
where $\xi^r = 0$ and hence $\xi^v / \xi^u = 1$,
the factor is just unity.
In a tetrad frame moving at radial velocity $V$ with respect
to the similarity frame,
so that $\xi^r / \xi^t = V$,
the factor
$( {\xi^v / \xi^u} )^{\mp s/2}$
is simply the $\mp s$'th power of the
special relativistic Doppler shift factor
\begin{equation}
\label{dopplerfactor}
  \left( {\xi^v \over \xi^u} \right)^{1/2}
  =
  \left[ {(1 + V) \over (1 - V)} \right]^{1/2}
  \ .
\end{equation}

Reduced spin raising and lowering operators
$\nd{s}_\vu$
can be defined which operate only on the radial part $\npsi{s}_{\omega l}(r)$
of the homogeneous solution $\Psi_s$,
in view of the relation~(\ref{Psi}),
\begin{equation}
\label{nds}
  \nd{s}_\vu
  \equiv
  \ee^{\im \omega t}
  A^{2}
  ( {\xi^v / \xi^u} )^{(s \pm 1)/2}
  \ncalD{s}_\vu \,
  ( {\xi^v / \xi^u} )^{- s/2}
  A^{-2}
  \ee^{- \im \omega t}
  \ ,
\end{equation}
which simplifies to
\begin{equation}
\label{nd}
  d_\vu
  \equiv
  \frac{1}{\sqrt{2 \Delta}}
  \Bigl(
    \pm \,
    \Dr{}{}
    -
    \im \omega
  \Bigr)
  \ ,
  \quad
  \nd{s}_\vu
  =
  \Delta^{\pm s/2} \,
  d_\vu \,
  \Delta^{\mp s/2}
  \ .
\end{equation}

\subsection{Asymptotic wave solutions at large radius}
\label{asymptotic}

Self-similar solutions to black holes
do not necessarily continue self-consistently
to asymptotically flat, empty space at large radius
\cite{HP05a,HP05b}.
Nevertheless,
it is convenient to suppose that the self-similar geometry
does asymptote at large radius
to a region where the Weyl scalar vanishes, $C = 0$,
since this makes it straightforward to identify
which solutions are to be identified as ingoing
and which as outgoing.

The solutions of equations~(\ref{kappa}) and (\ref{Dkappa})
with $C = 0$,
and with
$\kappa$
and
$\Delta$
tending to zero at large radius,
are
\begin{equation}
\label{kappaDeltaflat}
  \kappa = - {1 \over r^\ast}
  \ , \quad
  \Delta = {1 \over r^{\ast 2}}
  \ .
\end{equation}
In non-stationary self-similar solutions,
where the expansion velocity is non-zero, $\vel \neq 0$,
the Regge-Wheeler coordinate $r^\ast$
defined by equations~(\ref{partialra}) and (\ref{rast})
is related to the conformal radius $r$ at large radius by
$\vel r^\ast \rightarrow \ln r$.
This contrasts with the case of stationary solutions,
where $r^\ast \rightarrow r$ at large radius.

Denote the radial part $\npsi{s}_{\omega l}(r)$ of the solutions $\Psi_s$,
equation~(\ref{Psi}),
in the asymptotic region at large radius
by $\nz{s}_l$
\begin{equation}
  \npsi{s}_{\omega l}
  \sim
  \nz{s}_l
  \quad
  \mbox{at large radius}
  \ .
\end{equation}
The asymptotic solutions
$\nz{s}_l$
are the homogeneous solutions of the operator inside braces
in equation~(\ref{wave})
for the case where $C = 0$
and $\kappa$ and $\Delta$ are given by equations~(\ref{kappaDeltaflat}):
\begin{equation}
\label{DDz}
  \left[
  - \,
  \Dr{2}{}
  -
  \omega^2
  +
  {2 s \im \omega \over r^\ast}
  +
  {l ( l + 1 ) \over r^{\ast 2}}
  \right]
  \nz{s}_l
  =
  0
  \ .
\end{equation}
It evident from the form of equation~(\ref{DDz})
that the solutions can be taken to be functions
$\nz{s}_l(\omega r^\ast)$
of the product $\omega r^\ast$
of conformal frequency and radius.
To leading order at large radius $r^\ast$,
the two independent solutions of equation~(\ref{DDz}) are
\begin{equation}
\label{zasymptoticunnormalized}
  \begin{array}{rcll}
  \nz{s}^\textrm{in}_l
  &\sim&
  (\omega r^\ast)^s \,
  \ee^{- \im \omega r^\ast}
  &
  \quad
  (\mbox{propagating for $s \geq 0$})
  \ ,
\\
  \nz{s}^\textrm{out}_l
  &\sim&
  (\omega r^\ast)^{-s} \,
  \ee^{\im \omega r^\ast}
  &
  \quad
  (\mbox{propagating for $s \leq 0$})
  \ .
  \end{array}
\end{equation}
The modes are labeled in and out because
$\Psi_s \propto \ee^{- \im \omega t} \nz{s}^\textrm{in}_l
\propto
(\omega r^\ast)^s \, \ee^{- \im \omega ( t + r^\ast )}$
corresponds to ingoing modes,
while
$\Psi_s \propto \ee^{- \im \omega t} \nz{s}^\textrm{out}_l
\propto
(\omega r^\ast)^{-s} \, \ee^{- \im \omega ( t - r^\ast )}$
corresponds to outgoing modes.
For non-zero spins $s$,
only one of the two modes
of equations~(\ref{zasymptoticunnormalized})
is propagating,
the one that falls off most slowly at large radius $r^\ast$.
For positive spin $+s$,
the propagating mode is the ingoing mode,
$\nz{\plus s}^\textrm{in}_l \propto (\omega r^\ast)^s \ee^{- \im \omega r^\ast}$,
while for negative spin $-s$,
the propagating mode is the outgoing mode,
$\nz{\minus s}^\textrm{out}_l \propto (\omega r^\ast)^s \ee^{\im \omega r^\ast}$.
The other of the two modes, for non-zero spin,
corresponds to the short-range, non-propagating
spin-$(\mp s)$ component of a propagating spin-$(\pm s)$ wave.

Since changing the sign of the spin $s$ changes
the operator on the left hand side of equation~(\ref{DDz})
into its complex conjugate,
it follows that solutions
$\nz{s}_l$
of opposite spin
are complex conjugates of each other.
Specifically,
ingoing solutions
are complex conjugates of
outgoing solutions of opposite spin:
\begin{equation}
\label{zinout}
  \nz{s}^\textrm{in}_l
  =
  \left( \nz{\minus s}^\textrm{out}_l \right)^\ast
  \ .
\end{equation}

For vanishing Weyl scalar $C = 0$,
as is being considered in this subsection,
the radial operators
$\ncalD{s}_v$
and
$\ncalD{s}_u$
defined by equations~(\ref{Dvu})
form a consistent ladder of spin raising and lowering operators
[cf.\ the comments in the paragraphs containing
equations~(\ref{diffDDpm}) and (\ref{diffDDvu})].
As will now be shown,
these operators
can be used to construct exact analytic expressions for
the asymptotic solutions
$\nz{s}_l$
by successive raisings or lowerings
from the spin zero solutions.

The spin zero solutions
$\nz{\zero}_l(\omega r^\ast)$
of equation~(\ref{DDz})
are proportional to
spherical Hankel-Bessel functions
$h^{(\plusminus)}_l(\omega r^\ast) \equiv j_l(\omega r^\ast) \pm \im y_l(\omega r^\ast)
\equiv \sqrt{\pi / ( 2 \omega r^\ast )} \bigl[ J_{l+\frac{1}{2}}(\omega r^\ast) \pm \im Y_{l+\frac{1}{2}}(\omega r^\ast) \big]$,
\begin{equation}
\label{z0}
  \nz{\zero}^\textrm{out}_l(\omega r^\ast)
  =
  (\im)^{l + 1} \omega r^\ast h^{(\plus)}_l(\omega r^\ast)
  \ ,
  \quad
  \nz{\zero}^\textrm{in}_l(\omega r^\ast)
  =
  (- \im)^{l + 1} \omega r^\ast h^{(\minus)}_l(\omega r^\ast)
  \ ,
\end{equation}
which are complex conjugates of each other,
in accordance with equation~(\ref{zinout}).
The spin zero solutions
$\nz{\zero}l(\omega r^\ast)$, equations~(\ref{z0}),
are normalized so that
\begin{equation}
\label{z0asymptotic}
  \begin{array}{rcl}
    \nz{\zero}^\textrm{in}_l(\omega r^\ast)
    & \rightarrow &
    \ee^{- \im \omega r^\ast}
    \\
    \nz{\zero}^\textrm{out}_l(\omega r^\ast)
    & \rightarrow &
    \ee^{\im \omega r^\ast}
  \end{array}
  \quad
  \mbox{as } r^\ast \rightarrow \infty
  \ .
\end{equation}

The spin raising and lowering operators
$\ncalD{s}_\vu$
act
not on
$\nz{s}_l(\omega r^\ast)$
but rather
on the full temporal-radial part
$\nZ{s}_{\omega l}(t, r)$
of
$\Psi_s =
\nZ{s}_{\omega l}(t, r) \, \nY{s}_{lm}(\theta,\phi)$,
equation~(\ref{Psi}),
\begin{equation}
\label{Z}
  \nZ{s}_{\omega l}(t, r)
  \equiv
  ( {\xi^v / \xi^u} )^{- s/2} A^{-2} \, \ee^{- \im \omega t} \, \nz{s}_l(\omega r^\ast)
  \ .
\end{equation}
The functions
$\nZ{\plusminus s}_{\omega l}(t, r)$
are by definition eigenfunctions of the Hermitian operator
$\ncalD{\plusminus s}^{\dagger \uv} \, \ncalD{\plusminus s}_\uv$,
with eigenvalues
$- (l {-} s) ( l {+} s {+} 1) / 2$
in accordance with equation~(\ref{DDPsi}),
\begin{equation}
\label{DDZ}
  \left[
  \ncalD{\plusminus s}^{\dagger \uv} \, \ncalD{\plusminus s}_\uv
  +
  {(l - s) ( l + s + 1) \over 2}
  \right]
  \nZ{\plusminus s}_{\omega l}(t, r)
  =
  0
  \ .
\end{equation}
The radial spin raising and lowering operators
$\ncalD{s}_v$
and
$\ncalD{s}_u$
raise and lower the spin $s$ of
$\nZ{s}_{\omega l}(t, r)$
by one unit
\begin{equation}
\label{raiselowerZ}
  \ncalD{s}_\vu \,
  \nZ{s}_{\omega l}(t, r)
  =
  - \, \im
  \left[ {(l \mp s) ( l \pm s + 1) \over 2} \right]^{1/2}
  \!
  \nZ{s \plusminus \one}_{\omega l}(t, r)
\end{equation}
where the normalization factor follows,
modulo a phase factor,
from the fact that the raising and lowering operators
$\ncalD{s}_v$ and $\ncalD{s \plus \one}_u$
are Hermitian conjugates of each other,
equation~(\ref{Dvuh}),
and that their product
$\ncalD{s \plus \one}_u \, \ncalD{s}_v$
has eigenvalue
$- (l {-} s) ( l {+} s {+} 1) / 2$,
equation~(\ref{DDZ}).

The reduced radial spin raising and lowering operators
$\nd{s}_v$
and
$\nd{s}_u$
defined by equation~(\ref{nds}) or equivalently (\ref{nd})
simplify to
\begin{equation}
  d_\vu
  \equiv
  \frac{1}{\sqrt{2}}
  \left(
    \pm \,
    r^\ast \Dr{}{}
    -
    \im \omega r^\ast
  \right)
  \ ,
  \quad
  \nd{s}_\vu
  =
  ( r^\ast )^{\pm s} \,
  d_\vu \,
  ( r^\ast )^{\mp s}
  \ .
\end{equation}
Similar to equation~(\ref{raiselowerZ}),
the radial spin raising and lowering operators
$\nd{s}_\vu$
raise and lower the spin $s$ of
$\nz{s}_l(\omega r^\ast)$
by one unit
\begin{equation}
\label{raiselowerz}
  \nd{s}_\vu \,
  \nz{s}_l(\omega r^\ast)
  =
  - \, \im
  \left[ {(l \mp s) ( l \pm s + 1) \over 2} \right]^{1/2}
  \!
  \nz{s \plusminus \one}_l(\omega r^\ast)
  \ .
\end{equation}
There are of course both ingoing solutions
$\nz{s}^\textrm{in}_l$
and outgoing solutions
$\nz{s}^\textrm{out}_l$:
the operators
$\nd{s}_\vu$
raise and lower
ingoing solutions to other ingoing solutions,
and outgoing solutions to other outgoing solutions.
The asymptotic behavior of the spin zero modes
$\nz{\zero}_l$, equation~(\ref{z0asymptotic}),
combined with the raising and lowering equations~(\ref{raiselowerz}),
imply that the asymptotic behavior of
$\nz{s}_l$ at large radius is
\begin{equation}
\label{zasymptotic}
  \begin{array}{rcl}
    \nz{s}^\textrm{in}_l(\omega r^\ast)
    &\rightarrow&
    \displaystyle{
    \left[ {(l - s)! \over (l + s)!} \right]^{1/2}
    (2 \omega r^\ast)^s
    \ee^{- \im \omega r^\ast}
    }
    \\
    \nz{s}^\textrm{out}_l(\omega r^\ast)
    &\rightarrow&
    \displaystyle{
    \left[ {(l + s)! \over (l - s)!} \right]^{1/2}
    (2 \omega r^\ast)^{- s}
    \ee^{\im \omega r^\ast}
    }
  \end{array}
  \quad
  \mbox{as }
  r^\ast \rightarrow \infty
  \ ,
\end{equation}
in agreement with the earlier result~(\ref{zasymptoticunnormalized}).
The choice of phase factor $- \im$
in equation~(\ref{raiselowerZ})
and consequently (\ref{raiselowerz})
is such that
ingoing solutions
are complex conjugates of outgoing solutions
of opposite spin,
equation~(\ref{zinout}).

The Wronskian of the two solutions
$\nz{s}^\textrm{in}_l$
and
$\nz{s}^\textrm{out}_l$
must be constant,
in accordance with the usual rules
for homogeneous solutions
of second order differential equations,
and the value of the constant follows from
the asymptotic behavior~(\ref{zasymptotic}),
\begin{equation}
\label{wronskianz}
  \nz{s}^\textrm{in}_l
  \Dr{}{\nz{s}^\textrm{out}_l}
  -
  \nz{s}^\textrm{out}_l
  \Dr{}{\nz{s}^\textrm{in}_l}
  =
  2 \im \omega
  \ .
\end{equation}

%
%

\section{Perturbations to self-similar black holes}
\label{perturbations}

This section sets up the theory of
perturbations to spherically symmetric self-similar black hole spacetimes.
The approach follows the Newman-Penrose formalism
\cite{NP62,GHP73}.
The section marches successively through perturbations to the vierbein
(\S\ref{perturbedtetrad}),
tetrad connections
(\S\ref{perturbedconnections}),
Riemann tensor
(\S\ref{perturbedriemann}),
and
Einstein tensor
(\S\ref{perturbedeinstein}).
The results will be used in subsequent sections
\S\S\ref{gravitationalwaves}--\ref{scalarwaves}
on gravitational, electromagnetic, and scalar waves.

From the outset,
angular perturbations are required to
be expandable in spin-weighted spherical hamornics $\nY{s}$,
and conditions under which this can be done are obtained.
Initially,
from requiring that all tetrad connections be expandable in spin harmonics,
it is found that a certain set of gauge conditions is imposed,
the spherical gauge, equations~(\ref{gaugespherical}).
Subsequently,
it is found
(\S\ref{togauge})
that the tetrad connections that impose the spherical gauge conditions
are precisely those that do not actually appear in any of the wave equations,
so that it is consistent to expand vierbein perturbations in spin harmonics
with no gauge conditions whatsoever.
The general rule is established that the spin $s$ of any
quantity is equal to the sum of the $+$'s and $-$'s of
its covariant components in the Newman-Penrose formalism.
The result is so fundamental that it deserves stating on
a line by itself:
\begin{equation}
\label{spinrule}
  \mbox{spin } s
  =
  \mbox{sum of $+$'s and $-$'s of covariant indices}
  \ .
\end{equation}

The motivation for working with spin-weighted spherical harmonics,
aside from the familiar fact \cite{BP73,Chandrasekhar}
that the angular eigenmodes of
wave equations in the Newman-Penrose formalism
are spin harmonics,
is that spin harmonics, or more correctly
$\nY{s}_{lm}(\theta,\phi) \ee^{-\im s \chi}$,
are eigenmodes of the full 3-dimensional rotation group $O(3)$,
parameterized by three Euler angles
$\theta,\phi,\chi$
\cite{Goldberg67}.
Spherically symmetric spacetimes are symmetric
not merely with respect to rotations about the origin,
but with respect to the entire 3-dimensional rotation group.
Thus it is natural to expect that angular perturbations
of spherically symmetric spacetimes should be expandable
in spin harmonics.

In this section,
a ${\scriptstyle 0}$ (zero) overscript signifies an unperturbed quantity,
while a ${\scriptstyle 1}$ (one) overscript signifies a perturbation.
No overscript means the full quantity,
including both unperturbed and perturbed parts.
An overscript is attached only where necessary.
Thus if the unperturbed part of a quantity is zero,
then no overscript is needed, and none is attached.

\subsection{Perturbed tetrad}
\label{perturbedtetrad}

Define the vierbein perturbation $\varphi_{mn}$
so that the perturbed vierbein is
\begin{equation}
  {e_m}^\mu
  =
  \left(
  \delta_m^n - {\varphi_m}^n
  \right)
  \overset{\smallzero}{e}{{}_n}^\mu
  \ ,
\end{equation}
with corresponding inverse
\begin{equation}
  {e^m}_\mu
  =
  \left(
  \delta^m_n + {\varphi_n}^m
  \right)
  \overset{\smallzero}{e}{{}^n}_\mu
  \ .
\end{equation}
Since the vierbein perturbation $\varphi_{mn}$ is already of linear order,
its indices can be raised and lowered with the unperturbed metric,
and transformed between tetrad and coordinate frames
with the unperturbed vierbein.
The perturbation $\varphi_{mn}$ can thus be regarded as a tensor field
defined on the unperturbed background.
The perturbation $\varphi_{mn}$
has 16 degrees of freedom,
but only 6 of these degrees of freedom
correspond to real physical perturbations,
since 6 degrees of freedom are associated with arbitrary
infinitesimal changes in the choice of tetrad,
which is to say arbitrary infinitesimal Lorentz transformations,
and a further 4 degrees of freedom are associated with arbitrary
infinitesimal changes in the coordinates.

Under an infinitesimal tetrad transformation,
the covariant vierbein perturbations
$\varphi_{mn}$
transform as
\begin{equation}
  \varphi_{mn}
  \rightarrow
  \varphi_{mn}
  +
  \epsilon_{mn}
  \ ,
\end{equation}
where $\epsilon_{mn}$
is the generator of a Lorentz transformation,
which is to say an arbitrary antisymmetric tensor.
Thus the antisymmetric part
$\varphi_{mn} - \varphi_{nm}$
of the covariant perturbation $\varphi_{mn}$
is arbitrarily adjustable through an infinitesimal tetrad transformation,
while the symmetric part
$\varphi_{mn} + \varphi_{nm}$
is tetrad gauge-invariant.
It is easy to see when a perturbation to a quantity
is tetrad gauge-invariant:
it is tetrad gauge-invariant
if and only if it depends only on the symmetric part
of the vierbein perturbation,
not on the antisymmetric part.

Under an infinitesimal coordinate gauge transformation,
the coordinates $x^\mu$ change by an infinitesimal shift $A \epsilon^\mu$
(the factor $A$ is incorporated so that $\epsilon^\mu$ is dimensionless)
\begin{equation}
\label{gaugex}
  x^\mu
  \rightarrow
  x^\mu + A \epsilon^\mu
  \ .
\end{equation}
Because the dimensionless shift $\epsilon^\mu$ is,
like the vierbein perturbations $\varphi_{mn}$,
already of linear order,
its indices can be raised and lowered with the unperturbed metric,
and transformed between coordinate and tetrad frames
with the unperturbed vierbein.
Thus the dimensionless shift
$\epsilon^\mu$
can be regarded as a vector field
defined on the unperturbed background.
Under an infinitesimal coordinate gauge transformation~(\ref{gaugex}),
the vierbein perturbations
$\varphi_{mn}$
transform as
\begin{equation}
\label{gaugephi}
  \varphi_{mn}
  \rightarrow
  \varphi_{mn}
  -
  \left(
  \gamma_{nk} \partial_m
  + \Gamma_{nkm} - \Gamma_{nmk}
  \right)
  A
  \epsilon^k
  \ ,
\end{equation}
in which
$\epsilon^k = \overset{\smallzero}{e}{{}^k}_\mu \, \epsilon^\mu$
are the (dimensionless) tetrad components of the coordinate shift.
Explicitly,
in the Newman-Penrose tetrad
the perturbations
$\varphi_{mn}$
transform under infinitesimal coordinate gauge transformations~(\ref{gaugex}) as
\begin{subequations}
\label{gaugephis}
\begin{eqnarray}
  \varphi_{\overset{{\scriptstyle v v}}{{\scriptstyle u u}}}
  &\rightarrow&
  \varphi_{\overset{{\scriptstyle v v}}{{\scriptstyle u u}}}
  +
  A^2 \,
  \ncalD{\plusminus \one}_\vu \,
  A^{-2} \,
  \epsilon^\uv
  \ ,
\\
  \varphi_{\overset{{\scriptstyle v u}}{{\scriptstyle u v}}}
  &\rightarrow&
  \varphi_{\overset{{\scriptstyle v u}}{{\scriptstyle u v}}}
  +
  A \,
  \ncalD{\zero}_\vu \,
  A^{-1} \,
  \epsilon^\vu
  -
  h_\uv \, \epsilon^\uv
  \ ,
\\
  \varphi_{\overset{{\scriptstyle v} \plus}{{\scriptstyle u} \minus}}
  &\rightarrow&
  \varphi_{\overset{{\scriptstyle v} \plus}{{\scriptstyle u} \minus}}
  -
  A^2 \,
  \ncalD{\zero}_\vu \,
  A^{-2} \,
  \epsilon^\minusplus
  \ ,
\\
  \varphi_{\overset{{\scriptstyle v} \minus}{{\scriptstyle u} \plus}}
  &\rightarrow&
  \varphi_{\overset{{\scriptstyle v} \minus}{{\scriptstyle u} \plus}}
  -
  A^2 \,
  \ncalD{\zero}_\vu \,
  A^{-2} \,
  \epsilon^\plusminus
  \ ,
\\
\label{gaugephipv}
  \varphi_{\overset{\plus {\scriptstyle v}}{\minus {\scriptstyle u}}}
  &\rightarrow&
  \varphi_{\overset{\plus {\scriptstyle v}}{\minus {\scriptstyle u}}}
  +
  \ncalD{\zero}_\plusminus \,
  \epsilon^\uv
  \ ,
\\
\label{gaugephimv}
  \varphi_{\overset{\minus {\scriptstyle v}}{\plus {\scriptstyle u}}}
  &\rightarrow&
  \varphi_{\overset{\minus {\scriptstyle v}}{\plus {\scriptstyle u}}}
  +
  \ncalD{\zero}_\minusplus \,
  \epsilon^\uv
  \ ,
\\
\label{gaugephipm}
  \varphi_{\overset{\plus \minus}{\minus \plus}}
  &\rightarrow&
  \varphi_{\overset{\plus \minus}{\minus \plus}}
  -
  \Bigl(
  \beta_v \, \epsilon^v
  +
  \beta_u \, \epsilon^u
  +
  {\cot\theta \over \sqrt{2}} \,
  \epsilon^\minusplus
  +
  \ncalD{\zero}_\plusminus \, \epsilon^\plusminus
  \Bigr)
  \ ,
\\
\label{gaugephipp}
  \varphi_{\overset{\plus \plus}{\minus \minus}}
  &\rightarrow&
  \varphi_{\overset{\plus \plus}{\minus \minus}}
  -
  \ncalD{\plusminus \one}_\plusminus \, \epsilon^\minusplus
  \ .
\end{eqnarray}
\end{subequations}

Perturbations can be classified into polar modes,
which do not change sign,
and axial modes,
which change sign,
when the azimuthal angular tetrad axis
$\gamma_\phi$
is flipped in sign,
equation~(\ref{angularconjugate}).
In the Newman-Penrose formalism,
flipping the azimuthal axis $\gamma_\phi$ is equivalent to
swapping covariant angular indices
$+ \leftrightarrow -$,
equations~(\ref{npangularconjugate}),
which,
for quantities that are real
in an orthonormal tetrad,
is in turn equivalent
to taking the complex conjugate.
Thus in the Newman-Penrose formalism,
polar modes effectively correspond to real perturbations,
while axial modes correspond to imaginary perturbations.
It is a fundamental feature of the Newman-Penrose formalism
not to separate out the polar (real) and axial (imaginary) perturbations,
but rather to treat them together as a combined complex perturbation.
The advantage of the Newman-Penrose approach is that
the radial part of the wave operator in the wave equation
for non-zero spin components of the Weyl tensor
proves to be the same for both polar and axial modes,
revealing their fundamental unity
[see for example the discussion in the paragraph containing
equation~(\ref{waveCzstar})].


It should be commented that the general strategy for linearization,
which has already been invoked in obtaining equations~(\ref{gaugephis})
from equations~(\ref{gaugephi}),
and which is used throughout the rest of this paper,
is to discard products of quantities that are already first order,
since the result is of second order,
and to replace directed derivatives of first order quantities
by the spin raising and lowering operators $\ncalD{s}_m$,
equations~(\ref{Dpm}) and (\ref{Dvu}),
as follows:
\begin{subequations}
\label{linearize}
\begin{eqnarray}
\label{linearizevu}
  \partial_\vu
  +
  s \, \Gamma_{\overset{{\scriptstyle u v v}}{{\scriptstyle v u u}}}
  +
  a \, \Gamma_{\overset{\minus {\scriptstyle v} \plus}{\plus {\scriptstyle u} \minus}}
  +
  b \, \Gamma_{\overset{\plus {\scriptstyle v} \minus}{\minus {\scriptstyle u} \plus}}
  &\rightarrow&
  A^{1 + s - a - b}
  \ncalD{\plusminus s}_\vu \,
  A^{- 2 - s + a + b}
  \ ,
\\
\label{linearizepm}
  \partial_\plusminus
  -
  s \, \Gamma_{\overset{\minus \plus \plus}{\plus \minus \minus}}
  &\rightarrow&
  A^{-1}
  \ncalD{\plusminus s}_\plusminus
  \ .
\end{eqnarray}
\end{subequations}

\subsection{Perturbed tetrad connections}
\label{perturbedconnections}

This subsection presents expressions for the perturbed tetrad-frame connections
$\Gamma_{kmn}$
in terms of the vierbein perturbations
$\varphi_{mn}$,
equations~(\ref{Gamma}).
Much of the subsection is concerned with showing that
a certain set of gauge conditions,
spherical gauge, equations~(\ref{gaugespherical}),
emerges if it is required that the relations~(\ref{Gamma})
between the tetrad connections and vierbein perturbations
become algebraic upon resolution into spin-weighted harmonics,
that is to say, if each (perturbation to a) tetrad connection and each vierbein perturbation
is required to be individually expandable in spin-harmonics of some definite spin.
The next subsection, \S\ref{perturbedriemann},
continues the theme,
finding that spherical gauge emerges a second time
if it is required that the relations between the Riemann tensor
and the tetrad connections become algebraic on resolution into spin-weighted harmonics.
Ultimately, in \S\ref{togauge},
the need for spherical gauge is traced to just 4 offending tetrad connections
[the ones involving boxed terms in equations~(\ref{Gamma}) below],
and it is pointed out that these 4 connections are precisely the ones
that do not actually appear in the wave equations in the Newman-Penrose formalism.
Thus not only are the spherical gauge conditions not required,
but in fact the wave equations in the Newman-Penrose formalism
can be resolved into spin-weighted spherical
harmonics with no gauge conditions at all.

Three important lessons emerge from the analysis
of this and the next two subsections.
First, it is consistent to expand vierbein perturbations $\varphi_{mn}$
in spin-weighted harmonics from the outset:
there is no need for the more general system of tensor harmonics
\cite{Zerilli70a,NR05,MP05,FGR06}.
Second,
the wave equations in the Newman-Penrose formalism
become algebraic upon resolution into spin-weighted harmonics
without any gauge conditions;
that is, one is free to choose any gauge whatsoever.
However, if a gauge is called for,
then the spherical gauge conditions~(\ref{gaugespherical})
are, though not required, nevertheless natural,
since it is only in spherical gauge that each and every tetrad connection
can be expanded in spin harmonics of definite spin.
Subsequently, \S\ref{spin2gravitationalwaves},
it will be found that
spherical gauge proves to provide a particularly elegant gauge
for expressing the spin-$\pm 2$ wave equations
that govern the propagating components of gravitational waves.
Third,
the spin-weight of any object is given by the sum of the $+$'s and $-$'s
of its covariant angular indices in the Newman-Penrose formalism,
as already asserted by rule~(\ref{spinrule}).

To linear order in the vierbein perturbations
$\varphi_{mn}$,
the perturbed tetrad-frame connections $\Gamma_{kmn}$,
computed from the fundamental formula~(\ref{gammaGamma}),
are
(the boxed terms in the following equations
are terms that break algebraic resolution into spin-weighted harmonics,
as discussed below)
\begin{subequations}
\label{Gamma}
\begin{eqnarray}
\label{Gammauvv}
  \Gamma_{\overset{{\scriptstyle u v v}}{{\scriptstyle v u u}}}
  &=&
  {1 \over A}
  \left[
  h_\vu
  \bigl( 1 + \varphi_{\overset{{\scriptstyle v u}}{{\scriptstyle u v}}} \bigr)
  +
  A^2 \,
  \ncalD{\zero}_\vu \,
  A^{-2}
  \varphi_{\overset{{\scriptstyle u v}}{{\scriptstyle v u}}}
  -
  A \,
  \ncalD{\plusminus \one}_\uv \,
  A^{-1}
  \varphi_{\overset{{\scriptstyle v v}}{{\scriptstyle u u}}}
  \right]
\\
\label{Gammapvmp}
  \frac{1}{2}
  \Bigl(
  \Gamma_{\overset{\minus {\scriptstyle v} \plus}{\plus {\scriptstyle u} \minus}}
  +
  \Gamma_{\overset{\plus {\scriptstyle v} \minus}{\minus {\scriptstyle u} \plus}}
  \Bigr)
  &=&
  {1 \over 2 A}
  \left[
  2 \beta_\vu
  \bigl(
    1
    +
    \varphi_{\overset{{\scriptstyle v u}}{{\scriptstyle u v}}}
  \bigl)
  +
  2 \beta_\uv \,
  \varphi_{\overset{{\scriptstyle v v}}{{\scriptstyle u u}}}
  +
  A^2 \,
  \ncalD{\zero}_\vu \,
  A^{-2}
  \bigl(
    \varphi_{\overset{\plus \minus}{\minus \plus}}
    + \varphi_{\overset{\minus \plus}{\plus \minus}}
  \bigr)
  -
  \ncalD{\minusplus \one}_\plusminus \,
  \varphi_{\overset{{\scriptstyle v} \minus}{{\scriptstyle u} \plus}}
  -
  \ncalD{\plusminus \one}_\minusplus \,
  \varphi_{\overset{{\scriptstyle v} \plus}{{\scriptstyle u} \minus}}
  \right]
  \quad\quad\ 
\\
\label{Gammapvmm}
  \frac{1}{2}
  \Bigl(
  \Gamma_{\overset{\minus {\scriptstyle v} \plus}{\plus {\scriptstyle u} \minus}}
  -
  \Gamma_{\overset{\plus {\scriptstyle v} \minus}{\minus {\scriptstyle u} \plus}}
  \Bigr)
  &=&
  {1 \over 2 A}
  \left[
  \ncalD{\minusplus \one}_\plusminus
  \varphi_{\overset{\minus {\scriptstyle v}}{\plus {\scriptstyle u}}}
  -
  \ncalD{\plusminus \one}_\minusplus
  \varphi_{\overset{\plus {\scriptstyle v}}{\minus {\scriptstyle u}}}
  \right]
\\
\label{Gammampv}
  \Gamma_{\overset{\minus \plus {\scriptstyle v}}{\plus \minus {\scriptstyle u}}}
  -
  \frac{1}{2}
  \Bigl(
  \Gamma_{\overset{\minus {\scriptstyle v} \plus}{\plus {\scriptstyle u} \minus}}
  -
  \Gamma_{\overset{\plus {\scriptstyle v} \minus}{\minus {\scriptstyle u} \plus}}
  \Bigr)
  &=&
  {1 \over 2 A}
  \left[
  - \,
  A^2 \,
  \ncalD{\zero}_\vu \,
  A^{-2}
  \bigl(
    \varphi_{\overset{\plus \minus}{\minus \plus}}
    - \varphi_{\overset{\minus \plus}{\plus \minus}}
  \bigr)
  +
  \fbox{$
  \ncalD{\plusminus \one}_\plusminus \,
  \varphi_{\overset{{\scriptstyle v} \minus}{{\scriptstyle u} \plus}}
  $}
  -
  \fbox{$
  \ncalD{\minusplus \one}_\minusplus \,
  \varphi_{\overset{{\scriptstyle v} \plus}{{\scriptstyle u} \minus}}
  $}
 \right]
\\
\label{Gammampp}
  \Gamma_{\overset{\minus \plus \plus}{\plus \minus \minus}}
  &=&
  {1 \over A}
  \left[
  {\cot \theta \over \sqrt{2}}
  \bigl( 1 - \fbox{$
  \varphi_{\overset{\plus \minus}{\minus\plus}}
  $} \bigr)
  +
  \beta_\vu \,
  \varphi_{\overset{\plus {\scriptstyle u}}{\minus {\scriptstyle v}}}
  +
  \beta_\uv \,
  \varphi_{\overset{\plus {\scriptstyle v}}{\minus {\scriptstyle u}}}
  +
  \ncalD{\zero}_\plusminus \,
  \varphi_{\overset{\minus \plus}{\plus \minus}}
  -
  \fbox{$
  \ncalD{\plusminus \one}_\minusplus \,
  \varphi_{\overset{\plus \plus}{\minus \minus}}
  $}
  \right]
\\
\label{Gammapvup}
  \frac{1}{2}
  \Bigl(
  \Gamma_{\overset{\plus {\scriptstyle v} {\scriptstyle u}}{\minus {\scriptstyle u} {\scriptstyle v}}}
  +
  \Gamma_{\overset{\plus {\scriptstyle u} {\scriptstyle v}}{\minus {\scriptstyle v} {\scriptstyle u}}}
  \Bigr)
  &=&
  {1 \over 2 A}
  \left[
  \ncalD{\minusplus \one}_\vu \,
  \varphi_{\overset{\plus {\scriptstyle u}}{\minus {\scriptstyle v}}}
  +
  \ncalD{\plusminus \one}_\uv \,
  \varphi_{\overset{\plus {\scriptstyle v}}{\minus {\scriptstyle u}}}
  -
  \ncalD{\zero}_\plusminus \bigl(
    \varphi_{\overset{{\scriptstyle v u}}{{\scriptstyle u v}}}
    + \varphi_{\overset{{\scriptstyle u v}}{{\scriptstyle v u}}}
  \bigr)
  \right]
\\
\label{Gammapvum}
  \frac{1}{2}
  \Bigl(
  \Gamma_{\overset{\plus {\scriptstyle v} {\scriptstyle u}}{\minus {\scriptstyle u} {\scriptstyle v}}}
  -
  \Gamma_{\overset{\plus {\scriptstyle u} {\scriptstyle v}}{\minus {\scriptstyle v} {\scriptstyle u}}}
  \Bigr)
  &=&
  {1 \over 2 A}
  \left[
  A^2 \,
  \ncalD{\minusplus \one}_\vu \,
  A^{-2}
  \varphi_{\overset{{\scriptstyle u} \plus}{{\scriptstyle v} \minus}}
  -
  A^2 \,
  \ncalD{\plusminus \one}_\uv \,
  A^{-2}
  \varphi_{\overset{{\scriptstyle v} \plus}{{\scriptstyle u} \minus}}
  \right]
\\
\label{Gammauvp}
  \Gamma_{\overset{{\scriptstyle u v} \plus}{{\scriptstyle v u} \minus}}
  - \frac{1}{2}
  \Bigl(
  \Gamma_{\overset{\plus {\scriptstyle v} {\scriptstyle u}}{\minus {\scriptstyle u} {\scriptstyle v}}}
  -
  \Gamma_{\overset{\plus {\scriptstyle u} {\scriptstyle v}}{\minus {\scriptstyle v} {\scriptstyle u}}}
  \Bigr)
  &=&
  {1 \over 2 A}
  \left[
  A^2 \,
  \ncalD{\plusminus \one}_\vu \,
  A^{-2}
  \varphi_{\overset{\plus {\scriptstyle u}}{\minus {\scriptstyle v}}}
  -
  A^2 \,
  \ncalD{\minusplus \one}_\uv \,
  A^{-2}
  \varphi_{\overset{\plus {\scriptstyle v}}{\minus {\scriptstyle u}}}
  +
  \ncalD{\zero}_\plusminus \bigl(
    \varphi_{\overset{{\scriptstyle u v}}{{\scriptstyle v u}}}
    - \varphi_{\overset{{\scriptstyle v u}}{{\scriptstyle u v}}}
  \bigr)
  \right]
\\
\label{Gammapvv}
  \Gamma_{\overset{\plus {\scriptstyle v} {\scriptstyle v}}{\minus {\scriptstyle u} {\scriptstyle u}}}
  &=&
  {1 \over A}
  \left[
  A^2 \,
  \ncalD{\plusminus \one}_\vu \,
  A^{-2}
  \varphi_{\overset{\plus {\scriptstyle v}}{\minus {\scriptstyle u}}}
  -
  \ncalD{\zero}_\plusminus \,
  \varphi_{\overset{{\scriptstyle v v}}{{\scriptstyle u u}}}
  \right]
\\
\label{Gammamvv}
  \Gamma_{\overset{\minus {\scriptstyle v} {\scriptstyle v}}{\plus {\scriptstyle u} {\scriptstyle u}}}
  &=&
  {1 \over A}
  \left[
  A^2 \,
  \ncalD{\plusminus \one}_\vu \,
  A^{-2}
  \varphi_{\overset{\minus {\scriptstyle v}}{\plus {\scriptstyle u}}}
  -
  \ncalD{\zero}_\minusplus \,
  \varphi_{\overset{{\scriptstyle v v}}{{\scriptstyle u u}}}
  \right]
\\
\label{Gammapvp}
  \Gamma_{\overset{\plus {\scriptstyle v} \plus}{\minus {\scriptstyle u} \minus}}
  &=&
  {1 \over A}
  \left[
  A^2 \,
  \ncalD{\zero}_\vu \,
  A^{-2}
  \varphi_{\overset{\plus \plus}{\minus \minus}}
  -
  \ncalD{\plusminus \one}_\plusminus \,
  \varphi_{\overset{{\scriptstyle v} \plus}{{\scriptstyle u} \minus}}
  \right]
\\
\label{Gammamvm}
  \Gamma_{\overset{\minus {\scriptstyle v} \minus}{\plus {\scriptstyle u} \plus}}
  &=&
  {1 \over A}
  \left[
  A^2 \,
  \ncalD{\zero}_\vu \,
  A^{-2}
  \varphi_{\overset{\minus \minus}{\plus \plus}}
  -
  \ncalD{\minusplus \one}_\minusplus \,
  \varphi_{\overset{{\scriptstyle v} \minus}{{\scriptstyle u} \plus}}
  \right]
  \ .
\end{eqnarray}
\end{subequations}

In the unperturbed self-similar background,
the non-vanishing Newman-Penrose components of
the tetrad-frame connections $\Gamma_{kmn}$ reduce to
\begin{subequations}
\begin{eqnarray}
  \overset{\smallzero}{\Gamma}_{\overset{{\scriptstyle u v v}}{{\scriptstyle v u u}}}
  &=&
  A^{-1}
  h_\vu
  =
  {h_t \pm h_r \over A \sqrt{2}}
  \ ,
\\
  \overset{\smallzero}{\Gamma}_{\overset{\minus {\scriptstyle v} \plus}{\plus {\scriptstyle u} \minus}}
  =
  \overset{\smallzero}{\Gamma}_{\overset{\plus {\scriptstyle v} \minus}{\minus {\scriptstyle u} \plus}}
  &=&
  A^{-1}
  \beta_\vu
  =
  {\beta_t \pm \beta_r \over A \sqrt{2}}
  \ ,
\\
  \overset{\smallzero}{\Gamma}_{\overset{\minus \plus \plus}{\plus \minus \minus}}
  &=&
  {\cot\theta \over A \sqrt{2}}
  \ .
\end{eqnarray}
\end{subequations}

The tetrad connections $\Gamma_{kmn}$ are, by construction,
all coordinate gauge-invariant,
because they are defined, equation~(\ref{Gammadef}),
purely in terms of tetrad frame quantities,
independent of the choice of coordinates.
However,
for tetrad connections whose unperturbed part is finite
(non-zero),
the split between the unperturbed and perturbed parts
of $\Gamma_{kmn}$ is not coordinate gauge-invariant.
Moreover,
the tetrad connections $\Gamma_{kmn}$ are not
tetrad gauge-invariant.

It is a striking fact that
equations~(\ref{Gamma}) for the tetrad connections $\Gamma_{kmn}$
become algebraic upon resolution into spin-weighted harmonics
$\nY{s}$,
each perturbation $\varphi_{mn}$ and each connection $\Gamma_{kmn}$
being expanded in harmonics of appropriate spin $s$,
but only provided that certain gauge conditions are imposed,
namely that the perturbations
$\varphi_{m n}$ vanish whenever the right index $n$ is an angular index,
that is, $n = +$ or $-$:
\begin{equation}
\label{gaugespherical}
  \varphi_{\overset{{\scriptstyle v \plus}}{{\scriptstyle u \minus}}}
  =
  \varphi_{\overset{{\scriptstyle v \minus}}{{\scriptstyle u \plus}}}
  =
  \varphi_{\overset{{\scriptstyle \plus \plus}}{{\scriptstyle \minus \minus}}}
  =
  \varphi_{\overset{{\scriptstyle \plus \minus}}{{\scriptstyle \minus \plus}}}
  =
  0
  \ .
\end{equation}
These 8 gauge conditions can be termed ``spherical gauge'',
because physically
they correspond to choosing
a set of perturbed coordinate and tetrad frames
with respect to which
the spherical symmetry of the background is preserved to the greatest extent.
As argued immediately below, spherical gauge can be accomplished by
3 coordinate and 5 tetrad gauge transformations.


To see how the spherical gauge arises
mathematically from equations~(\ref{Gamma}),
consider for example
the perturbation $\varphi_{{\scriptstyle v} \plus}$,
which
appears operated on by
the angular operator
$\ncalD{\plus \one}_\minus$
in equation~(\ref{Gammapvmp}) for
$\frac{1}{2} \Bigl( \Gamma_{\overset{\minus {\scriptstyle v} \plus}{\plus {\scriptstyle u} \minus}} + \Gamma_{\overset{\plus {\scriptstyle v} \minus}{\minus {\scriptstyle u} \plus}} \Bigr)$,
and by
$\ncalD{\plus \one}_\plus$
in equation~(\ref{Gammapvp})
(the boxed term containing $\varphi_{{\scriptstyle v} \plus}$)
for
$\Gamma_{\overset{\plus {\scriptstyle v} \plus}{\minus {\scriptstyle u} \minus}}$,
but by
$\ncalD{\minus \one}_\minus$
in equation~(\ref{Gammampv}) for
$\Gamma_{\overset{\minus \plus {\scriptstyle v}}{\plus \minus {\scriptstyle u}}} - \frac{1}{2} \Bigl( \Gamma_{\overset{\minus {\scriptstyle v} \plus}{\plus {\scriptstyle u} \minus}} - \Gamma_{\overset{\plus {\scriptstyle v} \minus}{\minus {\scriptstyle u} \plus}} \Bigr)$.
Whereas $\ncalD{\plus \one}_\minus$ and $\ncalD{\plus \one}_\plus$ expect to operate on
a spin-harmonic
$\nY{s}$
of spin-weight $s = + 1$
(in accordance with the prefix $s$ on $\ncalD{s}_\minusplus$),
by contrast
$\ncalD{\minus \one}_\minus$ expects to operate on
a spin-harmonic of spin-weight $s = -1$.
This discrepancy in spin-weights can be accomodated
only if
$\varphi_{{\scriptstyle v} \plus}$
vanishes.
Similar discrepancies occur
between equations~(\ref{Gammapvmp}) or (\ref{Gammapvp}) and (\ref{Gammampv})
for
$\varphi_{{\scriptstyle v} \minus}$,
$\varphi_{{\scriptstyle u} \plus}$,
and
$\varphi_{{\scriptstyle u} \minus}$,
and again these can be accomodated
only if
all these perturbations vanish.
The 4 perturbations
$\varphi_{{\scriptstyle v} \plus}$,
$\varphi_{{\scriptstyle v} \minus}$,
$\varphi_{{\scriptstyle u} \plus}$,
and
$\varphi_{{\scriptstyle u} \minus}$,
can be made to vanish by 4 infinitesimal tetrad gauge transformations.
It is consistent to point the finger of blame
at the boxed terms
in equation~(\ref{Gammampv}) for
$\Gamma_{\overset{\minus \plus {\scriptstyle v}}{\plus \minus {\scriptstyle u}}} - \frac{1}{2} \Bigl( \Gamma_{\overset{\minus {\scriptstyle v} \plus}{\plus {\scriptstyle u} \minus}} - \Gamma_{\overset{\plus {\scriptstyle v} \minus}{\minus {\scriptstyle u} \plus}} \Bigr)$.
The general rule that emerges below,
equations~(\ref{phinY}) and (\ref{GammanY}),
is that the spin-weight $s$ of a quantity equals the sum
of the $+$'s and $-$'s of its covariant indices,
a rule already anticipated in equation~(\ref{spinrule}).
The boxed terms in equation~(\ref{Gammampv})
violate the rule.

Similar discrepancies occur for the perturbations
$\varphi_{\plus \plus}$, $\varphi_{\minus \minus}$,
$\varphi_{\plus \minus}$, and $\varphi_{\minus \plus}$.
For example,
$\varphi_{\plus \plus}$ appears operated on by
$\ncalD{\plus \one}_\minus$
in equation~(\ref{Gammampp})
(the boxed term containing
$\varphi_{\overset{{\scriptstyle \plus \plus}}{{\scriptstyle \minus \minus}}}$)
for
$\Gamma_{\minus \plus \plus}$,
which requires that $\varphi_{\plus \plus}$
be a harmonic of spin-weight $s = +1$,
whereas equation~(\ref{Gammapvp})
for $\Gamma_{\plus v \plus}$
[or alternatively equation~(\ref{Gammamvm}]
for
$\Gamma_{\plus u \plus}$]
requires that $\varphi_{\plus \plus}$
be a harmonic of spin-weight $s = +2$.
Consistency requires $\varphi_{\plus \plus}$,
and similarly $\varphi_{\minus \minus}$,
to vanish.
In the case of 
$\varphi_{\plus \minus}$ and $\varphi_{\minus \plus}$,
the $\cot\theta$ term multiplying these perturbations
in equation~(\ref{Gammampp})
(the boxed term containing
$\varphi_{\overset{{\scriptstyle \plus \minus}}{{\scriptstyle \minus \plus}}}$)
for
$\Gamma_{\overset{\minus \plus \plus}{\plus \minus \minus}}$
destroys algebraicity of the spherical transform
unless
$\varphi_{\plus \minus}$ and $\varphi_{\minus \plus}$ vanish.
Note that it is fine that
the unperturbed value of
$\Gamma_{\overset{\minus \plus \plus}{\plus \minus \minus}}$
be proportional to $\cot\theta$;
it is the perturbations, not the unperturbed values,
that are to be expanded in spin-weighted spherical harmonics.
Under an infinitesimal coordinate gauge transformation,
$\varphi_{\plus \plus}$
and
$\varphi_{\minus \minus}$
transform according to equation~(\ref{gaugephipp}),
and can thus be set to zero by a suitable choice of
$\epsilon^\minus$ and $\epsilon^\plus$.
Then
$\varphi_{\plus \minus} + \varphi_{\minus \plus}$,
which transforms, in accordance with equation~(\ref{gaugephipm}),
as
\begin{equation}
\label{gaugephipmp}
  \varphi_{\plus \minus}
  +
  \varphi_{\minus \plus}
  \rightarrow
  \varphi_{\plus \minus}
  +
  \varphi_{\minus \plus}
  -
  \Bigl(
  2 \, \beta_v \epsilon^v
  +
  2 \, \beta_u \epsilon^u
  +
  \ncalD{\plus \one}_\minus \, \epsilon^\minus
  +
  \ncalD{\minus \one}_\plus \, \epsilon^\plus
  \Bigr)
  \ ,
\end{equation}
can be set to zero by a suitable choice of
$\beta \cdot \epsilon \equiv \beta_v \epsilon^v + \beta_u \epsilon^u$.
Finally,
$\varphi_{\plus \minus} - \varphi_{\minus \plus}$
can be set to zero by an infinitesimal tetrad gauge transformation.

The conformal factor $A$ transforms
under an infinitesimal coordinate gauge transformation as
\begin{equation}
\label{gaugeA}
  A
  \rightarrow
  A + A \epsilon^m \partial_m A
  =
  A ( 1 + \beta \cdot \epsilon )
  \ .
\end{equation}
Thus a gauge choice of $\beta \cdot \epsilon$,
as accomplished by the vanishing of
$\varphi_{\plus \minus} + \varphi_{\minus \plus}$,
equation~(\ref{gaugephipmp}),
is equivalent to fixing the gauge of the conformal factor $A$.

Monopole and dipole harmonics,
$\nY{s}_{l m}$ with
$l = 0$ or $1$,
do not require special treatment.
For example,
equation~(\ref{gaugephipp})
constitutes an algebraic relation
between
the $l m$'th coefficient of
the expansion of
$\varphi_{\overset{\plus \plus}{\minus \minus}}$
in spin $\pm 2$ harmonics $\nY{\plusminus \two}_{l m}$,
and
the $l m$'th coefficient of
the expansion of
$\epsilon^\minusplus$
in spin $\pm 1$ harmonics $\nY{\plusminus \one}_{l m}$.
For the dipole harmonic, $l = 1$,
all terms in equation~(\ref{gaugephipp})
are identically zero,
since there is no spin $\pm 2$ dipole harmonic,
while
$\ncalD{\plusminus \one}_\plusminus$
yields zero when acting on any
spin $\pm 1$ dipole harmonic $\nY{\plusminus \one}_{1 m}$.
Thus for the dipole harmonic,
equation~(\ref{gaugephipp}) does not represent a
coordinate gauge freedom,
but rather it vanishes identically.

In spherical gauge,
two infinitesimal gauge freedoms remain,
1 coordinate freedom and 1 tetrad freedom.
The 1 coordinate freedom corresponds to an infinitesimal adjustment
of the conformal radius $r$
[the gauge of the conformal factor $A$ being fixed in spherical gauge,
in accordance with the comments following equation~(\ref{gaugeA})],
while the 1 tetrad freedom corresponds to an infinitesimal Lorentz boost
in the radial direction,
that is, to an adjustment of the antisymmetric part of $\varphi_{vu}$.

In spherical gauge,
equations~(\ref{gaugespherical}),
the tetrad-frame connections $\Gamma_{kmn}$, equations~(\ref{Gamma}), reduce to
\begin{subequations}
\label{linGamma}
\begin{eqnarray}
\label{linGammauvv}
  \Gamma_{\overset{{\scriptstyle u v v}}{{\scriptstyle v u u}}}
  &=&
  {1 \over A}
  \left[
  h_\vu
  \bigl( 1 + \varphi_{\overset{{\scriptstyle v u}}{{\scriptstyle u v}}} \bigr)
  +
  A^2 \,
  \ncalD{\zero}_\vu \,
  A^{-2}
  \varphi_{\overset{{\scriptstyle u v}}{{\scriptstyle v u}}}
  -
  \ncalD{\plusminus \one}_\uv \,
  \varphi_{\overset{{\scriptstyle v v}}{{\scriptstyle u u}}}
  \right]
\\
\label{linGammamvp}
  \frac{1}{2} \Bigl(
    \Gamma_{\overset{\minus {\scriptstyle v} \plus}{\plus {\scriptstyle u} \minus}}
    +
    \Gamma_{\overset{\plus {\scriptstyle v} \minus}{\minus {\scriptstyle u} \plus}}
  \Bigr)
  &=&
  {1 \over A}
  \left[
  \beta_\vu
  \bigl(
    1
    +
    \varphi_{\overset{{\scriptstyle v u}}{{\scriptstyle u v}}}
  \bigl)
  +
  \beta_\uv \,
  \varphi_{\overset{{\scriptstyle v v}}{{\scriptstyle u u}}}
  \right]
\\
\label{linGammampv}
  \Gamma_{\overset{\minus \plus {\scriptstyle v}}{\plus \minus {\scriptstyle u}}}
  \  = \ 
  \frac{1}{2} \Bigl(
    \Gamma_{\overset{\minus {\scriptstyle v} \plus}{\plus {\scriptstyle u} \minus}}
    -
    \Gamma_{\overset{\plus {\scriptstyle v} \minus}{\minus {\scriptstyle u} \plus}}
  \Bigr)
  &=&
  {1 \over 2 A}
  \left[
  \ncalD{\minusplus \one}_\plusminus \,
  \varphi_{\overset{\minus {\scriptstyle v}}{\plus {\scriptstyle u}}}
  -
  \ncalD{\plusminus \one}_\minusplus \,
  \varphi_{\overset{\plus {\scriptstyle v}}{\minus {\scriptstyle u}}}
  \right]
\\
\label{linGammampp}
  \Gamma_{\overset{\minus \plus \plus}{\plus \minus \minus}}
  &=&
  {1 \over A}
  \left[
  {\cot \theta \over \sqrt{2}}
  +
  \beta_\vu \,
  \varphi_{\overset{\plus {\scriptstyle u}}{\minus {\scriptstyle v}}}
  +
  \beta_\uv \,
  \varphi_{\overset{\plus {\scriptstyle v}}{\minus {\scriptstyle u}}}
  \right]
\\
\label{linGammapvu}
  \Gamma_{\overset{\plus {\scriptstyle v} {\scriptstyle u}}{\minus {\scriptstyle u} {\scriptstyle v}}}
  \  = \ 
  \Gamma_{\overset{\plus {\scriptstyle u} {\scriptstyle v}}{\minus {\scriptstyle v} {\scriptstyle u}}}
  &=&
  {1 \over 2 A}
  \left[
  \ncalD{\minusplus \one}_\vu \,
  \varphi_{\overset{\plus {\scriptstyle u}}{\minus {\scriptstyle v}}}
  +
  \ncalD{\plusminus \one}_\uv \,
  \varphi_{\overset{\plus {\scriptstyle v}}{\minus {\scriptstyle u}}}
  -
  \ncalD{\zero}_\plusminus \bigl(
    \varphi_{\overset{{\scriptstyle v u}}{{\scriptstyle u v}}}
    + \varphi_{\overset{{\scriptstyle u v}}{{\scriptstyle v u}}}
  \bigr)
  \right]
\\
\label{linGammauvp}
  \Gamma_{\overset{{\scriptstyle u v} \plus}{{\scriptstyle v u} \minus}}
  &=&
  {1 \over 2 A}
  \left[
  A^2 \,
  \ncalD{\plusminus \one}_\vu \,
  A^{-2}
  \varphi_{\overset{\plus {\scriptstyle u}}{\minus {\scriptstyle v}}}
  -
  A^2 \,
  \ncalD{\minusplus \one}_\uv \,
  A^{-2}
  \varphi_{\overset{\plus {\scriptstyle v}}{\minus {\scriptstyle u}}}
  +
  \ncalD{\zero}_\plusminus \bigl(
    \varphi_{\overset{{\scriptstyle u v}}{{\scriptstyle v u}}}
    - \varphi_{\overset{{\scriptstyle v u}}{{\scriptstyle u v}}}
  \bigr)
  \right]
\\
\label{linGammapvv}
  \Gamma_{\overset{\plus {\scriptstyle v} {\scriptstyle v}}{\minus {\scriptstyle u} {\scriptstyle u}}}
  &=&
  {1 \over A}
  \left[
  A^2 \,
  \ncalD{\plusminus \one}_\vu \,
  A^{-2}
  \varphi_{\overset{\plus {\scriptstyle v}}{\minus {\scriptstyle u}}}
  -
  \ncalD{\zero}_\plusminus \,
  \varphi_{\overset{{\scriptstyle v v}}{{\scriptstyle u u}}}
  \right]
\\
\label{linGammamvv}
  \Gamma_{\overset{\minus {\scriptstyle v} {\scriptstyle v}}{\plus {\scriptstyle u} {\scriptstyle u}}}
  &=&
  {1 \over A}
  \left[
  A^2 \,
  \ncalD{\plusminus \one}_\vu \,
  A^{-2}
  \varphi_{\overset{\minus {\scriptstyle v}}{\plus {\scriptstyle u}}}
  -
  \ncalD{\zero}_\minusplus \,
  \varphi_{\overset{{\scriptstyle v v}}{{\scriptstyle u u}}}
  \right]
\\
\label{linGammapvp}
  \Gamma_{\overset{\plus {\scriptstyle v} \plus}{\minus {\scriptstyle u} \minus}}
  &=&
  0
\\
\label{linGammamvm}
  \Gamma_{\overset{\minus {\scriptstyle v} \minus}{\plus {\scriptstyle u} \plus}}
  &=&
  0
  \ .
\end{eqnarray}
\end{subequations}
As intended,
these equations become algebraic upon spherical transform into spin-weighted harmonics,
with the perturbations $\varphi_{m n}$
and the perturbed connections $\overset{\smallone}{\Gamma}_{kmn}$
being proportional to spin-harmonics $\nY{s}$ as follows:
\begin{subequations}
\label{phinY}
\begin{eqnarray}
  \varphi_{\overset{{\scriptstyle v u}}{{\scriptstyle u v}}}
  \propto
  \varphi_{\overset{{\scriptstyle v v}}{{\scriptstyle u u}}}
  &\propto&
  \nY{\zero}
  \ ,
\\
  \varphi_{\overset{\plus {\scriptstyle v}}{\minus {\scriptstyle u}}}
  \propto
  \varphi_{\overset{\plus {\scriptstyle u}}{\minus {\scriptstyle v}}}
  &\propto&
  \nY{\plusminus \one}
  \ ,
\end{eqnarray}
\end{subequations}
and
\begin{subequations}
\label{GammanY}
\begin{eqnarray}
  \overset{\smallone}{\Gamma}_{\overset{{\scriptstyle u v v}}{{\scriptstyle v u u}}}
  \propto
  \overset{\smallone}{\Gamma}_{\overset{\minus {\scriptstyle v} \plus}{\plus {\scriptstyle u} \minus}}
  +
  \overset{\smallone}{\Gamma}_{\overset{\plus {\scriptstyle v} \minus}{\minus {\scriptstyle u} \plus}}
  \propto
  \Gamma_{\overset{\minus \plus {\scriptstyle v}}{\plus \minus {\scriptstyle u}}}
  &\propto&
  \nY{\zero}
  \ ,
\\
  \overset{\smallone}{\Gamma}_{\overset{\minus \plus \plus}{\plus \minus \minus}}
  \propto
  \Gamma_{\overset{\plus {\scriptstyle v} {\scriptstyle u}}{\minus {\scriptstyle u} {\scriptstyle v}}}
  \propto
  \Gamma_{\overset{{\scriptstyle u v} \plus}{{\scriptstyle v u} \minus}}
  \propto
  \Gamma_{\overset{\plus {\scriptstyle v} {\scriptstyle v}}{\minus {\scriptstyle u} {\scriptstyle u}}}
  \propto
  \Gamma_{\overset{\plus {\scriptstyle u} {\scriptstyle u}}{\minus {\scriptstyle v} {\scriptstyle v}}}
  &\propto&
  \nY{\plusminus \one}
  \ .
\end{eqnarray}
\end{subequations}
It is apparent
from equations~(\ref{phinY}) and (\ref{GammanY})
that the spin-weight $s$ of any term
is just equal to the sum of the $+$'s and $-$'s of its covariant indices,
in agreement with the previously asserted general rule~(\ref{spinrule}).

It is useful to record that
in spherical gauge
the directed derivatives of the conformal factor $A$ are,
to linear order,
\begin{subequations}
\label{dA}
\begin{eqnarray}
\label{dAa}
  \partial_\vu A
  &=&
  {\textstyle \frac{1}{2}}
  A
  \Bigl(
    \Gamma_{\overset{\minus {\scriptstyle v} \plus}{\plus {\scriptstyle u} \minus}}
    +
    \Gamma_{\overset{\plus {\scriptstyle v} \minus}{\minus {\scriptstyle u} \plus}}
  \Bigr)
  \ ,
\\
\label{dAb}
  \partial_\plusminus A
  &=&
  A \,
  \overset{\smallone}{\Gamma}_{\overset{\minus \plus \plus}{\plus \minus \minus}}
  \ .
\end{eqnarray}
\end{subequations}

\subsection{Perturbed Riemann tensor}
\label{perturbedriemann}

Much of this subsection is concerned with
demonstrating that the spherical gauge, equations~(\ref{gaugespherical}),
emerges a second time,
independent of the arguments of the previous subsection, \S\ref{perturbedconnections},
if it is required that
the linearized equations~(\ref{linnpR0})--(\ref{linnpR2})
for the tetrad components $R_{klmn}$ of the Riemann tensor
in terms of the tetrad connections $\Gamma_{kmn}$
become algebraic upon resolution into spin-weighted spherical harmonics.
Essential to the argument is that every tetrad connection
be individually expandable in spin harmonics.
In the next subsection, \S\ref{togauge},
it will be shown that indeed
it is the condition on the tetrad connections that imposes spherical gauge:
without the condition on the connections, no gauge conditions are imposed.

General expressions for the Newman-Penrose components
of the Riemann tensor
$R_{klmn}$
in terms of the tetrad connections
$\Gamma_{kmn}$
and their directed derivatives
are given
in the notation of the present paper
in the Appendix,
equations~(\ref{npR}).
To linear order, but in a general gauge,
not in spherical gauge,
the Newman-Penrose components of the Riemann tensor,
equations~(\ref{npR}),
reduce
in accordance with the linearization procedure described in the paragraph
containing equations~(\ref{linearize}) to
the expressions~(\ref{linnpR0})--(\ref{linnpR2})
given below.
For clarity,
the equations are subdivided into batches characterized
by the dependence of their perturbations on spin $s$
and on infinitesimal tetrad gauge (Lorentz) transformations.
As will be demonstrated below,
the spin $s$ of any component
$R_{klmn}$
of the Riemann tensor is equal
to the sum of the $+$'s and $-$'s of its covariant indices,
a statement that will be found in \S\ref{togauge}
to be true in any gauge, not just spherical gauge.
In the spherically symmetric unperturbed background,
all components of zero spin
except
$R_{v u \plus \minus}$
are are in general non-vanishing,
while components of non-zero spin necessarily vanish.

The boxed terms in
the expressions~(\ref{linnpR0})--(\ref{linnpR2})
are either (a) non-conforming terms that break algebraic resolution into spin harmonics
and hence lead to spherical gauge,
as discussed below,
or (b)
terms proportional to
$\Gamma_{\minus \plus {\scriptstyle v}}$
or
$\Gamma_{\plus \minus {\scriptstyle u}}$,
expressions for which in terms of vierbein potentials
were previously found,
equation~(\ref{Gammampv}),
to contain non-conforming terms.

Perturbations of the following 5 components of the Riemann tensor
are proportional to spin-$0$ harmonics $\nY{\zero}$,
and are tetrad gauge-invariant with respect to
all 6 arbitrary infinitesimal tetrad transformations.
Because of the Jacobi identity~(\ref{jacobi}),
the 5 components represent only 4 independent degrees of freedom.
The components
$G_{vu}$ and $G_{\plus \minus}$ of the Einstein tensor,
\S\ref{perturbedeinstein},
and
the spin-$0$ component
$\Cz_{\zero}$ of the complexified Weyl tensor,
\S\ref{perturbedweyl},
depend on these Riemann components:
\begin{subequations}
\label{linnpR0}
\begin{eqnarray}
\label{linnpR0a}
  R_{v u v u}
  &=&
  \bigl( \partial_{\scriptstyle u}
    - \Gamma_{{\scriptstyle v} {\scriptstyle u} {\scriptstyle u}}
  \bigr) \Gamma_{{\scriptstyle u} {\scriptstyle v} {\scriptstyle v}}
  + \bigl( \partial_{\scriptstyle v}
    - \Gamma_{{\scriptstyle u} {\scriptstyle v} {\scriptstyle v}}
  \bigr) \Gamma_{{\scriptstyle v} {\scriptstyle u} {\scriptstyle u}}
  \ ,
\\
\label{linnpR0b}
  R_{\plus \minus \plus \minus}
  &=&
  \fbox{$
  \bigl( \partial_\plus
    + \Gamma_{\minus \plus \plus}
  \bigr) \Gamma_{\plus \minus \minus}
  $}
  +
  \fbox{$
  \bigl( \partial_\minus
    + \Gamma_{\plus \minus \minus}
  \bigr) \Gamma_{\minus \plus \plus}
  $}
  -
  {\textstyle \frac{1}{2}} \bigl(
    \Gamma_{\minus {\scriptstyle v} \plus}
    +
    \Gamma_{\plus {\scriptstyle v} \minus}
  \bigr) \bigl(
    \Gamma_{\plus {\scriptstyle u} \minus}
    +
    \Gamma_{\minus {\scriptstyle u} \plus}
  \bigr)
  \ ,
\\
\label{linnpR0c}
  R_{v \plus u \minus}
  &=&
  - \, \bigl( \partial_{\scriptstyle v}
    - \Gamma_{{\scriptstyle u} {\scriptstyle v} {\scriptstyle v}}
    + \Gamma_{\minus {\scriptstyle v} \plus}
  \bigr)
  \Gamma_{\minus {\scriptstyle u} \plus}
  +
  A^{-1}
  \ncalD{\minus \one}_\plus \,
  \Gamma_{\minus {\scriptstyle u} {\scriptstyle v}}
  =
  - \, \bigl( \partial_{\scriptstyle u}
    - \Gamma_{{\scriptstyle v} {\scriptstyle u} {\scriptstyle u}}
    + \Gamma_{\plus {\scriptstyle u} \minus}
  \bigr)
  \Gamma_{\plus {\scriptstyle v} \minus}
  +
  A^{-1}
  \ncalD{\plus \one}_\minus \,
  \Gamma_{\plus {\scriptstyle v} {\scriptstyle u}}
  \ ,
\\
\label{linnpR0d}
  R_{v \minus u \plus}
  &=&
  - \, \bigl( \partial_{\scriptstyle v}
    - \Gamma_{{\scriptstyle u} {\scriptstyle v} {\scriptstyle v}}
    + \Gamma_{\plus {\scriptstyle v} \minus}
  \bigr)
  \Gamma_{\plus {\scriptstyle u} \minus}
  +
  A^{-1}
  \ncalD{\plus \one}_\minus \,
  \Gamma_{\plus {\scriptstyle u} {\scriptstyle v}}
  =
  - \, \bigl( \partial_{\scriptstyle u}
    - \Gamma_{{\scriptstyle v} {\scriptstyle u} {\scriptstyle u}}
    + \Gamma_{\minus {\scriptstyle u} \plus}
  \bigr)
  \Gamma_{\minus {\scriptstyle v} \plus}
  +
  A^{-1}
  \ncalD{\minus \one}_\plus \,
  \Gamma_{\minus {\scriptstyle v} {\scriptstyle u}}
  \ ,
\\
\label{linnpR0e}
  R_{v u \plus \minus}
  &=&
  \ncalD{\minus \one}_v \,
  A^{-1}
  \fbox{$
  \Gamma_{\plus \minus {\scriptstyle u}}
  $}
  +
  \ncalD{\plus \one}_u \,
  A^{-1}
  \fbox{$
  \Gamma_{\minus \plus {\scriptstyle v}}
  $}
  -
  \fbox{$
  \bigl(
    \Gamma_{\plus {\scriptstyle u} {\scriptstyle v}}
    -
    \Gamma_{\plus {\scriptstyle v} {\scriptstyle u}}
  \bigr)
  \Gamma_{\plus \minus \minus}
  $}
  -
  \fbox{$
  \bigl(
    \Gamma_{\minus {\scriptstyle v} {\scriptstyle u}}
    -
    \Gamma_{\minus {\scriptstyle u} {\scriptstyle v}}
  \bigr)
  \Gamma_{\minus \plus \plus}
  $}
\\
\nonumber
  &=&
  A^{-1}
  \ncalD{\minus \one}_\plus \,
  \Gamma_{{\scriptstyle v} {\scriptstyle u} \minus}
  +
  A^{-1}
  \ncalD{\plus \one}_\minus \,
  \Gamma_{{\scriptstyle u} {\scriptstyle v} \plus}
  -
  \bigl(
    \Gamma_{\minus {\scriptstyle v} \plus}
    -
    \Gamma_{\plus {\scriptstyle v} \minus}
  \bigr)
  \Gamma_{{\scriptstyle v} {\scriptstyle u} {\scriptstyle u}}
  -
  \bigl(
    \Gamma_{\plus {\scriptstyle u} \minus}
    -
    \Gamma_{\minus {\scriptstyle u} \plus}
  \bigr)
  \Gamma_{{\scriptstyle u} {\scriptstyle v} {\scriptstyle v}}
  +
  \Gamma_{\plus {\scriptstyle v} \minus} \,
  \Gamma_{\minus {\scriptstyle u} \plus}
  -
  \Gamma_{\minus {\scriptstyle v} \plus} \,
  \Gamma_{\plus {\scriptstyle u} \minus}
  \ .
\end{eqnarray}
\end{subequations}
The last three Riemann components
of equations~(\ref{linnpR0})
are related by
the only non-trivial occurrence of 
the Jacobi identity:
\begin{equation}
\label{jacobi}
  R_{v \plus \minus u}
  +
  R_{v \minus u \plus}
  +
  R_{v u \plus \minus}
  =
  0
  \ .
\end{equation}

Perturbations of the following 2 components of the Riemann tensor
are proportional to spin-$0$ harmonics $\nY{\zero}$,
but are not tetrad gauge-invariant,
since they vary under an infinitesimal tetrad transformation
associated with the antisymmetric part of $\varphi_{vu}$,
that is, under an infinitesimal Lorentz boost in the radial direction.
The components
$G_{vv}$ and $G_{uu}$
of the Einstein tensor are proportional to these Riemann components:
\begin{eqnarray}
  R_{\overset{{\scriptstyle v} \plus {\scriptstyle v} \minus}{{\scriptstyle u} \minus {\scriptstyle u} \plus}}
  &=&
  - \,
  \bigl( \partial_\vu
    + \Gamma_{\overset{{\scriptstyle u} {\scriptstyle v} {\scriptstyle v}}{{\scriptstyle v} {\scriptstyle u} {\scriptstyle u}}}
    + \Gamma_{\overset{\minus {\scriptstyle v} \plus}{\plus {\scriptstyle u} \minus}}
  \bigr) \Gamma_{\overset{\minus {\scriptstyle v} \plus}{\plus {\scriptstyle u} \minus}}
  +
  A^{-1}
  \ncalD{\minusplus \one}_\plusminus \,
  \Gamma_{\overset{\minus {\scriptstyle v} {\scriptstyle v}}{\plus {\scriptstyle u} {\scriptstyle u}}}
  =
  - \,
  \bigl( \partial_\vu
    + \Gamma_{\overset{{\scriptstyle u} {\scriptstyle v} {\scriptstyle v}}{{\scriptstyle v} {\scriptstyle u} {\scriptstyle u}}}
    + \Gamma_{\overset{\plus {\scriptstyle v} \minus}{\minus {\scriptstyle u} \plus}}
  \bigr) \Gamma_{\overset{\plus {\scriptstyle v} \minus}{\minus {\scriptstyle u} \plus}}
  +
  A^{-1}
  \ncalD{\plusminus \one}_\minusplus \,
  \Gamma_{\overset{\plus {\scriptstyle v} {\scriptstyle v}}{\minus {\scriptstyle u} {\scriptstyle u}}}
  \ .
  \quad
\end{eqnarray}

The following 8 components of the Riemann tensor
are proportional to spin-$\pm 1$ harmonics $\nY{\plusminus \one}$.
They are not tetrad gauge-invariant:
components proportional to $\nY{\plus \one}$
vary under transformation of the antisymmetric parts of
$\varphi_{v \plus}$ and $\varphi_{u \plus}$,
while
components proportional to $\nY{\minus \one}$
vary under transformation of the antisymmetric parts of
$\varphi_{v \minus}$ and $\varphi_{u \minus}$.
They are however tetrad gauge-invariant with respect to an infinitesimal radial Lorentz boost,
the one tetrad gauge freedom remaining in spherical gauge.
The components
$G_{v \plus}$,
$G_{v \minus}$,
$G_{u \plus}$,
and
$G_{u \minus}$
of the Einstein tensor,
and
the spin-$\pm 1$
components
$\Cz_{\plusminus \one}$
components of the complexified Weyl tensor,
depend on these Riemann components:
\begin{subequations}
\label{linnpR1}
\begin{eqnarray}
  R_{\overset{{\scriptstyle{v u v} \plus}}{{\scriptstyle{u v u} \minus}}}
  &=&
  - \,
  \ncalD{\zero}_\vu \,
  A^{-1}
  \Gamma_{\overset{\plus {\scriptstyle v} {\scriptstyle u}}{\minus {\scriptstyle u} {\scriptstyle v}}}
  +
  A^{-1}
  \ncalD{\plusminus \two}_\uv \,
  \Gamma_{\overset{\plus {\scriptstyle v} {\scriptstyle v}}{\minus {\scriptstyle u} {\scriptstyle u}}}
  +
  \Gamma_{\overset{\plus {\scriptstyle v} \minus}{\minus {\scriptstyle u} \plus}} \,
  \Gamma_{\overset{\plus {\scriptstyle u} {\scriptstyle v}}{\minus {\scriptstyle v} {\scriptstyle u}}}
\nonumber
\\
\label{linnpR1a}
  &=&
  - \,
  \ncalD{\zero}_\vu \,
  A^{-1}
  \Gamma_{\overset{{\scriptstyle u} {\scriptstyle v} \plus}{{\scriptstyle v} {\scriptstyle u} \minus}}
  +
  \bigl( \partial_\plusminus
    +
    \Gamma_{\overset{{\scriptstyle u} {\scriptstyle v} \plus}{{\scriptstyle v} {\scriptstyle u} \minus}}
    +
    \Gamma_{\overset{\plus {\scriptstyle u} {\scriptstyle v}}{\minus {\scriptstyle v} {\scriptstyle u}}}
  \bigr)
  \Gamma_{\overset{{\scriptstyle u} {\scriptstyle v} {\scriptstyle v}}{{\scriptstyle v} {\scriptstyle u} {\scriptstyle u}}}
  +
  \Gamma_{\overset{\minus {\scriptstyle v} \plus}{\plus {\scriptstyle u} \minus}} \,
  \Gamma_{\overset{\plus {\scriptstyle u} {\scriptstyle v}}{\minus {\scriptstyle v} {\scriptstyle u}}}
  -
  \bigl(
    \Gamma_{\overset{{\scriptstyle v} {\scriptstyle u} {\scriptstyle u}}{{\scriptstyle u} {\scriptstyle v} {\scriptstyle v}}}
    +
    \Gamma_{\overset{\minus {\scriptstyle u} \plus}{\plus {\scriptstyle v} \minus}}
  \bigr)
  \Gamma_{\overset{\plus {\scriptstyle v} {\scriptstyle v}}{\minus {\scriptstyle u} {\scriptstyle u}}}
  \ ,
\\
  R_{\overset{{\scriptstyle{u v u} \plus}}{{\scriptstyle{v u v} \minus}}}
  &=&
  - \,
  \ncalD{\zero}_\uv \,
  A^{-1}
  \Gamma_{\overset{\plus {\scriptstyle u} {\scriptstyle v}}{\minus {\scriptstyle v} {\scriptstyle u}}}
  +
  A^{-1}
  \ncalD{\plusminus \two}_\vu \,
  \Gamma_{\overset{\plus {\scriptstyle u} {\scriptstyle u}}{\minus {\scriptstyle v} {\scriptstyle v}}}
  +
  \Gamma_{\overset{\plus {\scriptstyle u} \minus}{\minus {\scriptstyle v} \plus}} \,
  \Gamma_{\overset{\plus {\scriptstyle v} {\scriptstyle u}}{\minus {\scriptstyle u} {\scriptstyle v}}}
\nonumber
\\
\label{linnpR1b}
  &=&
  - \,
  \ncalD{\zero}_\uv \,
  A^{-1}
  \Gamma_{\overset{{\scriptstyle v} {\scriptstyle u} \plus}{{\scriptstyle u} {\scriptstyle v} \minus}}
  +
  \bigl( \partial_\plusminus
    +
    \Gamma_{\overset{{\scriptstyle v} {\scriptstyle u} \plus}{{\scriptstyle u} {\scriptstyle v} \minus}}
    +
    \Gamma_{\overset{\plus {\scriptstyle v} {\scriptstyle u}}{\minus {\scriptstyle u} {\scriptstyle v}}}
  \bigr)
  \Gamma_{\overset{{\scriptstyle v} {\scriptstyle u} {\scriptstyle u}}{{\scriptstyle u} {\scriptstyle v} {\scriptstyle v}}}
  +
  \Gamma_{\overset{\minus {\scriptstyle u} \plus}{\plus {\scriptstyle v} \minus}} \,
  \Gamma_{\overset{\plus {\scriptstyle v} {\scriptstyle u}}{\minus {\scriptstyle u} {\scriptstyle v}}}
  -
  \bigl(
    \Gamma_{\overset{{\scriptstyle u} {\scriptstyle v} {\scriptstyle v}}{{\scriptstyle v} {\scriptstyle u} {\scriptstyle u}}}
    +
    \Gamma_{\overset{\minus {\scriptstyle v} \plus}{\plus {\scriptstyle u} \minus}}
  \bigr)
  \Gamma_{\overset{\plus {\scriptstyle u} {\scriptstyle u}}{\minus {\scriptstyle v} {\scriptstyle v}}}
  \ ,
\\
  R_{\overset{\plus \minus {\scriptstyle v} \plus}{\minus \plus {\scriptstyle u} \minus}}
  &=&
  - \,
  \bigl( \partial_\plusminus
    +
    \Gamma_{\overset{{\scriptstyle u} {\scriptstyle v} \plus}{{\scriptstyle v} {\scriptstyle u} \minus}}
    +
    \Gamma_{\overset{\plus {\scriptstyle v} {\scriptstyle u}}{\minus {\scriptstyle u} {\scriptstyle v}}}
  \bigr)
  \Gamma_{\overset{\plus {\scriptstyle v} \minus}{\minus {\scriptstyle u} \plus}}
  +
  A^{-1}
  \ncalD{\plusminus \two}_\minusplus \,
  \Gamma_{\overset{\plus {\scriptstyle v} \plus}{\minus {\scriptstyle u} \minus}}
  +
  \Gamma_{\overset{\minus {\scriptstyle v} \plus}{\plus {\scriptstyle u} \minus}} \,
  \Gamma_{\overset{\plus {\scriptstyle v} {\scriptstyle u}}{\minus {\scriptstyle u} {\scriptstyle v}}}
\nonumber
\\
\label{linnpR1c}
  &=&
  - \,
  \fbox{$
  \bigl( \partial_\vu
    -
    \Gamma_{\overset{\minus \plus {\scriptstyle v}}{\plus \minus {\scriptstyle u}}}
    +
    \Gamma_{\overset{\minus {\scriptstyle v} \plus}{\plus {\scriptstyle u} \minus}}
  \bigr)
  \Gamma_{\overset{\minus \plus \plus}{\plus \minus \minus}}
  $}
  +
  A^{-1}
  \ncalD{\zero}_\plusminus \,
  \fbox{$
  \Gamma_{\overset{\minus \plus {\scriptstyle v}}{\plus \minus {\scriptstyle u}}}
  $}
  +
  \Gamma_{\overset{\minus {\scriptstyle v} \plus}{\plus {\scriptstyle u} \minus}} \,
  \Gamma_{\overset{\plus {\scriptstyle u} {\scriptstyle v}}{\minus {\scriptstyle v} {\scriptstyle u}}}
  +
  \Gamma_{\overset{\minus {\scriptstyle u} \plus}{\plus {\scriptstyle v} \minus}} \,
  \Gamma_{\overset{\plus {\scriptstyle v} {\scriptstyle v}}{\minus {\scriptstyle u} {\scriptstyle u}}}
  +
  \fbox{$
  \Gamma_{\overset{\plus \minus \minus}{\minus \plus \plus}} \,
  \Gamma_{\overset{\plus {\scriptstyle v} \plus}{\minus {\scriptstyle u} \minus}}
  $}
  \ ,
\\
  R_{\overset{\plus \minus {\scriptstyle u} \plus}{\minus \plus {\scriptstyle v} \minus}}
  &=&
  - \,
  \bigl( \partial_\plusminus
    +
    \Gamma_{\overset{{\scriptstyle v} {\scriptstyle u} \plus}{{\scriptstyle u} {\scriptstyle v} \minus}}
    +
    \Gamma_{\overset{\plus {\scriptstyle u} {\scriptstyle v}}{\minus {\scriptstyle v} {\scriptstyle u}}}
  \bigr)
  \Gamma_{\overset{\plus {\scriptstyle u} \minus}{\minus {\scriptstyle v} \plus}}
  +
  A^{-1}
  \ncalD{\plusminus \two}_\minusplus \,
  \Gamma_{\overset{\plus {\scriptstyle u} \plus}{\minus {\scriptstyle v} \minus}}
  +
  \Gamma_{\overset{\minus {\scriptstyle u} \plus}{\plus {\scriptstyle v} \minus}} \,
  \Gamma_{\overset{\plus {\scriptstyle u} {\scriptstyle v}}{\minus {\scriptstyle v} {\scriptstyle u}}}
\nonumber
\\
\label{linnpR1d}
  &=&
  - \,
  \fbox{$
  \bigl( \partial_\uv
    -
    \Gamma_{\overset{\minus \plus {\scriptstyle u}}{\plus \minus {\scriptstyle v}}}
    +
    \Gamma_{\overset{\minus {\scriptstyle u} \plus}{\plus {\scriptstyle v} \minus}}
  \bigr)
  \Gamma_{\overset{\minus \plus \plus}{\plus \minus \minus}}
  $}
  +
  A^{-1}
  \ncalD{\zero}_\plusminus \,
  \fbox{$
  \Gamma_{\overset{\minus \plus {\scriptstyle u}}{\plus \minus {\scriptstyle v}}}
  $}
  +
  \Gamma_{\overset{\minus {\scriptstyle u} \plus}{\plus {\scriptstyle v} \minus}} \,
  \Gamma_{\overset{\plus {\scriptstyle v} {\scriptstyle u}}{\minus {\scriptstyle u} {\scriptstyle v}}}
  +
  \Gamma_{\overset{\minus {\scriptstyle v} \plus}{\plus {\scriptstyle u} \minus}} \,
  \Gamma_{\overset{\plus {\scriptstyle u} {\scriptstyle u}}{\minus {\scriptstyle v} {\scriptstyle v}}}
  +
  \fbox{$
  \Gamma_{\overset{\plus \minus \minus}{\minus \plus \plus}} \,
  \Gamma_{\overset{\plus {\scriptstyle u} \plus}{\minus {\scriptstyle v} \minus}}
  $}
  \ .
\end{eqnarray}
\end{subequations}

The following 6 components of the Riemann tensor
are proportional to spin-$\pm 2$ harmonics $\nY{\plusminus \two}$.
They are tetrad gauge-invariant with respect to
all 6 arbitrary infinitesimal tetrad transformations.
The components
$G_{\plus\plus}$ and $G_{\minus \minus}$ of the Einstein tensor,
and
the spin-$\pm 2$ components
$\Cz_{\plusminus \two}$ of the complexified Weyl tensor,
depend on these Riemann components:
\begin{subequations}
\label{linnpR2}
\begin{eqnarray}
  R_{\overset{{\scriptstyle v} \plus {\scriptstyle u} \plus}{{\scriptstyle u} \minus {\scriptstyle v} \minus}}
  &=&
  - \,
  A^{-1}
  \ncalD{\minusplus \one}_\vu \,
  \Gamma_{\overset{\plus {\scriptstyle u} \plus}{\minus {\scriptstyle v} \minus}}
  +
  A^{-1}
  \ncalD{\plusminus \one}_\plusminus \,
  \Gamma_{\overset{\plus {\scriptstyle u} {\scriptstyle v}}{\minus {\scriptstyle v} {\scriptstyle u}}}
\nonumber
\\
\label{linnpR2a}
  &=&
  - \,
  A^{-1}
  \ncalD{\plusminus \one}_\uv \,
  \Gamma_{\overset{\plus {\scriptstyle v} \plus}{\minus {\scriptstyle u} \minus}}
  +
  A^{-1}
  \ncalD{\plusminus \one}_\plusminus \,
  \Gamma_{\overset{\plus {\scriptstyle v} {\scriptstyle u}}{\minus {\scriptstyle u} {\scriptstyle v}}}
  \ ,
\\
\label{linnpR2b}
  R_{\overset{{\scriptstyle v} \plus {\scriptstyle v} \plus}{{\scriptstyle u} \minus {\scriptstyle u} \minus}}
  &=&
  - \,
  \ncalD{\plusminus \one}_\vu \,
  A^{-1}
  \Gamma_{\overset{\plus {\scriptstyle v} \plus}{\minus {\scriptstyle u} \minus}}
  +
  A^{-1}
  \ncalD{\plusminus \one}_\plusminus \,
  \Gamma_{\overset{\plus {\scriptstyle v} {\scriptstyle v}}{\minus {\scriptstyle u} {\scriptstyle u}}}
  \ ,
\\
\label{linnpR2c}
  R_{\overset{{\scriptstyle u} \plus {\scriptstyle u} \plus}{{\scriptstyle v} \minus {\scriptstyle v} \minus}}
  &=&
  - \,
  \ncalD{\plusminus \one}_\uv \,
  A^{-1}
  \Gamma_{\overset{\plus {\scriptstyle u} \plus}{\minus {\scriptstyle v} \minus}}
  +
  A^{-1}
  \ncalD{\plusminus \one}_\plusminus \,
  \Gamma_{\overset{\plus {\scriptstyle u} {\scriptstyle u}}{\minus {\scriptstyle v} {\scriptstyle v}}}
  \ .
\end{eqnarray}
\end{subequations}

In the unperturbed background,
the non-vanishing components of the tetrad frame Riemann tensor are
\begin{subequations}
\label{unperturbednpR}
\begin{eqnarray}
  \overset{\smallzero}{R}_{v u v u}
  &=&
  \ncalD{\minus \one}_v \,
  A^{-2}
  h_u
  +
  \ncalD{\plus \one}_u \,
  A^{-2}
  h_v
  \ ,
\\
\label{unperturbednpRb}
  \overset{\smallzero}{R}_{\plus \minus \plus \minus}
  &=&
  - \,
  2 \,
  A^{-2}
  M
  \ ,
\\
\label{unperturbednpRc}
  \overset{\smallzero}{R}_{v \plus u \minus}
  =
  \overset{\smallzero}{R}_{v \minus u \plus}
  &=&
  - \,
  A^{-1}
  \ncalD{\minus \one}_v \,
  A^{-1}
  \beta_u
  =
  - \,
  A^{-1}
  \ncalD{\plus \one}_u \,
  A^{-1}
  \beta_v
  \ ,
\\
\label{unperturbednpRd}
  \overset{\smallzero}{R}_{\overset{{\scriptstyle v} \plus {\scriptstyle v} \minus}{{\scriptstyle u} \minus {\scriptstyle u} \plus}}
  &=&
  - \,
  A \,
  \ncalD{\plusminus \one}_\vu \,
  A^{-3} \,
  \beta_\vu
  \ .
\end{eqnarray}
\end{subequations}
In terms of the dimensionless quantities
$R$, $P$, $P_\perp$, and $F$
defined by equations~(\ref{FRPPA})
and related
to the proper density, radial pressure, transverse pressure, and energy flux
through Einstein's equations~(\ref{einsteinRFPP}),
the unperturbed Riemann components are
\begin{subequations}
\label{unperturbednpRRFPP}
\begin{eqnarray}
  \overset{\smallzero}{R}_{v u v u}
  &=&
  ( R - P + 2 P_\perp - 2 M) / A^2
  \ ,
\\
\label{unperturbednpRRFPPb}
  \overset{\smallzero}{R}_{\plus \minus \plus \minus}
  &=&
  - 2 M / A^2
  \ ,
\\
\label{unperturbednpRRFPPc}
  \overset{\smallzero}{R}_{v \plus u \minus}
  =
  \overset{\smallzero}{R}_{v \minus u \plus}
  &=&
  -
  {\textstyle \frac{1}{2}}
  ( R - P - 2 M ) / A^2
  \ ,
\\
\label{unperturbednpRRFPPd}
  \overset{\smallzero}{R}_{\overset{{\scriptstyle v} \plus {\scriptstyle v} \minus}{{\scriptstyle u} \minus {\scriptstyle u} \plus}}
  &=&
  {\textstyle \frac{1}{2}}
  ( R + P \mp 2 F ) / A^2
  \ .
\end{eqnarray}
\end{subequations}

It is remarkable that
equations~(\ref{linnpR0})--(\ref{linnpR2})
for the perturbed Riemann tensor in terms of the tetrad connections
become algebraic upon resolution into spin-weighted harmonics,
each connection $\Gamma_{kmn}$ and each Riemann component $R_{klmn}$
being expanded in harmonics of appropriate spin $s$,
but only in a particular gauge,
and this is the same spherical gauge,
equations~(\ref{gaugespherical}),
as obtained in \S\ref{perturbedconnections}
from the requirement that the expressions~(\ref{Gamma})
for the connections in terms of the vierbein perturbations
become algebraic upon resolution into spin-weighted harmonics.

The argument
that equations~(\ref{linnpR0})--(\ref{linnpR2})
imply spherical gauge runs as follows.
First,
if a connection $\Gamma_{kmn}$ appears
operated on by a spin raising or lowering angular operator
$\ncalD{s}_\plusminus$
anywhere in equations~(\ref{linnpR0})--(\ref{linnpR2}),
then it follows immediately that
the connection $\Gamma_{kmn}$ must have spin $s$.
Inspection of equations~(\ref{linnpR0})--(\ref{linnpR2})
shows that all $16$ of the connections that vanish in the unperturbed background
do indeed appear somewhere acted on by an angular operator
$\ncalD{s}_\plusminus$,
and the spin $s$ of those $16$ connections then follows immediately.
As for the $8$ connections whose unperturbed values are finite (non-zero),
all $8$ of those connections appear
acted on by an angular directed derivative $\partial_\plusminus$
somewhere in equations~(\ref{linnpR0})--(\ref{linnpR2}).
In the case of the $6$ connections
whose unperturbed part is a conformal factor times a purely radial function,
that is,
$\Gamma_{vvu}$,
$\Gamma_{uvv}$,
$\Gamma_{\minus v \plus}$,
$\Gamma_{\plus v \minus}$,
$\Gamma_{\plus u \minus}$,
and
$\Gamma_{\minus u \plus}$
[every case except
$\Gamma_{\minus \plus \plus}$
and
$\Gamma_{\plus \minus \minus}$,
which will be considered below, equations~(\ref{DGammapm})],
these terms
can be split into parts depending
on the unperturbed connection
$\overset{\smallzero}{\Gamma}$
and its perturbation
$\overset{\smallone}{\Gamma}$
as follows
[an angular directed derivative $\partial_\plusminus$ acting on a first order quantity
can be replaced by $A^{-1} \ncalD{s}_\plusminus$ in accordance with the general rule
(\ref{linearizepm}),
and in every case in equations~(\ref{linnpR0})--(\ref{linnpR2})
where $\partial_\plusminus$ acts on a connection whose unperturbed value is finite,
the spin $s$ is zero]:
\begin{equation}
  \partial_\plusminus \,
  \Gamma
  =
  \bigl(
    \varphi_{\overset{\plus {\scriptstyle u}}{\minus {\scriptstyle v}}} \,
    \partial_\vu
    +
    \varphi_{\overset{\plus {\scriptstyle v}}{\minus {\scriptstyle u}}} \,
    \partial_\uv
  \bigr)
  \overset{\smallzero}{\Gamma}
  +
  A^{-1}
  \ncalD{\zero}_\plusminus \,
  \overset{\smallone}{\Gamma}
  \ .
\end{equation}
Here it is necessary to assume not only that
the perturbation
$\overset{\smallone}{\Gamma}$
has a definite spin $s$,
but also that each vierbein potential
$\varphi_{\plus v}$,
$\varphi_{\minus u}$,
$\varphi_{\minus v}$,
and
$\varphi_{\plus u}$
individually has definite spin $s$.
One way or another,
all $24$ tetrad connections appear acted on by
an angular operator somewhere in equations~(\ref{linnpR0})--(\ref{linnpR2}),
and the net consequence is that the spin $s$ of every connection
[with the possible exception of
$\Gamma_{\minus \plus \plus}$
and
$\Gamma_{\plus \minus \minus}$,
considered explicitly below, equations~(\ref{DGammapm})],
or of its perturbation if the unperturbed value is finite,
is equal to the sum of the $+$'s and $-$'s of its covariant indices,
in agreement with the earlier conclusion,
equations~(\ref{GammanY}).
The spin $s$ of each of the Riemann components $R_{klmn}$
then follows,
and again the result is that
the spin $s$ of the Riemann component,
or of its perturbation if the unperturbed value is finite,
is equal to the sum of the $+$'s and $-$'s of its covariant indices,
in agreement with the general rule~(\ref{spinrule}).
Radial directed derivatives $\partial_\vu$
acting on connections whose unperturbed values are finite
also appear variously in equations~(\ref{linnpR0})--(\ref{linnpR2}).
For the $6$ connections
whose unperturbed part is a conformal factor times a purely radial function,
the perturbations to the radial directed derivatives of these
can be split into parts depending
on the unperturbed connection
$\overset{\smallzero}{\Gamma}$
and its perturbation
$\overset{\smallone}{\Gamma}$
as follows:
\begin{equation}
  \overset{\smallone}{\overbrace{\partial_\vu \, \Gamma}}
  =
  \bigl(
    \varphi_{\overset{{\scriptstyle v u}}{{\scriptstyle u v}}} \,
    \partial_\vu
    +
    \varphi_{\overset{{\scriptstyle v v}}{{\scriptstyle u u}}} \,
    \partial_\uv
  \bigr)
  \overset{\smallzero}{\Gamma}
  +
  \partial_\vu \,
  \overset{\smallone}{\Gamma}
  \ .
\end{equation}
Here again it is necessary to assume that
the vierbein perturbations
$\varphi_{vu}$,
$\varphi_{uv}$,
$\varphi_{vv}$,
and
$\varphi_{uu}$
each have definite spins.
In combination with the constraints on
$\varphi_{\plus v}$,
$\varphi_{\minus u}$,
$\varphi_{\minus v}$,
and
$\varphi_{\plus u}$
one concludes that the spin of each vierbein perturbation
$\varphi_{mn}$ whose right index is $v$ or $u$
is equal to the sum of the $+$'s and $-$'s of its covariant indices,
again in agreement with equations~(\ref{phinY})
and in accordance with the general rule~(\ref{spinrule}).

So far it has been concluded that,
if it is assumed that (the perturbation of)
every tetrad connection $\Gamma_{kmn}$,
and every tetrad Riemann component $R_{klmn}$,
and in addition every vierbein potential $\varphi_{mn}$
whose right index $n$ is $v$ or $u$,
is expandable in spin harmonics $\nY{s}$
of definite spin $s$,
then equations~(\ref{linnpR0})--(\ref{linnpR2})
for the Riemann tensor in terms of the tetrad connections
require that the spin of each object
is equal to the sum of the $+$'s and $-$'s of its covariant indices.

The next step in establishing spherical gauge
is to identify in equations~(\ref{linnpR0})--(\ref{linnpR2})
non-conforming terms, those that do not have the correct spin.
These terms call attention to themselves by involving
$\Gamma_{\minus \plus \plus}$
or
$\Gamma_{\plus \minus \minus}$,
because the latter connections are proportional to
$\cot\theta$
in the unperturbed background,
and misfitting factors of $\cot\theta$
destroy algebraicity of the spherical transform into spin-weighted harmonics.
That is, $\cot\theta$ times a spin-weighted spherical harmonic $\nY{s}_{lm}$
is not another pure spin-weighted spherical harmonic.
Thus
the boxed terms
$\bigl( \Gamma_{\plus {\scriptstyle v} {\scriptstyle u}} - \Gamma_{\plus {\scriptstyle u} {\scriptstyle v}} \bigr) \Gamma_{\plus \minus \minus}$
and
$\bigl( \Gamma_{\minus {\scriptstyle u} {\scriptstyle v}} - \Gamma_{\minus {\scriptstyle v} {\scriptstyle u}} \bigr) \Gamma_{\minus \plus \plus}$
in equation~(\ref{linnpR0e})
require that
$\Gamma_{\plus {\scriptstyle v} {\scriptstyle u}} - \Gamma_{\plus {\scriptstyle u} {\scriptstyle v}}$
and
$\Gamma_{\minus {\scriptstyle u} {\scriptstyle v}} - \Gamma_{\minus {\scriptstyle v} {\scriptstyle u}}$
both be zero.
Similarly
the boxed terms
$\Gamma_{\overset{\plus \minus \minus}{\minus \plus \plus}} \, \Gamma_{\overset{\plus {\scriptstyle v} \plus}{\minus {\scriptstyle u} \minus}}$
in equation~(\ref{linnpR1c}),
and
$\Gamma_{\overset{\plus \minus \minus}{\minus \plus \plus}} \, \Gamma_{\overset{\plus {\scriptstyle u} \plus}{\minus {\scriptstyle v} \minus}}$
in equation~(\ref{linnpR1d}),
require that
$\Gamma_{\plus v \plus}$,
$\Gamma_{\minus u \minus}$,
$\Gamma_{\minus v \minus}$,
and
$\Gamma_{\plus u \plus}$
all be zero.
Other non-conforming terms in equations~(\ref{linnpR0})--(\ref{linnpR2}),
boxed for clarity,
are those where
$\Gamma_{\minus \plus \plus}$
or
$\Gamma_{\plus \minus \minus}$
is acted on by a directed derivative,
instances of which occur in equations~(\ref{linnpR0b}),
(\ref{linnpR1c}),
and
(\ref{linnpR1d}).
These terms are explicitly
\begin{subequations}
\label{DGammapm}
\begin{eqnarray}
\label{DGammapma}
  \bigl( \partial_\plusminus
    + \Gamma_{\overset{\minus \plus \plus}{\plus \minus \minus}}
  \bigr)
  \Gamma_{\overset{\plus \minus \minus}{\minus \plus \plus}}
  &=&
  {1 \over A^2}
  \Bigl[
  \frac{1}{2}
  \bigl(
    - \,
    1
    +
    \varphi_{\overset{\plus \plus}{\minus \minus}}
    +
    \varphi_{\overset{\plus \minus}{\minus \plus}}
  \bigr)
  +
  {\cot\theta \over \sqrt{2}}
  \bigl(
    \ncalD{\zero}_\plusminus \,
    \varphi_{\overset{\minus \plus}{\plus \minus}}
    -
    \ncalD{\zero}_\minusplus \,
    \varphi_{\overset{\plus \plus}{\minus \minus}}
  \bigr)
  \Bigr]
  +
  A^{-1}
  \ncalD{\minusplus \one}_\plusminus \,
  \overset{\smallone}{\Gamma}_{\overset{\plus \minus \minus}{\minus \plus \plus}}
  \ ,
\\
\label{DGammapmb}
  \bigl( \partial_\vu
    -
    \Gamma_{\overset{\minus \plus {\scriptstyle v}}{\plus \minus {\scriptstyle u}}}
    +
    \Gamma_{\overset{\minus {\scriptstyle v} \plus}{\plus {\scriptstyle u} \minus}}
  \bigr)
  \Gamma_{\overset{\minus \plus \plus}{\plus \minus \minus}}
  &=&
  {1 \over A^2}
  \Bigl[
  \frac{1}{2}
  \bigl(
    \varphi_{\overset{{\scriptstyle v} \plus}{{\scriptstyle u} \minus}}
    +
    \varphi_{\overset{{\scriptstyle v} \minus}{{\scriptstyle u} \plus}}
  \bigr)
  +
  {\cot\theta \over \sqrt{2}}
  \bigl(
    A^2 \,
    \ncalD{\zero}_\vu \,
    A^{-2} \,
    \varphi_{\overset{\plus \minus}{\minus \plus}}
    -
    \ncalD{\plusminus \one}_\plusminus \,
    \varphi_{\overset{{\scriptstyle v} \minus}{{\scriptstyle u} \plus}}
  \bigr)
  \Bigr]
  +
  \ncalD{\zero}_\vu \,
  A^{-1}
  \overset{\smallone}{\Gamma}_{\overset{\minus \plus \plus}{\plus \minus \minus}}
  \ ,
\\
\label{DGammapmc}
  \bigl( \partial_\uv
    -
    \Gamma_{\overset{\minus \plus {\scriptstyle u}}{\plus \minus {\scriptstyle v}}}
    +
    \Gamma_{\overset{\minus {\scriptstyle u} \plus}{\plus {\scriptstyle v} \minus}}
  \bigr)
  \Gamma_{\overset{\minus \plus \plus}{\plus \minus \minus}}
  &=&
  {1 \over A^2}
  \Bigl[
  \frac{1}{2}
  \bigl(
    \varphi_{\overset{{\scriptstyle u} \plus}{{\scriptstyle v} \minus}}
    +
    \varphi_{\overset{{\scriptstyle u} \minus}{{\scriptstyle v} \plus}}
  \bigr)
  +
  {\cot\theta \over \sqrt{2}}
  \bigl(
    A^2 \,
    \ncalD{\zero}_\uv \,
    A^{-2} \,
    \varphi_{\overset{\plus \minus}{\minus \plus}}
    -
    \ncalD{\plusminus \one}_\plusminus \,
    \varphi_{\overset{{\scriptstyle u} \minus}{{\scriptstyle v} \plus}}
  \bigr)
  \Bigr]
  +
  \ncalD{\zero}_\uv \,
  A^{-1}
  \overset{\smallone}{\Gamma}_{\overset{\minus \plus \plus}{\plus \minus \minus}}
  \ .
  \qquad\ \ 
\end{eqnarray}
\end{subequations}
Each of the terms multiplying $\cot\theta$ in equations~(\ref{DGammapm})
destroys algebraicity of the spherical transform into spin-weighted spherical harmonics,
and from these alone all of the spherical gauge conditions~(\ref{gaugespherical})
necessarily follow.
In spherical gauge,
$\Gamma_{\plus {\scriptstyle v} {\scriptstyle u}} - \Gamma_{\plus {\scriptstyle u} {\scriptstyle v}}$,
$\Gamma_{\minus {\scriptstyle u} {\scriptstyle v}} - \Gamma_{\minus {\scriptstyle v} {\scriptstyle u}}$,
$\Gamma_{\plus v \plus}$,
$\Gamma_{\minus u \minus}$,
$\Gamma_{\minus v \minus}$,
and
$\Gamma_{\plus u \plus}$
all vanish as required,
and
equations~(\ref{DGammapm}) reduce to
\begin{subequations}
\begin{eqnarray}
  \bigl( \partial_\plusminus
    + \Gamma_{\overset{\minus \plus \plus}{\plus \minus \minus}}
  \bigr)
  \Gamma_{\overset{\plus \minus \minus}{\minus \plus \plus}}
  &=&
  - \,
  {\textstyle {1 \over 2}}
  A^{-2}
  +
  A^{-1}
  \ncalD{\minusplus \one}_\plusminus \,
  \overset{\smallone}{\Gamma}_{\overset{\plus \minus \minus}{\minus \plus \plus}}
  \ ,
\\
  \bigl( \partial_\vu
    -
    \Gamma_{\overset{\minus \plus {\scriptstyle v}}{\plus \minus {\scriptstyle u}}}
    +
    \Gamma_{\overset{\minus {\scriptstyle v} \plus}{\plus {\scriptstyle u} \minus}}
  \bigr)
  \Gamma_{\overset{\minus \plus \plus}{\plus \minus \minus}}
  &=&
  \ncalD{\zero}_\vu \,
  A^{-1}
  \overset{\smallone}{\Gamma}_{\overset{\minus \plus \plus}{\plus \minus \minus}}
  \ ,
\\
  \bigl( \partial_\uv
    -
    \Gamma_{\overset{\minus \plus {\scriptstyle u}}{\plus \minus {\scriptstyle v}}}
    +
    \Gamma_{\overset{\minus {\scriptstyle u} \plus}{\plus {\scriptstyle v} \minus}}
  \bigr)
  \Gamma_{\overset{\minus \plus \plus}{\plus \minus \minus}}
  &=&
  \ncalD{\zero}_\uv \,
  A^{-1}
  \overset{\smallone}{\Gamma}_{\overset{\minus \plus \plus}{\plus \minus \minus}}
  \ ,
\end{eqnarray}
\end{subequations}
which now conform to being algebraic upon transform into spin-weighted harmonics,
provided that the spins of the perturbations
$\overset{\smallone}{\Gamma}_{\minus \plus \plus}$
and
$\overset{\smallone}{\Gamma}_{\plus \minus \minus}$
are $+1$ and $-1$ respectively,
in agreement with the general rule~(\ref{spinrule})
that the spin of an object
equals the sum of the $+$'s and $-$'s of its covariant indices.

This completes the demonstration that, as claimed,
equations~(\ref{linnpR0})--(\ref{linnpR2}) for the Riemann tensor in terms of the tetrad connections
become algebraic upon transformation into spin-weighted harmonics
in and only in spherical gauge, equations~(\ref{gaugespherical}).

\subsection{To gauge or not to gauge?}
\label{togauge}

Much of the previous two subsections,
\S\S\ref{perturbedconnections} and \ref{perturbedriemann},
was concerned with showing that a special gauge,
spherical gauge, equation~(\ref{gaugespherical}), emerges from demanding
either that every vierbein perturbation
$\varphi_{mn}$
and every tetrad connection
$\Gamma_{kmn}$
be individually expandable in spin-weighted harmonics,
or alternatively that
every tetrad connection
$\Gamma_{kmn}$
and every tetrad Riemann component
$R_{klmn}$
be individually expandable in spin-weighted harmonics.

But what if instead
one merely requires that each vierbein perturbation
$\varphi_{mn}$
and tetrad Riemann component
$R_{klmn}$
(but not necessarily each tetrad connection)
be individually expandable in spin-weighted harmonics?
As will now be shown,
this requirement is satisfied with no gauge conditions at all.

Inspection of equations~(\ref{Gamma})
for the tetrad connections in terms of the vierbein perturbations
shows that only 4 of the connections
contain non-conforming terms (boxed terms)
that break algebraic resolution into spin harmonics,
namely
$\Gamma_{\overset{\minus \plus {\scriptstyle v}}{\plus \minus {\scriptstyle u}}}$,
equation~(\ref{Gammampv}),
and
$\Gamma_{\overset{\minus \plus \plus}{\plus \minus \minus}}$,
equation~(\ref{Gammampp}).
In effect,
it is the requirement that these 4 connections
be expandable in spin harmonics that
imposes the spherical gauge conditions~(\ref{gaugespherical}).
The offending 4 tetrad connections
appear in just 6 components of the Riemann tensor,
the boxed terms in the expressions for
$R_{\plus \minus \plus \minus}$,
$R_{v u \plus \minus}$,
$R_{\overset{\plus \minus {\scriptstyle v} \plus}{\minus \plus {\scriptstyle u} \minus}}$,
and
$R_{\overset{\plus \minus {\scriptstyle u} \plus}{\minus \plus {\scriptstyle v} \minus}}$,
equations~(\ref{linnpR0b}),
(\ref{linnpR0e}),
(\ref{linnpR1c}),
and
(\ref{linnpR1d}).
However,
5 of these 6 Riemann components
have alternative expressions
in terms of tetrad connections and their derivatives,
and in each case the alternative expression
contains only conforming terms
(no boxed terms).
Since the alternative expressions in each case
must yield the same expression in terms of vierbein perturbations,
it follows that the more fundamental expressions
for the 5 Riemann components in terms of vierbein perturbations
contain only conforming terms.
This leaves only 1 potentially non-conforming Riemann component,
$R_{\plus \minus \plus \minus}$.
The non-conforming (boxed) terms in this component,
equation~(\ref{linnpR0b}),
are expanded in equation~(\ref{DGammapma}),
but the two parts combine to give
\begin{eqnarray}
\label{Nlightbulb}
  &&
  \bigl( \partial_\plus 
    +
    \Gamma_{\minus \plus \plus}
  \bigr)
  \Gamma_{\plus \minus \minus}
  +
  \bigl( \partial_\minus
    +
    \Gamma_{\plus \minus \minus}
  \bigr)
  \Gamma_{\minus \plus \plus}
  =
  A^{-2}
  \Bigl[
  - \,
  1
  +
  \varphi_{\plus \minus}
  +
  \varphi_{\minus \plus}
  +
  \ncalD{\plus \one}_\minus \,
  \ncalD{\zero}_\plus
  (
    \varphi_{\plus \minus}
    +
    \varphi_{\minus \plus}
  )
\nonumber
\\
  &&
  \quad
  + \,
  \ncalD{\plus \one}_\minus
  (
    \beta_u \,
    \varphi_{\plus v}
    +
    \beta_v \,
    \varphi_{\plus u}
  )
  +
  \ncalD{\minus \one}_\plus \,
  (
    \beta_u \,
    \varphi_{\minus v}
    +
    \beta_v \,
    \varphi_{\minus u}
  )
  -
  \ncalD{\plus \one}_\minus \,
  \ncalD{\plus \two}_\minus \,
  \varphi_{\plus \plus}
  -
  \ncalD{\minus \one}_\plus \,
  \ncalD{\minus \two}_\plus \,
  \varphi_{\minus \minus}
  \Bigr]
\end{eqnarray}
all terms of which now conform.
[For later reference,
the left hand side of equation~(\ref{Nlightbulb}) equals
$2 A^{-2} \gravN$
where $\gravN$ is defined by equation~(\ref{gravN})].

Thus
{\em no\/} gauge condition is imposed
if it is required only that each vierbein connection
$\varphi_{mn}$
and each tetrad Riemann component
$R_{klmn}$
be expandable in spin harmonics
(without any conditions on the tetrad connections).

The statement of the previous paragraph can be put another way.
If the vierbein perturbations
$\varphi_{mn}$
are expanded in spin-weighted harmonics
with spin given by the general rule~(\ref{spinrule}),
then it automatically follows that each Riemann component
$R_{klmn}$
will be a sum of spin-weighted harmonics
with spin given by the general rule~(\ref{spinrule}).

Since it is now evident that
the spherical gauge conditions~(\ref{gaugespherical})
follow precisely from requiring that the tetrad connections
be expandable in spin harmonics,
the question arises as to whether or not it is necessary
to impose these conditions in order to carry through the wave equations
in the Newman-Penrose formalism.
Recall that only 4 of the tetrad connections were non-conforming in the first place,
namely
$\Gamma_{\overset{\minus \plus {\scriptstyle v}}{\plus \minus {\scriptstyle u}}}$,
equation~(\ref{Gammampv}),
and
$\Gamma_{\overset{\minus \plus \plus}{\plus \minus \minus}}$,
equation~(\ref{Gammampp}).
Cunningly,
it turns out that
the linearized equations governing
gravitational, electromagnetic, and scalar waves
conspire to
involve all of the tetrad connections
{\em except\/}
the 4 non-conforming tetrad connections!
Notably,
the expressions~(\ref{linJz})
for the Weyl currents
$\Jz_{lmn}$
involve all but the 4 non-conforming tetrad connections.
The 4 non-conforming tetrad connections
seemingly appear in the
expressions~(\ref{Deltas}),
(\ref{Deltasprime}),
(\ref{frakDs}),
and
(\ref{frakDsprime})
for the differential operators
$\nDelta{s}_m$,
$\nDelta{s}{}^\prime_m$,
$\nfrakD{s}_m$,
and
$\nfrakD{s}{}^\prime_m$,
but everywhere that these operators actually
occur in the relevant linearized equations,
namely
equations~(\ref{Weyls}),
(\ref{waveWeyl}),
(\ref{Maxwells}),
(\ref{waveMaxwell}),
and
(\ref{wavescalar}),
the spin $s$ of the differential operator
is such that the contribution of non-conforming tetrad connections
vanishes.

Thus it can be concluded that
the wave equations in the Newman-Penrose formalism
can be expanded in spin-weighted harmonics,
and that this can be done without any gauge conditions.
Put another way,
one is free to choose any gauge that is convenient.

\subsection{Perturbed Einstein tensor}
\label{perturbedeinstein}

In a Newman-Penrose tetrad,
the Einstein tensor $G_{mn}$
is related to the Riemann tensor $R_{klmn}$ by
\begin{subequations}
\label{npEinstein}
\begin{eqnarray}
\label{npEinsteina}
  G_{\scriptstyle v u}
  &=&
  - \left(
  R_{\plus \minus \plus \minus}
  +
  R_{v \plus u \minus}
  +
  R_{v \minus u \plus}
  \right)
  \ ,
\\
\label{npEinsteinb}
  G_{\plus \minus}
  &=&
  R_{v u v u}
  +
  R_{v \plus u \minus}
  +
  R_{v \minus u \plus}
  \ ,
\\
\label{npEinsteinc}
  G_{\overset{{\scriptstyle v v}}{{\scriptstyle u u}}}
  &=&
  2 \,
  R_{\overset{{\scriptstyle v} \plus {\scriptstyle v} \minus}{{\scriptstyle u} \minus {\scriptstyle u} \plus}}
  \ ,
\\
\label{npEinsteind}
  G_{\overset{{\scriptstyle v} \plus}{{\scriptstyle u} \minus}}
  &=&
  R_{\overset{{\scriptstyle v u} {\scriptstyle v} \plus}{{\scriptstyle u v} {\scriptstyle u} \minus}}
  +
  R_{\overset{\plus \minus {\scriptstyle v} \plus}{\minus \plus {\scriptstyle u} \minus}}
  \ ,
\\
\label{npEinsteine}
  G_{\overset{{\scriptstyle v} \minus}{{\scriptstyle u} \plus}}
  &=&
  R_{\overset{{\scriptstyle v u} {\scriptstyle v} \minus}{{\scriptstyle u v} {\scriptstyle u} \plus}}
  +
  R_{\overset{\minus \plus {\scriptstyle v} \minus}{\plus \minus {\scriptstyle u} \plus}}
  \ ,
\\
\label{npEinsteinf}
  G_{\overset{\plus \plus}{\minus \minus}}
  &=&
  - \,
  2 \,
  R_{\overset{{\scriptstyle v} \plus {\scriptstyle u} \plus}{{\scriptstyle u} \minus {\scriptstyle v} \minus}}
  \ .
\end{eqnarray}
\end{subequations}
Because of the variety of expressions for the individual Riemann components,
equations~(\ref{linnpR0})--(\ref{linnpR2}),
there are several alternative ways to express each Einstein component
in terms of tetrad connections and their derivatives
(though only one way to write each of $G_{\plus \plus}$ and $G_{\minus \minus}$),
and there is no pressing reason to write out these expressions.

In the unperturbed self-similar background,
the non-vanishing Newman-Penrose components of the Einstein tensor $G_{mn}$ are,
in terms of the dimensionless quantities
$R$, $P$, $P_\perp$, and $F$
defined by equations~(\ref{FRPPA}),
\begin{subequations}
\label{npeinstein}
\begin{eqnarray}
  \overset{\smallzero}{G}_{vu}
  &=&
  ( R - P ) / A^2
  \ ,
\\
  \overset{\smallzero}{G}_{\plus\minus}
  &=&
  2 P_\perp / A^2
  \ ,
\\
  \overset{\smallzero}{G}_{vv}
  &=&
  ( R + P - 2 F ) / A^2
  \ ,
\\
  \overset{\smallzero}{G}_{uu}
  &=&
  ( R + P + 2 F ) / A^2
  \ .
\end{eqnarray}
\end{subequations}
It is convenient also to define an auxiliary dimensionless unperturbed quantity $E$ by
\begin{equation}
\label{E}
  E
  \equiv
  {\textstyle \frac{1}{2}}
  A^2
  ( \overset{\smallzero}{G}_{vu} + \overset{\smallzero}{G}_{\plus\minus} )
  =
  {\textstyle \frac{1}{2}}
  ( R - P + 2 P_\perp )
  \ ,
\end{equation}
since this quantity occurs several times in the wave equations,
notably equations~(\ref{waveCz}) and (\ref{waveCza}).

It follows from equations~(\ref{npEinstein})
that the perturbation of each component of the tetrad-frame Einstein tensor
follows the Riemann tensor
in conforming to the general rule~(\ref{spinrule})
that its spin-weight $s$
equals the sum of the $+$'s and $-$'s of its covariant indices:
\begin{subequations}
\label{GnY}
\begin{eqnarray}
  \overset{\smallone}{G}_{\overset{{\scriptstyle v u}}{\plus \minus}}
  \propto
  \overset{\smallone}{G}_{\overset{{\scriptstyle v v}}{{\scriptstyle u u}}}
  &\propto&
  \nY{\zero}
  \ ,
\\
  G_{\overset{{\scriptstyle v} \plus}{{\scriptstyle u} \minus}}
  \propto
  G_{\overset{{\scriptstyle u} \plus}{{\scriptstyle v} \minus}}
  &\propto&
  \nY{\plusminus \one}
  \ ,
\\
  G_{\overset{\plus \plus}{\minus \minus}}
  &\propto&
  \nY{\plusminus \two}
  \ .
\end{eqnarray}
\end{subequations}

Being tetrad-frame quantities,
all the components $G_{mn}$ of the tetrad-frame Einstein tensor
are automatically coordinate gauge-invariant,
although the split between the unperturbed and perturbed parts
of components whose unperturbed parts are non-zero,
namely
$G_{vu}$,
$G_{\plus\minus}$,
$G_{vv}$,
and
$G_{uu}$,
is not coordinate gauge-invariant.

The components
$G_{uv}$, $G_{\plus \minus}$,
$G_{\plus \plus}$,
and
$G_{\minus \minus}$
are tetrad gauge-invariant
with respect to all 6 arbitrary tetrad (Lorentz) transformations.
The remaining components
are not tetrad gauge-invariant.
The components
$G_{vv}$ and $G_{uu}$
vary under transformation of the antisymmetric part of
$\varphi_{v u}$,
that is,
under an infinitesimal radial Lorentz boost,
the one tetrad gauge freedom left in spherical gauge.
The components
$G_{v \plus}$
and
$G_{u \plus}$
each vary under transformation of the antisymmetric parts of
$\varphi_{v \plus}$ and $\varphi_{u \plus}$,
while
$G_{v \minus}$
and
$G_{u \minus}$
each vary under transformation of the antisymmetric parts of
$\varphi_{v \minus}$ and $\varphi_{u \minus}$.
The following quantities are tetrad gauge-invariant,
but not coordinate gauge-invariant:
\begin{subequations}
\label{G1tet}
\begin{eqnarray}
  &&
  G_{\overset{{\scriptstyle v v}}{{\scriptstyle u u}}}
  \bigl(
    1 + \varphi_{\overset{{\scriptstyle u v}}{{\scriptstyle v u}}} - \varphi_{\overset{{\scriptstyle v u}}{{\scriptstyle u v}}}
  \bigr)
  \ ,
\\
  &&
  G_{\overset{{\scriptstyle v} \plus}{{\scriptstyle u} \minus}}
  +
  \textstyle{\frac{1}{2}}
  \bigl(
    \varphi_{\overset{{\scriptstyle v} \plus}{{\scriptstyle u} \minus}}
    -
    \varphi_{\overset{\plus {\scriptstyle v}}{\minus {\scriptstyle u}}}
  \bigr)
  ( G_{vu} + G_{\plus\minus} )
  +
  \textstyle{\frac{1}{2}}
  \bigl(
    \varphi_{\overset{{\scriptstyle u} \plus}{{\scriptstyle v} \minus}}
    -
    \varphi_{\overset{\plus {\scriptstyle u}}{\minus {\scriptstyle v}}}
  \bigr)
  G_{\overset{{\scriptstyle v v}}{{\scriptstyle u u}}}
  \ ,
\\
  &&
  G_{\overset{{\scriptstyle v} \minus}{{\scriptstyle u} \plus}}
  +
  \textstyle{\frac{1}{2}}
  \bigl(
    \varphi_{\overset{{\scriptstyle v} \minus}{{\scriptstyle u} \plus}}
    -
    \varphi_{\overset{\minus {\scriptstyle v}}{\plus {\scriptstyle u}}}
  \bigr)
  ( G_{vu} + G_{\plus\minus} )
  +
  \textstyle{\frac{1}{2}}
  \bigl(
    \varphi_{\overset{{\scriptstyle u} \minus}{{\scriptstyle v} \plus}}
    -
    \varphi_{\overset{\minus {\scriptstyle u}}{\plus {\scriptstyle v}}}
  \bigr)
  G_{\overset{{\scriptstyle v v}}{{\scriptstyle u u}}}
  \ .
\end{eqnarray}
\end{subequations}
Note that the product
$G_{vv} G_{uu}$
is both coordinate and tetrad gauge-invariant.

There are just two components of the Einstein tensor,
the spin-$\pm 2$ components
$G_{\plus \plus}$
and
$G_{\minus \minus}$,
that are both coordinate and tetrad gauge-invariant,
and also vanishing in the background,
and therefore constitute
physical perturbations
that are unchanged by any gauge transformation.

In the limiting case of
the Reissner-Nordstr\"om geometry,
the only components of the Einstein tensor that are non-vanishing
in the unperturbed background are
$G_{vu} = G_{\plus \minus}$.
In this case the
components
$G_{vv}$ and $G_{uu}$
are also coordinate and tetrad gauge-invariant
and vanishing in the background.

Finally,
in the case of vanishing background energy-momentum tensor,
the Schwarzschild geometry,
all perturbations of the Einstein tensor
are coordinate and tetrad gauge-invariant.

\section{Gravitational waves}
\label{gravitationalwaves}

This section derives the equations that describe gravitational waves
in perturbed self-similar black hole spacetimes.
Subsection~\ref{perturbedweyl}
characterizes perturbations of the Weyl tensor,
\S\ref{evolutionweyl} presents the equations that govern the evolution of those perturbations,
and subsequent subsections
apply those equations to derive gravitational wave equations,
starting with \S\ref{spin2gravitationalwaves},
which presents the wave equations governing the propagating (spin-$\pm 2$)
components of gravitational waves.
Subsection~\ref{spinsgravitationalwaves}
discusses how the Weyl evolution equations
relate different spin components of gravitational waves to each other,
so that the different spin components fluctuate in harmony rather than independently.
Subsections~\ref{monopolegravitationalwaves} and \ref{dipolegravitationalwaves}
focus on the special cases of monopole and dipole modes.
In the final two subsections~\ref{axialgravitationalwaves} and \ref{polargravitationalwaves}
of this section,
the spin-$0$ wave equation in the Newman-Penrose formalism is used
to derive axial and polar spin-$0$ wave equations.

\subsection{Perturbed Weyl tensor}
\label{perturbedweyl}

Gravitational waves are most naturally described by the Weyl tensor
$C_{klmn}$,
the totally trace-free,
or tidal,
part of the Riemann tensor
$R_{klmn}$,
equation~(\ref{Weyl}).
Like the Riemann tensor,
the Weyl tensor is a symmetric bivector matrix,
that is, a symmetric matrix of antisymmetric tensors.

As is familiar from the geometric algebra~\cite{DL03},
a bivector is a 6-component object
which, in an orthonormal frame,
has a natural complex structure
(indeed, a natural complex quaternionic structure;
but the quaternionic structure goes beyond what is needed for the present paper).
The real, or ``electric'' ($E$),  part of the bivector
changes sign under spatial inversion,
while the imaginary, or ``magnetic'' ($B$), part
does not change sign under spatial inversion.
In the geometric algebra,
it is natural to treat a bivector as
a single complex object $E + \im B$.

In an orthonormal tetrad,
the Weyl tensor,
being a symmetric matrix of bivectors,
can be organized as a
$2 \times 2$ matrix of $3 \times 3$ blocks, with the structure
\begin{equation}
\label{blockweyl}
  \left(
  \begin{array}{cc}
  EE & EB \\
  EB & BB
  \end{array}
  \right)
\end{equation}
where
$EE$, $EB$, and $BB$ signify $3 \times 3$ matrices
with components of the indicated type,
$E$ for electric, and $B$ for magnetic.
In view of the bivector structure,
it is natural to define a complexified version
$\Cz_{klmn}$
of the Weyl tensor by
\begin{equation}
\label{Cz}
  \Cz_{klmn}
  \equiv
  \frac{1}{4}
  \Bigl(
  \delta^p_k \delta^q_l + \frac{\im}{2} {\varepsilon_{kl}}^{pq}
  \Bigr)
  \Bigl(
  \delta^r_m \delta^s_n + \frac{\im}{2} {\varepsilon_{mn}}^{rs}
  \Bigr)
  C_{pqrs}
  \ ,
\end{equation}
where
$\varepsilon_{klpq}$
is the
totally antisymmetric tensor,
here normalized so that
$\varepsilon_{tr\theta\phi} = -1$.
The overall factor of $1/4$ on the right hand side of equation~(\ref{Cz})
is introduced because then the operator
$P_{kl}^{pq} \equiv \frac{1}{2} \left( \delta_k^p \delta_l^q + \frac{\im}{2} {\varepsilon_{kl}}^{pq} \right)$
is a projection operator,
satisfying
$P^2 = P$.
The complexified Weyl tensor $\Cz_{klmn}$ is,
like the Weyl tensor, totally traceless.
If the Weyl tensor
$C_{klmn}$
is organized according to the structure~(\ref{blockweyl}),
then the complexified Weyl tensor
$\Cz_{klmn}$
has the corresponding structure
\begin{equation}
  \frac{1}{4}
  \left(
  \begin{array}{cc}
  1 & \im \\
  \im & -1
  \end{array}
  \right)
  ( EE - BB + 2 \im EB )
  \ .
\end{equation}
In tensor language, the complexified Weyl tensor
satisfies the symmetries
\begin{equation}
\label{dualweyl}
  \Cz_{klmn}
  =
  \frac{\im}{2} {\varepsilon_{kl}}^{pq} \Cz_{pqmn}
  =
  \frac{\im}{2} {\varepsilon_{mn}}^{rs} \Cz_{klrs}
  \ .
\end{equation}
Thus the independent components of the complexified Weyl tensor
$\Cz_{klmn}$
constitute a $3 \times 3$ complex symmetric traceless matrix
$EE {-} BB {+} 2 \im EB$,
with $5$ complex degrees of freedom.

The complexified Weyl tensor
$\Cz_{klmn}$
remains unchanged under spatial inversion,
because spatial inversion not only transforms
$E \rightarrow - E$ and $B \rightarrow B$,
but also changes the sign of the antisymmetric tensor
$\varepsilon_{klpq}$,
thereby effectively changing the sign of $\im$ in
$EE {-} BB {+} 2 \im EB$.

In a Newman-Penrose tetrad,
the non-vanishing components of
the complexified Weyl tensor
$\Cz_{klmn}$,
and their expressions in terms of the Weyl and Riemann tensors,
are
\begin{subequations}
\label{Czs}
\begin{eqnarray}
  \Cz_\zero
  &\equiv&
  \Cz_{u v u v}
  =
  \Cz_{\plus \minus \plus \minus}
  =
  \Cz_{u v \plus \minus}
  =
  \Cz_{v \plus \minus u}
  =
  \textstyle{\frac{1}{2}}
  \left( C_{u v u v} + C_{u v \plus \minus} \right)
  =
  \textstyle{\frac{1}{2}}
  \left( C_{\plus \minus \plus \minus} + C_{u v \plus \minus} \right)
  =
  C_{v \plus \minus u}
\label{Cz0}
\nonumber
\\
  &=&
  \textstyle{\frac{1}{6}}
  \left(
  R_{u v u v}
  + R_{\plus \minus \plus \minus}
  + 2 \, R_{v \plus \minus u}
  + 2 \, R_{u v \plus \minus}
  \right)
  \ ,
  \qquad
\\
\label{Cz1}
  \Cz_{\plusminus \one}
  &\equiv&
  \Cz_{\overset{{\scriptstyle v} {\scriptstyle u} {\scriptstyle v} \plus}{{\scriptstyle u} {\scriptstyle v} {\scriptstyle u} \minus}}
  =
  \Cz_{\overset{\minus \plus {\scriptstyle v} \plus}{\plus \minus {\scriptstyle u} \minus}}
  =
  C_{\overset{{\scriptstyle v} {\scriptstyle u} {\scriptstyle v} \plus}{{\scriptstyle u} {\scriptstyle v} {\scriptstyle u} \minus}}
  =
  C_{\overset{\minus \plus {\scriptstyle v} \plus}{\plus \minus {\scriptstyle u} \minus}}
  =
  \textstyle{\frac{1}{2}}
  \bigl(
    R_{\overset{{\scriptstyle v} {\scriptstyle u} {\scriptstyle v} \plus}{{\scriptstyle u} {\scriptstyle v} {\scriptstyle u} \minus}}
    +
    R_{\overset{\minus \plus {\scriptstyle v} \plus}{\plus \minus {\scriptstyle u} \minus}}
  \bigr)
  \ ,
\\
\label{Cz2}
  \Cz_{\plusminus \two}
  &\equiv&
  \Cz_{\overset{{\scriptstyle v} \plus {\scriptstyle v} \plus}{{\scriptstyle u} \minus {\scriptstyle u} \minus}}
  =
  C_{\overset{{\scriptstyle v} \plus {\scriptstyle v} \plus}{{\scriptstyle u} \minus {\scriptstyle u} \minus}}
  =
  R_{\overset{{\scriptstyle v} \plus {\scriptstyle v} \plus}{{\scriptstyle u} \minus {\scriptstyle u} \minus}}
  \ .
\end{eqnarray}
\end{subequations}
Notice that the only non-vanishing components of
the complexified Weyl tensor
$\Cz_{klmn}$
are those with bivector indices
$uv$, $\scriptstyle{+-}$, $v\scriptstyle{+}$, or $u\scriptstyle{-}$.
Components $\Cz_{klmn}$
with bivector indices $v-$ or $u+$ all vanish:
\begin{subequations}
\begin{eqnarray}
\label{Czzeroa}
  &&
  \Cz_{v \minus \plus u}
  =
  \Cz_{\overset{{\scriptstyle v} {\scriptstyle u} {\scriptstyle v} \minus}{{\scriptstyle u} {\scriptstyle v} {\scriptstyle u} \plus}}
  =
  \Cz_{\overset{\plus \minus {\scriptstyle v} \minus}{\minus \plus {\scriptstyle u} \plus}}
  =
  \Cz_{\overset{{\scriptstyle v} \minus {\scriptstyle v} \minus}{{\scriptstyle u} \plus {\scriptstyle u} \plus}}
  =
  0
  \ ,
\\
\label{Czzerob}
  &&
  \Cz_{\overset{{\scriptstyle v} \minus {\scriptstyle v} \plus}{{\scriptstyle u} \plus {\scriptstyle u} \minus}}
  =
  \Cz_{\overset{{\scriptstyle v} \minus {\scriptstyle u} \minus}{{\scriptstyle u} \plus {\scriptstyle v} \plus}}
  =
  C_{\overset{{\scriptstyle v} \minus {\scriptstyle v} \plus}{{\scriptstyle u} \plus {\scriptstyle u} \minus}}
  =
  C_{\overset{{\scriptstyle v} \minus {\scriptstyle u} \minus}{{\scriptstyle u} \plus {\scriptstyle v} \plus}}
  =
  0
  \ ,
\end{eqnarray}
\end{subequations}
in which the top and bottom lines, equations~(\ref{Czzeroa}) and (\ref{Czzerob}),
collect components $\Cz_{klmn}$
for which the parent Weyl tensor $C_{klmn}$ respectively does not, and does,
also vanish.

In the unperturbed self-similar background, only the spin-$0$ component
${\Cz}_\zero$
of the complexified Weyl tensor is non-zero,
and it equals
\begin{equation}
\label{Czunperturbed}
  \overset{\smallzero}{\Cz}_\zero
  =
  C / A^2
\end{equation}
where $C$ is the unperturbed Weyl scalar given by equation~(\ref{C}).

The $\Cz_s$
defined by equations~(\ref{Czs})
can be referred to as
the spin-$s$ components of the complexified Weyl tensor,
since $\Cz_s$
(or in the case $s = 0$ the perturbation
$\overset{\smallone}{\Cz}_\zero$)
is proportional to a harmonic of spin-weight $s$
\begin{equation}
\label{CzsY}
  \overset{\smallone}{\Cz}_s
  \propto
  \nY{s}
\end{equation}
as follows from equations~(\ref{Czs})
and the fact, demonstrated in \S\ref{perturbedriemann},
that each component of the Riemann tensor
conforms to the general rule~(\ref{spinrule})
that its spin equals the sum of the $+$'s and $-$'s of its covariant indices,
As expounded in \S\ref{spin2gravitationalwaves} below,
the spin-$\pm 2$ components $\Cz_{\plusminus \two}$
of the complexified Weyl tensor
describe propagating components of gravitational waves:
$\Cz_{\plus \two}$
describes the propagating component of ingoing gravitational waves,
while
$\Cz_{\minus \two}$
describes the propagating component of outgoing gravitational waves.
In more traditional notation
\cite{NP62,T72,T73,Chandrasekhar},
$\Cz_s = \Psi_{2 - s}$,
so that
$\Cz_{\plus \two}$ is $\Psi_0$,
while
$\Cz_{\minus \two}$ is $\Psi_4$
(Chandrasekhar's \cite{Chandrasekhar} metric signature
${+}{-}{-}{-}$
is opposite to the present paper's
${-}{+}{+}{+}$,
so his Riemann tensor is opposite in sign to the present paper's).

The complexified Weyl tensor
$\Cz_{klmn}$,
even though defined, equation~(\ref{Cz}),
as a projection of the Weyl tensor
$C_{klmn}$,
nevertheless retains all the degrees of freedom of the Weyl tensor.
This can be demonstrated by
resolving $\Cz_s$ into polar $(p)$ and axial $(a)$ parts
\begin{equation}
\label{Czspa}
  \Cz_s
  =
  \Cz^{(p)}_s + \Cz^{(a)}_s
  \ ,
\end{equation}
defined by the property that polar and axial parts respectively
do not and do change sign
when the azimuthal angular tetrad axis
$\gamma_\phi$
is flipped in sign, equation~(\ref{angularconjugate}).
The angular conjugate of
$\Cz_s$
is
\begin{equation}
\label{Czsconj}
  \Cz^\star_s
  =
  \Cz^{(p)}_s - \Cz^{(a)}_s
  \ .
\end{equation}
As remarked after equation~(\ref{npangularconjugate}),
the effect of angular conjugation
is to swap angular indices $+ \leftrightarrow -$
on tensors (such as the Riemann tensor) and other natively real objects
in a Newman-Penrose tetrad (such as the tetrad connections);
but angular conjugation does {\em not\/} swap angular indices on a complexified object such as
the complexified Weyl tensor
$\Cz_{klmn}$.
Thus the angular conjugates
$\Cz^\star_s$
of the spin components of the complexified Weyl tensor are
given
in terms of the Weyl and Riemann tensors
by the angular conjugates of equations~(\ref{Czs}):
\begin{subequations}
\label{Czstar}
\begin{eqnarray}
\label{Cz0star}
  \Cz^\star_\zero
  &=&
  \textstyle{\frac{1}{2}}
  \left( C_{u v u v} - C_{u v \plus \minus} \right)
  =
  \textstyle{\frac{1}{2}}
  \left( C_{\plus \minus \plus \minus} - C_{u v \plus \minus} \right)
  =
  C_{v \minus \plus u}
  =
  \textstyle{\frac{1}{6}}
  \left(
  R_{u v u v}
  + R_{\plus \minus \plus \minus}
  + 2 \, R_{v \minus \plus u}
  - 2 \, R_{u v \plus \minus}
  \right)
  \ ,
  \qquad
\\
\label{Cz1star}
  \Cz^\star_{\plusminus \one}
  &=&
  C_{\overset{{\scriptstyle v} {\scriptstyle u} {\scriptstyle v} \minus}{{\scriptstyle u} {\scriptstyle v} {\scriptstyle u} \plus}}
  =
  C_{\overset{\plus \minus {\scriptstyle v} \minus}{\minus \plus {\scriptstyle u} \plus}}
  =
  \textstyle{\frac{1}{2}}
  \bigl(
    R_{\overset{{\scriptstyle v} {\scriptstyle u} {\scriptstyle v} \minus}{{\scriptstyle u} {\scriptstyle v} {\scriptstyle u} \plus}}
    +
    R_{\overset{\plus \minus {\scriptstyle v} \minus}{\minus \plus {\scriptstyle u} \plus}}
  \bigr)
  \ ,
\\
\label{Cz2star}
  \Cz^\star_{\plusminus \two}
  &=&
  C_{\overset{{\scriptstyle v} \minus {\scriptstyle v} \minus}{{\scriptstyle u} \plus {\scriptstyle u} \plus}}
  =
  R_{\overset{{\scriptstyle v} \minus {\scriptstyle v} \minus}{{\scriptstyle u} \plus {\scriptstyle u} \plus}}
  \ .
\end{eqnarray}
\end{subequations}
Equations~(\ref{Czs}) and (\ref{Czstar})
together account for all the non-zero components of the Weyl tensor
$C_{klmn}$
in terms of
$\Cz_s$ and its angular conjugate $\Cz^\star_s$.

The angular conjugates
$\Cz^\star_s$
are proportional to spin harmonics of spin-weight $-s$,
as follows from the fact that angular conjugation
swaps angular indices $+ \leftrightarrow -$ on any Riemann component,
and the rule~(\ref{spinrule}) that the spin of any Riemann component equals
the sum of the $+$'s and $-$'s of its covariant indices:
\begin{equation}
\label{CzsstarY}
  \overset{\smallone}{\Cz}{}^\star_s
  \propto
  \nY{\minus s}
  \ .
\end{equation}

Linearized expressions for the
spin-$s$ components
$\Cz_s$
of the complexified Weyl tensor
in a Newman-Penrose tetrad
follow from their general expressions~(\ref{Czs})
in terms of the Riemann tensor,
and from the linearized expressions~(\ref{linnpR0})--(\ref{linnpR2})
for the Riemann tensor in terms of the tetrad connections
and their directed derivatives.
Because of the variety of expressions for the Riemann tensor,
there are several ways to write
the spin-$0$ and spin-$\pm 1$ components $\Cz_\zero$
and $\Cz_{\plusminus \one}$,
though only one way to write the spin-$\pm 2$ components $\Cz_{\plusminus \two}$.

Being tetrad-frame quantities,
all components $\Cz_s$ of the complexified Weyl tensor
(and their angular conjugates)
are coordinate gauge-invariant,
although the split between the unperturbed and perturbed parts
of the polar part of
$\Cz_\zero$,
the only component whose unperturbed value is non-vanishing,
is not coordinate gauge-invariant.
The spin-$0$ and spin-$\pm 2$ components
$\Cz_{\zero}$
and
$\Cz_{\plusminus \two}$
(and their angular conjugates)
are not only coordinate gauge-invariant
but tetrad gauge-invariant with respect to all 6
arbitrary infinitesimal tetrad (Lorentz) transformations.
The spin-$\pm 1$ components
$\Cz_{\plusminus \one}$
(and their angular conjugates)
are not tetrad gauge-invariant with respect to arbitrary
infinitesimal tetrad transformations,
being changed by a tetrad transformation that varies
the antisymmetric part of
$\varphi_{\overset{{\scriptstyle v} \plus}{{\scriptstyle u} \minus}}$
(or its angular conjugate
$\varphi_{\overset{{\scriptstyle v} \minus}{{\scriptstyle u} \plus}}$).
The following combinations
of
$\Cz_{\plusminus \one}$
or $\Cz^\star_{\plusminus \one}$
and $\Cz_{\zero}$
are tetrad gauge-invariant,
but not coordinate gauge-invariant:
\begin{subequations}
\label{C1tet}
\begin{eqnarray}
\label{C1teta}
  &&
  \Cz_{\plusminus \one}
  +
  {\textstyle \frac{3}{2}}
  \bigl(
    \varphi_{\overset{{\scriptstyle v} \plus}{{\scriptstyle u} \minus}}
    -
    \varphi_{\overset{\plus {\scriptstyle v}}{\minus {\scriptstyle u}}}
  \bigr)
  \Cz_\zero
  \ ,
\\
\label{C1tetb}
  &&
  \Cz^\star_{\plusminus \one}
  +
  {\textstyle \frac{3}{2}}
  \bigl(
    \varphi_{\overset{{\scriptstyle v} \minus}{{\scriptstyle u} \plus}}
    -
    \varphi_{\overset{\minus {\scriptstyle v}}{\plus {\scriptstyle u}}}
  \bigr)
  \Cz^\star_\zero
  \ .
\end{eqnarray}
\end{subequations}

Special interest attaches to quantities
that are both coordinate and tetrad gauge-invariant,
and also vanishing in the background,
because such quantities constitute
physical perturbations
that are unchanged by any gauge transformation.
There are five components of the
complexified Weyl tensor
that are coordinate and tetrad gauge-invariant
and vanishing in the background,
namely
the axial spin-$0$ component
$\Cz^{(a)}_\zero$,
\begin{equation}
\label{Cza}
  \Cz^{(a)}_\zero
  =
  {\textstyle \frac{1}{2}}
  ( \Cz_\zero - \Cz^\star_\zero )
  =
  \textstyle{\frac{1}{2}}
  C_{u v \plus \minus}
  =
  \textstyle{\frac{1}{2}}
  R_{u v \plus \minus}
  \ ,
\end{equation}
and
the spin-$\pm 2$ components
$\Cz_{\plusminus \two}$
and their angular conjugates
$\Cz^\star_{\plusminus \two}$.

It follows from equations~(\ref{G1tet}) and (\ref{C1tet}) that
the following combinations of spin-$\pm 1$
Weyl and Einstein components
are both coordinate and tetrad gauge-invariant,
and vanishing in the background:
\begin{subequations}
\label{CG1tet}
\begin{eqnarray}
  &&
  \bigl[
    ( G_{vu} + G_{\plus\minus} )^2
    -
    G_{vv} G_{uu}
  \bigr]
  \Cz_{\plusminus \one}
  -
  3 \Cz_\zero
  ( G_{vu} + G_{\plus\minus} )
  G_{\overset{{\scriptstyle v} \plus}{{\scriptstyle u} \minus}}
  +
  3 \Cz_\zero
  G_{\overset{{\scriptstyle v v}}{{\scriptstyle u u}}}
  G_{\overset{{\scriptstyle u} \plus}{{\scriptstyle v} \minus}}
  \ ,
\\
  &&
  \bigl[
    ( G_{vu} + G_{\plus\minus} )^2
    -
    G_{vv} G_{uu}
  \bigr]
  \Cz^\star_{\plusminus \one}
  -
  3 \Cz^\star_\zero
  ( G_{vu} + G_{\plus\minus} )
  G_{\overset{{\scriptstyle v} \minus}{{\scriptstyle u} \plus}}
  +
  3 \Cz^\star_\zero
  G_{\overset{{\scriptstyle v v}}{{\scriptstyle u u}}}
  G_{\overset{{\scriptstyle u} \minus}{{\scriptstyle v} \plus}}
  \ .
\end{eqnarray}
\end{subequations}

\subsection{Evolution of the Weyl tensor}
\label{evolutionweyl}

Equations governing the evolution of gravitational waves
are obtained by applying the Bianchi identities
to the Weyl tensor.
The central result of this subsection is the set of
linearized Weyl evolution equations~(\ref{Weyls}),
which describe the coupled evolution of the various spin
$s = -2$ to $+ 2$
components
$\Cz_s$
of the complexified Weyl tensor
in a spherically symmetric self-similar unperturbed background.
These Weyl evolution equations
can be combined in pairs to yield sourced wave equations~(\ref{waveWeyl})
for each of the spin components
$\Cz_s$.
Since the different spin components are related to each other by the
Weyl evolution equations~(\ref{Weyls}),
the wave equations do not describe independent modes of vibration,
but nevertheless it is useful to have the full suite of wave equations available.
The next subsection, \S\ref{spin2gravitationalwaves},
considers the case of the spin-$\pm 2$ components
$\Cz_{\plusminus \two}$,
which describe the propagating components of gravitational waves,
while the final subsections, \S\ref{axialgravitationalwaves} and \S\ref{polargravitationalwaves},
consider the axial and polar spin-$0$ components
$\Cz^{(a)}_{\zero}$
and
$\Cz^{(p)}_{\zero}$.

The Bianchi identities
\begin{equation}
  D_k R_{lmnp}
  +
  D_l R_{mknp}
  +
  D_m R_{klnp}
  =
  0
\end{equation}
yield evolution equations for the Weyl tensor
\begin{equation}
\label{DC}
  D^k C_{klmn}
  =
  J_{lmn}
\end{equation}
with Weyl current
\begin{equation}
  J_{lmn}
  =
  {\textstyle \frac{1}{2}} \left(
  D_m G_{ln} - D_n G_{lm}
  \right)
  - {\textstyle \frac{1}{6}} \left(
  \gamma_{ln} D_m G - \gamma_{lm} D_n G
  \right)
  \ ,
\end{equation}
which
satisfies the conservation law
$D^l J_{lmn} = 0$,
in view of equation~(\ref{DC})
and the antisymmetry of $C_{klmn}$ with respect to the indices $kl$.
The corresponding evolution equations for
the complexified Weyl tensor are
\begin{equation}
\label{DCz}
  D^k \Cz_{klmn}
  =
  \Jz_{lmn}
\end{equation}
where $\Jz_{lmn}$ is the complexified Weyl current
\begin{equation}
  \Jz_{lmn}
  =
  \frac{1}{2}
  \left(
  \delta^r_m \delta^s_n + \frac{\im}{2} {\varepsilon_{mn}}^{rs}
  \right)
  J_{lrs}
  \ ,
\end{equation}
which satisfies the conservation law
$D^l \Jz_{lmn} = 0$.
The complexified currents
$\Jz_{lmn}$
are related to the currents
$J_{lmn}$
by
\begin{subequations}
\label{npJz}
\begin{eqnarray}
  \Jz_{\overset{{\scriptstyle v} {\scriptstyle v} {\scriptstyle u}}{{\scriptstyle u} {\scriptstyle u} {\scriptstyle v}}}
  =
  \Jz_{\overset{{\scriptstyle v} \minus \plus}{{\scriptstyle u} \plus \minus}}
  =
  \Jz_{\overset{\minus {\scriptstyle v} \plus}{\plus {\scriptstyle u} \minus}}
  &=&
  {\textstyle \frac{1}{2}}
  \bigl(
    J_{\overset{{\scriptstyle v} {\scriptstyle v} {\scriptstyle u}}{{\scriptstyle u} {\scriptstyle u} {\scriptstyle v}}}
    +
    J_{\overset{{\scriptstyle v} \minus \plus}{{\scriptstyle u} \plus \minus}}
  \bigr)
  =
  J_{\overset{\minus {\scriptstyle v} \plus}{\plus {\scriptstyle u} \minus}}
  \ ,
\\
  \Jz_{\overset{\plus {\scriptstyle v} {\scriptstyle u}}{\minus {\scriptstyle u} {\scriptstyle v}}}
  =
  \Jz_{\overset{\plus \minus \plus}{\minus \plus \minus}}
  =
  \Jz_{\overset{{\scriptstyle u} {\scriptstyle v} \plus}{{\scriptstyle v} {\scriptstyle u} \minus}}
  &=&
  {\textstyle \frac{1}{2}}
  \bigl(
    J_{\overset{\plus {\scriptstyle v} {\scriptstyle u}}{\minus {\scriptstyle u} {\scriptstyle v}}}
    +
    J_{\overset{\plus \minus \plus}{\minus \plus \minus}}
  \bigr)
  =
  J_{\overset{{\scriptstyle u} {\scriptstyle v} \plus}{{\scriptstyle v} {\scriptstyle u} \minus}}
  \ ,
\\
  \Jz_{\overset{{\scriptstyle v} {\scriptstyle v} \plus}{{\scriptstyle u} {\scriptstyle u} \minus}}
  &=&
  J_{\overset{{\scriptstyle v} {\scriptstyle v} \plus}{{\scriptstyle u} {\scriptstyle u} \minus}}
  \ ,
\\
  \Jz_{\overset{\plus {\scriptstyle v} \plus}{\minus {\scriptstyle u} \minus}}
  &=&
  J_{\overset{\plus {\scriptstyle v} \plus}{\minus {\scriptstyle u} \minus}}
  \ .
\end{eqnarray}
\end{subequations}
Complexified Weyl currents
$\Jz_{lmn}$
with right bivector index $mn$ equal to $v-$ or $u+$
are zero
\begin{equation}
  \Jz_{\overset{\minus {\scriptstyle v} \minus}{\plus {\scriptstyle u} \plus}}
  =
  \Jz_{\overset{{\scriptstyle u} {\scriptstyle v} \minus}{{\scriptstyle v} {\scriptstyle u} \plus}}
  =
  \Jz_{\overset{{\scriptstyle v} {\scriptstyle v} \minus}{{\scriptstyle u} {\scriptstyle u} \plus}}
  =
  \Jz_{\overset{\plus {\scriptstyle v} \minus}{\minus {\scriptstyle u} \plus}}
  =
  0
  \ .
\end{equation}
The angular conjugates
$\Jz^\star_{lmn}$
of the complexified currents
are given by the angular conjugates of equations~(\ref{npJz}),
which swaps angular indices $+ \leftrightarrow -$ on the currents
$J_{lmn}$
on the right hand sides
(but does not swap angular indices on the complexified currents
$\Jz_{lmn}$):
\begin{subequations}
\begin{eqnarray}
  \Jz^\star_{\overset{{\scriptstyle v} {\scriptstyle v} {\scriptstyle u}}{{\scriptstyle u} {\scriptstyle u} {\scriptstyle v}}}
  =
  \Jz^\star_{\overset{{\scriptstyle v} \minus \plus}{{\scriptstyle u} \plus \minus}}
  =
  \Jz^\star_{\overset{\minus {\scriptstyle v} \plus}{\plus {\scriptstyle u} \minus}}
  &=&
  {\textstyle \frac{1}{2}}
  \bigl(
    J_{\overset{{\scriptstyle v} {\scriptstyle v} {\scriptstyle u}}{{\scriptstyle u} {\scriptstyle u} {\scriptstyle v}}}
    -
    J_{\overset{{\scriptstyle v} \minus \plus}{{\scriptstyle u} \plus \minus}}
  \bigr)
  =
  J_{\overset{\plus {\scriptstyle v} \minus}{\minus {\scriptstyle u} \plus}}
  \ ,
\\
  \Jz^\star_{\overset{\plus {\scriptstyle v} {\scriptstyle u}}{\minus {\scriptstyle u} {\scriptstyle v}}}
  =
  \Jz^\star_{\overset{\plus \minus \plus}{\minus \plus \minus}}
  =
  \Jz^\star_{\overset{{\scriptstyle u} {\scriptstyle v} \plus}{{\scriptstyle v} {\scriptstyle u} \minus}}
  &=&
  {\textstyle \frac{1}{2}}
  \bigl(
    - \,
    J_{\overset{\minus {\scriptstyle u} {\scriptstyle v}}{\plus {\scriptstyle v} {\scriptstyle u}}}
    +
    J_{\overset{\minus \plus \minus}{\plus \minus \plus}}
  \bigr)
  =
  J_{\overset{{\scriptstyle u} {\scriptstyle v} \minus}{{\scriptstyle v} {\scriptstyle u} \plus}}
  \ ,
\\
  \Jz^\star_{\overset{{\scriptstyle v} {\scriptstyle v} \plus}{{\scriptstyle u} {\scriptstyle u} \minus}}
  &=&
  J_{\overset{{\scriptstyle v} {\scriptstyle v} \minus}{{\scriptstyle u} {\scriptstyle u} \plus}}
  \ ,
\\
  \Jz^\star_{\overset{\plus {\scriptstyle v} \plus}{\minus {\scriptstyle u} \minus}}
  &=&
  J_{\overset{\minus {\scriptstyle v} \minus}{\plus {\scriptstyle u} \plus}}
  \ .
\end{eqnarray}
\end{subequations}

In a general Newman-Penrose tetrad, the equations~(\ref{DCz})
governing the evolution of the complexified Weyl tensor $\Cz_s$ are explicitly
\begin{subequations}
\label{npWeyl}
\begin{eqnarray}
  - \, 2 \, \Gamma_{\overset{\plus {\scriptstyle v} {\scriptstyle v}}{\minus {\scriptstyle u} {\scriptstyle u}}} \, A^2 \, \Cz_{\minusplus \one}
  + \nDelta{\zero}_\vu \, A \, \Cz_\zero
  - \nDelta{\plusminus \one}_\minusplus \, A \, \Cz_{\plusminus \one}
  + \Gamma_{\overset{\minus {\scriptstyle u} \minus}{\plus {\scriptstyle v} \plus}} \, A^2 \, \Cz_{\plusminus \two}
  &=&
  A^2 \,
  \Jz_{\overset{\minus {\scriptstyle v} \plus}{\plus {\scriptstyle u} \minus}}
  \ ,
\\
  - \, 2 \, \Gamma_{\overset{\plus {\scriptstyle v} \plus}{\minus {\scriptstyle u} \minus}} \, A^2 \, \Cz_{\minusplus \one}
  + \nDelta{\zero}_\plusminus \, A \, \Cz_\zero
  - \nDelta{\plusminus \one}_\uv \, A \, \Cz_{\plusminus \one}
  + \Gamma_{\overset{\minus {\scriptstyle u u}}{\plus {\scriptstyle v v}}} \, A^2 \, \Cz_{\plusminus \two}
  &=&
  A^2 \,
  \Jz_{\overset{{\scriptstyle u} {\scriptstyle v} \plus}{{\scriptstyle v} {\scriptstyle u} \minus}}
  \ ,
\\
  - \, 3 \, \Gamma_{\overset{\plus {\scriptstyle v} {\scriptstyle v}}{\minus {\scriptstyle u} {\scriptstyle u}}} \, A^2 \, \Cz_\zero
  + \nDelta{\plusminus \one}_\vu \, A \, \Cz_{\plusminus \one}
  - \nDelta{\plusminus \two}_\minusplus \, A \, \Cz_{\plusminus \two}
  &=&
  A^2 \,
  \Jz_{\overset{{\scriptstyle v} {\scriptstyle v} \plus}{{\scriptstyle u} {\scriptstyle u} \minus}}
  \ ,
\\
  - \, 3 \, \Gamma_{\overset{\plus {\scriptstyle v} \plus}{\minus {\scriptstyle u} \minus}} \, A^2 \, \Cz_\zero
  + \nDelta{\plusminus \one}_\plusminus \, A \, \Cz_{\plusminus \one}
  - \nDelta{\plusminus \two}_\uv \, A \, \Cz_{\plusminus \two}
  &=&
  A^2 \,
  \Jz_{\overset{\plus {\scriptstyle v} \plus}{\minus {\scriptstyle u} \minus}}
  \ ,
\end{eqnarray}
\end{subequations}
where the differential operators
$\nDelta{s}_m$
are
\begin{subequations}
\label{Deltas}
\begin{eqnarray}
  \nDelta{\plusminus s}_\vu
  &\equiv&
  A^2 \Bigl[
  \partial_\vu
  +
  s \, \Gamma_{\overset{{\scriptstyle u v v}}{{\scriptstyle v u u}}}
  -
  s \, \Gamma_{\overset{\minus \plus {\scriptstyle v}}{\plus \minus {\scriptstyle u}}}
  +
  ( s + 3 ) \, \Gamma_{\overset{\plus {\scriptstyle v} \minus}{\minus {\scriptstyle u} \plus}}
  \Bigr] A^{-1}
  \ ,
\\
  \nDelta{\plusminus s}_\plusminus
  &\equiv&
  A^2 \Bigl[
  \partial_\plusminus
  -
  s \, \Gamma_{\overset{\minus \plus \plus}{\plus \minus \minus}}
  +
  s \, \Gamma_{\overset{{\scriptstyle u v} \plus}{{\scriptstyle v u} \minus}}
  +
  ( s + 3 ) \, \Gamma_{\overset{\plus {\scriptstyle v u}}{\minus {\scriptstyle u v}}}
  \Bigr] A^{-1}
  \ ,
\end{eqnarray}
\end{subequations}
which reduce in the unperturbed background to
\begin{equation}
  \nDeltazero{s}_m
  =
  \ncalD{s}_m
  \ .
\end{equation}
Explicit expressions for the Weyl currents
$\Jz_{lmn}$
on the right hand sides of equations~(\ref{npWeyl})
are given in the Appendix, equations~(\ref{Jz}).

To linear order of perturbations on the self-similar background,
the Weyl evolution equations~(\ref{npWeyl}) reduce
(in a general gauge)
to
\begin{subequations}
\label{Weyls}
\begin{eqnarray}
\label{Weylsa}
  A^2 \left(
  \partial_\vu
  +
  3 \, \Gamma_{\overset{\plus {\scriptstyle v} \minus}{\minus {\scriptstyle u} \plus}}
  \right)
  A^{-2}
  C
  +
  \ncalD{\zero}_\vu \, A \, \overset{\smallone}{\Cz}_\zero
  -
  \ncalD{\plusminus \one}_\minusplus \, A \, \Cz_{\plusminus \one}
  &=&
  A^2 \,
  \Jz_{\overset{\minus {\scriptstyle v} \plus}{\plus {\scriptstyle u} \minus}}
  \ ,
\\
\label{Weylsb}
  A^2 \left(
  \partial_\plusminus
  +
  3 \, \Gamma_{\overset{\plus {\scriptstyle v u}}{\minus {\scriptstyle u v}}}
  \right)
  A^{-2}
  C
  +
  \ncalD{\zero}_\plusminus \, A \, \overset{\smallone}{\Cz}_\zero
  -
  \ncalD{\plusminus \one}_\uv \, A \, \Cz_{\plusminus \one}
  &=&
  A^2 \,
  \Jz_{\overset{{\scriptstyle u} {\scriptstyle v} \plus}{{\scriptstyle v} {\scriptstyle u} \minus}}
  \ ,
\\
\label{Weylsc}
  - \,
  3 \, \Gamma_{\overset{\plus {\scriptstyle v} {\scriptstyle v}}{\minus {\scriptstyle u} {\scriptstyle u}}} \,
  C
  +
  \ncalD{\plusminus \one}_\vu \, A \, \Cz_{\plusminus \one}
  -
  \ncalD{\plusminus \two}_\minusplus \, A \, \Cz_{\plusminus \two}
  &=&
  A^2 \,
  \Jz_{\overset{{\scriptstyle v} {\scriptstyle v} \plus}{{\scriptstyle u} {\scriptstyle u} \minus}}
  \ ,
\\
\label{Weylsd}
  - \,
  3 \, \Gamma_{\overset{\plus {\scriptstyle v} \plus}{\minus {\scriptstyle u} \minus}} \,
  C
  +
  \ncalD{\plusminus \one}_\plusminus \, A \, \Cz_{\plusminus \one}
  -
  \ncalD{\plusminus \two}_\uv \, A \, \Cz_{\plusminus \two}
  &=&
  A^2 \,
  \Jz_{\overset{\plus {\scriptstyle v} \plus}{\minus {\scriptstyle u} \minus}}
  \ ,
\end{eqnarray}
\end{subequations}
where the linearized Weyl currents
$\Jz_{lmn}$
on the right hand sides of equations~(\ref{Weyls})
are
(again in a general gauge),
from equations~(\ref{Jz}),
\begin{subequations}
\label{linJz}
\begin{eqnarray}
\label{linJza}
  \Jz_{\overset{{\scriptstyle v} {\scriptstyle v} {\scriptstyle u}}{{\scriptstyle u} {\scriptstyle u} {\scriptstyle v}}}
  \  = \ 
  \Jz_{\overset{{\scriptstyle v} \minus \plus}{{\scriptstyle u} \plus \minus}}
  &=&
  \frac{1}{4} \Bigl[
  \partial_\vu
  \bigl(
    {\textstyle \frac{1}{3}} G_{{\scriptstyle v} {\scriptstyle u}}
    +
    {\textstyle \frac{2}{3}} G_{\plus \minus}
  \bigr)
  -
  \bigl( \partial_\uv
    - 2 \, \Gamma_{\overset{{\scriptstyle v} {\scriptstyle u} {\scriptstyle u}}{{\scriptstyle u} {\scriptstyle v} {\scriptstyle v}}}
    + \Gamma_{\overset{\plus {\scriptstyle u} \minus}{\minus {\scriptstyle v} \plus}}
    - \Gamma_{\overset{\minus {\scriptstyle u} \plus}{\plus {\scriptstyle v} \minus}}
  \bigr) G_{\overset{{\scriptstyle v} {\scriptstyle v}}{{\scriptstyle u} {\scriptstyle u}}}
  -
  A^{-1}
  \ncalD{\minusplus \one}_\plusminus \,
  G_{\overset{{\scriptstyle v} \minus}{{\scriptstyle u} \plus}}
  +
  A^{-1}
  \ncalD{\plusminus \one}_\minusplus \,
  G_{\overset{{\scriptstyle v} \plus}{{\scriptstyle u} \minus}}
\nonumber
\\
  &&
  \quad
  + \,
  \bigl(
    \Gamma_{\overset{\minus {\scriptstyle v} \plus}{\plus {\scriptstyle u} \minus}}
    - \Gamma_{\overset{\plus {\scriptstyle v} \minus}{\minus {\scriptstyle u} \plus}}
  \bigr) \bigl(
    \overset{\smallzero}{G}_{{\scriptstyle v} {\scriptstyle u}}
    +
    \overset{\smallzero}{G}_{\plus \minus}
  \bigr)
  \Bigr]
\\
\label{linJzb}
  = \ 
  \Jz_{\overset{\minus {\scriptstyle v} \plus}{\plus {\scriptstyle u} \minus}}
  &=&
  \frac{1}{2} \Bigl[
  \partial_\vu
  \bigl(
    {\textstyle \frac{2}{3}} G_{{\scriptstyle v} {\scriptstyle u}}
    +
    {\textstyle \frac{1}{3}} G_{\plus \minus}
  \bigr)
  -
  A^{-1}
  \ncalD{\minusplus \one}_\plusminus \,
  G_{\overset{{\scriptstyle v} \minus}{{\scriptstyle u} \plus}}
  +
  \Gamma_{\overset{\minus {\scriptstyle v} \plus}{\plus {\scriptstyle u} \minus}}
  \bigl(
    G_{{\scriptstyle v} {\scriptstyle u}}
    +
    G_{\plus \minus}
  \bigr)
  +
  \Gamma_{\overset{\minus {\scriptstyle u} \plus}{\plus {\scriptstyle v} \minus}}
  \, G_{\overset{{\scriptstyle v} {\scriptstyle v}}{{\scriptstyle u} {\scriptstyle u}}}
  \Bigr]
  \ ,
\\
\label{linJzc}
  \Jz_{\overset{\plus {\scriptstyle v} {\scriptstyle u}}{\minus {\scriptstyle u} {\scriptstyle v}}}
  \  = \ 
  \Jz_{\overset{\plus \minus \plus}{\minus \plus \minus}}
  &=&
  \frac{1}{4} \Bigl[
  - \,
  \partial_\plusminus
  \bigl(
    {\textstyle \frac{2}{3}} G_{{\scriptstyle v} {\scriptstyle u}}
    +
    {\textstyle \frac{1}{3}} G_{\plus \minus}
  \bigr)
  +
  A^{-1}
  \ncalD{\minusplus \one}_\vu \,
  G_{\overset{{\scriptstyle u} \plus}{{\scriptstyle v} \minus}}
  -
  A^{-1}
  \ncalD{\plusminus \one}_\uv \,
  G_{\overset{{\scriptstyle v} \plus}{{\scriptstyle u} \minus}}
  +
  A^{-1}
  \ncalD{\plusminus \two}_\minusplus \,
  G_{\overset{\plus \plus}{\minus \minus}}
\nonumber
\\
  &&
  \quad
  + \,
  \bigl( 
    \Gamma_{\overset{\plus {\scriptstyle v} {\scriptstyle u}}{\minus {\scriptstyle u} {\scriptstyle v}}}
    - \Gamma_{\overset{\plus {\scriptstyle u} {\scriptstyle v}}{\minus {\scriptstyle v} {\scriptstyle u}}}
  \bigr) \bigl(
    \overset{\smallzero}{G}_{{\scriptstyle v} {\scriptstyle u}} + \overset{\smallzero}{G}_{\plus \minus}
  \bigr)
  +
  \Gamma_{\overset{\plus {\scriptstyle u} {\scriptstyle u}}{\minus {\scriptstyle v} {\scriptstyle v}}} \,
  \overset{\smallzero}{G}_{\overset{{\scriptstyle v} {\scriptstyle v}}{{\scriptstyle u} {\scriptstyle u}}}
  -
  \Gamma_{\overset{\plus {\scriptstyle v} {\scriptstyle v}}{\minus {\scriptstyle u} {\scriptstyle u}}} \,
  \overset{\smallzero}{G}_{\overset{{\scriptstyle u} {\scriptstyle u}}{{\scriptstyle v} {\scriptstyle v}}}
  \Bigr]
  \qquad
\\
\label{linJzd}
  = \ 
  \Jz_{\overset{{\scriptstyle u} {\scriptstyle v} \plus}{{\scriptstyle v} {\scriptstyle u} \minus}}
  &=&
  \frac{1}{2} \Bigl[
  - \,
  \partial_\plusminus
  \bigl(
    {\textstyle \frac{1}{3}} G_{{\scriptstyle v} {\scriptstyle u}}
    +
    {\textstyle \frac{2}{3}} G_{\plus \minus}
  \bigr)
  +
  A^{-1}
  \ncalD{\minusplus \one}_\vu \,
  G_{\overset{{\scriptstyle u} \plus}{{\scriptstyle v} \minus}}
  -
  \Gamma_{\overset{\plus {\scriptstyle u} {\scriptstyle v}}{\minus {\scriptstyle v} {\scriptstyle u}}}
  \bigl(
    \overset{\smallzero}{G}_{{\scriptstyle v} {\scriptstyle u}} + \overset{\smallzero}{G}_{\plus \minus}
  \bigr)
  -
  \Gamma_{\overset{\plus {\scriptstyle v} {\scriptstyle v}}{\minus {\scriptstyle u} {\scriptstyle u}}} \,
  \overset{\smallzero}{G}_{\overset{{\scriptstyle u} {\scriptstyle u}}{{\scriptstyle v} {\scriptstyle v}}}
  +
  \Gamma_{\overset{\minus {\scriptstyle u} \plus}{\plus {\scriptstyle v} \minus}} \,
  G_{\overset{{\scriptstyle v} \plus}{{\scriptstyle u} \minus}}
  \Bigr]
  \ ,
  \qquad\quad
\\
\label{linJze}
  \Jz_{\overset{{\scriptstyle v} {\scriptstyle v} \plus}{{\scriptstyle u} {\scriptstyle u} \minus}}
  &=&
  \frac{1}{2} \Bigl[
  \ncalD{\plusminus \one}_\vu \,
  A^{-1}
  G_{\overset{{\scriptstyle v} \plus}{{\scriptstyle u} \minus}}
  -
  \bigl( \partial_\plusminus
    + 2 \, \Gamma_{\overset{{\scriptstyle u} {\scriptstyle v} \plus}{{\scriptstyle v} {\scriptstyle u} \minus}}
    + \Gamma_{\overset{\plus {\scriptstyle u} {\scriptstyle v}}{\minus {\scriptstyle v} {\scriptstyle u}}}
  \bigr)
  G_{\overset{{\scriptstyle v} {\scriptstyle v}}{{\scriptstyle u} {\scriptstyle u}}}
  -
  \Gamma_{\overset{\plus {\scriptstyle v} {\scriptstyle v}}{\minus {\scriptstyle u} {\scriptstyle u}}}
  \bigl(
    \overset{\smallzero}{G}_{{\scriptstyle v} {\scriptstyle u}} + \overset{\smallzero}{G}_{\plus \minus}
  \bigr)
  \Bigr]
  \ ,
\\
\label{linJzf}
  \Jz_{\overset{\plus {\scriptstyle v} \plus}{\minus {\scriptstyle u} \minus}}
  &=&
  \frac{1}{2} \Bigl[
  \ncalD{\zero}_\vu \,
  A^{-1}
  G_{\overset{\plus \plus}{\minus \minus}}
  -
  A^{-1}
  \ncalD{\plusminus \one}_\plusminus \,
  G_{\overset{{\scriptstyle v} \plus}{{\scriptstyle u} \minus}}
  +
  \Gamma_{\overset{\plus {\scriptstyle v} \plus}{\minus {\scriptstyle u} \minus}}
  \bigl(
    \overset{\smallzero}{G}_{{\scriptstyle v} {\scriptstyle u}} + \overset{\smallzero}{G}_{\plus \minus}
  \bigr)
  +
  \Gamma_{\overset{\plus {\scriptstyle u} \plus}{\minus {\scriptstyle v} \minus}} \,
  \overset{\smallzero}{G}_{\overset{{\scriptstyle v} {\scriptstyle v}}{{\scriptstyle u} {\scriptstyle u}}}
  \Bigr]
  \ .
\end{eqnarray}
\end{subequations}
The equality of the expressions~(\ref{linJza}) and (\ref{linJzb}),
and of (\ref{linJzc}) and (\ref{linJzd}),
expresses the vanishing of the covariant derivative of the Einstein tensor,
$D^m G_{mn} = 0$,
which implies the conservation of energy-momentum.

In the unperturbed background,
the Weyl evolution equations~(\ref{Weyls}) simplify to
\begin{subequations}
\label{npWeylunperturbed}
\begin{eqnarray}
\label{npWeylunperturbedvu}
  \ncalD{\zero}_\vu \, A^{-1} C
  &=&
  A^2 \,
  \overset{\smallzero}{\Jz}_{\overset{\minus {\scriptstyle v} \plus}{\plus {\scriptstyle u} \minus}}
  \ ,
\\
\label{npWeylunperturbedpm}
  \ncalD{\zero}_\plusminus \, A^{-1} C
  &=&
  0
  \ ,
\end{eqnarray}
\end{subequations}
where the only non-vanishing component of the unperturbed Weyl current is
\begin{eqnarray}
  \overset{\smallzero}{\Jz}_{\overset{{\scriptstyle v} {\scriptstyle v} {\scriptstyle u}}{{\scriptstyle u} {\scriptstyle u} {\scriptstyle v}}}
  \  = \ 
  \overset{\smallzero}{\Jz}_{\overset{{\scriptstyle v} \minus \plus}{{\scriptstyle u} \plus \minus}}
  &=&
  \frac{1}{4} \Bigl[
  A \,
  \ncalD{\zero}_\vu \,
  A^{-2}
  \bigl(
    {\textstyle \frac{1}{3}} \overset{\smallzero}{G}_{{\scriptstyle v} {\scriptstyle u}}
    +
    {\textstyle \frac{2}{3}} \overset{\smallzero}{G}_{\plus \minus}
  \bigr)
  -
  A^{-1}
  \ncalD{\plusminus \two}_\uv \,
  \overset{\smallzero}{G}_{\overset{{\scriptstyle v} {\scriptstyle v}}{{\scriptstyle u} {\scriptstyle u}}}
  \Bigr]
\nonumber
\\
\label{unperturbedJz}
  = \ 
  \overset{\smallzero}{\Jz}_{\overset{\minus {\scriptstyle v} \plus}{\plus {\scriptstyle u} \minus}}
  &=&
  \frac{1}{2} \Bigl[
  {\textstyle \frac{2}{3}}
  A^{-1/2}
  \ncalD{\zero}_\vu \,
  A^{-1/2}
  \overset{\smallzero}{G}_{{\scriptstyle v} {\scriptstyle u}}
  +
  {\textstyle \frac{1}{3}}
  A^{-2}
  \ncalD{\zero}_\vu \,
  A \,
  \overset{\smallzero}{G}_{\plus \minus}
  +
  A^{-1}
  \beta_\uv
  \, \overset{\smallzero}{G}_{\overset{{\scriptstyle v} {\scriptstyle v}}{{\scriptstyle u} {\scriptstyle u}}}
  \Bigr]
  \ ,
\end{eqnarray}
in which the equality of the two expressions
expresses
$D^m G_{mn} = 0$ in the unperturbed background,
which implies energy-momentum conservation in the unperturbed background.
A ``nice'' expression for
$\overset{\smallzero}{\Jz}_{\overset{\minus {\scriptstyle v} \plus}{\plus {\scriptstyle u} \minus}}$,
obtained by combining the first and second expressions of equations~(\ref{unperturbedJz})
in the proportions $\frac{2}{3}$ and $\frac{1}{3}$, is
\begin{equation}
  \overset{\smallzero}{\Jz}_{\overset{\minus {\scriptstyle v} \plus}{\plus {\scriptstyle u} \minus}}
  =
  \frac{1}{6}
  \Bigl[
  \ncalD{\zero}_\vu \, A^{-1}
  \bigl(
    \overset{\smallzero}{G}_{{\scriptstyle v} {\scriptstyle u}} + \overset{\smallzero}{G}_{\plus \minus}
  \bigr)
  -
  \ncalD{\plusminus \two}_\uv \,
  A^{-1}
  \overset{\smallzero}{G}_{\overset{{\scriptstyle v} {\scriptstyle v}}{{\scriptstyle u} {\scriptstyle u}}}
  \Bigr]
  \ .
\end{equation}

In a general Newman-Penrose tetrad,
the Weyl current conservation equations
$D^l \Jz_{lmn} = 0$
are explicitly
\begin{subequations}
\label{DJz}
\begin{eqnarray}
\label{DJzuv}
  A^3 D^l \Jz_{l u v}
  &=&
  \nDelta{\plus \one}^\prime_u \,
  A^2
  \Jz_{v v u}
  -
  \nDelta{\minus \one}^\prime_v \,
  A^2
  \Jz_{u u v}
  -
  \nDelta{\plus \one}^\prime_\minus \,
  A^2
  \Jz_{\plus v u}
  +
  \nDelta{\minus \one}^\prime_\plus \,
  A^2
  \Jz_{\minus u v}
\nonumber
\\
  &&
  - \, \Gamma_{\minus u u}
  \, \Jz_{v v \plus}
  + \Gamma_{\plus v v}
  \, \Jz_{u u \minus}
  + \Gamma_{\minus u \minus}
  \, \Jz_{\plus v \plus}
  - \Gamma_{\plus v \plus}
  \, \Jz_{\minus u \minus}
  \  = \ 
  0
  \ ,
\\
\label{DJzvp}
  A^3 D^l \Jz_{\overset{{\scriptstyle l} {\scriptstyle v} \plus}{{\scriptstyle l} {\scriptstyle u} \minus}}
  &=&
  \nDelta{\zero}^\prime_\plusminus \,
  A^2
  \Jz_{\overset{{\scriptstyle {\scriptstyle v}} {\scriptstyle v} {\scriptstyle u}}{{\scriptstyle {\scriptstyle u}} {\scriptstyle u} {\scriptstyle v}}}
  -
  \nDelta{\zero}^\prime_\vu \,
  A^2
  \Jz_{\overset{\plus {\scriptstyle v} {\scriptstyle u}}{\minus {\scriptstyle u} {\scriptstyle v}}}
  -
  \nDelta{\plusminus \two}^\prime_\uv \,
  A^2
  \Jz_{\overset{{\scriptstyle v} {\scriptstyle v} \plus}{{\scriptstyle u} {\scriptstyle u} \minus}}
  +
  \nDelta{\plusminus \two}^\prime_\minusplus \,
  A^2
  \Jz_{\overset{\plus {\scriptstyle v} \plus}{\minus {\scriptstyle u} \minus}}
  - 2 \, \Gamma_{\overset{\plus {\scriptstyle v} {\scriptstyle v}}{\minus {\scriptstyle u} {\scriptstyle u}}}
  \, \Jz_{\overset{{\scriptstyle u} {\scriptstyle u} {\scriptstyle v}}{{\scriptstyle v} {\scriptstyle v} {\scriptstyle u}}}
  + 2 \, \Gamma_{\overset{\plus {\scriptstyle v} \plus}{\minus {\scriptstyle u} \minus}}
  \, \Jz_{\overset{\minus {\scriptstyle u} {\scriptstyle v}}{\plus {\scriptstyle v} {\scriptstyle u}}}
  \  = \ 
  0
  \ ,
  \quad\quad\quad
\end{eqnarray}
\end{subequations}
where the differential operators
$\nDelta{\zero}^\prime_m$
are
\begin{subequations}
\label{Deltasprime}
\begin{eqnarray}
  \nDelta{\plusminus s}^\prime_\vu
  &\equiv&
  A^3 \Bigl[
  \partial_\vu
  +
  s \, \Gamma_{\overset{{\scriptstyle u v v}}{{\scriptstyle v u u}}}
  -
  ( s + 1 ) \, \Gamma_{\overset{\minus \plus {\scriptstyle v}}{\plus \minus {\scriptstyle u}}}
  +
  \Gamma_{\overset{\minus {\scriptstyle v} \plus}{\plus {\scriptstyle u} \minus}}
  +
  ( s + 3 ) \, \Gamma_{\overset{\plus {\scriptstyle v} \minus}{\minus {\scriptstyle u} \plus}}
  \Bigr] A^{-2}
  \ ,
\\
  \nDelta{\plusminus s}^\prime_\plusminus
  &\equiv&
  A^3 \Bigl[
  \partial_\plusminus
  -
  s \, \Gamma_{\overset{\minus \plus \plus}{\plus \minus \minus}}
  +
  ( s + 1 ) \, \Gamma_{\overset{{\scriptstyle u v} \plus}{{\scriptstyle v u} \minus}}
  +
  \Gamma_{\overset{\plus {\scriptstyle u v}}{\minus {\scriptstyle v u}}}
  +
  ( s + 3 ) \, \Gamma_{\overset{\plus {\scriptstyle v u}}{\minus {\scriptstyle u v}}}
  \Bigr] A^{-2}
  \ ,
\end{eqnarray}
\end{subequations}
which reduce in the unperturbed background to
\begin{equation}
  \nDeltazero{s}{}^\prime_m
  =
  \ncalD{s}_m
  \ .
\end{equation}
Linearized,
the Weyl current conservation equations~(\ref{DJz}) reduce
(in a general gauge)
to
\begin{subequations}
\label{linDJz}
\begin{eqnarray}
\label{linDJzuv}
  &\!\!\!\!\!\!&
  \nDelta{\plus \one}^\prime_u \,
  A^2
  \Jz_{v v u}
  -
  \nDelta{\minus \one}^\prime_v \,
  A^2
  \Jz_{u u v}
  -
  \ncalD{\plus \one}_\minus \,
  A^2
  \Jz_{\plus v u}
  +
  \ncalD{\minus \one}_\plus \,
  A^2
  \  = \ 
  0
  \ ,
\\
\label{linDJzvp}
  &\!\!\!\!\!\!&
  \nDelta{\zero}^\prime_\plusminus \,
  A^2
  \Jz_{\overset{{\scriptstyle {\scriptstyle v}} {\scriptstyle v} {\scriptstyle u}}{{\scriptstyle {\scriptstyle u}} {\scriptstyle u} {\scriptstyle v}}}
  -
  \ncalD{\zero}_\vu \,
  A^2
  \Jz_{\overset{\plus {\scriptstyle v} {\scriptstyle u}}{\minus {\scriptstyle u} {\scriptstyle v}}}
  -
  \ncalD{\plusminus \two}_\uv \,
  A^2
  \Jz_{\overset{{\scriptstyle v} {\scriptstyle v} \plus}{{\scriptstyle u} {\scriptstyle u} \minus}}
  +
  \ncalD{\plusminus \two}_\minusplus \,
  A^2
  \Jz_{\overset{\plus {\scriptstyle v} \plus}{\minus {\scriptstyle u} \minus}}
  - 2 \, \Gamma_{\overset{\plus {\scriptstyle v} {\scriptstyle v}}{\minus {\scriptstyle u} {\scriptstyle u}}}
  \, \Jz_{\overset{{\scriptstyle u} {\scriptstyle u} {\scriptstyle v}}{{\scriptstyle v} {\scriptstyle v} {\scriptstyle u}}}
  \  = \ 
  0
  \ .
\end{eqnarray}
\end{subequations}

Combining the linearized Weyl evolution equations~(\ref{Weyls}) in pairs,
by taking one of the operators $\nDelta{s}^\prime_m$ of equations~(\ref{Deltasprime}) times one equation,
minus another of the operators $\nDelta{s}^\prime_m$ times an adjacent equation,
the choice of operators being guided by the Weyl current conservation
equations~(\ref{linDJz}),
yields equations that look like sourced wave equations
for each of the (perturbations of the) spin components
$\Cz_s$ of the complexified Weyl tensor
(in a general gauge):
\begin{subequations}
\label{waveWeyl}
\begin{eqnarray}
\label{waveWeyla}
  \Bigl(
  \ncalD{\zero}^{\dagger \vu} \,
  \ncalD{\zero}_\vu
  +
  \ncalD{\zero}^{\dagger \plusminus} \,
  \ncalD{\zero}_\plusminus
  \Bigl)
  A \, \overset{\smallone}{\Cz}_\zero
  &=&
  \nDelta{\plusminus \one}^\prime_\uv
  \Bigl(
    - \,
    \nDelta{\zero}_\vu \,
    A^{-1}
    C
    +
    A^2 \,
    \Jz_{\overset{\minus {\scriptstyle v} \plus}{\plus {\scriptstyle u} \minus}}
  \Bigr)
  -
  \ncalD{\plusminus \one}_\minusplus
  \Bigl(
    - \,
    \nDelta{\zero}_\plusminus \,
    A^{-1}
    C
    +
    A^2 \,
    \Jz_{\overset{{\scriptstyle u} {\scriptstyle v} \plus}{{\scriptstyle v} {\scriptstyle u} \minus}}
  \Bigr)
  \ ,
  \qquad
\\
\label{waveWeylb}
  \Bigl(
  \ncalD{\plusminus \one}^{\dagger \uv} \,
  \ncalD{\plusminus \one}_\uv
  +
  \ncalD{\plusminus \one}^{\dagger \minusplus} \,
  \ncalD{\plusminus \one}_\minusplus
  \Bigl)
  A \, \Cz_{\plusminus \one}
  &=&
  \nDelta{\zero}^\prime_\plusminus
  \Bigl(
    - \,
    \nDelta{\zero}_\vu \,
    A^{-1}
    C
    +
    A^2 \,
    \Jz_{\overset{\minus {\scriptstyle v} \plus}{\plus {\scriptstyle u} \minus}}
  \Bigr)
  -
  \ncalD{\zero}_\vu
  \Bigl(
    - \,
    \nDelta{\zero}_\plusminus \,
    A^{-1}
    C
    +
    A^2 \,
    \Jz_{\overset{{\scriptstyle u} {\scriptstyle v} \plus}{{\scriptstyle v} {\scriptstyle u} \minus}}
  \Bigr)
  \ ,
\\
\label{waveWeylc}
  \Bigl(
  \ncalD{\plusminus \one}^{\dagger \vu} \,
  \ncalD{\plusminus \one}_\vu
  +
  \ncalD{\plusminus \one}^{\dagger \plusminus} \,
  \ncalD{\plusminus \one}_\plusminus
  \Bigl)
  A \, \Cz_{\plusminus \one}
  &=&
  \ncalD{\plusminus \two}_\uv
  \Bigl(
    3 \, \Gamma_{\overset{\plus {\scriptstyle v} {\scriptstyle v}}{\minus {\scriptstyle u} {\scriptstyle u}}} \,
    C
    +
    A^2 \,
    \Jz_{\overset{{\scriptstyle v} {\scriptstyle v} \plus}{{\scriptstyle u} {\scriptstyle u} \minus}}
  \Bigr)
  -
  \ncalD{\plusminus \two}_\minusplus
  \Bigl(
    3 \, \Gamma_{\overset{\plus {\scriptstyle v} \plus}{\minus {\scriptstyle u} \minus}} \,
    C
    +
    A^2 \,
    \Jz_{\overset{\plus {\scriptstyle v} \plus}{\minus {\scriptstyle u} \minus}}
  \Bigr)
  \ ,
\\
\label{waveWeyld}
  \Bigl(
  \ncalD{\plusminus \two}^{\dagger \uv} \,
  \ncalD{\plusminus \two}_\uv
  +
  \ncalD{\plusminus \two}^{\dagger \minusplus} \,
  \ncalD{\plusminus \two}_\minusplus
  \Bigl)
  A \, \Cz_{\plusminus \two}
  &=&
  \ncalD{\plusminus \one}_\plusminus
  \Bigl(
    3 \, \Gamma_{\overset{\plus {\scriptstyle v} {\scriptstyle v}}{\minus {\scriptstyle u} {\scriptstyle u}}} \,
    C
    +
    A^2 \,
    \Jz_{\overset{{\scriptstyle v} {\scriptstyle v} \plus}{{\scriptstyle u} {\scriptstyle u} \minus}}
  \Bigr)
  -
  \ncalD{\plusminus \one}_\vu
  \Bigl(
    3 \, \Gamma_{\overset{\plus {\scriptstyle v} \plus}{\minus {\scriptstyle u} \minus}} \,
    C
    +
    A^2 \,
    \Jz_{\overset{\plus {\scriptstyle v} \plus}{\minus {\scriptstyle u} \minus}}
  \Bigr)
  \ .
\end{eqnarray}
\end{subequations}
On the left hand side of equations~(\ref{waveWeyl}),
the leftmost of each pair of differential operators
has been replaced with an equivalent Hermitian conjugate operator,
per equations~(\ref{Dvuh}) and (\ref{Dpmh}),
which makes manifest the Hermitian character of the
wave operators on the left hand side.
It should be emphasized that
these are not independent wave equations:
the various spin components
$\Cz_s$
are not independently adjustable,
but rather fluctuate in harmony with each other
in accordance with the Weyl evolution equations~(\ref{Weyls}).

Equation~(\ref{waveWeyla})
apparently constitutes two separate equations for
$\Cz_\zero$,
while equations~(\ref{waveWeylb}) and (\ref{waveWeylc})
apparently provide two separate equations for each of
$\Cz_{\plusminus \one}$.
It is a straightforward if somewhat lengthy exercise
to confirm directly,
using
the commutation relations~(\ref{partialcommute}),
the operator relations~(\ref{diffDDpm}) and (\ref{diffDDvu}),
and the Weyl current conservation equations~(\ref{linDJz}),
that the separate equations for each
$\Cz_s$
are in fact equivalent.
For
$\Cz_\zero$,
the confirmation also invokes the equivalence of the
two expressions~(\ref{npRc}) for
$R_{v \plus u \minus}$,
while for
$\Cz_{\plusminus \one}$,
it is necessary to use
expression~(\ref{Cz1})
for $\Cz_{\plusminus \one}$
with the top lines of equations~(\ref{npRg}) and (\ref{npRi}) substituted.

It should be commented that
it is legitimate to replace
the differential operators
$\nDelta{\plusminus \one}^\prime_\uv$
and
$\nDelta{\zero}^\prime_\plusminus$
on the right hand sides of equations~(\ref{waveWeyla}) and (\ref{waveWeylb})
with their unperturbed limits
$\ncalD{\plusminus \one}_\uv$
and
$\ncalD{\zero}_\plusminus$,
since the combined quantities in parentheses that they operate on
are already of first order
(although the separate terms inside the parentheses
are not individually of first order).
However,
it is convenient to retain the operators in the unperturbed form,
since this makes application of the Weyl current conservation equations~(\ref{linDJz}),
as in the previous paragraph,
transparent.

One of the more complicated steps
in confirming that equations~(\ref{waveWeylb}) and (\ref{waveWeylc})
are equivalent is to derive
equation~(\ref{DeltaC}) given in the Appendix,
which linearizes (in a general gauge) to
\begin{equation}
\label{linDeltaC}
  \Bigl(
  \nDelta{\zero}^\prime_\plusminus \, \nDelta{\zero}_\vu
  -
  \nDelta{\zero}^\prime_\vu \, \nDelta{\zero}_\plusminus
  \Bigr)
  A^{-1} C
  =
  3 C
  \Bigl(
  2 A \, \Cz_{\plusminus \one}
  -
  A \,
  \ncalD{\plusminus \two}_\uv \,
  A^{-1}
  \Gamma_{\overset{\plus {\scriptstyle v} {\scriptstyle v}}{\minus {\scriptstyle u} {\scriptstyle u}}}
  +
  \ncalD{\plusminus \two}_\minusplus \,
  \Gamma_{\overset{\plus {\scriptstyle v} \plus}{\minus {\scriptstyle u} \minus}}
  \Bigr)
  -
  A \,
  \Gamma_{\overset{\plus {\scriptstyle v} {\scriptstyle v}}{\minus {\scriptstyle u} {\scriptstyle u}}} \,
  \ncalD{\zero}_\uv \,
  A^{-1} C
  \ ,
\end{equation}
the last term on the right hand side of which
can be converted to unperturbed Weyl currents
$\overset{\smallzero}{\Jz}_{lmn}$
using equation~(\ref{npWeylunperturbedvu}).

Another relation that proves useful
in reducing the wave equation
for the axial spin-$0$ component
$\Cz^{(a)}_\zero$,
\S\ref{axialgravitationalwaves} below,
is equation~(\ref{DeltaCa}) in the Appendix,
which linearizes (in a general gauge) to
\begin{eqnarray}
\label{linDeltaCa}
\lefteqn{
  \Bigl(
  \nDelta{\plus \one}^\prime_u \, \nDelta{\zero}_v
  -
  \nDelta{\plus \one}^\prime_\minus \, \nDelta{\zero}_\plus
  -
  \nDelta{\plus \one}^{\prime \star}_u \, \nDelta{\zero}^\star_v
  +
  \nDelta{\plus \one}^{\prime \star}_\minus \, \nDelta{\zero}^\star_\plus
  \Bigr)
  A^{-1} C
}
  &&
\nonumber
\\
  &=&
  6 C
  \, A \, \Cz^{(a)}_\zero
  -
  2 A
  \Bigl[
  (
    \Gamma_{\minus v \plus}
    -
    \Gamma_{\plus v \minus}
  ) \,
  \ncalD{\zero}_u
  +
  (
    \Gamma_{\plus u \minus}
    -
    \Gamma_{\minus u \plus}
  ) \,
  \ncalD{\zero}_v
  \Bigr]
  A^{-1} C
  \ ,
\end{eqnarray}
the last set of terms on the right hand side of which
can be converted to unperturbed Weyl currents
$\overset{\smallzero}{\Jz}_{lmn}$
using equation~(\ref{npWeylunperturbedvu}).

\subsection{Spin-two (propagating) components of gravitational waves}
\label{spin2gravitationalwaves}

The spin-$\pm 2$ components
$\Cz_{\plusminus \two}$
of the complexified Weyl tensor,
and their angular conjugates
$\Cz^\star_{\plusminus \two}$,
are of particular interest because
they describe the ingoing and outgoing propagating components of gravitational waves.
They are gauge-invariant with respect to
infinitesimal coordinate and tetrad transformations,
and vanishing in the background,
and therefore represent physical perturbations.
The main result of this subsection is the sourced spin-$\pm 2$ wave equation~(\ref{waveCz})
[and its angular conjugate~(\ref{waveCzstar})],
whose homogeneous solutions are separable as given by equations~(\ref{Cz2homogeneous}),
the radial part satisfying the generalized spin-$\pm 2$ Teukolsky equation~(\ref{DD2}).

The wave equation~(\ref{waveWeyld})
governing the evolution of each of
the spin-$\pm 2$ components
$\Cz_{\plusminus \two}$
is,
like $\Cz_{\plusminus \two}$ itself,
coordinate and tetrad gauge-invariant.
However,
there is an inherent ambiguity in deciding how to split the wave equation
into a left and a right hand side
-- a ``homogeneous'' part and a ``source'' part --
an ambiguity that is not resolved simply by writing the equation
in a form such as equation~(\ref{waveWeyld}).
The problem is that, although
$\Cz_{\plusminus \two}$
is itself coordinate and tetrad gauge-invariant,
the space of perturbations ``orthogonal'' to
$\Cz_{\plusminus \two}$
is not.
Indeed, as noted in \S\ref{perturbedeinstein},
the only coordinate and tetrad gauge-invariant perturbations
to the Einstein tensor are its spin-$\pm 2$ components
$G_{\overset{\plus \plus}{\minus \minus}}$.
For example,
the Einstein component
$G_{v \plus}$
[which appears in the Weyl currents
$\Jz_{\overset{{\scriptstyle v} {\scriptstyle v} \plus}{{\scriptstyle u} {\scriptstyle u} \minus}}$
and
$\Jz_{\overset{\plus {\scriptstyle v} \plus}{\minus {\scriptstyle u} \minus}}$,
equations~(\ref{linJze}) and (\ref{linJzf}),
on the right hand side of the wave equation~(\ref{waveWeyld})]
is not tetrad gauge-invariant,
so by a suitable infinitesimal tetrad gauge transformation
(that is, by a suitable choice of tetrad gauge)
it is possible in effect to add
to $G_{v \plus}$
arbitrary amounts of
$\ncalD{\plus \two}_\minus \Cz_{\plus \two}$
(the angular lowering operator
$\ncalD{\plus \two}_\minus$
being required to convert the spin-$2$ object
$\Cz_{\plus \two}$
into a spin-$1$ object like
$G_{v \plus}$).
The word ``arbitrary'' in the previous sentence is a bit too general:
it is desirable that the homogeneous part of the wave equation
should be (coordinate and tetrad) gauge-invariant,
so the amount of
$\ncalD{\plus \two}_\minus \Cz_{\plus \two}$
added to
$G_{v \plus}$
should be restricted to being gauge-invariant;
but this still allows a broad range of choices.
It would be possible to split
$G_{v \plus}$
unambigously into homogeneous and source parts respectively
proportional and orthogonal to
$\Cz_{\plusminus \two}$
if there were an unambiguous notion of orthogonality for
$\Cz_{\plusminus \two}$,
which would be true if
$\Cz_{\plusminus \two}$
were an eigenmode of some operator.
However,
$\Cz_{\plusminus \two}$,
whose evolution is governed by the
coupled Weyl evolution equations~(\ref{Weyls}),
is not an eigenmode of those equations,
except in the case of vanishing background energy-momentum tensor,
the Schwarzschild geometry.

The argument of the previous paragraph
shows that splitting the wave equation for
the spin-$\pm 2$ components
$\Cz_{\plusminus \two}$
into homogeneous and source parts
effectively involves choosing a gauge.
Intriguingly,
spherical gauge,
equations~(\ref{gaugespherical}),
which emerged
in \S\S\ref{perturbedconnections}
and \ref{perturbedriemann}
from the (not necessary, but nevertheless ``natural'')
requirement that all tetrad connections
be expandable in spin harmonics,
provides a split that proves particularly elegant.
In spherical gauge,
the connections
$\Gamma_{\overset{\plus {\scriptstyle v} \plus}{\minus {\scriptstyle u} \minus}}$
and
$\Gamma_{\overset{\plus {\scriptstyle u} \plus}{\minus {\scriptstyle v} \minus}}$
vanish,
and
$\Cz_{\plusminus \two}$
and
$\Cz^\star_{\plusminus \two}$
can be expressed in terms of tetrad connections as,
equations~(\ref{Cz2}) and (\ref{Cz2star}) with (\ref{linnpR2b}) and (\ref{linnpR2c}),
\begin{subequations}
\label{Cz2spherical}
\begin{eqnarray}
\label{Cz2aspherical}
  \Cz_{\plusminus \two}
  &=&
  A^{-1}
  \ncalD{\plusminus \one}_\plusminus \,
  \Gamma_{\overset{\plus {\scriptstyle v} {\scriptstyle v}}{\minus {\scriptstyle u} {\scriptstyle u}}}
  \ ,
\\
\label{Cz2bspherical}
  \Cz^\star_{\plusminus \two}
  &=&
  A^{-1}
  \ncalD{\minusplus \one}_\minusplus \,
  \Gamma_{\overset{\minus {\scriptstyle v} {\scriptstyle v}}{\plus {\scriptstyle u} {\scriptstyle u}}}
  \ .
\end{eqnarray}
\end{subequations}
For subsequent reference it is useful to record here that in spherical gauge
the spin-$\pm 2$ components
$G_{\overset{\plus \plus}{\minus \minus}}$
of the Einstein tensor are,
equations~(\ref{npEinsteinf}) and (\ref{linnpR2a}),
\begin{equation}
\label{G2spherical}
  G_{\overset{\plus \plus}{\minus \minus}}
  \  = \ 
  - \,
  2 \,
  A^{-1}
  \ncalD{\plusminus \one}_\plusminus \,
  \Gamma_{\overset{\plus {\scriptstyle v} {\scriptstyle u}}{\minus {\scriptstyle u} {\scriptstyle v}}}
  \ .
\end{equation}
Although the right hand sides of equations~(\ref{Cz2aspherical}), (\ref{Cz2bspherical}), and (\ref{G2spherical})
are not (tetrad) gauge-invariant,
they are evidently equivalent
to gauge-invariant quantities.
For non-dipole spherical harmonics, $\nY{s}_{lm}$ with $l \neq 1$,
equations~(\ref{Cz2aspherical}), (\ref{Cz2bspherical}), and (\ref{G2spherical})
invert to yield expressions for
$\Gamma_{\overset{\plus {\scriptstyle v} {\scriptstyle v}}{\minus {\scriptstyle u} {\scriptstyle u}}}$,
$\Gamma_{\overset{\minus {\scriptstyle v} {\scriptstyle v}}{\plus {\scriptstyle u} {\scriptstyle u}}}$,
and
$\Gamma_{\overset{\plus {\scriptstyle v} {\scriptstyle u}}{\minus {\scriptstyle u} {\scriptstyle v}}}$
(in spherical gauge)
in terms of
$\Cz_{\plusminus \two}$,
$\Cz^\star_{\plusminus \two}$,
and
$G_{\overset{\plus \plus}{\minus \minus}}$:
\begin{subequations}
\label{Gammaspherical}
\begin{eqnarray}
\label{Gammapvvspherical}
  \Gamma_{\overset{\plus {\scriptstyle v} {\scriptstyle v}}{\minus {\scriptstyle u} {\scriptstyle u}}}
  &=&
  - \,
  {2 A \over ( l - 1 ) ( l + 2 )} \,
  \ncalD{\plusminus \two}_\minusplus \,
  \Cz_{\plusminus \two}
  \ ,
\\
\label{Gammamvvspherical}
  \Gamma_{\overset{\minus {\scriptstyle v} {\scriptstyle v}}{\plus {\scriptstyle u} {\scriptstyle u}}}
  &=&
  - \,
  {2 A \over ( l - 1 ) ( l + 2 )} \,
  \ncalD{\minusplus \two}_\plusminus \,
  \Cz^\star_{\plusminus \two}
  \ ,
\\
\label{Gammamvuspherical}
  \Gamma_{\overset{\plus {\scriptstyle v} {\scriptstyle u}}{\minus {\scriptstyle u} {\scriptstyle v}}}
  &=&
  {A \over ( l - 1 ) ( l + 2 )} \,
  \ncalD{\plusminus \two}_\minusplus \,
  G_{\overset{\plus \plus}{\minus \minus}}
  \ .
\end{eqnarray}
\end{subequations}

The right hand side of the wave equation~(\ref{waveWeyld}) for
$\Cz_{\plusminus \two}$
contains the connections
$\Gamma_{\overset{\plus {\scriptstyle v} {\scriptstyle v}}{\minus {\scriptstyle u} {\scriptstyle u}}}$
in two places,
in a form that converts them precisely into
$\Cz_{\plusminus \two}$,
and it makes sense to take these terms over to the left hand side of the wave equation,
where they become incorporated into the wave operator.
The first of the two terms is proportional to the Weyl scalar $C$,
while the second occurs in the current
$\Jz_{\overset{{\scriptstyle v} {\scriptstyle v} \plus}{{\scriptstyle u} {\scriptstyle u} \minus}}$,
equation~(\ref{linJze}),
and is proportional to
$E \equiv \frac{1}{2} A^2 \bigl( \overset{\smallzero}{G}_{{\scriptstyle v} {\scriptstyle u}} + \overset{\smallzero}{G}_{\plus \minus} \bigr)$.
The two terms combine in the proportions
\begin{equation}
\label{waveCzc}
  - \,
  3 C
  +
  E
  \ .
\end{equation}

The resulting wave equation for
the spin-$\pm 2$ Weyl tensor
$\Cz_{\plusminus \two}$ is
\begin{equation}
\label{waveCz}
  \Bigl(
  \ncalD{\plusminus \two}^{\dagger \uv} \,
  \ncalD{\plusminus \two}_\uv
  +
  \ncalD{\plusminus \two}^{\dagger \minusplus} \,
  \ncalD{\plusminus \two}_\minusplus
  -
  3 C
  +
  E
  \Bigr)
  A \, \Cz_{\plusminus \two}
  =
  S_{\plusminus \two}
  \ ,
\end{equation}
where the spin-$\pm 2$ source term
$S_{\plusminus \two}$
on the right hand side
is, in spherical gauge,
\begin{equation}
\label{S2spherical}
  S_{\plusminus \two}
  =
  \frac{A}{2}
  \Bigl[
  - \,
  A^{-1}
  \ncalD{\plusminus \one}_\vu \,
  A^2 \,
  \ncalD{\zero}_\vu \,
  A^{-1}
  G_{\overset{\plus \plus}{\minus \minus}}
  +
  2 \,
  \ncalD{\plusminus \one}_\vu \,
  \ncalD{\plusminus \one}_\plusminus \,
  G_{\overset{{\scriptstyle v} \plus}{{\scriptstyle u} \minus}}
  -
  A \,
  \ncalD{\plusminus \one}_\plusminus
  \bigl( \partial_\plusminus
    + 2 \, \Gamma_{\overset{{\scriptstyle u} {\scriptstyle v} \plus}{{\scriptstyle v} {\scriptstyle u} \minus}}
    + \Gamma_{\overset{\plus {\scriptstyle u} {\scriptstyle v}}{\minus {\scriptstyle v} {\scriptstyle u}}}
  \bigr) G_{\overset{{\scriptstyle v} {\scriptstyle v}}{{\scriptstyle u} {\scriptstyle u}}}
  \Bigr]
  \ .
\end{equation}
Although spherical gauge has been invoked to split the wave equation~(\ref{waveCz})
into homogeneous and source parts,
the left hand (homogeneous) side of equation~(\ref{waveCz})
is coordinate and tetrad gauge-invariant,
and therefore the spin-$\pm 2$ source term
$S_{\plusminus \two}$
constituting the right hand side
must be equivalent to a
coordinate and tetrad gauge-invariant object.
In fact the full (gauge-invariant) expression for the source term
$S_{\plusminus \two}$
is
(in a general gauge, not spherical gauge)
\begin{eqnarray}
\label{S2}
  S_{\plusminus \two}
  &=&
  - \,
  \ncalD{\plusminus \one}_\vu \,
  A^2
  \bigl[
    \Gamma_{\overset{\plus {\scriptstyle v} \plus}{\minus {\scriptstyle u} \minus}}
    \bigl(
      \overset{\smallzero}{G}_{vu} + \overset{\smallzero}{G}_{\plus \minus}
    \bigr)
    +
    {\textstyle \frac{1}{2}} \,
    \Gamma_{\overset{\plus {\scriptstyle u} \plus}{\minus {\scriptstyle v} \minus}} \,
    \overset{\smallzero}{G}_{\overset{{\scriptstyle v} {\scriptstyle v}}{{\scriptstyle u} {\scriptstyle u}}}
  \bigr]
  +
  {\textstyle \frac{1}{2}} \,
  A^3 \,
  \Gamma_{\overset{\plus {\scriptstyle v} \plus}{\minus {\scriptstyle u} \minus}} \,
  \ncalD{\plusminus \two}_\uv \,
  A^{-1}
  \overset{\smallzero}{G}_{\overset{{\scriptstyle v} {\scriptstyle v}}{{\scriptstyle u} {\scriptstyle u}}}
\nonumber
\\
  &&
  + \,
  \frac{A}{2}
  \Bigl[
  - \,
  A^{-1}
  \ncalD{\plusminus \one}_\vu \,
  A^2 \,
  \ncalD{\zero}_\vu \,
  A^{-1}
  G_{\overset{\plus \plus}{\minus \minus}}
  +
  2 \,
  \ncalD{\plusminus \one}_\vu \,
  \ncalD{\plusminus \one}_\plusminus \,
  G_{\overset{{\scriptstyle v} \plus}{{\scriptstyle u} \minus}}
  -
  A \,
  \ncalD{\plusminus \one}_\plusminus
  \bigl( \partial_\plusminus
    + 2 \, \Gamma_{\overset{{\scriptstyle u} {\scriptstyle v} \plus}{{\scriptstyle v} {\scriptstyle u} \minus}}
    + \Gamma_{\overset{\plus {\scriptstyle u} {\scriptstyle v}}{\minus {\scriptstyle v} {\scriptstyle u}}}
  \bigr) G_{\overset{{\scriptstyle v} {\scriptstyle v}}{{\scriptstyle u} {\scriptstyle u}}}
  \Bigr]
  \ .
\end{eqnarray}
It is evident from equation~(\ref{S2}) that spherical gauge,
where
$\Gamma_{\overset{\plus {\scriptstyle v} \plus}{\minus {\scriptstyle u} \minus}}$
and
$\Gamma_{\overset{\plus {\scriptstyle u} \plus}{\minus {\scriptstyle v} \minus}}$
vanish,
indeed leads to a significant simplification
of the source term.

Spherical gauge, equation~(\ref{gaugespherical}),
does not fix the gauge uniquely,
since it still leaves two infinitesimal gauge freedoms,
corresponding to 1 coordinate freedom in the choice of radial coordinate $r$,
and 1 tetrad gauge freedom in the antisymmetric part of $\varphi_{vu}$
(a Lorentz boost in the radial direction).
It is worth pointing out
that these gauge freedoms are precisely sufficient
to set the contribution from the unperturbed part of
$G_{\overset{{\scriptstyle v} {\scriptstyle v}}{{\scriptstyle u} {\scriptstyle u}}}$
in the source term
$S_{\plusminus \two}$,
equation~(\ref{S2spherical}),
to zero
(although in the next paragraph it will be argued
that this may not be a preferred gauge choice).
That is,
the contribution to the source term
$S_{\plusminus \two}$
from
$G_{\overset{{\scriptstyle v} {\scriptstyle v}}{{\scriptstyle u} {\scriptstyle u}}}$
splits into parts depending on the unperturbed Einstein component
$\overset{\smallzero}{G}_{\overset{{\scriptstyle v} {\scriptstyle v}}{{\scriptstyle u} {\scriptstyle u}}}$
and its perturbation
$\overset{\smallone}{G}_{\overset{{\scriptstyle v} {\scriptstyle v}}{{\scriptstyle u} {\scriptstyle u}}}$
as
\begin{equation}
\label{S2Gvv}
  \bigl( \partial_\plusminus
    + 2 \, \Gamma_{\overset{{\scriptstyle u} {\scriptstyle v} \plus}{{\scriptstyle v} {\scriptstyle u} \minus}}
    + \Gamma_{\overset{\plus {\scriptstyle u} {\scriptstyle v}}{\minus {\scriptstyle v} {\scriptstyle u}}}
  \bigr)
  G_{\overset{{\scriptstyle v} {\scriptstyle v}}{{\scriptstyle u} {\scriptstyle u}}}
  =
  \bigl(
    \varphi_{\overset{\plus {\scriptstyle u}}{\minus {\scriptstyle v}}}
    \partial_\vu
    +
    \varphi_{\overset{\plus {\scriptstyle v}}{\minus {\scriptstyle u}}}
    \partial_\uv
    + 2 \, \Gamma_{\overset{{\scriptstyle u} {\scriptstyle v} \plus}{{\scriptstyle v} {\scriptstyle u} \minus}}
    + \Gamma_{\overset{\plus {\scriptstyle u} {\scriptstyle v}}{\minus {\scriptstyle v} {\scriptstyle u}}}
  \bigr)
  \overset{\smallzero}{G}_{\overset{{\scriptstyle v} {\scriptstyle v}}{{\scriptstyle u} {\scriptstyle u}}}
  +
  A^{-1}
  \ncalD{\zero}_\plus \,
  \overset{\smallone}{G}_{\overset{{\scriptstyle v} {\scriptstyle v}}{{\scriptstyle u} {\scriptstyle u}}}
  \ .
\end{equation}
Since one of
$\varphi_{\overset{\plus {\scriptstyle u}}{\minus {\scriptstyle v}}}$
or
$\varphi_{\overset{\plus {\scriptstyle v}}{\minus {\scriptstyle u}}}$
can be adjusted with the coordinate gauge freedom in
$\epsilon^v$ or $\epsilon^u$,
equations~(\ref{gaugephipv}) and (\ref{gaugephimv}),
while one of
$\Gamma_{\overset{{\scriptstyle u} {\scriptstyle v} \plus}{{\scriptstyle v} {\scriptstyle u} \minus}}$
can be adjusted with the tetrad gauge freedom in $\varphi_{vu}$,
equation~(\ref{Gammauvp}),
the two gauge freedoms available in spherical gauge
are just such as to make it possible to set
\begin{equation}
\label{gaugeGvv}
  \bigl(
    \varphi_{\overset{\plus {\scriptstyle u}}{\minus {\scriptstyle v}}}
    \partial_\vu
    +
    \varphi_{\overset{\plus {\scriptstyle v}}{\minus {\scriptstyle u}}}
    \partial_\uv
    + 2 \, \Gamma_{\overset{{\scriptstyle u} {\scriptstyle v} \plus}{{\scriptstyle v} {\scriptstyle u} \minus}}
    + \Gamma_{\overset{\plus {\scriptstyle u} {\scriptstyle v}}{\minus {\scriptstyle v} {\scriptstyle u}}}
  \bigr)
  \overset{\smallzero}{G}_{\overset{{\scriptstyle v} {\scriptstyle v}}{{\scriptstyle u} {\scriptstyle u}}}
  =
  0
  \ .
\end{equation}
If this gauge choice is made, then the spin-$\pm 2$ source term becomes
\begin{equation}
\label{S2total}
  S_{\plusminus \two}
  =
  \frac{A}{2}
  \Bigl[
  - \,
  A^{-1}
  \ncalD{\plusminus \one}_\vu \,
  A^2 \,
  \ncalD{\zero}_\vu \,
  A^{-1}
  G_{\overset{\plus \plus}{\minus \minus}}
  +
  2 \,
  \ncalD{\plusminus \one}_\vu \,
  \ncalD{\plusminus \one}_\plusminus \,
  G_{\overset{{\scriptstyle v} \plus}{{\scriptstyle u} \minus}}
  -
  \ncalD{\plusminus \one}_\plusminus \,
  \ncalD{\zero}_\plus \,
  \overset{\smallone}{G}_{\overset{{\scriptstyle v} {\scriptstyle v}}{{\scriptstyle u} {\scriptstyle u}}}
  \Bigr]
  \ ,
\end{equation}
which is precisely the form that the spin-$\pm 2$ source term
takes in the case of the Schwarzschild geometry,
where all unperturbed components of the Einstein tensor vanish,
and all perturbations to the tetrad frame Einstein tensor are
coordinate and tetrad gauge-invariant.

However,
the gauge conditions~(\ref{gaugeGvv}),
though feasible, are ad hoc,
unlike the spherical gauge conditions~(\ref{gaugespherical}),
which have a ``natural'' origin
(that all tetrad connections be expandable in spin harmonics).
Moreover the gauge conditions~(\ref{gaugeGvv})
represent a rather complicated condition on the vierbein perturbations $\varphi_{mn}$,
which may not be ideal for numerical calculations.
Indeed,
in a hypothetical asymptotically flat empty region far from the black hole,
the unperturbed Einstein component
$\overset{\smallzero}{G}_{\overset{{\scriptstyle v} {\scriptstyle v}}{{\scriptstyle u} {\scriptstyle u}}}$
vanishes asymptotically,
so that the conditions~(\ref{gaugeGvv})
are satisfied automatically rather than being gauge conditions,
and then how the conditions~(\ref{gaugeGvv}) translate into conditions on
the vierbein perturbations $\varphi_{mn}$
may become a numerically delicate issue.


Equation~(\ref{S2spherical})
shows that the tensor (spin-$\pm 2$) components of gravitational waves
couple not only to tensor sources ($G_{\overset{\plus \plus}{\minus \minus}}$)
but also to vector
($G_{\overset{{\scriptstyle v} \plus}{{\scriptstyle u} \minus}}$)
and scalar
($G_{\overset{{\scriptstyle v} {\scriptstyle v}}{{\scriptstyle u} {\scriptstyle u}}}$)
sources.
If there is a non-vanishing background energy-momentum,
then gravitational waves can excite other (non-gravitational) kinds of waves in the energy-momentum,
which in turn provide a source for gravitational waves.
The coupling of modes in black holes
is unlike the case of the Friedmann-Robertson-Walker metric of cosmology,
where
spatial translation and rotation symmetries
ensure that
scalar, vector, and tensor modes decouple completely from each other
\cite{Bardeen80,MFB92,Brandenberger05}.

The angular conjugates
$\Cz^\star_{\plusminus \two}$
of the spin-$\pm 2$ components of the complexified Weyl tensor
satisfy the angular conjugate of equation~(\ref{waveCz})
\begin{equation}
\label{waveCzstar}
  \Bigl(
  \ncalD{\plusminus \two}^{\dagger \uv} \,
  \ncalD{\plusminus \two}_\uv
  +
  \ncalD{\minusplus \two}^{\dagger \plusminus} \,
  \ncalD{\minusplus \two}_\plusminus
  -
  3 C
  +
  E
  \Bigr)
  A \, \Cz^\star_{\plusminus \two}
  =
  S^\star_{\plusminus \two}
  \ .
\end{equation}
The angular Hermitian operators
$\ncalD{\plus \two}^{\dagger \minus} \ncalD{\plus \two}_\minus$
and
$\ncalD{\minus \two}^{\dagger \plus} \ncalD{\minus \two}_\plus$
inside the wave operators
of equation~(\ref{waveCz}) and its angular conjugate~(\ref{waveCzstar})
have the same eigenvalues
(with angular eigenfunctions that are complex conjugates of each other),
in view of equations~(\ref{DpmhDpm}),
and it follows that
the temporal-radial (i.e.\ non-angular) part of the eigenfunctions of the wave operators
are the same for both
$\Cz_{\plusminus \two}$
and its angular conjugate
$\Cz^\star_{\plusminus \two}$.
However,
the temporal-radial parts of
the source terms
$S_{\plusminus \two}$
are in general not equal to their angular conjugates
$S^\star_{\plusminus \two}$.

The wave operator on the left hand side of equation~(\ref{waveCz})
differs from the wave operator discussed in
\S\ref{spinradialoperators}
only by the addition of the terms
$- \, 3 C + E$,
which are functions only of conformal radius $r$.
The homogeneous solutions
(when the source terms vanish)
of the wave equation~(\ref{waveCz})
and its angular conjugate~(\ref{waveCzstar})
are, similarly to equation~(\ref{Psi}),
\begin{equation}
\label{Cz2homogeneous}
  \left.
  \begin{array}{r}
    \Cz_{\plusminus \two} \\
    \Cz^\star_{\plusminus \two}
  \end{array}
  \right\}
  =
  ( {\xi^v / \xi^u} )^{\mp 1}
  A^{-3} \,
  \ee^{- \im \omega t}
  \ncz{\plusminus \two}_{\omega l}(r) \,
  \left\{
  \begin{array}{l}
    \nY{\plusminus \two}_{lm}(\theta,\phi) \\
    \nY{\plusminus \two}^\ast_{lm}(\theta,\phi)
  \end{array}
  \right.
\end{equation}
in which
$\Cz_{\plusminus \two}$
and its angular conjugate
$\Cz^\star_{\plusminus \two}$
differ only in that their angular parts
are complex conjugates of each other.
Note that the complex conjugate of a harmonic of spin $s$
has opposite spin $-s$,
in accordance with equation~(\ref{Yconjugate})
and in agreement with equation~(\ref{CzsstarY}).
The polar and axial parts of the eigenfunctions~(\ref{Cz2homogeneous})
have the same temporal-radial dependence, and
simply project out the real and imaginary parts of the angular factor,
the spin-weighted spherical harmonic
$\nY{\plusminus \two}_{lm}(\theta,\phi)$.
The radial function
$\ncz{\plusminus \two}_{\omega l}(r)$
in equation~(\ref{Cz2homogeneous})
depends only on conformal radius $r$,
and satisfies the second order ordinary differential equation
\begin{equation}
\label{DD2}
  \left\{
  - \,
  \Dr{2}{}
  +
  \bigl(
    \im \omega
    \, \pm \, 2 \kappa
  \bigr)^2
  +
  \bigl[
    l ( l + 1 )
    - 4
    - 18 C + 2 E
  \bigr]
  \Delta
  \right\}
  \ncz{\plusminus \two}_{\omega l}
  =
  0
  \ ,
\end{equation}
which agrees with the generalized Teukolsky equation~(\ref{teukolsky})
given in the Introduction, since,
equations~(\ref{C}) and (\ref{E}),
\begin{equation}
  - 18 C + 2 E
  =
  - 5 ( R - P ) - 4 P_\perp + 18 M
  \ .
\end{equation}

Equation~(\ref{DD2})
has two linearly independent eigensolutions,
which may be identified as ingoing and outgoing
based on their properties
in a hypothetical asymptotically flat region at large radius,
as already discussed in \S\ref{asymptotic}.
In a hypothetical aysmptotically flat region at large radius,
the radial eigensolutions
$\ncz{s}_{\omega l}$
(with $s = \pm 2$)
go over to the asymptotic solutions
$\nz{s}_l$
discussed in \S\ref{asymptotic},
equation~(\ref{zasymptotic}),
\begin{equation}
\label{casymptotic}
  \begin{array}{rcccl}
    \ncz{s}^\textrm{in}_{\omega l}
    &\rightarrow&
    \nz{s}^\textrm{in}_l(\omega r^\ast)
    &\rightarrow&
    \displaystyle{
    \left[ {(l - s)! \over (l + s)!} \right]^{1/2}
    (2 \omega r^\ast)^s
    \ee^{- \im \omega r^\ast}
    }
    \\
    \ncz{s}^\textrm{out}_{\omega l}
    &\rightarrow&
    \nz{s}^\textrm{out}_l(\omega r^\ast)
    &\rightarrow&
    \displaystyle{
    \left[ {(l + s)! \over (l - s)!} \right]^{1/2}
    (2 \omega r^\ast)^{- s}
    \ee^{\im \omega r^\ast}
    }
  \end{array}
  \quad
  \mbox{as }
  r^\ast \rightarrow \infty
  \ .
\end{equation}
Because the wave operators for modes of opposite spin are
complex conjugates of each other,
equation~(\ref{DD2}),
it follows that
the ingoing and outgoing radial eigensolutions
$\ncz{s}_{\omega l}$
of opposite spin can be taken to be
complex conjugates of each other,
in complete analogy to equation~(\ref{zinout}),
\begin{equation}
\label{cinout}
  \ncz{s}^\textrm{in}_{\omega l}
  =
  \left( \ncz{\minus s}^\textrm{out}_{\omega l} \right)^\ast
  \ .
\end{equation}
The propagating components of
$\Cz_s \propto A^{-3} \ee^{- \im \omega t} \ncz{s}_{\omega l}$
are those that fall off most slowly at large radius,
which identifies the positive spin component
$\Cz_{\plus \two}$
as describing the propagating component of ingoing gravitational waves,
and the negative spin component
$\Cz_{\minus \two}$
as describing the propagating component of outgoing gravitational waves:
\begin{subequations}
\begin{eqnarray}
  \Cz^\textrm{in}_{\plus \two}
  \  \propto \ 
  A^{-3} \ee^{- \im \omega t}
  \ncz{\plus \two}^\textrm{in}_{\omega l}
  \  \sim \ 
  A^{-3} (r^\ast)^2 \ee^{- \im \omega ( t + r^\ast )}
  &\quad&
  ( \mbox{propagating, ingoing} )
  \ ,
\\
  \Cz^\textrm{out}_{\minus \two}
  \  \propto \ 
  A^{-3} \ee^{- \im \omega t}
  \ncz{\minus \two}^\textrm{out}_{\omega l}
  \  \sim \ 
  A^{-3} (r^\ast)^2 \ee^{- \im \omega ( t - r^\ast )}
  &\quad&
  ( \mbox{propagating, outgoing} )
  \ .
\end{eqnarray}
\end{subequations}
The other of the two modes for each spin corresponds to the non-propagating,
short-range spin-$(-s)$ partner of a propagating mode of spin $s$:
\begin{subequations}
\begin{eqnarray}
  \Cz^\textrm{in}_{\minus \two}
  \  \propto \ 
  A^{-3} \ee^{- \im \omega t}
  \ncz{\minus \two}^\textrm{in}_{\omega l}
  \  \sim \ 
  A^{-3} (r^\ast)^{-2} \ee^{- \im \omega ( t + r^\ast )}
  &\quad&
  ( \mbox{non-propagating, ingoing} )
  \ ,
\\
  \Cz^\textrm{out}_{\plus \two}
  \  \propto \ 
  A^{-3} \ee^{- \im \omega t}
  \ncz{\plus \two}^\textrm{out}_{\omega l}
  \  \sim \ 
  A^{-3} (r^\ast)^{-2} \ee^{- \im \omega ( t - r^\ast )}
  &\quad&
  ( \mbox{non-propagating, outgoing} )
  \ .
\end{eqnarray}
\end{subequations}

\subsection{From spin-$\pm 2$ to general spin-$s$ components of gravitational waves}
\label{spinsgravitationalwaves}

The Weyl evolution equations~(\ref{Weyls})
relate the five spin
$s = -2$ to $+ 2$
components
$\Cz_s$
of the complexified Weyl tensor to each other,
so that once one component has been specified,
then the others follow.
Thus,
except for monopole and dipole harmonics,
$l = 0$ or $1$,
for which all spin-$\pm 2$ quantities vanish identically,
and which will be considered specially in
\S\S\ref{monopolegravitationalwaves} and \ref{dipolegravitationalwaves},
either of the spin-$\pm 2$ components
$\Cz_{\plusminus \two}$
of the complexified Weyl tensor
is sufficient to determine all the other Weyl components.
%
Starting with either of the
components
$\Cz_{\plus \two}$
or
$\Cz_{\minus \two}$,
the Weyl evolution equations,
reordered in the sequence
(\ref{Weylsd}), (\ref{Weylsb}), (\ref{Weylsa}), and (\ref{Weylsc}),
yield in succession
\begin{equation}
\label{Weylsequence}
  \Cz_{\plusminus \two}
  \rightarrow
  \Cz_{\plusminus \one}
  \rightarrow
  \overset{\smallone}{\Cz}_{\zero}
  \rightarrow
  \Cz_{\minusplus \one}
  \rightarrow
  \Cz_{\minusplus \two}
  \ .
\end{equation}
The angular raising/lowering operators
$\ncalD{s}_\plusminus$
involved in extracting each successive spin-$s$ component
can be canceled with corresponding angular lowering/raising operators
$\ncalD{s \plus \one}_\minusplus$,
and then
equation~(\ref{Weylsd})
yields
$\Cz_{\plusminus \one}$ in terms of
$\Cz_{\plusminus \two}$ (plus other things),
equation~(\ref{Weylsb})
yields
$\overset{\smallone}{\Cz}_{\zero}$ in terms of
$\Cz_{\plusminus \one}$ (plus other things),
and so on.

It is not worth writing out explicitly all of the resulting set of expressions
for the various spin-$s$ components
$\Cz_s$ of the Weyl tensor,
because the expressions for the most part
offer little insight beyond that already in the Weyl evolution equations~(\ref{Weyls}).

However,
the first rung on the ladder,
equation~(\ref{Weylsd}),
which yields
$\Cz_{\plusminus \one}$ in terms of
$\Cz_{\plusminus \two}$ plus other things,
does deserve comment,
because it reveals explicitly how
the spin-$\pm 1$ and spin-$\pm 2$ components of the Weyl tensor
fluctuate together rather than varying independently.
%
%
%
In spherical gauge, and for non-dipole modes,
the Weyl evolution
equation~(\ref{Weylsd})
integrates to
\begin{equation}
\label{Czp1}
    \Cz_{\plusminus \one}
    +
    {\textstyle \frac{1}{2}}
    G_{\overset{{\scriptstyle v} \plus}{{\scriptstyle u} \minus}}
  =
  - \,
  {2 \over ( l - 1 ) ( l + 2 )} \,
  \ncalD{\plusminus \two}_\minusplus \,
  \Bigl(
    A^{-1}
    \ncalD{\plusminus \two}_\uv \,
    A \,
    \Cz_{\plusminus \two}
    +
    {\textstyle \frac{1}{2}} \,
    A \,
    \ncalD{\zero}_\vu \,
    A^{-1}
    G_{\overset{\plus \plus}{\minus \minus}}
  \Bigr)
  \quad
  (l \neq 1)
  \ ,
\end{equation}
which expresses the combination
$\Cz_{\plusminus \one} + {\textstyle \frac{1}{2}} G_{\overset{{\scriptstyle v} \plus}{{\scriptstyle u} \minus}}$
of spin-$\pm 1$
components of the Weyl and Einstein tensors
in terms of the
spin-$\pm 2$
components
$\Cz_{\plusminus \two}$
and
$G_{\overset{\plus \plus}{\minus \minus}}$
of the Weyl and Einstein tensors.
Equation~(\ref{Czp1}) thus shows that,
as asserted,
the spin-$\pm 1$ and spin-$\pm 2$ components of the Weyl tensor
fluctuate together rather than varying independently.
In spherical gauge,
the spin-$\pm 2$ components of the Weyl and Einstein tensors
are related to tetrad connections by equations~(\ref{Cz2spherical}) and (\ref{G2spherical}),
and equation~(\ref{Czp1}) can be rewritten as
\begin{equation}
\label{Czp1integrated}
  \Cz_{\plusminus \one}
  +
  {\textstyle \frac{1}{2}}
  G_{\overset{{\scriptstyle v} \plus}{{\scriptstyle u} \minus}}
  =
  A^{-1}
  \Bigl(
  \ncalD{\plusminus \two}_\uv \,
  \Gamma_{\overset{\plus {\scriptstyle v} {\scriptstyle v}}{\minus {\scriptstyle u} {\scriptstyle u}}}
  -
  \ncalD{\zero}_\vu \,
  \Gamma_{\overset{\plus {\scriptstyle v} {\scriptstyle u}}{\minus {\scriptstyle u} {\scriptstyle v}}}
  \Bigr)
  \ ,
\end{equation}
which can be recognized as the same as the Riemann equation~(\ref{linnpR1a}),
expressed in spherical gauge
(where
$\Gamma_{\overset{\plus {\scriptstyle v} {\scriptstyle u}}{\minus {\scriptstyle u} {\scriptstyle v}}} = \Gamma_{\overset{\plus {\scriptstyle u} {\scriptstyle v}}{\minus {\scriptstyle v} {\scriptstyle u}}}$),
and is therefore valid for dipole modes $l = 1$ as well as non-dipole modes.

\subsection{Gravitational monopole modes}
\label{monopolegravitationalwaves}

For monopole or dipole modes, $l = 0$ or $1$,
all spin-$\pm 2$ quantities vanish identically,
and therefore the spin-$\pm 2$ components
$\Cz_{\plusminus \two}$
of the complexified Weyl tensor
cannot describe these modes.
Thus monopole and dipole modes,
which as emphasized by \cite{WP05}
are essential to a complete description of gravitational perturbations,
require special attention.
The present subsection covers monopole modes,
while the following subsection, \S\ref{dipolegravitationalwaves}, will deal with dipole modes.
The main results of this subsection are:
the polar monopole equation~(\ref{DNmonopole}),
which simplifies in spherical gauge to equation~(\ref{Nmonopole});
and the axial monopole equation~(\ref{Czamonopole}),
which states simply that the axial monopole vanishes identically.

An equation for the monopole follows from the first Weyl evolution equation~(\ref{Weylsa}).
After some manipulation,
the Weyl evolution equation~(\ref{Weylsa})
can be recast
in the following form
(in a general gauge, and for arbitrary harmonics),
with the feature that every term is individually of first order:
\begin{eqnarray}
\label{DvuN}
  &&
  A^2 
  \bigl(
    \partial_\vu
    +
    \Gamma_{\overset{\minus {\scriptstyle v} \plus}{\plus {\scriptstyle u} \minus}}
    +
    \Gamma_{\overset{\plus {\scriptstyle v} \minus}{\minus {\scriptstyle u} \plus}}
  \bigr)
  A^{-2} \gravN
  +
  \ncalD{\zero}_\vu \,
  A \, \Cz^{(a)}_\zero
  -
  \ncalD{\plusminus \one}_\minusplus \, A \, \Cz_{\plusminus \one}
  \  = \ 
  {\textstyle \frac{1}{2}}
  \bigl(
    \Gamma_{\overset{\minus {\scriptstyle v} \plus}{\plus {\scriptstyle u} \minus}}
    -
    \Gamma_{\overset{\plus {\scriptstyle v} \minus}{\minus {\scriptstyle u} \plus}}
  \bigr)
  ( 3 C + E )
  -
  {\textstyle \frac{1}{4}}
  A^2
  \bigl(
    \Gamma_{\overset{\plus {\scriptstyle u} \minus}{\minus {\scriptstyle v} \plus}}
    -
    \Gamma_{\overset{\minus {\scriptstyle u} \plus}{\plus {\scriptstyle v} \minus}}
  \bigr)
  \overset{\smallzero}{G}_{\overset{{\scriptstyle v v}}{{\scriptstyle u u}}}
\nonumber
\\
  &&
  \qquad
  + \,
  {\textstyle \frac{1}{2}} \,
  \beta_\uv
  \bigl(
    \ncalD{\plusminus \one}_\minusplus \,
    \Gamma_{\overset{\plus {\scriptstyle v v}}{\minus {\scriptstyle u u}}}
    +
    \ncalD{\minusplus \one}_\plusminus \,
    \Gamma_{\overset{\minus {\scriptstyle v v}}{\plus {\scriptstyle u u}}}
  \bigr)
  +
  {\textstyle \frac{1}{2}} \,
  \beta_\vu
  \bigl(
    \ncalD{\plusminus \one}_\minusplus \,
    \Gamma_{\overset{\plus {\scriptstyle v u}}{\minus {\scriptstyle u v}}}
    +
    \ncalD{\minusplus \one}_\plusminus \,
    \Gamma_{\overset{\minus {\scriptstyle v u}}{\plus {\scriptstyle u v}}}
  \bigr)
  -
  {\textstyle \frac{1}{2}}
  A \,
  \ncalD{\minusplus \one}_\plusminus \,
  G_{\overset{{\scriptstyle v} \minus}{{\scriptstyle u} \plus}}
  \ ,
\end{eqnarray}
where
$\Cz^{(a)}_\zero$
is the axial spin-$0$ component of the Weyl tensor,
and
$\gravN$
is defined to be the following dimensionless combination
of polar spin-$0$ components of
the Weyl tensor, Einstein tensor, and tetrad connections
\begin{equation}
\label{gravN}
  \gravN
  \equiv
  A^2
  \big[
  \Cz^{(p)}_\zero
  -
  {\textstyle \frac{1}{3}}
  G_{vu}
  -
  {\textstyle \frac{1}{6}}
  G_{\plus \minus}
  +
  {\textstyle \frac{1}{4}}
  \bigl(
    \Gamma_{\minus v \plus}
    +
    \Gamma_{\plus v \minus}
  \bigr)
  \bigl(
    \Gamma_{\plus u \minus}
    +
    \Gamma_{\minus u \plus}
  \bigr)
  \bigr]
  \ .
\end{equation}
The combination of 
polar spin-$0$ components of the Weyl and Einstein tensors in the definition~(\ref{gravN}) of $\gravN$
is proportional to one of the Riemann components,
$A^2 \bigl( \Cz^{(p)}_\zero - {\textstyle \frac{1}{3}} G_{vu} - {\textstyle \frac{1}{6}} G_{\plus \minus} \bigr) = \frac{1}{2} A^2 R_{\plus \minus \plus \minus}$,
which in the unperturbed background
reduces to minus the dimensionless interior mass, $- M$,
equation~(\ref{unperturbednpRb}).
The polar spin-$0$ quantity $\gravN$ is just
$\frac{1}{2} A^2$ times the object on the left hand side of
equation~(\ref{Nlightbulb}),
so the expression for $\gravN$
in terms of vierbein perturbations $\varphi_{mn}$
is given by
$\frac{1}{2} A^2$ times
the right hand side of equation~(\ref{Nlightbulb}).
The quantity $\gravN$ is not coordinate gauge-invariant,
although $A^{-2} \gravN$ is coordinate gauge-invariant,
and $\gravN$ is not tetrad gauge-invariant,
being changed by infinitesimal tetrad gauge transformations
of any of the antisymmetric parts of
$\varphi_{\plus v}$,
$\varphi_{\plus u}$,
$\varphi_{\minus v}$,
or
$\varphi_{\minus u}$,
equation~(\ref{Nlightbulb}).
However, the monopole part of $\gravN$ is tetrad gauge-invariant,
since
$\varphi_{\plus v}$,
$\varphi_{\plus u}$,
$\varphi_{\minus v}$,
and
$\varphi_{\minus u}$
are spin-$\pm 1$ objects,
whose monopole parts therefore vanish identically.
The notable asset of $\gravN$
is that the split between its unperturbed and perturbed parts
is coordinate and tetrad gauge-invariant,
the unperturbed part being simply a constant,
a coordinate and tetrad gauge-invariant object,
\begin{equation}
\label{unperturbedN}
  \overset{\smallzero}{\gravN}
  =
  - \,
  {\textstyle \frac{1}{2}}
  \ .
\end{equation}


For the monopole mode, $l = 0$,
all non-zero spin objects in equation~(\ref{DvuN})
vanish identically
[including
$\Gamma_{\minus {\scriptstyle v} \plus} - \Gamma_{\plus {\scriptstyle v} \minus}$
and
$\Gamma_{\plus {\scriptstyle u} \minus} - \Gamma_{\minus {\scriptstyle u} \plus}$,
whose expressions~(\ref{Gammapvmm}) in terms of vierbein components $\varphi_{mn}$
contain only spin-$\pm 1$ terms],
and the equation turns into one that determines
the monopole to linear order:
\begin{equation}
\label{Czmonopole}
  A^2
  \Bigl(
    \partial_\vu
    +
    \Gamma_{\overset{\minus {\scriptstyle v} \plus}{\plus {\scriptstyle u} \minus}}
    +
    \Gamma_{\overset{\plus {\scriptstyle v} \minus}{\minus {\scriptstyle u} \plus}}
  \Bigr)
  A^{-2} \gravN
  +
  \ncalD{\zero}_\uv \,
  A \, \Cz^{(a)}_\zero
  =
  0
  \quad
  ( l = 0 )
  \ ,
\end{equation}
a coordinate and tetrad gauge-invariant equation.

The polar part of the monopole equation~(\ref{Czmonopole}) is,
in a general gauge,
\begin{equation}
\label{DNmonopole}
  A^2
  \Bigl(
    \partial_\vu
    +
    \Gamma_{\overset{\minus {\scriptstyle v} \plus}{\plus {\scriptstyle u} \minus}}
    +
    \Gamma_{\overset{\plus {\scriptstyle v} \minus}{\minus {\scriptstyle u} \plus}}
  \Bigr)
  A^{-2} \gravN
  =
  0
  \quad
  ( l = 0 )
  \ .
\end{equation}
In spherical gauge,
where equation~(\ref{dAa}) holds,
equation~(\ref{DNmonopole})
is equivalent to
$\partial_\vu \gravN = 0$,
which integrates to
\begin{equation}
\label{Nmonopole}
  \gravN
  =
  - \,
  {\textstyle \frac{1}{2}}
  \quad
  ( l = 0 )
  \ ,
\end{equation}
the constant of integration being determined
by the unperturbed value~(\ref{unperturbedN}).
The fact that
in spherical gauge
the monopole component of the polar spin-$0$ quantity $\gravN$
equals $- \frac{1}{2}$
not only to unperturbed order but also to linear order
is a pretty result,
and definitely another feather in the cap of spherical gauge.
This feature will be exploited in \S\ref{polargravitationalwaves} below
to derive a wave equation for polar spin-$0$ modes.

The axial part of the monopole equation~(\ref{Czmonopole}) is,
in a general gauge,
$\ncalD{\zero}_\vu \, A \, \Cz^{(a)}_\zero = 0$,
which integrates immediately to
\begin{equation}
\label{Czamonopole}
  \Cz^{(a)}_\zero
  =
  0
  \quad
  ( l = 0 )
  \ ,
\end{equation}
that is, the monopole axial spin-$0$ component
of the Weyl tensor is identically zero,
a coordinate and tetrad gauge-invariant statement.

\subsection{Gravitational dipole modes}
\label{dipolegravitationalwaves}

An equation for the dipole follows from the third Weyl evolution equation~(\ref{Weylsc})
with
$\Cz_{\plusminus \two}$
set to zero,
which gives
(in a general gauge)
\begin{equation}
\label{Czdipole}
  \ncalD{\plusminus \one}_\vu \,
  A \,
  \Cz_{\plusminus \one}
  =
  ( 3 C - E ) \,
  \Gamma_{\overset{\plus {\scriptstyle v} {\scriptstyle v}}{\minus {\scriptstyle u} {\scriptstyle u}}}
  +
  {\textstyle \frac{1}{2}} \,
  A^{2} \,
  \ncalD{\plusminus \one}_\vu \,
  A^{-1}
  G_{\overset{{\scriptstyle v} \plus}{{\scriptstyle u} \minus}}
  -
  {\textstyle \frac{1}{2}} \,
  A^{2} \,
  \bigl( \partial_\plusminus
    + 2 \, \Gamma_{\overset{{\scriptstyle u} {\scriptstyle v} \plus}{{\scriptstyle v} {\scriptstyle u} \minus}}
    + \Gamma_{\overset{\plus {\scriptstyle u} {\scriptstyle v}}{\minus {\scriptstyle v} {\scriptstyle u}}}
  \bigr)
  G_{\overset{{\scriptstyle v} {\scriptstyle v}}{{\scriptstyle u} {\scriptstyle u}}}
  \quad
  ( l = 1 )
  \ .
\end{equation}
In contrast to equation~(\ref{Czp1integrated}),
which constitutes an identity between Riemann components
and tetrad connections,
equation~(\ref{Czdipole}) is a genuine evolutionary equation,
a first order differential equation governing
the dipole spin-$\pm 1$ components
$\Cz_{\plusminus \one}$
of the Weyl tensor.
It is tempting to think that the tetrad connections
$\Gamma_{\overset{\plus {\scriptstyle v} {\scriptstyle v}}{\minus {\scriptstyle u} {\scriptstyle u}}}$
in the dipole evolution equation~(\ref{Czdipole})
would vanish in spherical gauge,
in view of the relations~(\ref{Gammapvvspherical})
between these tetrad connections
and the spin-$\pm 2$ components
$\Cz_{\plusminus \two}$
of the Weyl tensor,
but for dipole modes the relations~(\ref{Gammapvvspherical}) fails,
and so the tetrad connections do not necessarily vanish.

\subsection{Axial spin-zero component of gravitational waves}
\label{axialgravitationalwaves}

The previous four subsections, \S\S\ref{spin2gravitationalwaves}--\ref{dipolegravitationalwaves},
have given a complete description of gravitational waves
(perturbations to the Weyl tensor)
in spherically symmetric self-similar black holes,
including monopole and dipole modes,
so it might seem superfluous to explore the equations any further.

However,
there are two good reasons
not to stop at this point,
but rather
to consider wave equations for the polar and axial parts of the spin-$0$ component
$\Cz_\zero$
of the complexified Weyl tensor.
The first good reason
is that
the potentials in the wave operators
for the polar and axial spin-$0$ components are real,
not complex.
For real potentials,
theorems familiar from quantum mechanics apply,
such as whether or not there exists a discrete spectrum of bound states
(which happens if the potential has one or more minima).
And the second reason is that
it is desirable to attempt to make contact
with the traditional description of perturbations
in terms of polar and axial perturbations and
the associated Regge-Wheeler and Zerilli potentials
\cite{RW57,Zerilli70a,Zerilli74,Chandrasekhar,CF91}.

As matters stand,
the connection between the Newman-Penrose formalism
and the traditional polar-axial description of perturbations
remains incompletely understood.
Chandrasekhar
\cite{Chandrasekhar},
by an ingenious suite of manipulations,
is able to work from the spin-$\pm 2$ wave equations
in the Newman-Penrose formalism
to derive a non-linear differential equation,
his equation~(311) on page 186,
which is satisfied
by the polar and axial potentials
of the Schwarzschild and Reissner-Nordstr\"om geometries,
and which is such as to ensure that
the potentials yield the same transmission and reflection coefficients.
Chandrasekhar's primary motivation
was to recover
the Regge-Wheeler and Zerilli decoupled wave equations
for the polar and axial perturbations
of the Schwarzschild and Reissner-Nordstr\"om geometries,
and in this he was successful.
However,
it seems unlikely that the gravito-electromagnetic wave
equations would decouple in the case
of general spherically symmetric self-similar solutions,
and it is unclear how Chandrasekhar's procedure
might provide a way forward in this more general case.

On the other hand,
the Newman-Penrose formalism does volunteer a spin-$0$ wave equation,
namely equation~(\ref{Weylsa}),
and it would seem good practice to explore where this equation leads.
Chandrasekhar
\cite{Chandrasekhar,CF91}
did not consider this spin-$0$ wave equation.

For non-zero spin $s$,
the two distinct degrees of freedom of a Weyl component $\Cz_s$
are represented by ingoing and outgoing modes,
which satisfy wave equations with complex conjugate potentials,
and the potentials are the same for both polar and axial modes.
For zero spin,
by contrast, 
the two distinct degrees of freedom of $\Cz_\zero$
are represented by polar and axial modes,
which satisfy wave equations with two different real potentials,
and the potentials are the same for both ingoing and outgoing modes.
The present subsection derives the wave equation~(\ref{waveCza})
governing the axial spin-$0$ component
$\Cz^{(a)}_\zero$
of the Weyl tensor,
a much easier task than for the polar spin-$0$ component
$\Cz^{(p)}_\zero$,
which is considered in the next subsection, \S\ref{polargravitationalwaves}.

The reason that the axial spin-$0$ component
$\Cz^{(a)}_\zero$,
equation~(\ref{Cza}),
is straightforward to address is that it is
coordinate and tetrad gauge-invariant and vanishing in the background,
the only other component of the Weyl tensor that has this happy property
besides the spin-$\pm 2$ components
$\Cz_{\plusminus \two}$
and
$\Cz^\star_{\plusminus \two}$
considered in
\S\ref{spin2gravitationalwaves}.
A wave equation for the axial spin-$0$ component
$\Cz^{(a)}_\zero$
follows from the wave equation~(\ref{waveWeyla})
minus its angular conjugate.

Somewhat surprisingly,
spherical gauge,
which proved to be a reliable guide
in the case of the spin-$\pm 2$ wave equations,
turns out not to provide the most elegant gauge
for splitting the axial spin-$0$ wave equation
into homogeneous and source parts.
In fact,
as will be seen below,
there is a gauge
in which the axial spin-$0$ source term takes its (apparently) simplest form,
but not only is this gauge not spherical gauge,
but it is incompatible with spherical gauge.
The gauge, which can be termed ``axial gauge'',
consists of the two conditions
\begin{equation}
\label{gaugeaxial}
  \Gamma_{\minus v \plus} - \Gamma_{\plus v \minus}
  =
  \Gamma_{\plus u \minus} - \Gamma_{\minus u \plus}
  =
  0
  \ ,
\end{equation}
which can be accomplished for example
by 2 infinitesimal tetrad transformations
of
$\varphi_{\plus {\scriptstyle v}}$
and
$\varphi_{\minus {\scriptstyle u}}$,
in accordance with equations~(\ref{Gammapvmm}).
As discussed in \S\ref{togauge},
the requirement that perturbations be expandable in spin-weighted harmonics
of itself imposes no gauge conditions,
so the axial gauge conditions~(\ref{gaugeaxial})
are entirely legitimate.
But it is curious that the spin-$\pm 2$
and axial spin-$0$ wave equations should
``prefer''
two mutually incompatible gauges.
In the next subsection, \S\ref{polargravitationalwaves},
it will be found that the polar spin-$0$ wave equation
``prefers''
spherical gauge.


In carrying out the somewhat lengthy calculation
needed to evaluate the source term for the axial spin-$0$ wave equation,
one finds that
(much as occurred earlier for spin-$\pm 2$ waves)
there are two sets of terms that naturally convert themselves
into quantities proportional to
$\Cz^{(a)}_\zero$,
and which are therefore incorporated naturally
into the wave operator on the left hand side.
One of these two sets of terms,
proportional to the unperturbed Weyl scalar $C$,
yields
$\Cz^{(a)}_\zero$
via equation~(\ref{linDeltaCa}).
The other set of terms arises from
the terms proportional to
$E \equiv \frac{1}{2} A^2 \bigl( \overset{\smallzero}{G}_{{\scriptstyle v} {\scriptstyle u}} + \overset{\smallzero}{G}_{\plus \minus} \bigr)$
in the Weyl currents
$\Jz_{\overset{\minus {\scriptstyle v} \plus}{\plus {\scriptstyle u} \minus}}$
and
$\Jz_{\overset{{\scriptstyle u} {\scriptstyle v} \plus}{{\scriptstyle v} {\scriptstyle u} \minus}}$,
equations~(\ref{linJzb}) and (\ref{linJzd}).
In axial gauge, the terms prove to combine in the proportions
\begin{equation}
\label{waveCzca}
  3 C
  -
  E
  \ ,
\end{equation}
which is opposite in sign to the
spin-$\pm 2$ expression~(\ref{waveCzc}).

The resulting wave equation for
the axial spin-$0$ component
$\Cz^{(a)}_\zero$
of the complexified Weyl tensor is
\begin{equation}
\label{waveCza}
  \Bigl(
  \ncalD{\zero}^{\dagger \vu} \,
  \ncalD{\zero}_\vu
  +
  \ncalD{\zero}^{\dagger \plusminus} \,
  \ncalD{\zero}_\plusminus
  + 3 C - E
  \Bigl)
  A \, \Cz^{(a)}_\zero
  =
  S^{(a)}_\zero
  \ ,
\end{equation}
where the axial spin-$0$ source term
$S^{(a)}_\zero$ is, in axial gauge,
\begin{equation}
\label{Saaxial}
  S^{(a)}_\zero
  =
  \frac{A}{4}
  \bigl(
    \ncalD{\plus \one}_u \, \ncalD{\plus \one}_\minus \,
    G_{v \plus}
    -
    \ncalD{\plus \one}_u \, \ncalD{\minus \one}_\plus \,
    G_{v \minus}
    +
    \ncalD{\minus \one}_v \, \ncalD{\minus \one}_\plus \,
    G_{u \minus}
    -
    \ncalD{\minus \one}_v \, \ncalD{\plus \one}_\minus \,
    G_{u \plus}
  \bigr)
  \ .
\end{equation}
Equation~(\ref{Saaxial})
is precisely the form that the axial spin-$0$ source term
takes in the case of the Schwarzschild geometry,
where all unperturbed components of the Einstein tensor vanish,
and all perturbations to the tetrad-frame Einstein tensor are
coordinate and tetrad gauge-invariant.
The full (gauge-invariant) expression for the source term
$S^{(a)}_\zero$
is
(in a general gauge, not axial gauge)
\begin{eqnarray}
\label{Sa}
  S^{(a)}_\zero
  &=&
  \frac{A}{4}
  \Bigl\{
  \ncalD{\plus \one}_u \,
  A
  \bigl[
  ( \Gamma_{\minus v \plus} - \Gamma_{\plus v \minus} )
  \bigl(
    \overset{\smallzero}{G}_{vu} + \overset{\smallzero}{G}_{\plus \minus}
  \bigr)
  -
  ( \Gamma_{\plus u \minus} - \Gamma_{\minus u \plus} )
  \overset{\smallzero}{G}_{vv}
  \bigr]
\nonumber
\\
  &&
  \quad
  + \,
  \ncalD{\minus \one}_v \,
  A
  \bigl[
  ( \Gamma_{\plus u \minus} - \Gamma_{\minus u \plus} )
  \bigl(
    \overset{\smallzero}{G}_{vu} + \overset{\smallzero}{G}_{\plus \minus}
  \bigr)
  -
  ( \Gamma_{\minus v \plus} - \Gamma_{\plus v \minus} )
  \overset{\smallzero}{G}_{uu}
  \bigr]
\nonumber
\\
  &&
  \quad
  + \,
  \ncalD{\plus \one}_u \, \ncalD{\plus \one}_\minus \,
  G_{v \plus}
  -
  \ncalD{\plus \one}_u \, \ncalD{\minus \one}_\plus \,
  G_{v \minus}
  +
  \ncalD{\minus \one}_v \, \ncalD{\minus \one}_\plus \,
  G_{u \minus}
  -
  \ncalD{\minus \one}_v \, \ncalD{\plus \one}_\minus \,
  G_{u \plus}
  \Bigr\}
  \ ,
\end{eqnarray}
and it is evident that,
as asserted above,
the axial spin-$0$ source term
$S^{(a)}_\zero$
takes its apparently simplest form in axial gauge,
equations~(\ref{gaugeaxial}).

The special case of the monopole mode, $l = 0$,
was considered already in \S\ref{monopolegravitationalwaves},
where it was found that the axial monopole mode vanishes identically,
equation~(\ref{Czamonopole}).
For the monopole mode,
the axial gauge conditions~(\ref{gaugeaxial})
are automatically satisfied,
and do not constitute additional gauge conditions,
since
$\Gamma_{\minus v \plus} - \Gamma_{\plus v \minus}$
and
$\Gamma_{\plus u \minus} - \Gamma_{\minus u \plus}$
depend only on spin-$\pm 1$ components of the vierbein perturbations $\varphi_{mn}$,
equations~(\ref{Gammapvmm}),
and therefore automatically vanish identically for the monopole.


The wave operator on the left hand side of equation~(\ref{waveCza})
differs from the wave operator discussed in
\S\ref{spinradialoperators}
only by the addition of the terms
$3 C - E$,
which are functions only of conformal radius $r$.
The homogeneous solutions
(when the source terms vanish)
are, similarly to equation~(\ref{Psi}),
\begin{equation}
\label{Czahomogeneous}
  \Cz^{(a)}_{\zero}
  =
  A^{-3}
  \ee^{- \im \omega t}
  \ncz{\zero}^{(a)}_{\omega l}(r) \,
  \nY{\zero}_{lm}(\theta,\phi)
  \ .
\end{equation}
Previously,
in the case of the spin-$\pm 2$ components
$\Cz_{\plusminus \two}$
of the Weyl tensor,
the polar and axial parts of the eigenfunctions~(\ref{Cz2homogeneous})
projected out the real and imaginary parts of the spin-$\pm 2$ spherical harmonic
$\nY{\plusminus \two}_{lm}$.
The two parts could be distinguished fundamentally
because the complex conjugate of a harmonic of spin $s$ has opposite spin $-s$,
equation~(\ref{Yconjugate}).
In the present case of the axial spin-$0$ component
$\Cz^{(a)}_{\zero}$,
the complex conjugate of a spin-$0$ spherical harmonic
is just another spin-$0$ spherical harmonic,
so the real and imaginary parts are not fundamentally distinguishable.
Thus the axial spin-$0$ eigenfunctions~(\ref{Czahomogeneous})
encompass both real and imaginary parts of the spin-$0$ spherical harmonic
$\nY{\zero}_{lm}$,
not just the imaginary part.
The radial function $\ncz{\zero}^{(a)}_{\omega l}(r)$
in equation~(\ref{Czahomogeneous})
depends only on conformal radius $r$,
and satisfies the second order ordinary differential equation
\begin{equation}
\label{DDa}
  \left\{
  - \,
  \Dr{2}{}
  -
  \omega^2
  +
  \bigl[
    l ( l + 1 )
  + 6 C - 2 E
  \bigr]
  \Delta
  \right\}
  \ncz{\zero}^{(a)}_{\omega l}
  =
  0
  \ .
\end{equation}
The effective potential
$\bigl[ l ( l + 1 ) + 6 C - 2 E \bigr] \Delta$
of the wave operator
for
$\ncz{\zero}^{(a)}_{\omega l}$
is real,
in contrast to the effective potential of the wave operator for
$\ncz{\plusminus \two}_{\omega l}$,
equation~(\ref{DD2}),
which is complex.
In terms of the dimensionless interior mass $M$, equation~(\ref{M}),
and dimensionless proper density $R$ and radial pressure $P$, equations~(\ref{einsteinRFPP}),
the factor $6 C - 2 E$ in the potential is, equations~(\ref{C}) and (\ref{E}),
\begin{equation}
  6 C - 2 E
  =
  R - P - 6 M
  \ .
\end{equation}

Equation~(\ref{DDa})
has two linearly independent eigensolutions,
which may be taken to be complex conjugates of each other,
and which may be identified as ingoing and outgoing
based on their properties
in a hypothetical asymptotically flat region at large radius,
as discussed in \S\ref{asymptotic}.
In a hypothetical aysmptotically flat region at large radius,
the radial eigensolutions
$\ncz{\zero}^{(a)}_{\omega l}$
go over to the asymptotic solutions
$\nz{\zero}_l$
discussed in \S\ref{asymptotic},
equation~(\ref{z0}),
\begin{equation}
\label{caasymptotic}
  \begin{array}{rcccl}
    \ncz{\zero}^{(a) \, \textrm{in}}_{\omega l}
    &\rightarrow&
    \nz{\zero}^\textrm{in}_l(\omega r^\ast)
    &\rightarrow&
    \ee^{- \im \omega r^\ast}
    \\
    \ncz{\zero}^{(a) \, \textrm{out}}_{\omega l}
    &\rightarrow&
    \nz{\zero}^\textrm{out}_l(\omega r^\ast)
    &\rightarrow&
    \ee^{\im \omega r^\ast}
  \end{array}
  \quad
  \mbox{as }
  r^\ast \rightarrow \infty
  \ .
\end{equation}

In the case of the Schwarzschild geometry,
the potential in the wave operator of the wave equation~(\ref{DDa})
for the axial spin-$0$ modes
is precisely the familiar Regge-Wheeler \cite{RW57,Chandrasekhar}
potential $V^{(-)}$,
\begin{equation}
  \bigl[ l ( l + 1 ) + 6 C - 2 E \bigr] \Delta
  =
  \bigl[ l ( l + 1 ) - 6 M \bigr] ( 1 - 2 M )
  =
  V^{(-)}
  \ ,
\end{equation}
where the dimensionless interior mass
$M$ of the Schwarzschild black hole is
related to its constant mass $m_\bullet$ by
$M = m_\bullet / A = m_\bullet / r$.
The axial potential
$\bigl[ l ( l + 1 ) + 6 C - 2 E \bigr] \Delta$
also agrees with that obtained by \cite{CF91} [their equation~(149)]
for spherical stationary ideal-fluid uncharged ``stars'',
and by
\cite{Karlovini02} [his equations~(91)--(93) for $s = 2$]
for general spherical stationary distributions.

In the case of the Reissner-Nordstr\"om geometry,
the axial potential
$\bigl[ l ( l + 1 ) + 6 C - 2 E \bigr] \Delta$
in the wave operator of equation~(\ref{DDa})
is not the same as the
Zerilli axial potential $V^{(-)}$
\cite{Zerilli74,Moncrief75,Chandrasekhar}.
This is because
the wave equation~(\ref{DDa})
does not describe decoupled
eigenmodes of gravito-electromagnetic waves,
but rather it describes homogeneous solutions
of a wave equation~(\ref{waveCza})
whose source
$S^{(a)}_\zero$
includes electromagnetic perturbations
that do not vanish
for the Reissner-Nordstr\"om geometry.
Physically,
if the black hole is charged,
then electromagnetic waves
constitute a source for gravitational waves,
and correspondingly gravitational waves
constitute a source for electromagnetic waves
[as will become evident below when electromagnetic waves are treated,
for example equation~(\ref{Sema})].
In order to obtain a decoupled wave equation,
it would be necessary to find axial eigensolutions of the
coupled wave equations for
gravitational waves
and electromagnetic waves.
Thanks to the work of Zerilli and Moncrief
\cite{Zerilli74,Moncrief75}
it is known that such a decoupling can be accomplished for the
Reissner-Nordstr\"om geometry,
but for the case of general self-similar solutions
there is no reason to expect that such a decoupling would occur.
The present paper does not pursue this issue further.

\subsection{Polar spin-zero component of gravitational waves}
\label{polargravitationalwaves}

The derivation of a wave equation for the polar spin-$0$ component
of gravitational waves
presents a formidable challenge.
This is largely because
the polar spin-$0$ component
$\Cz^{(p)}_\zero$
of the Weyl tensor
does not vanish in the unperturbed background
(even in the Schwarzschild geometry),
and consequently issues of gauge freedom are especially delicate.
Nevertheless,
for the reasons described at the beginning of the previous subsection,
\S\ref{axialgravitationalwaves},
it seems wise to pursue the polar part of the
spin-$0$ wave equation~(\ref{waveWeyla})
that arises in the Newman-Penrose formalism,
to see where it leads.
Where it leads is, apparently,
to the sourced polar spin-$0$ wave equation~(\ref{waveN}).
The present paper does not address
the problem of
how this equation and the other equations to which it couples
might be decoupled so as to recover the Zerilli-Moncrief equations
\cite{Zerilli74,Moncrief75}
in the Reissner-Nordstr\"om geometry.

The polar spin-$0$ component
$\Cz^{(p)}_\zero = \frac{1}{2} ( \Cz_\zero + \Cz^\star_\zero )$
of the complexified Weyl tensor
is gauge-invariant with respect to infinitesimal coordinate and tetrad transformations,
but the split between
the unperturbed and perturbed parts is not coordinate gauge-invariant.
One might think (for an instant) that,
since the unperturbed part
of $\Cz^{(p)}_\zero$
is pure monopole ($l = 0$),
at least the non-monopole parts of the expansion
of $\Cz^{(p)}_\zero$
in spherical harmonics would be gauge-invariant,
but this is not true.
Under an infinitesimal coordinate gauge transformation
$x^\mu \rightarrow x^{\prime \mu} = x^\mu + A \epsilon^\mu$,
the unperturbed part
$\overset{\smallzero}{\Cz}{}^{(p)}_\zero$
of the unperturbed spin-$0$ Weyl tensor
transforms as
\begin{equation}
\label{gaugeCzp0}
  \overset{\smallzero}{\Cz}{}^{(p)}_\zero(t, r)
  \rightarrow
  \overset{\smallzero}{\Cz}{}^{(p)}_\zero(t^\prime, r^\prime)
  =
  \overset{\smallzero}{\Cz}{}^{(p)}_\zero
  +
  A (
    \epsilon^v \partial_v
    +
    \epsilon^u \partial_u
  )
  \overset{\smallzero}{\Cz}{}{}^{(p)}_\zero
  \ ,
\end{equation}
so that,
since
$\Cz^{(p)}_\zero = \overset{\smallzero}{\Cz}{}{}^{(p)}_\zero + \overset{\smallone}{\Cz}{}{}^{(p)}_\zero$
as a whole is gauge-invariant,
the perturbation
$\overset{\smallone}{\Cz}{}^{(p)}_\zero$
transforms as
\begin{equation}
\label{gaugeCzp1}
  \overset{\smallone}{\Cz}{}^{(p)}_\zero
  \rightarrow
  \overset{\smallone}{\Cz}{}^{(p)}_\zero
  -
  A (
    \epsilon^v \partial_v
    +
    \epsilon^u \partial_u
  )
  \overset{\smallzero}{\Cz}{}{}^{(p)}_\zero
  \ .
\end{equation}
The dimensionless coordinate perturbations
$\epsilon^v$
and
$\epsilon^u$
in the transformation~(\ref{gaugeCzp1})
will in general contain arbitrary non-monopole harmonics,
changing the spherical harmonic expansion of
the perturbation
$\overset{\smallone}{\Cz}{}^{(p)}_\zero$
arbitrarily.
Indeed,
as long as
the unperturbed spin-$0$ Weyl tensor
$\overset{\smallzero}{\Cz}{}^{(p)}_\zero$
is non-constant,
which is generally true except
in a hypothetical asymptotically flat region at large radius,
it is possible in principle to eliminate the perturbation
$\overset{\smallone}{\Cz}{}^{(p)}_\zero$
to the polar spin-$0$ Weyl tensor
altogether by a coordinate gauge transformation arranged to satisfy
\begin{equation}
\label{badCzgauge}
  \overset{\smallone}{\Cz}{}^{(p)}_\zero
  -
  A (
    \epsilon^v \partial_v
    +
    \epsilon^u \partial_u
  )
  \overset{\smallzero}{\Cz}{}^{(p)}_\zero
  =
  0
  \ .
\end{equation}
However,
it is not possible to eliminate simultaneously all polar spin-$0$ perturbations.
Specifically, it is not possible to set
the polar spin-$0$ vierbein perturbations
$\varphi_{vv}$,
$\varphi_{uu}$,
and
$\varphi_{vu} + \varphi_{uv}$
simultaneously to zero,
because each of these 3 vierbein perturbations
varies only under a coordinate transformation
of $\epsilon^v$ or $\epsilon^u$
(not
under a coordinate transformation
of $\epsilon^\plus$ or $\epsilon^\minus$,
nor
under a tetrad transformation),
equations~(\ref{gaugephis}),
and 3 gauge conditions cannot be achieved with just 2 freedoms.
Thus there is reason to expect that
the polar part of the spin-$0$ wave equation~(\ref{waveWeyla})
is not vacuous,
but may in fact encode the evolution of polar spin-$0$ perturbations.

A clue as to how to proceed
is provided by the monopole mode of the polar spin-$0$ perturbation,
which was found in \S\ref{monopolegravitationalwaves}
to be described most naturally not by
the polar spin-$0$ Weyl component
$\Cz^{(p)}_\zero$
itself,
but rather by the related polar spin-$0$ quantity $\gravN$
defined by equation~(\ref{gravN}).
Although
$\gravN$
is itself
neither coordinate nor tetrad gauge-invariant,
the equation~(\ref{DNmonopole})
governing the monopole part of $\gravN$
is coordinate and tetrad gauge-invariant.
Therefore,
if the monopole equation~(\ref{DNmonopole}) is subtracted from the
general equation~(\ref{DvuN}) governing
$\gravN$,
then the result,
being a difference of two
coordinate and tetrad gauge-invariant equations,
will be coordinate and tetrad gauge-invariant,
and moreover vanishing in the background
[that is to say, equation~(\ref{DvuN}) as a whole is
coordinate and tetrad gauge-invariant,
though the split between its left and right hand sides
is in general not coordinate and tetrad gauge-invariant].

As was found in \S\ref{monopolegravitationalwaves},
the monopole part of $\gravN$ takes a particularly simple form
in spherical gauge,
being equal to its unperturbed constant value
$- \frac{1}{2}$
not only to unperturbed order but also to linear order,
equation~(\ref{Nmonopole}).
Thus in spherical gauge,
subtracting the monopole equation from the general equation governing $\gravN$
amounts to replacing
$\gravN$ by its perturbation
$\overset{\smallone}{\gravN}$,
a trivial operation.
Although the resulting equation for the perturbation
$\overset{\smallone}{\gravN}$
is no longer gauge-invariant
(being expressed in a particular gauge, spherical gauge),
nevertheless the equation is equivalent to
a gauge-invariant equation,
and therefore has real physical significance
independent of any gauge.
The perturbation
$\overset{\smallone}{\gravN}$
is coordinate and tetrad gauge-invariant
with respect to the two gauge freedoms left in spherical gauge.


The two Weyl evolution equations that combine to produce
the spin-$0$ wave equation~(\ref{waveWeyla}) are
equations~(\ref{Weylsa}) and (\ref{Weylsb}).
The first of Weyl evolution equation~(\ref{Weylsa}) was already
recast in terms of $\gravN$ in equation~(\ref{DvuN}).
Similarly recasting the second Weyl evolution equation~(\ref{Weylsb})
in terms of $\gravN$ gives
(in a general gauge, and for arbitrary harmonics)
\begin{eqnarray}
\label{DpmN}
  &&
  \partial_\plusminus \gravN
  +
  \ncalD{\zero}_\plusminus \,
  A \,
  \Cz^{(a)}_\zero
  -
  \ncalD{\plusminus \one}_\uv \, A \, \Cz_{\plusminus \one}
  =
  - \,
  A^{-1} \partial_\plusminus A
  - 3 \,
  \Gamma_{\overset{\plus {\scriptstyle v u}}{\minus {\scriptstyle u v}}} \,
  C
  -
  {\textstyle \frac{1}{2}} \,
  A^2
  \bigl(
    \partial_\plusminus
    +
    \Gamma_{\overset{\plus {\scriptstyle u v}}{\minus {\scriptstyle v u}}}
  \bigr)
  ( G_{vu} + G_{\plus\minus} )
\nonumber
\\ 	
  &&
  \qquad
  + \,
  {\textstyle \frac{1}{4}} \,
  A^2 \,
  \partial_\plusminus
  \bigl[
  \bigl(
  \Gamma_{\overset{\minus {\scriptstyle v} \plus}{\plus {\scriptstyle u} \minus}}
  +
  \Gamma_{\overset{\plus {\scriptstyle v} \minus}{\minus {\scriptstyle u} \plus}}
  \bigr)
  \bigl(
  \Gamma_{\overset{\plus {\scriptstyle u} \minus}{\minus {\scriptstyle v} \plus}}
  +
  \Gamma_{\overset{\minus {\scriptstyle u} \plus}{\plus {\scriptstyle v} \minus}}
  \bigr)
  \bigr]
  -
  {\textstyle \frac{1}{2}} \,
  A^2 \,
  \Gamma_{\overset{\plus {\scriptstyle v v}}{\minus {\scriptstyle u u}}} \,
  \overset{\smallzero}{G}_{\overset{{\scriptstyle u u}}{{\scriptstyle v v}}}
  +
  {\textstyle \frac{1}{2}} \,
  A \,
  \ncalD{\minusplus \one}_\vu \,
  G_{\overset{{\scriptstyle u} \plus}{{\scriptstyle v} \minus}}
  +
  {\textstyle \frac{1}{2}} \,
  A \,
  \beta_\uv \,
  G_{\overset{{\scriptstyle v} \plus}{{\scriptstyle u} \minus}}
  \ .
\end{eqnarray}
The monopole part of equation~(\ref{DpmN}) vanishes identically,
since all the terms of the equation are spin-$\pm 1$,
so there is no need to subtract the monopole part of equation~(\ref{DpmN}).

As discussed above,
it is helpful from this point to work in spherical gauge,
equations~(\ref{gaugespherical}),
because the subtraction of the monopole is trivial in this gauge,
amounting to a replacement of
$\gravN$
by its perturbation
$\overset{\smallone}{\gravN}$.
In spherical gauge,
the perturbation
$\overset{\smallone}{\gravN}$ is,
from $\frac{1}{2} A^2$ times equation~(\ref{Nlightbulb}),
\begin{equation}
  \overset{\smallone}{\gravN}
  =
  {\textstyle \frac{1}{2}}
  \bigl[
  \ncalD{\plus \one}_\minus
  (
    \beta_u \,
    \varphi_{\plus v}
    +
    \beta_v \,
    \varphi_{\plus u}
  )
  +
  \ncalD{\minus \one}_\plus \,
  (
    \beta_u \,
    \varphi_{\minus v}
    +
    \beta_v \,
    \varphi_{\minus u}
  )
  \bigl]
  \ ,
\end{equation}
which may also be written
\begin{equation}
\label{perturbationNspherical}
  \overset{\smallone}{\gravN}
  =
  {\textstyle \frac{1}{2}}
  \bigl(
    \ncalD{\plus \one}_\minus \,
    \partial_\plus \, A
    +
    \ncalD{\minus \one}_\plus \,
    \partial_\minus \, A
  \bigl)
  \ .
\end{equation}
In spherical gauge,
the term on the right hand side of equation~(\ref{DpmN})
involving the directed angular derivative of products of tetrad connections
can be rewritten,
using equations~(\ref{dAa})
and the commutation relations~(\ref{partialcommute}),
\begin{equation}
\label{DGG}
  {\textstyle \frac{1}{4}}
  A^2
  \partial_\plusminus
  \bigl[
  \bigl(
  \Gamma_{\overset{\minus {\scriptstyle v} \plus}{\plus {\scriptstyle u} \minus}}
  +
  \Gamma_{\overset{\plus {\scriptstyle v} \minus}{\minus {\scriptstyle u} \plus}}
  \bigr)
  \bigl(
  \Gamma_{\overset{\plus {\scriptstyle u} \minus}{\minus {\scriptstyle v} \plus}}
  +
  \Gamma_{\overset{\minus {\scriptstyle u} \plus}{\plus {\scriptstyle v} \minus}}
  \bigr)
  \bigr]
  =
  \bigl(
  \beta_u
  \partial_v
  +
  \beta_v
  \partial_u
  \bigr)
  \partial_\plusminus
  A
  - \,
  \beta_u^2 \,
  \Gamma_{\overset{\plus {\scriptstyle v v}}{\minus {\scriptstyle u u}}}
  -
  \beta_v^2 \,
  \Gamma_{\overset{\plus {\scriptstyle u u}}{\minus {\scriptstyle v v}}}
  -
  2 \,
  \beta_u \beta_v \,
  \Gamma_{\overset{\plus {\scriptstyle v u}}{\minus {\scriptstyle u v}}}
  \ .
\end{equation}
When equations~(\ref{DvuN}) and (\ref{DpmN}) are combined
into a polar wave equation for the perturbation
$\overset{\smallone}{\gravN}$,
terms proportional to $\partial_\plusminus A$,
namely the term
$- A^{-1} \partial_\plusminus A$ on the right hand side of equation~(\ref{DpmN}),
and the terms proportional to $\partial_\plusminus A$
on the right hand hand side of equation~(\ref{DGG}),
convert into quantities proportional to
$\overset{\smallone}{\gravN}$,
equation~(\ref{perturbationNspherical}),
and it is natural to incorporate these terms into the
wave operator on the left hand side.

The resulting wave equation for the perturbation
$\overset{\smallone}{\gravN}$ is
\begin{eqnarray}
\label{waveNpre}
  \Bigl(
    \ncalD{\plus \one}_u \,
    \partial_v
    +
    \ncalD{\minus \one}_v \,
    \partial_u
    -
    \ncalD{\plus \one}_\minus \,
    \partial_\plus
    +
    \ncalD{\minus \one}_\plus \,
    \partial_\minus
    +
    2 \,
    \beta_u \partial_v
    +
    2 \,
    \beta_v \partial_u
    - 2 A^{-1}
  \Bigr)
  \overset{\smallone}{\gravN}
  =
  2 S^{(p)}_\zero
  \ ,
\end{eqnarray}
where
the factor of $2$ on the right hand side is introduced
so that it disappears from the wave equation~(\ref{waveN}) below.
The polar spin-$0$ source term
$S^{(p)}_\zero$
on the right hand side of equation~(\ref{waveNpre})
is,
in spherical gauge,
the unprepossessingly complicated object
\begin{eqnarray}
\label{Sp}
  S^{(p)}_\zero
  &=&
  \frac{1}{4}
  \Bigl\{
  A^2
  \bigl(
    \ncalD{\plus \one}_\minus \,
    \partial_\plus
    +
    \ncalD{\minus \one}_\plus \,
    \partial_\minus
  \bigr)
  \bigl(
    G_{vu} + G_{\plus \minus}
  \bigr)
\nonumber
\\
  &&
  \quad
  + \,
  A^{-3}
  \beta_u \,
  \ncalD{\plus \two}_u \,
  A^3
  \bigl(
    \ncalD{\plus \one}_\minus \, \Gamma_{\plus v v}
    +
    \ncalD{\minus \one}_\plus \, \Gamma_{\minus v v}
  \bigr)
  +
  A^{-3}
  \beta_v \,
  \ncalD{\minus \two}_v \,
  A^3
  \bigl(
    \ncalD{\plus \one}_\minus \, \Gamma_{\plus u u}
    +
    \ncalD{\minus \one}_\plus \, \Gamma_{\minus u u}
  \bigr)
\nonumber
\\
  &&
  \quad
  + \,
  A^{-1}
  \bigl(
  \beta_u \,
  \ncalD{\zero}_v \,
  +
  \beta_v \,
  \ncalD{\zero}_u \,
  \bigr)
  A
  \bigl(
    \ncalD{\plus \one}_\minus \, \Gamma_{\plus v u}
    +
    \ncalD{\minus \one}_\plus \, \Gamma_{\minus v u}
  \bigr)
  +
  4 ( R - P + P_\perp - 2 M )
  \bigl(
    \ncalD{\plus \one}_\minus \, \Gamma_{\plus v u}
    +
    \ncalD{\minus \one}_\plus \, \Gamma_{\minus v u}
  \bigr)
\nonumber
\\
  &&
  \quad
  - \,
  \ncalD{\plus \one}_u \, A \, \ncalD{\plus \one}_\minus \,
  G_{v \plus}
  -
  \ncalD{\plus \one}_u \, A \, \ncalD{\minus \one}_\plus \,
  G_{v \minus}
  -
  \ncalD{\minus \one}_v \, A \, \ncalD{\minus \one}_\plus \,
  G_{u \minus}
  -
  \ncalD{\minus \one}_v \, A \, \ncalD{\plus \one}_\minus \,
  G_{u \plus}
  \Bigr\}
  \ .
\end{eqnarray}
The wave operator on the left hand side
of equation~(\ref{waveNpre})
is
(barring the $- 2 A^{-1}$ term)
$A^{-2} \ncalD{\zero}^{\dagger m} \, A^3 \ncalD{\zero}_m \, A^{-2}$,
which may be recast using the identity
(with $a$ an arbitrary constant)
\begin{equation}
\label{ADAD}
  A^{-2a} \, \ncalD{\zero}^{\dagger m} \, A^{2a} \ncalD{\zero}_m
  =
  A^{-a} \, \ncalD{\zero}^{\dagger m} \, \ncalD{\zero}_m \, A^a
  -
  a
  ( 1 + R - P - 4 M )
  +
  a^2
  ( 1 - 2 M)
  \ .
\end{equation}
The identity~(\ref{ADAD}),
along with the operator equalities
$\ncalD{\zero}^{\dagger v} \, \ncalD{\zero}_v = \ncalD{\zero}^{\dagger u} \, \ncalD{\zero}_u$
and
$\ncalD{\zero}^{\dagger \plus} \, \ncalD{\zero}_\plus = \ncalD{\zero}^{\dagger \minus} \, \ncalD{\zero}_\minus$
valid for zero spin,
equations~(\ref{diffDDvu}) and (\ref{diffDDpm}),
brings the wave equation~(\ref{waveNpre})
to a form similar to that already encountered
for spin-$\pm 2$ and axial spin-$0$ modes,
equations~(\ref{waveCz}) and (\ref{waveCza}):
\begin{equation}
\label{waveN}
  A^{-1/2}
  \Bigl(
  \ncalD{\zero}^{\dagger \vu} \,
  \ncalD{\zero}_\vu
  +
  \ncalD{\zero}^{\dagger \plusminus} \,
  \ncalD{\zero}_\plusminus
  -
  {\textstyle \frac{5}{8}}
  -
  {\textstyle \frac{3}{4}} R
  +
  {\textstyle \frac{3}{4}} P
  +
  {\textstyle \frac{3}{4}} M
  \Bigl)
  A^{-1/2}
  \overset{\smallone}{\gravN}
  =
  S^{(p)}_\zero
  \ .
\end{equation}

As written,
the expression~(\ref{Sp})
for the polar spin-$0$ source term
$S^{(p)}_\zero$
is valid for all harmonics,
including dipole harmonics.
For quadrupole and higher harmonics, $l \geq 2$,
the tetrad connections in the source term
$S^{(p)}_\zero$
can be replaced by spin-$\pm 2$ components
of the Weyl and Einstein tensors,
in accordance with equations~(\ref{Gammaspherical}),
valid in spherical gauge:
\begin{subequations}
\label{DGammaspherical}
\begin{eqnarray}
\label{DGammavvspherical}
  \ncalD{\plusminus \one}_\minusplus \,
  \Gamma_{\overset{\plus {\scriptstyle v} {\scriptstyle v}}{\minus {\scriptstyle u} {\scriptstyle u}}}
  +
  \ncalD{\minusplus \one}_\plusminus \,
  \Gamma_{\overset{\minus {\scriptstyle v} {\scriptstyle v}}{\plus {\scriptstyle u} {\scriptstyle u}}}
  &=&
  - \,
  {2 A \over ( l - 1 ) ( l + 2 )}
  \bigl(
    \ncalD{\plusminus \one}_\minusplus \,
    \ncalD{\plusminus \two}_\minusplus \,
    \Cz_{\plusminus \two}
    +
    \ncalD{\minusplus \one}_\plusminus \,
    \ncalD{\minusplus \two}_\plusminus \,
    \Cz^\star_{\plusminus \two}
  \bigr)
  \ ,
\\
\label{DGammavuspherical}
  \ncalD{\plus \one}_\minus \, \Gamma_{\plus v u}
  +
  \ncalD{\minus \one}_\plus \, \Gamma_{\minus v u}
  &=&
  {A \over ( l - 1 ) ( l + 2 )}
  \bigl(
    \ncalD{\plus \one}_\minus \,
    \ncalD{\plus \two}_\minus \,
    G_{\plus \plus}
    +
    \ncalD{\minus \one}_\plus \,
    \ncalD{\minus \two}_\plus \,
    G_{\minus \minus}
  \bigr)
  \ .
\end{eqnarray}
\end{subequations}
Physically,
the right hand sides of equations~(\ref{DGammaspherical})
are proportional to the polar parts of
$\Cz_{\plusminus \two}$
and
$G_{\overset{\plus \plus}{\minus \minus}}$
angularly lowered and raised to zero spin.

As discussed above,
taken as a whole
the wave equation~(\ref{waveN})
is equivalent to a
(coordinate and tetrad)
gauge-invariant equation,
being the difference of two
gauge-invariant equations
(a general equation and a monopole equation).
However,
the split between the left and right hand sides
of equation~(\ref{waveN})
is in general not gauge-invariant.
In the case of the Schwarzschild geometry,
where the background energy-momentum vanishes,
and all perturbations to the Einstein tensor are
coordinate and tetrad gauge-invariant,
the polar spin-$0$ source term
$S^{(p)}_\zero$
with the replacements~(\ref{DGammaspherical})
becomes a coordinate and tetrad gauge-invariant object.
Therefore,
at least in the case of the Schwarzschild geometry,
and for quadrupole and higher harmonics,
the left and right hand sides of the wave equation~(\ref{waveN})
are individually equivalent to coordinate and tetrad gauge-invariant objects.

The polar spin-$0$ wave equation~(\ref{waveN})
has two unusual features
compared to the other wave equations derived in this paper.
The first is that
(for quadrupole and higher harmonics)
the spin-$\pm 2$ Weyl tensor
$\Cz_{\plusminus \two}$
and its angular conjugate
$\Cz^\star_{\plusminus \two}$,
in the combination given by equation~(\ref{DGammavvspherical}),
appear in the source term
$S^{(p)}_\zero$,
equation~(\ref{Sp}).
That is,
the wave equation for polar spin-$0$ perturbation
$\overset{\smallone}{\gravN}$
appears to be sourced by spin-$\pm 2$ modes of the Weyl tensor
(even in the case of the Schwarzschild geometry).

The second unusual feature of the wave equation~(\ref{waveN})
is that its homogeneous solutions
decay with conformal factor as
$\overset{\smallone}{\gravN} \propto A^{-3/2}$
(so that the dimensional perturbation
$A^{-2} \overset{\smallone}{\gravN}$
with the same dimension as the spin-$0$ Weyl tensor
$\Cz_{\zero}$
goes as
$A^{-2} \overset{\smallone}{\gravN} \propto A^{-7/2}$),
which is a factor of $A^{-1/2}$
faster than expected
for the non-propagating spin-$0$
companion of a
propagating spin-$\pm 2$ mode
[such as the axial component
$\Cz^{(a)}_{\zero}$,
whose homogeneous solutions fall off as
$A^{-3}$,
equation~(\ref{Czahomogeneous})].
This anomalous behavior is related to the presence of
the spin-$\pm 2$ Weyl tensor as a source of the wave equation.
Since at least one of the spin-$\pm 2$ components
$\Cz_{\plusminus \two}$
is propagating,
it remains an active source for the polar spin-$0$ perturbation
$\overset{\smallone}{\gravN}$
even in a hypothetically flat empty region
far from the black hole.
Dimensional analysis
indicates that the part of
$\overset{\smallone}{\gravN}$
sourced by the driving spin-$\pm 2$ Weyl component is expected to
go with conformal factor $A$ as
$\overset{\smallone}{\gravN} \propto A S^{(p)}_\zero \propto A^2 \Cz_{\plusminus \two}$,
which according to the homogeneous solutions~(\ref{Cz2homogeneous})
for
$\Cz_{\plusminus \two}$
goes as
$\overset{\smallone}{\gravN} \sim A^{-1}$
in asymptotically flat empty space far from the black hole.
Thus the sourced part of
$\overset{\smallone}{\gravN}$
is expected to dominate the homogeneous part
far from the black hole.

The homogeneous solutions of the wave equation~(\ref{waveN})
are, similarly to equation~(\ref{Psi}),
\begin{equation}
\label{Nhomogeneous}
  \overset{\smallone}{\gravN}
  =
  A^{-3/2}
  \ee^{- \im \omega t}
  n_{\omega l}(r) \,
  \nY{\zero}_{lm}(\theta,\phi)
  \ ,
\end{equation}
where the radial eigenmodes
$n_{\omega l}(r)$
satisfy
\begin{equation}
\label{DDN}
  \left\{
  - \,
  \Dr{2}{}
  -
  \omega^2
  +
  \bigl[
    l ( l + 1 )
    -
    {\textstyle \frac{5}{4}}
    -
    {\textstyle \frac{3}{2}} R
    +
    {\textstyle \frac{3}{2}} P
    +
    {\textstyle \frac{3}{2}} M
  \bigr]
  \Delta
  \right\}
  n_{\omega l}
  =
  0
  \ .
\end{equation}
As discussed in the previous paragraph,
in a hypothetical asymptotically flat empty region far from the black hole,
the behavior of the polar spin-$0$ perturbation
$\overset{\smallone}{\gravN}$
is expected to be dominated not by a homogeneous solution~(\ref{Nhomogeneous}),
but rather by the driving spin-$\pm 2$ Weyl component
in the source
$S^{(p)}_\zero$.

The polar potential
$\bigl[ l ( l + 1 ) - {\textstyle \frac{5}{4}} - {\textstyle \frac{3}{2}} R + {\textstyle \frac{3}{2}} P + {\textstyle \frac{3}{2}} M \bigr] \Delta$
in the wave operator of the homogeneous polar spin-$0$ wave equation~(\ref{DDN})
is not the same as the
Zerilli polar potential $V^{(+)}$
even for the Schwarzschild geometry
\cite{Zerilli70a},
let alone for the Reissner-Nordstr\"om geometry
\cite{Zerilli74,Moncrief75}.
There is no contradiction because
equation~(\ref{DDN})
does not describe decoupled eigenmodes,
but rather it
describes the radial part of homogeneous solutions~(\ref{Nhomogeneous})
of a wave equation~(\ref{waveN})
whose source
$S^{(p)}_\zero$
does not vanish even for the Schwarzschild geometry.

The polar part of the spin-$0$ wave equation~(\ref{Weylsa})
in the Newman-Penrose formalism has (apparently)
led to the wave equation~(\ref{DDN}),
but it has not led all the way to an independent derivation of the
Zerilli-Moncrief polar potential $V^{(+)}$
\cite{Zerilli74,Moncrief75}
for decoupled gravito-elecromagnetic eigenmodes
in either the Schwarzschild or Reissner-Nordstr\"om geometries.
It would be good to accomplish such a derivation,
but it may not be trivial to do so,
and here I take the matter no further.


\section{Electromagnetic waves}
\label{electromagneticwaves}

This section derives the equations that describe electromagnetic waves
in perturbed self-similar black hole spacetimes.
Subsections~\ref{electromagneticfield} and \ref{maxwellsequations}
characterize perturbations of the electromagnetic field
and the Maxwell's equations that govern their evolution.
Subsequent subsections
consider the propagating (spin-$\pm 1$) components of electromagnetic waves
(\S\ref{spin1electromagneticwaves}),
monopole electromagnetic modes
(\S\ref{monopoleelectromagneticwaves}),
and the axial (\S\ref{axialelectromagneticwaves}) and polar (\S\ref{polarelectromagneticwaves})
parts of the electromagnetic spin-$0$ wave equation.

\subsection{Electromagnetic field}
\label{electromagneticfield}

The electromagnetic field $F_{mn}$ is a bivector,
an antisymmetric tensor.
As earlier discussed
in \S\ref{perturbedweyl}
with regard to the bivector structure of the Weyl tensor,
this means that the electromagnetic field
is a 6-component object with a natural complex structure.
The real part of the eletcromagnetic bivector field is the electric 3-vector $E$,
which changes sign under spatial inversion,
while the imaginary part is the magnetic 3-vector $B$,
which does not change sign under spatial inversion.

The natural complex structure motivates writing Maxwell's equations
as the single complex equation
\begin{equation}
\label{Maxwell}
  \left( \delta_k^m \delta_l^n + \frac{\im}{2} {\varepsilon_{kl}}^{mn}
  \right) D^k F_{mn}
  = 4 \pi j_l
  \ ,
\end{equation}
whose real (vector) and imaginary (trivector) parts
represent respectively the source and source-free
parts of Maxwell's equations.
Absent magnetic monopoles,
the current $j_l$ is purely electric, a real vector.
Maxwell's equations~(\ref{Maxwell}) can immediately be recast as
\begin{equation}
\label{Maxwellz}
  D^k
  \Fz_{kl}
  = 2 \pi j_l
  \ ,
\end{equation}
where the complexified electromagnetic field $\Fz$ is
\begin{equation}
\label{Fz}
  \Fz_{kl}
  \equiv
  \frac{1}{2}
  \left( \delta_k^m \delta_l^n + \frac{\im}{2} {\varepsilon_{kl}}^{mn}
  \right)
  F_{mn}
\end{equation}
[as earlier in the definition~(\ref{Cz}) of the complexified Weyl tensor,
the factor of $1/2$ on the right hand side of equation~(\ref{Fz})
is introduced because then the operator
$P_{kl}^{mn} \equiv \frac{1}{2} \left( \delta_k^m \delta_l^n + \frac{\im}{2} {\varepsilon_{kl}}^{mn} \right)$
is a projection operator,
satisfying
$P^2 = P$.]
In an orthonormal tetrad basis,
the electromagnetic field $F_{kl}$
can be represented as a pair of electric and magnetic $3$-vectors
$E$ and $B$
\begin{equation}
  F
  =
  \left(
  \begin{array}{cc}
  E & B \\
  \end{array}
  \right)
  \ .
\end{equation}
The complexified electromagnetic field $\Fz_{kl}$ then has the structure
\begin{equation}
  \Fz
  =
  \frac{1}{2}
  \left(
  \begin{array}{cc}
  1 & i \\
  \end{array}
  \right)
  ( E + \im B )
  \ .
\end{equation}
Thus the magnetic part of the complexified electromagnetic field
is $\im$ times the electric part.
In tensor language,
the complexified electromagnetic field satisfies the symmetry
\begin{equation}
  \Fz_{kl}
  =
  \frac{\im}{2} {\varepsilon_{kl}}^{mn} \Fz_{mn}
  \ ,
\end{equation}
as follows immediately from the definition~(\ref{Fz}).
Thanks to this symmetry,
the independent components of
$\Fz_{kl}$
constitute a single complex $3$-vector,
containing
the electric field $E$
and the magnetic field $B$
in the complex combination $E + \im B$.

The complexified electromagnetic tensor
$\Fz_{kl}$
changes sign under spatial inversion,
because spatial inversion not only transforms
$E \rightarrow - E$ and $B \rightarrow B$,
but also changes the sign of the antisymmetric tensor
$\varepsilon_{klmn}$,
thereby effectively changing the sign of $\im$ in
$E + \im B$.

With respect to a general Newman-Penrose tetrad,
the non-vanishing components
of the complexified electromagnetic field $\Fz_{kl}$,
and their expressions in terms of the electromagnetic field tensor $F_{kl}$,
and the electric and magnetic fields $E$ and $B$,
are
\begin{subequations}
\begin{eqnarray}
  \Fz_\zero
  &\equiv&
  \Fz_{uv}
  =
  \Fz_{\plus \minus}
  =
  {\textstyle \frac{1}{2}} \left(
    F_{uv}
    +
    F_{\plus\minus}
  \right)
  =
  {\textstyle \frac{1}{2}} \left(
  E_r + \im B_r
  \right)
  \ ,
\\
  \Fz_{\plusminus \one}
  &\equiv&
  \Fz_{\overset{{\scriptstyle v} \plus}{{\scriptstyle u} \minus}}
  =
  F_{\overset{{\scriptstyle v} \plus}{{\scriptstyle u} \minus}}
  =
  {\textstyle \frac{1}{2}} \bigl[
  E_\theta + \im B_\theta \mp \im (E_\phi + \im B_\phi)
  \bigr]
  \ .
\end{eqnarray}
\end{subequations}
As with the Weyl tensor, the only non-vanishing components of
the complexified electromagnetic field tensor
$\Fz_{kl}$
are those with bivector indices
$uv$, $\scriptstyle{+-}$, $v\scriptstyle{+}$, or $u\scriptstyle{-}$.
Components $\Fz_{kl}$
with bivector indices $v-$ or $u+$ vanish:
\begin{equation}
\label{Fzzero}
  \Fz_{v \minus}
  =
  \Fz_{u \plus}
  =
  0
  \ .
\end{equation}

In the unperturbed self-similar background,
only the polar spin-$0$ component
${\Fz}_\zero$
of the electromagnetic field is non-zero,
and it equals
$\frac{1}{2}$
the unperturbed radial electric field
$\overset{\smallzero}{E}_r$
\begin{equation}
\label{Fzunperturbed}
  \overset{\smallzero}{\Fz}_\zero
  =
  {\textstyle \frac{1}{2}} \,
  \overset{\smallzero}{F}_\zero
  =
  {\textstyle \frac{1}{2}} \,
  \overset{\smallzero}{E}_r
  \ .
\end{equation}
The unperturbed value of the electric field
can be taken to define what is meant by the
electric charge $q$ interior to $r$ in the unperturbed background,
and the associated dimensionless charge
$Q \equiv q / A$:
\begin{equation}
\label{Q}
  \overset{\smallzero}{\Fz}_\zero
  \equiv
  {\textstyle \frac{1}{2}} \,
  q / A^2
  \equiv
  {\textstyle \frac{1}{2}} \,
  Q / A
  \ .
\end{equation}

Like the complexified Weyl tensor,
equation~(\ref{Czspa}),
the complexified electromagnetic field tensor $\Fz_{kl}$
can be resolved into polar $(p)$ and axial $(a)$ parts
which respectively do not and do change sign
when the azimuthal angular tetrad axis
$\gamma_\phi$
is flipped in sign, equation~(\ref{angularconjugate}),
\begin{equation}
\label{Fzspa}
  \Fz_s
  =
  \Fz^{(p)}_s + \Fz^{(a)}_s
  \ .
\end{equation}
The angular conjugate of
$\Fz_{kl}$
is by definition
\begin{equation}
\label{Fzsconj}
  \Fz^\star_s
  =
  \Fz^{(p)}_s - \Fz^{(a)}_s
  \ .
\end{equation}
The angular conjugates
$\Fz^\star_{kl}$
of the spin components of
the complexified electromagnetic field tensor are,
in terms of the electromagnetic field tensor $F_{kl}$,
and also in terms of the electric and magnetic fields $E$ and $B$,
\begin{subequations}
\begin{eqnarray}
  \Fz^\star_\zero
  =
  {\textstyle \frac{1}{2}} \left(
    F_{uv}
    -
    F_{\plus\minus}
  \right)
  =
  {\textstyle \frac{1}{2}} \left(
  E_r - \im B_r
  \right)
  \ ,
\\
  \Fz^\star_{\plusminus \one}
  =
  F_{\overset{{\scriptstyle v} \minus}{{\scriptstyle u} \plus}}
  =
  {\textstyle \frac{1}{2}} \bigl[
  E_\theta - \im B_\theta \pm \im (E_\phi - \im B_\phi)
  \bigr]
  \ .
\end{eqnarray}
\end{subequations}

Being tetrad-frame quantities,
all components $\Fz_s$ of the complexified electromagnetic field tensor
(and their angular conjugates)
are coordinate gauge-invariant,
although the split between the unperturbed and perturbed parts
of the polar part of
$\Fz_\zero$,
the only component whose unperturbed value is non-vanishing,
is not coordinate gauge-invariant.
The spin-$0$ component
$\Fz_{\zero}$
(and its angular conjugate)
are not only coordinate gauge-invariant
but tetrad gauge-invariant with respect to all 6
arbitrary infinitesimal tetrad transformations,
including those tetrad freedoms that are fixed in spherical gauge,
equations~(\ref{gaugespherical}).
The spin-$\pm 1$ components
$\Fz_{\plusminus \one}$
(and their angular conjugates)
are not tetrad gauge-invariant with respect to arbitrary
infinitesimal tetrad transformations,
being changed by a tetrad transformation that varies
the antisymmetric part of
$\varphi_{\overset{{\scriptstyle v} \plus}{{\scriptstyle u} \minus}}$
(or its angular conjugate
$\varphi_{\overset{{\scriptstyle v} \minus}{{\scriptstyle u} \plus}}$).
The following combinations
of
$\Fz_{\plusminus \one}$
or
$\Fz^\star_{\plusminus \one}$
and $\Fz_{\zero}$
are tetrad gauge-invariant,
but not coordinate gauge-invariant:
\begin{subequations}
\label{F1tet}
\begin{eqnarray}
\label{F1teta}
  &&
  \Fz_{\plusminus \one}
  \mp
  \bigl(
    \varphi_{\overset{{\scriptstyle v} \plus}{{\scriptstyle u} \minus}}
    -
    \varphi_{\overset{\plus {\scriptstyle v}}{\minus {\scriptstyle u}}}
  \bigr)
  \Fz_\zero
  \ ,
\\
\label{F1tetb}
  &&
  \Fz^\star_{\plusminus \one}
  \mp
  \bigl(
    \varphi_{\overset{{\scriptstyle v} \minus}{{\scriptstyle u} \plus}}
    -
    \varphi_{\overset{\minus {\scriptstyle v}}{\plus {\scriptstyle u}}}
  \bigr)
  \Fz_\zero
  \ .
\end{eqnarray}
\end{subequations}

The one component of the complexified electromagnetic field tensor that is
coordinate and tetrad gauge-invariant and vanishing in the background
is the axial spin-$0$ component
$\Fz^{(a)}_\zero$
\begin{equation}
\label{Fza}
  \Fz^{(a)}_\zero
  =
  {\textstyle \frac{1}{2}}
  ( \Fz_\zero - \Fz^\star_\zero )
  =
  {\textstyle \frac{1}{2}}
  F_{\plus\minus}
  =
  {\textstyle \frac{1}{2}}
  \im B_r
  \ ,
\end{equation}
which is $\frac{1}{2} \im$ times the radial component $B_r$ of the magnetic field.

The electromagnetic energy-momentum tensor is given by
$4 \pi T_{mn} = F_{mk} {F_{n}}^k - \frac{1}{4} \gamma_{mn} F_{kl} F^{kl}$.
The components of the electromagnetic energy-momentum tensor
in a Newman-Penrose tetrad are given in the Appendix, equations~(\ref{npTem}).
To linear order,
equations~(\ref{npTem})
reduce to
\begin{subequations}
\label{linnpTem}
\begin{eqnarray}
  4 \pi T_{uv}
  =
  4 \pi T_{\plus \minus}
  &=&
  {\textstyle \frac{1}{2}}
  F_{uv}^2
  =
  2 \,
  \Fz_\zero
  \Fz^\star_\zero
  \ ,
\\
  4 \pi T_{\overset{{\scriptstyle v} \plus}{{\scriptstyle u} \minus}}
  &=&
  F_{\overset{{\scriptstyle v u}}{{\scriptstyle u v}}}
  F_{\overset{{\scriptstyle v} \plus}{{\scriptstyle u} \minus}}
  =
  \mp \,
  ( Q / A )
  \Fz_{\plusminus \one}
  \ ,
\\
  4 \pi T_{\overset{{\scriptstyle v} \minus}{{\scriptstyle u} \plus}}
  &=&
  F_{\overset{{\scriptstyle v u}}{{\scriptstyle u v}}}
  F_{\overset{{\scriptstyle v} \minus}{{\scriptstyle u} \plus}}
  =
  \mp \,
  ( Q / A )
  \Fz^\star_{\plusminus \one}
  \,
\\
  4 \pi T_{\overset{{\scriptstyle v v}}{{\scriptstyle u u}}}
  =
  4 \pi T_{\overset{\plus \plus}{\minus \minus}}
  &=&
  0
  \ .
\end{eqnarray}
\end{subequations}
In the unperturbed background,
the non-vanishing components of the electromagnetic energy-momentum tensor simplify to
\begin{equation}
  4 \pi \overset{\smallzero}{T}_{uv}
  =
  4 \pi \overset{\smallzero}{T}_{\plus \minus}
  =
  {\textstyle \frac{1}{2}}
  ( Q / A )^2
  \ .
\end{equation}

\subsection{Maxwell's equations}
\label{maxwellsequations}

The central result of this subsection is the set of
linearized Maxwell's equations~(\ref{Maxwells})
which describe the evolution of electromagnetic perturbations
in spherically symmetric self-similar black hole spacetimes.
The resulting electromagnetic wave equations~(\ref{waveMaxwell})
will be applied in subsequent subsections.

In a general Newman-Penrose tetrad, Maxwell's equations~(\ref{Maxwellz}) are explicitly,
\begin{subequations}
\label{npMaxwell}
\begin{eqnarray}
  - \, \Gamma_{\overset{\plus {\scriptstyle v} {\scriptstyle v}}{\minus {\scriptstyle u} {\scriptstyle u}}} \, A \, \Fz_{\minusplus \one}
  \, \mp \, \nfrakD{\zero}_\vu \, \Fz_\zero
  - \nfrakD{\plusminus \one}_\minusplus \, \Fz_{\plusminus \one}
  &=&
  2 \pi \, A \,
  j_\vu
  \ ,
  \quad\quad
\\
  \Gamma_{\overset{\plus {\scriptstyle v} \plus}{\minus {\scriptstyle u} \minus}} \, A \, \Fz_{\minusplus \one}
  \, \mp \, \nfrakD{\zero}_\plusminus \, \Fz_\zero
  - \nfrakD{\plusminus \one}_\uv \, \Fz_{\plusminus \one}
  &=&
  2 \pi \, A \,
  j_\plusminus
  \ ,
\end{eqnarray}
\end{subequations}
where the differential operators
$\nfrakD{s}_m$
are defined by
\begin{subequations}
\label{frakDs}
\begin{eqnarray}
  \nfrakD{\plusminus s}_\vu
  &\equiv&
  A \Bigl[
  \partial_\vu
  +
  s \, \Gamma_{\overset{{\scriptstyle u v v}}{{\scriptstyle v u u}}}
  -
  s \, \Gamma_{\overset{\minus \plus {\scriptstyle v}}{\plus \minus {\scriptstyle u}}}
  +
  ( s + 2 ) \, \Gamma_{\overset{\plus {\scriptstyle v} \minus}{\minus {\scriptstyle u} \plus}}
  \Bigr]
  \ ,
\\
  \nfrakD{\plusminus s}_\plusminus
  &\equiv&
  A \Bigl[
  \partial_\plusminus
  -
  s \, \Gamma_{\overset{\minus \plus \plus}{\plus \minus \minus}}
  +
  s \, \Gamma_{\overset{{\scriptstyle u v} \plus}{{\scriptstyle v u} \minus}}
  +
  ( s + 2 ) \, \Gamma_{\overset{\plus {\scriptstyle v u}}{\minus {\scriptstyle u v}}}
  \Bigr]
  \ ,
\end{eqnarray}
\end{subequations}
which reduce in the unperturbed background to
\begin{equation}
  \nfrakDzero{s}_m
  =
  \ncalD{s}_m
  \ .
\end{equation}

To linear order of perturbations on the self-similar background,
Maxwell's equations~(\ref{npMaxwell}) reduce
(in a general gauge) to
\begin{subequations}
\label{Maxwells}
\begin{eqnarray}
\label{Maxwellsa}
  \mp \, {\textstyle \frac{1}{2}} \, \nfrakD{\zero}_\vu \, A^{-1} Q
  \, \mp \, \ncalD{\zero}_\vu \, \overset{\smallone}{\Fz}_\zero
  - \ncalD{\plusminus \one}_{\minusplus} \, \Fz_{\plusminus \one}
  &=&
  2 \pi \, A \,
  j_\vu
  \ ,
\\
\label{Maxwellsb}
  \mp \, {\textstyle \frac{1}{2}} \, \nfrakD{\zero}_\plusminus \, A^{-1} Q
  \, \mp \, \ncalD{\zero}_\plusminus \, \overset{\smallone}{\Fz}_\zero
  - \ncalD{\plusminus \one}_\uv \, \Fz_{\plusminus \one}
  &=&
  2 \pi \, A \,
  j_\plusminus
  \ .
\end{eqnarray}
\end{subequations}
In the unperturbed self-similar background,
Maxwell's equations~(\ref{Maxwells}) simplify to
\begin{subequations}
\label{Maxwellsunperturbed}
\begin{eqnarray}
  \mp \, {\textstyle \frac{1}{2}} \, \ncalD{\zero}_\vu \, A^{-1} Q
  &=&
  2 \pi \, A \,
  \overset{\smallzero}{j}_\vu
  \ ,
\\
  \mp \, {\textstyle \frac{1}{2}} \, \ncalD{\zero}_\plusminus \, A^{-1} Q
  &=&
  0
  \ .
\end{eqnarray}
\end{subequations}

From Maxwell's equations~(\ref{Maxwellz})
and the antisymmetry of the electromagnetic field tensor $\Fz_{kl}$
it follows 
that the electric current $j_l$ satisfies the charge conservation law
\begin{equation}
\label{chargeconservation}
  D^l j_l
  =
  0
  \ .
\end{equation}
In a general Newman-Penrose tetrad,
the electric current conservation equation~(\ref{chargeconservation})
is explicitly
\begin{equation}
\label{Dj}
  A^2 D^l j_l
  =
  - \,
  \nfrakD{\plus \one}^\prime_u \,
  A \,
  j_v
  -
  \nfrakD{\minus \one}^\prime_v \,
  A \,
  j_u
  +
  \nfrakD{\minus \one}^\prime_\plus \,
  A \,
  j_\minus
  +
  \nfrakD{\plus \one}^\prime_\minus \,
  A \,
  j_\plus
  =
  0
  \ ,
\end{equation}
where the differential operators
$\nfrakD{s}^\prime_m$
are
\begin{subequations}
\label{frakDsprime}
\begin{eqnarray}
  \nfrakD{\plusminus s}^\prime_\vu
  &\equiv&
  A^2 \Bigl[
  \partial_\vu
  +
  s \, \Gamma_{\overset{{\scriptstyle u v v}}{{\scriptstyle v u u}}}
  -
  ( s + 1 ) \, \Gamma_{\overset{\minus \plus {\scriptstyle v}}{\plus \minus {\scriptstyle u}}}
  +
  \Gamma_{\overset{\minus {\scriptstyle v} \plus}{\plus {\scriptstyle u} \minus}}
  +
  ( s + 2 ) \, \Gamma_{\overset{\plus {\scriptstyle v} \minus}{\minus {\scriptstyle u} \plus}}
  \Bigr] A^{-1}
  \ ,
\\
  \nfrakD{\plusminus s}^\prime_\plusminus
  &\equiv&
  A^2 \Bigl[
  \partial_\plusminus
  -
  s \, \Gamma_{\overset{\minus \plus \plus}{\plus \minus \minus}}
  +
  ( s + 1 ) \, \Gamma_{\overset{{\scriptstyle u v} \plus}{{\scriptstyle v u} \minus}}
  +
  \Gamma_{\overset{\plus {\scriptstyle u v}}{\minus {\scriptstyle v u}}}
  +
  ( s + 2 ) \, \Gamma_{\overset{\plus {\scriptstyle v u}}{\minus {\scriptstyle u v}}}
  \Bigr] A^{-1}
  \ ,
\end{eqnarray}
\end{subequations}
which reduce in the unperturbed background to
\begin{equation}
  \nfrakDzero{s}{}^\prime_m
  =
  \ncalD{s}_m
  \ .
\end{equation}
Linearized, the electric current conservation equation~(\ref{Dj})
reduces to
\begin{equation}
\label{linDj}
  - \,
  \nfrakD{\plus \one}^\prime_u \,
  A \,
  j_v
  -
  \nfrakD{\minus \one}^\prime_v \,
  A \,
  j_u
  +
  \ncalD{\minus \one}_\plus \,
  A \,
  j_\minus
  +
  \ncalD{\plus \one}_\minus \,
  A \,
  j_\plus
  =
  0
  \ .
\end{equation}

The pair of Maxwell's equations~(\ref{Maxwellsa})
can be combined into wave equations for
the complexified electromagnetic tensor
$\Fz_s$
in much the same way as was done earlier
for the complexified Weyl tensor
$\Cz_s$,
equations~(\ref{waveWeyl}).
Taking one of the operators $\nfrakD{s}^\prime_m$ of equations~(\ref{frakDsprime})
times equation~(\ref{Maxwellsa})
minus another of the operators $\nfrakD{s}^\prime_m$
times equation~(\ref{Maxwellsb}),
the choice of operators being guided by the electric current conservation
equation~(\ref{linDj}),
yields equations that look like sourced wave equations
for each of the (perturbations of the) spin components
$\Fz_s$ of the complexified electromagnetic tensor
(in a general gauge):
\begin{subequations}
\label{waveMaxwell}
\begin{eqnarray}
\label{waveMaxwella}
  \Bigl(
  \ncalD{\zero}^{\dagger \vu} \,
  \ncalD{\zero}_\vu
  +
  \ncalD{\zero}^{\dagger \plusminus} \,
  \ncalD{\zero}_\plusminus
  \Bigl)
  \overset{\smallone}{\Fz}_\zero
  &=&
  - \,
  {\textstyle \frac{1}{2}} \,
  \nfrakD{\plusminus \one}^\prime_\uv
  \Bigl(
    \nfrakD{\zero}_\vu \,
    A^{-1}
    Q
    \pm
    4 \pi \,
    A \,
    j_\vu
  \Bigr)
  +
  {\textstyle \frac{1}{2}} \,
  \ncalD{\plusminus \one}_\minusplus
  \Bigl(
    \nfrakD{\zero}_\plusminus \,
    A^{-1}
    Q
    \pm
    4 \pi \,
    A \,
    j_\plusminus
  \Bigr)
  \ ,
  \qquad
\\
\label{waveMaxwellb}
  \Bigl(
  \ncalD{\plusminus \one}^{\dagger\uv}
  \,
  \ncalD{\plusminus \one}_\uv
  +
  \ncalD{\plusminus \one}^{\dagger\minusplus}
  \,
  \ncalD{\plusminus \one}_\minusplus
  \Bigr)
  \Fz_{\plusminus \one}
  &=&
  {\textstyle \frac{1}{2}} \,
  \nfrakD{\zero}^\prime_\plusminus
  \Bigl(
    \pm \,
    \nfrakD{\zero}_\vu \,
    A^{-1}
    Q
    +
    4 \pi \,
    A \,
    j_\vu
  \Bigr)
  -
  {\textstyle \frac{1}{2}} \,
  \ncalD{\zero}_\vu
  \Bigl(
    \pm \,
    \nfrakD{\zero}_\plusminus \,
    A^{-1}
    Q
    +
    4 \pi \,
    A \,
    j_\plusminus
  \Bigr)
  \ .
\end{eqnarray}
\end{subequations}
As with the wave equations~(\ref{waveWeyl}) for the Weyl tensor
$\Cz_s$,
the wave equations~(\ref{waveMaxwell}) for the electromagnetic tensor
$\Fz_s$
are not independent wave equations:
the spin components $\Fz_s$ are not independently adjustable,
but rather fluctuate in harmony with each other
in accordance with Maxwell's equations~(\ref{Maxwells}).

Equation~(\ref{waveMaxwella})
appears to constitute two separate equations for
$\overset{\smallone}{\Fz}_\zero$,
but the two equations are in fact equivalent.
The mathematical derivation of the equivalence
is almost identical to that which demonstrates
the equivalence of the two wave equations~(\ref{waveWeyla}) for
$\overset{\smallone}{\Cz}_\zero$,
except that
it is necessary to use the electric current conservation equation~(\ref{linDj})
instead of the Weyl current conservation equations~(\ref{linDJz}).

It should be remarked that the electric current conservation
equation~(\ref{linDj})
is insufficient to imply the definition~(\ref{frakDsprime})
of the operators
$\nfrakD{s}{}^\prime_m$
for arbitrary spin $s$,
in particular for the zero spin operators
$\nfrakD{\zero}{}^\prime_m$
that appear in the wave equation~(\ref{waveMaxwellb}) for
$\Fz_{\plusminus \one}$.
Actually,
it is fine to replace the differential operators
$\nfrakD{s}{}^\prime_m$
on the right hand sides of both wave equations~(\ref{waveMaxwell})
with their unperturbed limits
$\ncalD{\plusminus \one}_m$,
since the combined quantities in parentheses that they operate on
are already of first order
(although the separate terms inside the parentheses
are not individually of first order).
The advantage of retaining
$\nfrakD{\plusminus \one}{}^\prime_\uv$
in equation~(\ref{waveMaxwella})
is that it makes application of the electric current conservation equation~(\ref{linDj}),
as in the previous paragraph,
transparent.
As regards equation~(\ref{waveMaxwellb}),
the advantage of
$\nfrakD{\zero}{}^\prime_\plusminus$
as given by equations~(\ref{frakDsprime}),
besides the resulting symmetrical relation between the
expressions~(\ref{Deltasprime}) and (\ref{frakDsprime}) for $\nDelta{s}{}^\prime_m$ and $\nfrakD{s}{}^\prime_m$
and the
expressions~(\ref{Deltas}) and (\ref{frakDs}) for $\nDelta{s}_m$ and $\nfrakD{s}_m$,
is that this choice elimates terms equal to first order quantities times
$\bigl( \partial_\vu + 2 \, \Gamma_{\overset{\plus {\scriptstyle v} \minus}{\minus {\scriptstyle u} \plus}} \bigr) A^{-1} Q$
and
$\bigl( \partial_\plusminus + 2 \, \Gamma_{\overset{\plus {\scriptstyle v u}}{\minus {\scriptstyle u v}}} \bigr) A^{-1} Q$
that would otherwise
appear on the last line of equation~(\ref{frakDQ}),
and that would then
need to be converted to (unperturbed) currents
$\overset{\smallzero}{j}_m$
using equation~(\ref{Maxwellsunperturbed}).
The operators
$\nfrakD{\zero}{}^\prime_m$
defined by equations~(\ref{frakDsprime}) are,
modulo the conformal factors $A$,
the same as those
used by Chandrasekhar \cite{Chandrasekhar} before equation~(204) on page 238.

The tricky parts of the derivations
of the wave equations
for the spin-$\pm 1$ components
$\Fz_{\plusminus \one}$
in \S\ref{spin1electromagneticwaves},
and for the axial spin-$0$ component
$\Fz^{(a)}_{\plusminus \one}$
in \S\ref{axialelectromagneticwaves},
are the parts that depend
on the unperturbed dimensionless interior charge $Q$
on the right hand side of equations~(\ref{waveMaxwell}).
In the case of
$\Fz_{\plusminus \one}$,
this tricky part of the right hand side of
the wave equation~(\ref{waveMaxwellb})
is given as equation~(\ref{frakDQ}) in the Appendix,
which linearizes (in a general gauge) to
\begin{equation}
\label{linfrakDQ}
  \Bigl(
  \nfrakD{\zero}^\prime_\plusminus \, \nfrakD{\zero}_\vu
  -
  \nfrakD{\zero}^\prime_\vu \, \nfrakD{\zero}_\plusminus
  \Bigr)
  A^{-1} Q
  =
  2 Q
  \Bigl(
  2 A \, \Cz_{\plusminus \one}
  -
  A \,
  \ncalD{\plusminus \two}_\uv \,
  A^{-1}
  \Gamma_{\overset{\plus {\scriptstyle v} {\scriptstyle v}}{\minus {\scriptstyle u} {\scriptstyle u}}}
  +
  \ncalD{\plusminus \two}_\minusplus \,
  \Gamma_{\overset{\plus {\scriptstyle v} \plus}{\minus {\scriptstyle u} \minus}}
  \Bigr)
  -
  A \,
  \Gamma_{\overset{\plus {\scriptstyle v} {\scriptstyle v}}{\minus {\scriptstyle u} {\scriptstyle u}}} \,
  \ncalD{\zero}_\uv \,
  A^{-1} Q
  \ ,
\end{equation}
the last term on the right hand side of which
can be converted to unperturbed electric currents
$\overset{\smallzero}{j}_l$
using equation~(\ref{Maxwellsunperturbed}).

In the case of the axial spin-$0$ component
$\Fz^{(a)}_\zero$,
the tricky part of the right hand side of
the wave equation~(\ref{waveMaxwella})
is given as equation~(\ref{frakDQa}) in the Appendix,
which linearizes (in a general gauge) to
\begin{equation}
\label{linfrakDQa}
  \Bigl(
  \nfrakD{\plus \one}^\prime_u \, \nfrakD{\zero}_v
  -
  \nfrakD{\plus \one}^\prime_\minus \, \nfrakD{\zero}_\plus
  -
  \nfrakD{\plus \one}^{\prime \star}_u \, \nfrakD{\zero}^\star_v
  +
  \nfrakD{\plus \one}^{\prime \star}_\minus \, \nfrakD{\zero}^\star_\plus
  \Bigr)
  A^{-1} Q
  =
  4 \, Q
  \, A \, \Cz^{(a)}_\zero
  -
  A
  \Bigl[
  (
    \Gamma_{\minus v \plus}
    -
    \Gamma_{\plus v \minus}
  ) \,
  \ncalD{\zero}_u
  +
  (
    \Gamma_{\plus u \minus}
    -
    \Gamma_{\minus u \plus}
  ) \,
  \ncalD{\zero}_v
  \Bigr]
  A^{-1} Q
  \ ,
\end{equation}
the last set of terms on the right hand side of which
can be converted to unperturbed electric currents
$\overset{\smallzero}{j}_l$
using equation~(\ref{Maxwellsunperturbed}).

\subsection{Spin-one (propagating) components of electromagnetic waves}
\label{spin1electromagneticwaves}

The spin-$\pm 1$ components
$\Fz_{\plusminus \one}$
of the complexified electromagnetic field
are of particular interest because they describe the propagating
components of electromagnetic waves.
The main result of this subsection is the
sourced spin-$\pm 1$ electromagnetic wave equation~(\ref{waveFz}),
whose homogeneous solutions~(\ref{Fz1homogeneous})
have radial part satisfying the generalized spin-$\pm 1$ Teukolsky equation~(\ref{DD1}).

If the unperturbed black hole is uncharged, $Q = 0$,
then the spin-$\pm 1$ components
$\Fz_{\plusminus \one}$
are both coordinate and tetrad gauge-invariant
and vanishing in the background,
and therefore represent real physical perturbations.
However, in the more general case where the black hole is charged,
$Q \neq 0$,
then the spin-$\pm 1$ components
$\Fz_{\plusminus \one}$
are not tetrad gauge-invariant,
being changed by an infinitesimal tetrad transformation of
the antisymmetric part of
$\varphi_{\overset{{\scriptstyle v} \plus}{{\scriptstyle u} \minus}}$.
For $Q \neq 0$,
it is in principle possible to use the tetrad gauge freedom
to set
$\Fz_{\plusminus \one} = 0$.
Indeed,
in application to the Reissner-Nordstr\"om geometry,
Chandrasekhar \cite{Chandrasekhar} page 240
adopts precisely such a gauge,
calling it the ``phantom gauge''.
However,
from a physical perspective
it is unnatural to adopt a gauge
where the propagating component of a field is set to zero.
As noted previously,
expression~(\ref{F1teta}),
the combination
$\Fz_{\plusminus \one} \mp \bigl( \varphi_{\overset{{\scriptstyle v} \plus}{{\scriptstyle u} \minus}} - \varphi_{\overset{\plus {\scriptstyle v}}{\minus {\scriptstyle u}}} \bigr) \Fz_\zero$
is tetrad gauge-invariant.
In situations where
the propagating components of the electromagnetic field
are much larger than the static component,
$\bigl| \Fz_{\plusminus \one} \bigr| \gg \bigl| \Fz_\zero \bigr|$,
which an observer would interpret confidently as light waves,
setting
$\Fz_{\plusminus \one}$
to zero
would require setting the supposedly small Lorentz transformation
$\varphi_{\overset{{\scriptstyle v} \plus}{{\scriptstyle u} \minus}} - \varphi_{\overset{\plus {\scriptstyle v}}{\minus {\scriptstyle u}}}$
to some large value,
which would be quite unnatural.
Thus
the spin-$\pm 1$ components
$\Fz_{\plusminus \one}$
of the electromagnetic field
have physical content
notwithstanding the tetrad gauge ambiguity,
and it makes sense to consider the wave equation
that governs their propagation.


The linearized wave equation
for the spin-$\pm 1$ components $\Fz_{\plusminus \one}$
of the electromagnetic field is,
from equations~(\ref{waveMaxwellb}) and (\ref{linfrakDQ}),
\begin{equation}
\label{waveFz}
  \Bigl(
  \ncalD{\plusminus \one}^{\dagger\uv}
  \,
  \ncalD{\plusminus \one}_\uv
  +
  \ncalD{\plusminus \one}^{\dagger\minusplus}
  \,
  \ncalD{\plusminus \one}_\minusplus
  \Bigr)
  \Fz_{\plusminus \one}
  =
  \emS_{\plusminus \one}
  \ ,
\end{equation}
where the source term
$\emS_{\plusminus \one}$ is,
in a general gauge,
\begin{equation}
\label{Sem}
  \emS_{\plusminus \one}
  =
  \pm \,
  A
  Q
  \Bigl(
  2 \, \Cz_{\plusminus \one}
  -
  \ncalD{\plusminus \two}_\uv \,
  A^{-1}
  \Gamma_{\overset{\plus {\scriptstyle v} {\scriptstyle v}}{\minus {\scriptstyle u} {\scriptstyle u}}}
  +
  \ncalD{\plusminus \two}_\minusplus \,
  A^{-1}
  \Gamma_{\overset{\plus {\scriptstyle v} \plus}{\minus {\scriptstyle u} \minus}}
  \Bigr)
  +
  2 \pi
  A^2 \,
  \Gamma_{\overset{\plus {\scriptstyle v} {\scriptstyle v}}{\minus {\scriptstyle u} {\scriptstyle u}}} \,
  \overset{\smallzero}{j}_\uv
  +
  2 \pi \Bigl(
  \nfrakD{\zero}{}^\prime_\plusminus
  \,
  A \,
  j_\vu
  -
  \ncalD{\zero}_\vu \,
  A \,
  j_\plusminus
  \Bigr)
  \ .
\end{equation}
In spherical gauge,
where
$\Gamma_{\overset{\plus {\scriptstyle v} \plus}{\minus {\scriptstyle u} \minus}}$
vanishes,
the source term
$\emS_{\plusminus \one}$,
equation~(\ref{Sem}),
reduces to
\begin{equation}
\label{Semspherical}
  \emS_{\plusminus \one}
  =
  \pm \,
  A
  Q
  \Bigl(
  2 \, \Cz_{\plusminus \one}
  -
  \ncalD{\plusminus \two}_\uv \,
  A^{-1}
  \Gamma_{\overset{\plus {\scriptstyle v} {\scriptstyle v}}{\minus {\scriptstyle u} {\scriptstyle u}}}
  \Bigr)
  +
  2 \pi
  A^2 \,
  \Gamma_{\overset{\plus {\scriptstyle v} {\scriptstyle v}}{\minus {\scriptstyle u} {\scriptstyle u}}} \,
  \overset{\smallzero}{j}_\uv
  +
  2 \pi \Bigl(
  \nfrakD{\zero}{}^\prime_\plusminus
  \,
  A \,
  j_\vu
  -
  \ncalD{\zero}_\vu \,
  A \,
  j_\plusminus
  \Bigr)
  \ .
\end{equation}
In spherical gauge,
and for non-dipole modes, $l \neq 2$,
the connection
$\Gamma_{\overset{\plus {\scriptstyle v} {\scriptstyle v}}{\minus {\scriptstyle u} {\scriptstyle u}}}$
is determined by the spin-$\pm 2$ components
$\Cz_{\plusminus \two}$
of the complexified Weyl tensor,
equation~(\ref{Cz2aspherical}).
Equation~(\ref{Semspherical})
for
$\emS_{\plusminus \one}$
thus shows that
electromagnetic waves are sourced by two separate effects,
first by gravitational waves,
the first sets of terms on the right hand side of equation~(\ref{Semspherical})
(including the term
proportional to the unperturbed electric current
$\overset{\smallzero}{j}_l$),
and second by
perturbations to the electric currents $j_l$,
the last set of terms on the right hand side of equation~(\ref{Semspherical}).
If the unperturbed black hole is uncharged, $Q = 0$
[which among other things implies that
the unperturbed electric current vanishes,
$\overset{\smallzero}{j}_l$ = 0,
in accordance with equations~(\ref{Maxwellsunperturbed})],
then the gravitational wave source vanishes,
but if the black hole is charged,
then gravitational waves are an inevitable source of electromagnetic waves.

Much of the discussion in \S\ref{spin2gravitationalwaves}
concerning the propagating components
$\Cz_{\plusminus \two}$
of gravitational waves,
starting from the paragraph containing equation~(\ref{waveCzstar})
and on to the end of the subsection,
carries through with little change
for the case of the propagating components
$\Fz_{\plusminus \one}$
of electromagnetic waves.
The angular conjugates
$\Fz^\star_{\plusminus \one}$
satisfy the angular conjugate of equation~(\ref{waveFz})
\begin{equation}
\label{waveFzstar}
  \Bigl(
  \ncalD{\plusminus \one}^{\dagger \uv} \,
  \ncalD{\plusminus \one}_\uv
  +
  \ncalD{\minusplus \one}^{\dagger \plusminus} \,
  \ncalD{\minusplus \one}_\plusminus
  \Bigr)
  \Fz^\star_{\plusminus \one}
  =
  \emS^\star_{\plusminus \one}
  \ .
\end{equation}
The angular Hermitian operators
$\ncalD{\plus \one}^{\dagger \minus} \ncalD{\plus \one}_\minus$
and
$\ncalD{\minus \one}^{\dagger \plus} \ncalD{\minus \one}_\plus$
inside the wave operators
of equation~(\ref{waveFz}) and its angular conjugate~(\ref{waveFzstar})
have the same eigenvalues
(and complex conjugate angular eigenfunctions),
and it follows that
the temporal-radial (i.e.\ non-angular) part of the eigenfunctions of the wave operators
are the same for both
$\Fz_{\plusminus \one}$
and its angular conjugate
$\Fz^\star_{\plusminus \one}$.

The wave operator on the left hand side of equation~(\ref{waveFz})
is identical to the wave operator discussed in
\S\ref{spinradialoperators},
equation~(\ref{wave}),
for the case of spin one, $s = 1$.
The homogeneous solutions
(when the source terms vanish)
of the wave equation~(\ref{waveFz})
and its angular conjugate~(\ref{waveFzstar})
are, similarly to equation~(\ref{Psi}),
\begin{equation}
\label{Fz1homogeneous}
  \left.
  \begin{array}{r}
    \Fz_{\plusminus \one} \\
    \Fz^\star_{\plusminus \one}
  \end{array}
  \right\}
  =
  ( {\xi^v / \xi^u} )^{\mp 1/2}
  A^{-2} \,
  \ee^{- \im \omega t}
  \nfz{\plusminus \one}_l(r) \,
  \left\{
  \begin{array}{l}
    \nY{\plusminus \one}_{lm}(\theta,\phi) \\
    \nY{\plusminus \one}^\ast_{lm}(\theta,\phi)
  \end{array}
  \right.
\end{equation}
where $\nfz{\plusminus \one}_l(r)$ depends only on conformal radius $r$,
and satisfies the second order ordinary differential equation
\begin{equation}
\label{DD1}
  \left\{
  - \,
  \Dr{2}{}
  +
  \bigl(
    \im \omega
    \, \pm \, 2 \kappa
  \bigr)^2
  +
  \bigl[
    l ( l + 1 )
    - 1
    - 6 C
  \bigr]
  \Delta
  \right\}
  \nfz{\plusminus \one}_{\omega l}
  =
  0
  \ ,
\end{equation}
which agrees with the generalized Teukolsky equation~(\ref{teukolsky})
given in the Introduction, since,
equation~(\ref{C}),
\begin{equation}
  - 6 C
  =
  - 2 ( R - P ) - 2 P_\perp + 6 M
  \ .
\end{equation}

Equation~(\ref{DD1})
has two linearly independent eigensolutions,
which may be identified as ingoing and outgoing
based on their properties
in a hypothetical asymptotically flat region at large radius,
as already discussed in \S\ref{asymptotic}.
In a hypothetical aysmptotically flat region at large radius,
the radial eigensolutions
$\nfz{s}_{\omega l}$
(with $s = \pm 1$)
go over to the asymptotic solutions
$\nz{s}_l$
discussed in \S\ref{asymptotic},
equation~(\ref{zasymptotic}).
Because the wave operators for modes of opposite spin are
complex conjugates of each other,
equation~(\ref{DD1}),
it follows that
the ingoing and outgoing radial eigensolutions
$\nfz{s}_{\omega l}$
of opposite spin can be taken to be
complex conjugates of each other,
$\nfz{s}^\textrm{in}_{\omega l} = \bigl( \nfz{\minus s}^\textrm{out}_{\omega l} \bigr)^\ast$,
in complete analogy to equation~(\ref{zinout}).
The propagating components of
$\Fz_s \propto A^{-2} \ee^{- \im \omega t} \nfz{s}_{\omega l}$
are those that fall off most slowly at large radius,
which identifies the positive spin component
$\Fz_{\plus \one}$
as describing the propagating component of ingoing electromagnetic waves,
and the negative spin component
$\Fz_{\minus \one}$
as describing the propagating component of outgoing electromagnetic waves:
\begin{subequations}
\begin{eqnarray}
  \Fz^\textrm{in}_{\plus \one}
  \  \propto \ 
  A^{-2} \ee^{- \im \omega t}
  \nfz{\plus \one}^\textrm{in}_{\omega l}
  \  \sim \ 
  A^{-2} r^\ast \ee^{- \im \omega ( t + r^\ast )}
  &\quad&
  ( \mbox{propagating, ingoing} )
  \ ,
\\
  \Fz^\textrm{out}_{\minus \one}
  \  \propto \ 
  A^{-2} \ee^{- \im \omega t}
  \nfz{\minus \one}^\textrm{out}_{\omega l}
  \  \sim \ 
  A^{-2} r^\ast \ee^{- \im \omega ( t - r^\ast )}
  &\quad&
  ( \mbox{propagating, outgoing} )
  \ .
\end{eqnarray}
\end{subequations}
The other of the two modes for each spin corresponds to the non-propagating,
short-range spin-$(-s)$ partner of a propagating mode of spin $s$:
\begin{subequations}
\begin{eqnarray}
  \Fz^\textrm{in}_{\minus \one}
  \  \propto \ 
  A^{-2} \ee^{- \im \omega t}
  \nfz{\minus \one}^\textrm{in}_{\omega l}
  \  \sim \ 
  A^{-2} (r^\ast)^{-1} \ee^{- \im \omega ( t + r^\ast )}
  &\quad&
  ( \mbox{non-propagating, ingoing} )
  \ ,
\\
  \Fz^\textrm{out}_{\plus \one}
  \  \propto \ 
  A^{-2} \ee^{- \im \omega t}
  \nfz{\plus \one}^\textrm{out}_{\omega l}
  \  \sim \ 
  A^{-2} (r^\ast)^{-1} \ee^{- \im \omega ( t - r^\ast )}
  &\quad&
  ( \mbox{non-propagating, outgoing} )
  \ .
\end{eqnarray}
\end{subequations}

\subsection{Electromagnetic monopole modes}
\label{monopoleelectromagneticwaves}

For monopole modes, $l = 0$,
all spin-$\pm 1$ quantities vanish identically,
and therefore the spin-$\pm 1$ components
$\Fz_{\plusminus \one}$
of the complexified electromagnetic field tensor
cannot describe monopole modes.
The bottom line of this subsection is
the unsurprising result
that the equation~(\ref{DFzmonopole})
for the polar part of the electromagnetic monopole mode
simply enforces conservation of interior electric charge,
and does not describe any electromagnetic wave.
The axial electromagnetic monopole mode vanishes identically,
equation~(\ref{Fzamonopole})
(at least in the absence of magnetic monopoles).

An equation for the monopole follows from the
linearized Maxwell equation~(\ref{Maxwellsa}),
which can be written
(in a general gauge, and for arbitrary harmonics)
\begin{equation}
\label{DvuQ}
  A^2
  \bigl(
    \partial_\vu
    +
    \Gamma_{\overset{\minus {\scriptstyle v} \plus}{\plus {\scriptstyle u} \minus}}
    +
    \Gamma_{\overset{\plus {\scriptstyle v} \minus}{\minus {\scriptstyle u} \plus}}
  \bigr)
  A^{-2} \emQ
  +
  2 A \,
  \ncalD{\zero}_\vu \, \Fz^{(a)}_\zero
  \, \pm \,
  2 A \,
  \ncalD{\plusminus \one}_{\minusplus} \, \Fz_{\plusminus \one}
  =
  A Q
  \bigl(
    \Gamma_{\overset{\minus {\scriptstyle v} \plus}{\plus {\scriptstyle u} \minus}}
    -
    \Gamma_{\overset{\plus {\scriptstyle v} \minus}{\minus {\scriptstyle u} \plus}}
  \bigr)
  \, \mp \,
  4 \pi \, A^2 \,
  j_\vu
  \ ,
\end{equation}
where
$\emQ$
is the generalized interior electric charge
defined by
(notation:
script $\emQ$ is the generalized interior electric charge,
to be distinguished from the
italic $Q$ dimensionless unperturbed interior electric charge
related to the unperturbed electric charge $q$ by $Q = q / A$)
\begin{equation}
\label{emQ}
  \emQ
  \equiv
  2 A^2 \Fz^{(p)}_\zero
  \ ,
\end{equation}
whose unperturbed value is just
the electric charge $q$
interior to radius $r$ in the unperturbed background,
equation~(\ref{Q}),
\begin{equation}
\label{emQunperturbed}
  \overset{\smallzero}{\emQ}
  =
  q
  \ .
\end{equation}
The generalized interior charge $\emQ$
is tetrad gauge-invariant,
but it is not coordinate gauge-invariant,
and moreover the split between its unperturbed and perturbed parts
is not coordinate gauge-invariant.
However,
the unperturbed part
$\overset{\smallzero}{\emQ} = q$,
which transforms under a coordinate gauge transformation as
\begin{equation}
  \overset{\smallzero}{\emQ}
  \rightarrow
  \overset{\smallzero}{\emQ}
  +
  A (
    \epsilon^v \partial_v
    +
    \epsilon^u \partial_u
  )
  \overset{\smallzero}{\emQ}
  =
  \overset{\smallzero}{\emQ}
  -
  4 \pi A
  \bigl(
    \epsilon^v
    \overset{\smallzero}{j}_v
    -
    \epsilon^u
    \overset{\smallzero}{j}_u
  \bigr)
  \ ,
\end{equation}
does become coordinate (and tetrad) gauge-invariant
in any region where the unperturbed background electric charge and current vanish,
$\overset{\smallzero}{j}_l = 0$,
such as in a hypothetical asymptotically flat empty region far from the black hole.
In such a region where the unperturbed background charge and current vanish,
the unperturbed Maxwell's equations~(\ref{Maxwellsunperturbed})
imply that
the unperturbed interior electric charge
$\overset{\smallzero}{\emQ}$
goes over to a constant,
the charge $q_\bullet$ of the black hole,
\begin{equation}
\label{emQunperturbedconstant}
  \overset{\smallzero}{\emQ}
  \rightarrow
  q_\bullet
  =
  \mbox{constant}
  \ ,
\end{equation}
a coordinate and tetrad gauge-invariant quantity.

For the monopole mode, $l = 0$,
all non-zero spin objects in equation~(\ref{DvuQ})
vanish identically
[including
$\Gamma_{\minus {\scriptstyle v} \plus} - \Gamma_{\plus {\scriptstyle v} \minus}$
and
$\Gamma_{\plus {\scriptstyle u} \minus} - \Gamma_{\minus {\scriptstyle u} \plus}$,
whose expressions~(\ref{Gammapvmm}) in terms of vierbein components $\varphi_{mn}$
contain only spin-$\pm 1$ terms],
and equation~(\ref{DvuQ}) simplifies to
\begin{equation}
\label{Fzmonopole}
  A^2
  \bigl(
    \partial_\vu
    +
    \Gamma_{\overset{\minus {\scriptstyle v} \plus}{\plus {\scriptstyle u} \minus}}
    +
    \Gamma_{\overset{\plus {\scriptstyle v} \minus}{\minus {\scriptstyle u} \plus}}
  \bigr)
  A^{-2}
  \emQ
  +
  2 A \,
  \ncalD{\zero}_\vu \, \Fz^{(a)}_\zero
  =
  \mp \,
  4 \pi \, A^2 \,
  j_\vu
  \quad
  ( l = 0 )
  \ .
\end{equation}

The polar part of the electromagnetic monopole equation~(\ref{Fzmonopole}) is,
in a general gauge,
\begin{equation}
\label{DFzmonopole}
  A^2
  \bigl(
    \partial_\vu
    +
    \Gamma_{\overset{\minus {\scriptstyle v} \plus}{\plus {\scriptstyle u} \minus}}
    +
    \Gamma_{\overset{\plus {\scriptstyle v} \minus}{\minus {\scriptstyle u} \plus}}
  \bigr)
  A^{-2}
  \emQ
  =
  \mp \,
  4 \pi \, A^2 \,
  j_\vu
  \quad
  ( l = 0 )
  \ .
\end{equation}
In spherical gauge,
where equation~(\ref{dAa}) holds,
equation~(\ref{DFzmonopole})
simplifies to
\begin{equation}
\label{DFzmonopolespherical}
  \partial_\vu \,
  \emQ
  =
  \mp \,
  4 \pi \, A^2 \,
  j_\vu
  \quad
  ( l = 0 )
  \ .
\end{equation}
Equation~(\ref{DFzmonopole}),
or its spherical gauge version~(\ref{DFzmonopolespherical}),
simply expresses conservation of (the monopole part of the)
generalized interior electric charge
$\emQ$
to linear order.

%

The axial part of the monopole equation~(\ref{Fzmonopole}) is,
in a general gauge,
$\ncalD{\zero}_\vu \, \Fz^{(a)}_\zero = 0$,
which integrates to
\begin{equation}
\label{Fzamonopole}
  \Fz^{(a)}_\zero
  =
  0
  \quad
  ( l = 0 )
  \ ,
\end{equation}
that is, the monopole axial spin-$0$ component
of the electromagnetic field tensor is identically zero,
a coordinate and tetrad gauge-invariant statement.

\subsection{Axial spin-zero component of electromagnetic waves}
\label{axialelectromagneticwaves}

The previous two subsections,
\S\S\ref{spin1electromagneticwaves} and \ref{monopoleelectromagneticwaves}
have given a complete description of electromagnetic perturbations:
\S\ref{spin1electromagneticwaves}
covered spin-$\pm 1$ electromagnetic waves,
which encompass all electromagnetic perturbations except monopole modes;
while \S\ref{spin1electromagneticwaves} treated the special case of monopole modes.
Thus in principle there is no need to probe
the electromagnetic equations any further.

However,
for the reasons enunciated at the beginning of \S\ref{axialgravitationalwaves},
it is desirable to consider wave equations for the polar and axial parts
of the spin-$0$ component
$\Fz_\zero$
of the electromagnetic field tensor,
and since
the Newman-Penrose formalism does provide a spin-$0$ wave equation
for electromagnetic perturbations,
equation~(\ref{waveMaxwella}),
it seems a good idea to explore where this equation leads.
This subsection gives the wave equation~(\ref{waveFza})
for axial spin-$0$ electromagnetic modes,
while the next subsection \S\ref{polarelectromagneticwaves}
considers the wave equation for polar spin-$0$ electromagnetic modes.

The axial spin-$0$ component
$\Fz^{(a)}_\zero$,
equation~(\ref{Fza}),
of the complexified electromagnetic field tensor,
which equals $\frac{1}{2} \im$ times the radial component $B_r$ of the magnetic field,
is the only component of the electromagnetic field tensor
that is
coordinate and tetrad gauge-invariant and vanishing in the background
(an assertion that includes the propagating spin-$\pm 1$ components,
which are not tetrad gauge-invariant).
The linearized wave equation for
$\Fz^{(a)}_\zero$
is,
from equations~(\ref{waveMaxwella}) and (\ref{linfrakDQa}),
\begin{equation}
\label{waveFza}
  \Bigl(
  \ncalD{\zero}^{\dagger \vu} \,
  \ncalD{\zero}_\vu
  +
  \ncalD{\zero}^{\dagger \plusminus} \,
  \ncalD{\zero}_\plusminus
  \Bigl)
  \Fz^{(a)}_\zero
  =
  \emS^{(a)}_\zero
  \ ,
\end{equation}
where the source term
$\emS^{(a)}_{\zero}$ is,
in a general gauge,
\begin{equation}
\label{Sema}
  \emS^{(a)}_{\zero}
  =
  - \,
  Q
  \, A \, \Cz^{(a)}_\zero
  +
  \pi
  A^2
  \bigl[
  (
    \Gamma_{\minus v \plus}
    -
    \Gamma_{\plus v \minus}
  ) \,
  \overset{\smallzero}{j}_u
  -
  (
    \Gamma_{\plus u \minus}
    -
    \Gamma_{\minus u \plus}
  ) \,
  \overset{\smallzero}{j}_v
  \bigr]
  -
  \pi
  A
  \bigl(
    \ncalD{\minus \one}_\plus \,
    j_\minus
    -
    \ncalD{\plus \one}_\minus \,
    j_\plus
  \bigr)
  \ .
\end{equation}
Axial gauge,
equation~(\ref{gaugeaxial}),
simplifies the the source term
$\emS^{(a)}_{\zero}$
for axial spin-$0$ electromagnetic waves,
as was previously found in the source term
$S^{(a)}_\zero$
for axial spin-$0$ gravitational waves.
In axial gauge,
the source term
$\emS^{(a)}_{\zero}$
reduces to
\begin{equation}
\label{Semaaxial}
  \emS^{(a)}_{\zero}
  =
  - \,
  Q
  \, A \, \Cz^{(a)}_\zero
  -
  \pi
  A
  \bigl(
    \ncalD{\minus \one}_\plus \,
    j_\minus
    -
    \ncalD{\plus \one}_\minus \,
    j_\plus
  \bigr)
  \ .
\end{equation}
Equation~(\ref{Semaaxial}) for
$\emS^{(a)}_\zero$
shows that
the axial spin-$0$ component of electromagnetic waves is,
like the spin-$\pm 1$ components,
sourced by two separate effects,
first by gravitational waves,
the first term on the right hand side of equation~(\ref{Semaaxial}),
and second by
perturbations to the electric currents $j_l$,
the last set of terms on the right hand side of equation~(\ref{Semaaxial}).

Much of the discussion in \S\ref{axialgravitationalwaves}
concerning the axial spin-$0$ component
$\Cz^{(a)}_{\zero}$
of gravitational waves,
starting from the paragraph containing equation~(\ref{Czahomogeneous})
and on to the end of the subsection,
carries through with little change
for the case of the axial spin-$0$ component
$\Fz^{(a)}_{\zero}$
of electromagnetic waves.
The homogeneous solutions
(when the source terms vanish)
of the wave equation~(\ref{waveFza})
are, similarly to equation~(\ref{Psi}),
\begin{equation}
\label{Fzahomogeneous}
  \Fz^{(a)}_{\zero}
  =
  A^{-2}
  \ee^{- \im \omega t}
  \nfz{\zero}^{(a)}_{\omega l}(r) \,
  \nY{\zero}_{lm}(\theta,\phi)
  \ ,
\end{equation}
where the radial function $\nfz{\zero}^{(a)}_{\omega l}(r)$
in equation~(\ref{Fzahomogeneous})
depends only on conformal radius $r$,
and satisfies the second order ordinary differential equation
\begin{equation}
\label{DDFa}
  \left[
  - \,
  \Dr{2}{}
  -
  \omega^2
  +
  l ( l + 1 )
  \Delta
  \right]
  \nfz{\zero}^{(a)}_{\omega l}
  =
  0
  \ .
\end{equation}
The scalar potential
$l ( l + 1) \Delta$
in equation~(\ref{DDFa}) agrees with that given by
\cite{Karlovini02} [his equations~(91)--(93) for $s = 1$].

Equation~(\ref{DDFa})
has two linearly independent eigensolutions,
which may be taken to be complex conjugates of each other,
and which may be identified as ingoing and outgoing
based on their properties
in a hypothetical asymptotically flat region at large radius,
as discussed in \S\ref{asymptotic}.
In a hypothetical aysmptotically flat region at large radius,
the radial eigensolutions
$\nfz{\zero}^{(a)}_{\omega l}$
go over to the asymptotic solutions
$\nz{\zero}_l$
discussed in \S\ref{asymptotic},
equation~(\ref{z0}),
\begin{equation}
\label{faasymptotic}
  \begin{array}{rcccl}
    \nfz{\zero}^{(a) \, \textrm{in}}_{\omega l}
    &\rightarrow&
    \nz{\zero}^\textrm{in}_l(\omega r^\ast)
    &\rightarrow&
    \ee^{- \im \omega r^\ast}
    \\
    \nfz{\zero}^{(a) \, \textrm{out}}_{\omega l}
    &\rightarrow&
    \nz{\zero}^\textrm{out}_l(\omega r^\ast)
    &\rightarrow&
    \ee^{\im \omega r^\ast}
  \end{array}
  \quad
  \mbox{as }
  r^\ast \rightarrow \infty
  \ .
\end{equation}

\subsection{Polar spin-zero component of electromagnetic waves}
\label{polarelectromagneticwaves}

The main result of this subsection is the sourced wave equation~(\ref{waveFzp})
for the polar spin-$0$ component of electromagnetic waves.
The polar wave equation~(\ref{waveFzp})
is the same as its axial counterpart~(\ref{Fza}),
aside from the difference in source terms.

The polar spin-$0$ component
$\Fz^{(p)}_\zero = \frac{1}{2} ( \Fz_\zero + \Fz^\star_\zero ) = \frac{1}{2} E_r$
of the electromagnetic field,
which equals $\frac{1}{2}$ the radial component $E_r$ of the electric field,
is both coordinate and tetrad gauge-invariant.
If the unperturbed black hole is uncharged, $Q = 0$,
then
$\Fz^{(p)}_\zero$
is also vanishing in the background,
and its perturbation then represents a real physical perturbation.
However, in the more general case where the black hole is charged,
$Q \neq 0$,
the polar spin-$0$ electromagnetic field
$\Fz^{(p)}_\zero$
does not vanish in the background,
and then the split between its unperturbed and perturbed parts
is not coordinate gauge-invariant.
Indeed,
as long as
the unperturbed spin-$0$ electromagnetic field
$\overset{\smallzero}{\Fz}{}^{(p)}_\zero$
is non-constant,
it is possible in principle to eliminate the perturbation
$\overset{\smallone}{\Fz}{}^{(p)}_\zero$
altogether by a coordinate gauge transformation arranged to satisfy
\begin{equation}
\label{badFzgauge}
  \overset{\smallone}{\Fz}{}^{(p)}_\zero
  -
  A (
    \epsilon^v \partial_v
    +
    \epsilon^u \partial_u
  )
  \overset{\smallzero}{\Fz}{}^{(p)}_\zero
  =
  0
  \ ,
\end{equation}
in much the same way that
it is possible in principle to eliminate the perturbation
$\overset{\smallone}{\Cz}{}^{(p)}_\zero$
to the polar spin-$0$ Weyl tensor,
equation~(\ref{badCzgauge}).
However,
as with the Weyl tensor,
there is reason to expect
that the polar part of
the electromagnetic spin-$0$ wave equation~(\ref{waveMaxwella})
is not vacuous, but genuinely encodes the evolution
of polar spin-$0$ electromagnetic perturbations,
as is certainly the case
if the unperturbed black hole is uncharged, $Q = 0$.

In a sense it is straightforward to obtain
a wave equation for the polar spin-$0$ electromagnetic perturbation:
the Maxwell equation~(\ref{DvuQ})
is (coordinate and tetrad) gauge-invariant
(that is,
the split between the left and right hand sides of the equation is not gauge-invariant,
but the equation taken as a whole is gauge-invariant),
and the monopole equation~(\ref{Fzmonopole})
is similarly gauge-invariant,
and therefore the difference between the general and monopole equations
is gauge-invariant,
and moreover vanishing in the background.
When this equation is combined with equation~(\ref{DpmQ}) below,
the result will be the desired electromagnetic polar spin-$0$ wave equation,
which as a whole will be gauge-invariant
(though the split between left and right hand sides will in general not be gauge-invariant).
However,
subtracting the monopole equation
does not eliminate the contribution of the monopole component
of the electromagnetic field from the wave equation,
because the wave equation involves perturbed derivatives
$\overset{\smallone}{\partial}_m \overset{\smallzero}{\emQ}$
of the unperturbed monopole,
which in general do not vanish
because in general the unperturbed monopole is not constant.

In the case of gravitational polar spin-$0$ waves,
\S\ref{polargravitationalwaves},
it was found most natural to cast the wave equation
in terms of
the polar spin-$0$ quantity $\gravN$ defined by equation~(\ref{gravN}),
which had the property that its unperturbed part was a constant,
equation~(\ref{unperturbedN}),
a coordinate and tetrad gauge-invariant object.
For electromagnetic polar spin-$0$ waves,
there does not appear to be a polar spin-$0$ quantity
whose unperturbed part is always constant.
The closest analogous quantity is the
generalized interior electric charge $\emQ$, equation~(\ref{emQ}),
whose unperturbed part is constant,
equation~(\ref{emQunperturbedconstant}),
over any region where the unperturbed background
electric charge and current vanish,
$\overset{\smallzero}{j}_l = 0$,
such as in a hypothetical asymptotically flat empty region far from the black hole.

The first Maxwell equation~(\ref{Maxwellsa}) was already
recast in terms of the generalized interior electric charge
$\emQ$ in equation~(\ref{DvuQ}).
Similarly recasting the second Maxwell equation~(\ref{Maxwellsb})
in terms of $\emQ$ gives
(in a general gauge, and for arbitrary harmonics)
\begin{equation}
\label{DpmQ}
  \partial_\plusminus
  \emQ
  +
  2 A \,
  \ncalD{\zero}_\plusminus \Fz^{(a)}_\zero
  \, \pm \,
  2 A \,
  \ncalD{\plusminus \one}_\uv \Fz_{\plusminus \one}
  =
  2 Q \,
  \partial_\plusminus A
  - 2 A Q \,
  \Gamma_{\overset{\plus {\scriptstyle v u}}{\minus {\scriptstyle u v}}}
  \, \mp \,
  4 \pi
  A^2
  j_\plusminus
  \ .
\end{equation}
Combining the polar parts of equations~(\ref{DvuQ}) and (\ref{DpmQ})
yields the following prototype wave equation for
the generalized interior charge
$\emQ$
(in a general gauge)
\begin{align}
\label{waveQpre}
  &
  \Bigl[
    \nfrakD{\plus \one}^\prime_u \,
    A
    \bigl(
      \partial_v
      +
      \Gamma_{\minus v \plus}
      +
      \Gamma_{\plus v \minus}
    \bigr)
    A^{-2}
    +
    \nfrakD{\minus \one}^\prime_v \,
    A
    \bigl(
      \partial_u
      +
      \Gamma_{\plus u \minus}
      +
      \Gamma_{\minus u \plus}
    \bigr)
    A^{-2}
    -
    A^{-1}
    \ncalD{\plus \one}_\minus \,
    \partial_\plus
    -
    A^{-1}
    \ncalD{\minus \one}_\plus \,
    \partial_\minus
  \Bigr]
  \emQ
\nonumber
\\
  &
  \quad\quad\quad
  = \,
  2
  A^{-1}
  Q
  \bigl(
    \ncalD{\plus \one}_\minus \,
    \partial_\plus \, A
    +
    \ncalD{\minus \one}_\plus \,
    \partial_\minus \, A
  \bigl)
  \  - \ 
  Q
  \bigl(
    \ncalD{\plus \one}_\minus \, \Gamma_{\plus v u}
    +
    \ncalD{\minus \one}_\plus \, \Gamma_{\minus v u}
    +
    \ncalD{\plus \one}_\minus \, \Gamma_{\plus u v}
    +
    \ncalD{\minus \one}_\plus \, \Gamma_{\minus u v}
  \bigr)
\nonumber
\\
  &
  \quad\quad\quad
  \ \ \ \
  - \,
  4 \pi
  \bigl(
    \nfrakD{\plus \one}^\prime_u \,
    A \,
    j_v
    -
    \nfrakD{\minus \one}^\prime_v \,
    A \,
    j_u
  \bigr)
  \ ,
\end{align}
where the differential operators
$\nfrakD{s}^\prime_m$
are given by equations~(\ref{frakDsprime}).
Equation~(\ref{waveQpre})
has not yet had its unperturbed part subtracted.
In spherical gauge,
where equations~(\ref{dAa})
and
(\ref{perturbationNspherical})
hold,
and
$\Gamma_{\overset{\plus {\scriptstyle v} {\scriptstyle u}}{\minus {\scriptstyle u} {\scriptstyle v}}} = \Gamma_{\overset{\plus {\scriptstyle u} {\scriptstyle v}}{\minus {\scriptstyle v} {\scriptstyle u}}}$
according to
equation~(\ref{linGammapvu}),
the prototype wave equation~(\ref{waveQpre}) becomes
\begin{align}
\label{waveQprespherical}
  &
  \Bigl(
    \nfrakD{\plus \one}^\prime_u \,
    A^{-1}
    \partial_v
    +
    \nfrakD{\minus \one}^\prime_v \,
    A^{-1}
    \partial_u
    -
    A^{-1}
    \ncalD{\plus \one}_\minus \,
    \partial_\plus
    -
    A^{-1}
    \ncalD{\minus \one}_\plus \,
    \partial_\minus
  \Bigl)
  \emQ
\nonumber
\\
  &
  = \,
  4 A^{-1}
  Q \,
  \overset{\smallone}{\gravN}
  -
  2
  Q
  \bigl(
    \ncalD{\plus \one}_\minus \, \Gamma_{\plus v u}
    +
    \ncalD{\minus \one}_\plus \, \Gamma_{\minus v u}
  \bigr)
  -
  4 \pi
  \bigl(
    \nfrakD{\plus \one}^\prime_u \,
    A \,
    j_v
    -
    \nfrakD{\minus \one}^\prime_v \,
    A \,
    j_u
  \bigr)
  \ ,
\end{align}
which again has not yet had its unperturbed part subtracted.
It is seen that if the unperturbed part
$\overset{\smallzero}{\emQ}$
of the generalized interior electric charge were constant,
so that all its derivatives vanished,
$\partial_m \overset{\smallzero}{\emQ} = 0$,
then
$\emQ$
on the left hand side of equation~(\ref{waveQprespherical})
could be replaced by its perturbation
$\overset{\smallone}{\emQ}$.
However,
the unperturbed interior electric charge
$\overset{\smallzero}{\emQ}$
is in general not constant.
Therefore,
when the unperturbed part of equation~(\ref{waveQprespherical}) is subtracted,
the residual contributions from perturbed derivatives of
$\overset{\smallzero}{\emQ}$
must be taken over to the right hand side,
where they become incorporated into the source term.

Acting on the perturbed part
$\overset{\smallone}{\emQ}$
of the generalized interior electric charge,
the wave operator on the left hand side of equation~(\ref{waveQprespherical}) is
$\ncalD{\zero}^{\dagger m} \, \ncalD{\zero}_m \, A^{-2}$.
This,
along with the operator equalities
$\ncalD{\zero}^{\dagger v} \, \ncalD{\zero}_v = \ncalD{\zero}^{\dagger u} \, \ncalD{\zero}_u$
and
$\ncalD{\zero}^{\dagger \plus} \, \ncalD{\zero}_\plus = \ncalD{\zero}^{\dagger \minus} \, \ncalD{\zero}_\minus$
valid for zero spin,
brings the electromagnetic polar spin-$0$ wave equation~(\ref{waveQprespherical})
in spherical gauge to the form
\begin{equation}
\label{waveFzp}
  \Bigl(
  \ncalD{\zero}^{\dagger \vu} \,
  \ncalD{\zero}_\vu
  +
  \ncalD{\zero}^{\dagger \plusminus} \,
  \ncalD{\zero}_\plusminus
  \Bigl)
  \overset{\smallone}{\Fz}{}^{(p)}_\zero
  =
  \emS^{(p)}_\zero
  \ ,
\end{equation}
in which the perturbation
$\overset{\smallone}{\emQ}$
of the generalized interior electric charge
has been replaced by the equivalent perturbation
$\overset{\smallone}{\Fz}{}^{(p)}_\zero = \frac{1}{2} A^{-2} \overset{\smallone}{\emQ}$
of the polar spin-$0$ electromagnetic field,
so as to bring out the similarity between the polar spin-$0$ wave equation~(\ref{waveFzp})
and its axial counterpart~(\ref{waveFza}).
The polar spin-$0$ source term
$\emS^{(p)}_\zero$
on the right hand side of the wave equation~(\ref{waveFzp}) is,
in spherical gauge,
\begin{align}
\label{Sempspherical}
  \emS^{(p)}_\zero
  = \ 
  &
  A^{-1}
  Q \,
  \overset{\smallone}{\gravN}
  -
  {\textstyle \frac{1}{2}}
  Q
  \bigl(
    \ncalD{\plus \one}_\minus \, \Gamma_{\plus v u}
    +
    \ncalD{\minus \one}_\plus \, \Gamma_{\minus v u}
  \bigr)
  -
  \pi
  \bigl(
    \ncalD{\plus \one}_u \,
    A \,
    \overset{\smallone}{j}_v
    -
    \ncalD{\minus \one}_v \,
    A \,
    \overset{\smallone}{j}_u
  \bigr)
\nonumber
\\
  &
  - \,
  {\textstyle \frac{1}{4}}
  \bigl(
    \ncalD{\plus \one}_u \,
    A^{-1}
    \overset{\smallone}{\partial}_v
    +
    \ncalD{\minus \one}_v \,
    A^{-1}
    \overset{\smallone}{\partial}_u
    -
    A^{-1}
    \ncalD{\plus \one}_\minus \,
    \overset{\smallone}{\partial}_\plus
    -
    A^{-1}
    \ncalD{\minus \one}_\plus \,
    \overset{\smallone}{\partial}_\minus
  \bigl)
  \overset{\smallzero}{\emQ}
  \ ,
\end{align}
where the perturbations
$\overset{\smallone}{\partial}_m \, \overset{\smallzero}{\emQ}$
to the directed derivatives of the unperturbed interior electric charge
$\overset{\smallzero}{\emQ}$
are
\begin{equation}
  \overset{\smallone}{\partial}_m \,
  \overset{\smallzero}{\emQ}
  =
  \bigl(
    \varphi_{mu} \,
    \partial_v
    +
    \varphi_{mv} \,
    \partial_u
  \bigr)
  \overset{\smallzero}{\emQ}
  \ ,
\end{equation}
which would vanish if the unperturbed interior electric charge
$\overset{\smallzero}{\emQ}$
were constant.

The expression~(\ref{Sempspherical})
for the source term
$\emS^{(p)}_\zero$
in spherical gauge
is valid for all harmonics,
including monopole and dipole harmonics.
For quadrupole and higher harmonics, $l \geq 2$,
the combination
$\ncalD{\plus \one}_\minus \, \Gamma_{\plus v u} + \ncalD{\minus \one}_\plus \, \Gamma_{\minus v u}$
of tetrad connections
in
$\emS^{(p)}_\zero$
can be replaced by the combination of spin-$\pm 2$ Einstein components
$G_{\plus \plus}$ and $G_{\minus \minus}$
given by equation~(\ref{DGammavuspherical}),
valid in spherical gauge.
Equation~(\ref{Sempspherical}) for
$\emS^{(p)}_\zero$
shows that,
like the other components
(spin-$\pm 1$ and axial spin-$0$)
of electromagnetic waves,
the polar spin-$0$ component
of electromagnetic waves
is sourced both by gravitational waves,
the first two sets of terms on the right hand side of equation~(\ref{Sempspherical}),
and by perturbations to electric currents,
the last two sets of terms on the right hand side of equation~(\ref{Sempspherical})
including the terms involving derivatives
of the unperturbed interior electric charge $\overset{\smallzero}{\emQ}$.
As with the other spin components of electromagnetic waves,
if the unperturbed black hole is uncharged, $Q = 0$,
then the gravitational wave contribution to the electromagnetic source vanishes,
but if the unperturbed black hole is charged, $Q \neq 0$,
then gravitational waves are an inevitable source of electromagnetic waves.

The electromagnetic polar spin-$0$ wave equation~(\ref{waveFzp})
is the same as its axial spin-$0$ counterpart~(\ref{waveFza})
apart from the difference in the polar and axial source terms
$\emS^{(p)}_\zero$
and
$\emS^{(a)}_\zero$.
Therefore the homogeneous solutions for
the polar spin-$0$ perturbation
$\overset{\smallone}{\Fz}{}^{(p)}_\zero$
are essentially the same as those~(\ref{Fzahomogeneous})
for the axial spin-$0$ component
$\Fz^{(a)}_\zero$.
The polar radial eigenfunction
$\nfz{\zero}^{(p)}_{\omega l}(r)$
satisfies the same differential equation~(\ref{DDFa})
as its axial counterpart,
and its asymptotic behavior is likewise the same~(\ref{faasymptotic})
as its axial counterpart.

\section{Scalar waves}
\label{scalarwaves}

This section derives the equations that describe massless scalar waves
in perturbed self-similar black hole spacetimes.

\subsection{Massless scalar field}
\label{scalarfield}

No massless scalar field is known in nature.
Nevertheless, massless scalar fields are of interest
because they describe hypothetical waves of spin zero
that propagate at the speed of light,
and as such provide a theoretical counterpart to the
known waves of spin one (electromagnetic) and two (gravitational)
that propagate at the speed of light.
The wave equation for a massless scalar field $\psi$ is
the massless Klein-Gordon equation
\begin{equation}
\label{scalar}
  D^m \partial_m
  \psi
  =
  0
  \ .
\end{equation}


Without loss of generality,
the scalar field $\psi$ can be taken to be real in an orthonormal tetrad.
In quantum field theory,
it is natural to consider a complex scalar field $\psi$,
in which the field $\psi$
and its complex conjugate $\psi^\ast$
describe particles and anti-particles respectively,
but this is not needed for the present paper.

The scalar field $\psi$
is by assumption a scalar,
a coordinate and tetrad gauge-invariant object.
However, the split between the unperturbed and perturbed parts
of the scalar field is not
coordinate gauge-invariant.

Being scalar,
the field $\psi$ is unchanged by angular conjugation,
which flips the sign of the azimuthal tetrad axis $\gamma_\phi$,
equation~(\ref{angularconjugate}).
The scalar field $\psi$ is thus a purely polar quantity.
For a scalar field there only polar waves, no axial waves.

The energy-momentum tensor of the scalar field is given by
$4 \pi T_{mn} = \pi_m \pi_n - \frac{1}{2} \gamma_{mn} \pi_k \pi^k$,
page 483 of \cite{MTW},
where $\pi_m$ is defined to be the gradient,
or momentum, of $\psi$,
\begin{equation}
\label{pi}
  \pi_m
  \equiv
  \partial_m
  \psi
  \ .
\end{equation}
The components of the scalar energy-momentum tensor
in a general Newman-Penrose tetrad are given in the Appendix, equations~(\ref{npTscalar}).
To linear order,
equations~(\ref{npTscalar})
reduce to
\begin{subequations}
\label{linnpTscalar}
\begin{eqnarray}
  4 \pi T_{\plus \minus}
  &=&
  \pi_v \pi_u
  \ ,
\\
  4 \pi T_{\overset{{\scriptstyle v v}}{{\scriptstyle u u}}}
  &=&
  \pi_\vu \pi_\vu
  \ ,
\\
  4 \pi T_{\overset{{\scriptstyle v} \plus}{{\scriptstyle u} \minus}}
  &=&
  \overset{\smallzero}{\pi}_\vu \pi_\plusminus
  \ ,
\\
  4 \pi T_{\overset{{\scriptstyle v} \minus}{{\scriptstyle u} \plus}}
  &=&
  \overset{\smallzero}{\pi}_\vu \pi_\minusplus
  \ ,
\\
  4 \pi T_{uv}
  \  = \ 
  4 \pi T_{\overset{\plus \plus}{\minus \minus}}
  &=&
  0
  \ .
\end{eqnarray}
\end{subequations}

\subsection{Massless scalar waves}
\label{spin0scalarwaves}

The main result of this subsection
is the sourced massless scalar wave equation~(\ref{wavepsi}),
whose homogeneous solutions~(\ref{psihomogeneous}) have radial part
satisfying the generalized spin-$0$ Teukolsky equation~(\ref{DD0}).

In a Newman-Penrose tetrad, the massless scalar wave equation~(\ref{scalar}) is explicitly
\begin{equation}
\label{npscalar}
  \bigl(
  \nfrakD{\plus \one}^\prime_u \,
  A \,
  \partial_v
  +
  \nfrakD{\minus \one}^\prime_v \,
  A \,
  \partial_u
  -
  \nfrakD{\plus \one}^\prime_\minus \,
  A \,
  \partial_\plus
  -
  \nfrakD{\minus \one}^\prime_\plus \,
  A \,
  \partial_\minus
  \bigr)
  \psi
  =
  0
  \ ,
\end{equation}
where the differential operators
$\nfrakD{s}^\prime_m$
are defined by equation~(\ref{frakDsprime}).
The wave equation~(\ref{npscalar}) linearizes to
\begin{equation}
\label{wavescalar}
  \bigl(
  \nfrakD{\plus \one}^\prime_u \,
  A \,
  \partial_v
  +
  \nfrakD{\minus \one}^\prime_v \,
  A \,
  \partial_u
  -
  \ncalD{\plus \one}_\minus \,
  A \,
  \partial_\plus
  -
  \ncalD{\minus \one}_\plus \,
  A \,
  \partial_\minus
  \bigr)
  \psi
  =
  0
  \ .
\end{equation}
If the unperturbed wave equation is subtracted from the
linearized wave equation~(\ref{wavescalar}),
and the residual contribution from perturbed derivatives of the unperturbed field
$\overset{\smallzero}{\psi}$
taken over to the right hand side,
then the result is a wave equation for the perturbation
$\overset{\smallone}{\psi}$
(it is worth remarking here that
$\partial^{\dagger m} \partial_m = A^{-2} \, \ncalD{\zero}^{\dagger m} \, A^2 \, \ncalD{\zero}_m \, A^{-2}$)
\begin{equation}
\label{linnpscalar}
  \ncalD{\zero}^{\dagger m} \, A^2 \, \ncalD{\zero}_m \, A^{-2} \,
  \overset{\smallone}{\psi}
  =
  2 \sigma
  \ ,
\end{equation}
with source $\sigma$
[the factor of $2$ on the right hand side of equation~(\ref{linnpscalar})
is inserted so that it disappears from equation~(\ref{wavepsi}) below]
given by the following combination of derivatives
of the unperturbed scalar field:
\begin{eqnarray}
\label{Sscalar}
  \sigma
  &=&
  \frac{1}{2}
  \Bigl\{
    - \,
    \ncalD{\plus \one}_u \,
    A \,
    \overset{\smallone}{\partial}_v
    -
    \ncalD{\minus \one}_v \,
    A \,
    \overset{\smallone}{\partial}_u
    +
    \ncalD{\plus \one}_\minus \,
    A \,
    \overset{\smallone}{\partial}_\plus
    +
    \ncalD{\minus \one}_\plus \,
    A \,
    \overset{\smallone}{\partial}_\minus
\nonumber
\\
  &&
  + \,
  A^2
  \bigl[
    - \,
    \bigl(
      \overset{\smallone}{\partial}_u
      -
      \overset{\smallone}{\Gamma}_{vuu}
      +
      \overset{\smallone}{\Gamma}_{\plus u \minus}
      +
      \overset{\smallone}{\Gamma}_{\minus u \plus}
    \bigl)
    \overset{\smallzero}{\partial}_v
    -
    \bigl(
      \overset{\smallone}{\partial}_v
      -
      \overset{\smallone}{\Gamma}_{uvv}
      +
      \overset{\smallone}{\Gamma}_{\minus v \plus}
      +
      \overset{\smallone}{\Gamma}_{\plus v \minus}
    \bigl)
    \overset{\smallzero}{\partial}_u
  \Bigr\}
  \,
  \overset{\smallzero}{\psi}
  \ ,
\end{eqnarray}
where the perturbations
$\overset{\smallone}{\partial}_m \, \overset{\smallzero}{\psi}$
to the directed derivatives of the unperturbed scalar field are
\begin{equation}
  \overset{\smallone}{\partial}_m \,
  \overset{\smallzero}{\psi}
  =
  \bigl(
    \varphi_{mu} \,
    \partial_v
    +
    \varphi_{mv} \,
    \partial_u
  \bigr)
  \overset{\smallzero}{\psi}
  \ .
\end{equation}

If the unperturbed scalar field
$\overset{\smallzero}{\psi}$
is constant,
as for example occurs asymptotically
in a hypothetical asymptotically flat empty spacetime
far from the black hole,
then the source $\sigma$,
equation~(\ref{Sscalar}),
vanishes identically,
and the wave equation~(\ref{npscalar})
is (coordinate and tetrad) gauge-invariant.
However,
in the more general case that the unperturbed scalar field
$\overset{\smallzero}{\psi}$
is non-constant,
then the split between the unperturbed and perturbed parts
of the scalar field
is not coordinate gauge-invariant,
and the wave equation~(\ref{linnpscalar})
is then likewise not coordinate gauge-invariant.
Indeed,
as long as the unperturbed scalar field is non-constant,
it is in principle possible to eliminate the perturbation
$\overset{\smallone}{\psi}$
altogether,
by a coordinate gauge transformation arranged to satisfy
\begin{equation}
\label{badscalargauge}
  A \epsilon^m \partial_m
  \overset{\smallzero}{\psi}
  =
  - \,
  \overset{\smallone}{\psi}
  \ .
\end{equation}
However,
if the perturbation
$\overset{\smallone}{\psi}$
to the scalar field varies on scales much smaller
than the scale over which the unperturbed field
$\overset{\smallzero}{\psi}$
is varying,
then the gauge choice~(\ref{badscalargauge})
can be accomplished only by making the supposedly small
coordinate perturbation $\epsilon^m$ large.
Thus the gauge choice~(\ref{badscalargauge})
is not natural,
and it is increasingly less natural for perturbations of higher frequency
in regions where the unperturbed scalar field is more slowly varying.

It can be argued that
a more natural gauge choice is simply to demand that the
source term $\sigma$,
equation~(\ref{Sscalar}),
vanishes
\begin{equation}
\label{scalargauge}
  \sigma
  =
  0
  \ ,
\end{equation}
a condition that is automatically satisfied
in any region where the unperturbed scalar field is constant,
such as in a hypothetical asymptotically flat
empty region far from the black hole.
However,
in the general case that the unperturbed scalar field is not constant,
the gauge condition~(\ref{scalargauge})
represents a rather complicated condition on the vierbein perturbations $\varphi_{mn}$,
which may not be ideal for numerical calculations.
Moreover,
in the asymptotically flat region far from the black hole,
the question of how the condition~(\ref{scalargauge}) translates into a condition on
the vierbein perturbations $\varphi_{mn}$
may become numerically delicate,
because the gauge condition~(\ref{scalargauge})
is satisfied automatically rather than being a gauge condition.

On balance,
it may be preferable to adopt whatever gauge happens to be
most numerically convenient,
and to accept that the scalar source term~(\ref{Sscalar})
may be complicated.


The identity~(\ref{ADAD}),
along with the operator equalities
$\ncalD{\zero}^{\dagger v} \, \ncalD{\zero}_v = \ncalD{\zero}^{\dagger u} \, \ncalD{\zero}_u$
and
$\ncalD{\zero}^{\dagger \plus} \, \ncalD{\zero}_\plus = \ncalD{\zero}^{\dagger \minus} \, \ncalD{\zero}_\minus$
valid for zero spin,
equations~(\ref{diffDDvu}) and (\ref{diffDDpm}),
brings the linearized scalar wave equation~(\ref{linnpscalar}) to a form
similar to those previously encountered for gravitational waves,
equations~(\ref{waveCz}), (\ref{waveCza}), and (\ref{waveN}),
and electromagnetic waves,
equations~(\ref{waveFz}), (\ref{waveFza}), and (\ref{waveFzp}),
\begin{equation}
\label{wavepsi}
  \Bigl[
  \ncalD{\zero}^{\dagger \uv} \,
  \ncalD{\zero}_\uv
  +
  \ncalD{\zero}^{\dagger \minusplus} \,
  \ncalD{\zero}_\minusplus
  +
  {\textstyle \frac{1}{2}}
  ( - \, R + P + 2 M )
  \Bigr]
  A^{-1} \,
  \overset{\smallone}{\psi}
  =
  \sigma
  \ .
\end{equation}
The wave operator on the left hand side of equation~(\ref{wavepsi})
differs from the wave operator discussed in
\S\ref{spinradialoperators}
only by the addition of the terms
${\textstyle \frac{1}{2}} ( - \, R + P + 2 M )$,
which are functions only of conformal radius $r$.
The homogeneous solutions
of the scalar wave equation~(\ref{wavepsi})
(when the source term vanishes)
are, similarly to equation~(\ref{Psi}),
\begin{equation}
\label{psihomogeneous}
  \overset{\smallone}{\psi}
  =
  A^{-1}
  \ee^{- \im \omega t} \psi_{\omega l}(r) Y_{lm}(\theta, \phi)
  \ ,
\end{equation}
whose radial part $\psi_{\omega l}(r)$
depends only on conformal radius $r$,
and satisfies the second order differential equation
\begin{equation}
\label{DD0}
  \left\{
  - \,
  {\dd^2 \over \dd r^{\ast 2}}
  -
  \omega^2
  +
  \bigl[
  l ( l + 1)
  - R + P + 2 M
  \bigr]
  \Delta
  \right\}
  \psi_{\omega l}(r)
  =
  0
  \ ,
\end{equation}
which agrees with the generalized Teukolsky equation~(\ref{teukolsky})
given in the Introduction.
The scalar potential
$\bigl[ l ( l + 1) - R + P + 2 M \bigr] \Delta$
in equation~(\ref{DD0}) agrees with that given by
\cite{Karlovini02} [his equations~(91)--(93) for $s = 0$].

Equation~(\ref{DD0})
has two linearly independent eigensolutions,
which may be taken to be complex conjugates of each other,
and which may be identified as ingoing and outgoing
based on their properties
in a hypothetical asymptotically flat region at large radius,
as discussed in \S\ref{asymptotic}.
In a hypothetical aysmptotically flat region at large radius,
the radial eigensolutions
$\psi_{\omega l}$
go over to the asymptotic solutions
$\nz{\zero}_l$
discussed in \S\ref{asymptotic},
equation~(\ref{z0}),
\begin{equation}
\label{psiasymptotic}
  \begin{array}{rcccl}
    \psi^{\textrm{in}}_{\omega l}
    &\rightarrow&
    \nz{\zero}^\textrm{in}_l(\omega r^\ast)
    &\rightarrow&
    \ee^{- \im \omega r^\ast}
    \\
    \psi^{\textrm{out}}_{\omega l}
    &\rightarrow&
    \nz{\zero}^\textrm{out}_l(\omega r^\ast)
    &\rightarrow&
    \ee^{\im \omega r^\ast}
  \end{array}
  \quad
  \mbox{as }
  r^\ast \rightarrow \infty
  \ .
\end{equation}

\section{Summary}
\label{summary}

This paper has presented the foundations of the theory
of perturbations of spherically symmetric self-similar black holes.
The theory encompasses the limiting cases of
the Schwarzschild and Reissner-Nordstr\"om geometries,
and of more general stationary spherical spacetimes.
No additional condition on the nature of the energy-momentum has been imposed
beyond those of spherical symmetry and self-similarity
(for example, the energy-momentum is not required to be an ideal fluid).

The primary motivation has been to explore the mass inflation instability
inside black holes,
where the geometry departs hugely from any stationary geometry,
but given the length and technicality of the present paper,
it has seemed best to defer application,
including all issues of numerical computation,
to subsequent work.

Wave equations have been derived for
the propagating components of
gravitational, electromagnetic, and scalar waves,
and it has been shown how the non-propagating,
short-range components are related to the propagating components.
The special cases of the (non-propagating) monopole and dipole components
of gravitational perturbations,
and monopole components of electromagnetic perturbations,
have been treated.
The homogeneous solutions of the wave equations
are separable with respect to
a suitable set of conformal coordinates:
a conformal time $t$ and conformal radius $r$,
and the usual polar and azimuthal angular coordinates $\theta$ and $\phi$.
A generalization of the Teukolsky equation is given as
equation~(\ref{teukolsky})
in the Introduction.

Although separable, the wave equations are not decoupled.
In the presence of non-vanishing background energy-momentum
(that is, except in the limiting case of the Schwarzschild geometry),
waves of each spin -- tensor (spin-$2$), vector (spin-$1$), or scalar (spin-$0$) --
are sourced by perturbations of other kinds,
so that waves of different spin are inevitably coupled to each other,
and convert from one kind to another.
This is unlike the case of the Friedmann-Robertson-Walker metric of cosmology,
where spatial translation and rotation symmetries
ensure that tensor, vector, and scalar modes
are decoupled completely from each other
\cite{Bardeen80,MFB92,Brandenberger05}.

The paper follows the Newman-Penrose null tetrad formalism
\cite{NP62,GHP73}.
From the outset,
each vierbein perturbation is required
to be expandable in spin-weighted harmonics of definite spin
(as opposed to the more general system of tensor harmonics),
and it is shown that this requirement can be imposed consistently
without any gauge conditions.
If each vierbein perturbation is expanded in spin-weighted harmonics,
then it automatically follows that
each component of the tetrad-frame Riemann tensor,
and consequently Einstein and Weyl tensors,
is expandable in spin-weighted harmonics of definite spin.
The spin of each perturbation follows the general rule that it
equals the sum of the $+$'s and $-$'s of its covariant angular indices
in the Newman-Penrose formalism.

Although the Newman-Penrose wave equations
resolve into spin-weighted harmonics
with no gauge conditions,
if in addition it is required that
each tetrad connection (Ricci rotation coefficient)
be expandable in spin-weighted harmonics of definite spin,
then this imposes a set of 8 gauge conditions,
here termed ``spherical'' gauge,
equations~(\ref{gaugespherical}).
Spherical gauge proves to provide a particularly nice
gauge for expressing the source terms of propagating gravitational waves,
and also for exploring polar (but not axial, apparently)
spin-$0$ perturbations,
including monopole modes.

The Newman-Penrose formalism
provides wave equations not only for propagating components of waves,
but also for non-propagating components.
Wave equations for polar and axial spin-$0$ modes are derived in this paper,
partly in an attempt to make some contact
with the traditional description
of perturbations to the Schwarzschild and Reissner-Nordstr\"om geometries
in terms of polar and axial perturbations
\cite{RW57,Zerilli70a,Zerilli74,Moncrief75,Chandrasekhar,CF91}.
The Newman-Penrose spin-$0$ wave equation
does correctly reproduce the Regge-Wheeler axial potential
for the Schwarzschild geometry,
but this is the only potential that it recovers without further manipulation.
The Newman-Penrose formalism
yields not decoupled wave equations
for polar and axial spin-$0$ perturbations,
but rather sourced wave equations.
It would be useful in future work
to see how the Newman-Penrose spin-$0$ wave equations
for case of the Reissner-Nordstr\"om geometry
might be decoupled so as to recover the
Zerilli-Moncrief
polar and axial potentials.

\begin{acknowledgements}
This work was supported in part by
NSF award ESI-0337286.
I thank Jim Hartle for emphasizing
the advantage of working in a Newman-Penrose tetrad,
Roger Penrose for insights
into singularity theorems and what really happens
inside real black holes,
Pedro Avel\'ino,
who spent several months on sabbatical
as a Visiting Fellow at JILA,
for wise advice that helped bring this paper into being,
and
Ji\v{r}\'i Bi\v{c}\'ak, also a JILA Visiting Fellow,
for helpful conversations.
\end{acknowledgements}


\appendix

\section{Newman-Penrose components}
\label{npcomponents}

This appendix gives
expressions for the Newman-Penrose components of
various quantities,
in the notation of the present paper.

\subsection{Riemann tensor}
\label{npRiemann}

The Newman-Penrose components of the Riemann tensor
$R_{klmn}$
in terms of the tetrad connections
$\Gamma_{kmn}$
and their directed derivatives
are,
from equation~(\ref{gammaRiemann}),
\begin{subequations}
\label{npR}
\begin{eqnarray}
  R_{v u v u}
  &=&
  \bigl( \partial_{\scriptstyle v}
    - \Gamma_{{\scriptstyle u} {\scriptstyle v} {\scriptstyle v}}
  \bigr)
  \Gamma_{{\scriptstyle v} {\scriptstyle u} {\scriptstyle u}}
  +
  \bigl( \partial_{\scriptstyle u}
    - \Gamma_{{\scriptstyle v} {\scriptstyle u} {\scriptstyle u}}
  \bigr)
  \Gamma_{{\scriptstyle u} {\scriptstyle v} {\scriptstyle v}}
  -
  \bigl(
    \Gamma_{\plus {\scriptstyle u} {\scriptstyle v}}
    -
    \Gamma_{\plus {\scriptstyle v} {\scriptstyle u}}
  \bigr)
  \Gamma_{{\scriptstyle v} {\scriptstyle u} \minus}
  -
  \bigl(
    \Gamma_{\minus {\scriptstyle v} {\scriptstyle u}}
    -
    \Gamma_{\minus {\scriptstyle u} {\scriptstyle v}}
  \bigr)
  \Gamma_{{\scriptstyle u} {\scriptstyle v} \plus}
\nonumber
\\
  &&
  + \,
  \Gamma_{\plus {\scriptstyle v} {\scriptstyle u}} \,
  \Gamma_{\minus {\scriptstyle u} {\scriptstyle v}}
  +
  \Gamma_{\plus {\scriptstyle u} {\scriptstyle v}} \,
  \Gamma_{\minus {\scriptstyle v} {\scriptstyle u}}
  -
  \Gamma_{\plus {\scriptstyle v} {\scriptstyle v}} \,
  \Gamma_{\minus {\scriptstyle u} {\scriptstyle u}}
  -
  \Gamma_{\minus {\scriptstyle v} {\scriptstyle v}} \,
  \Gamma_{\plus {\scriptstyle u} {\scriptstyle u}}
  \ ,
\label{npRa}
\\
  R_{\plus \minus \plus \minus}
  &=&
  \bigl( \partial_\plus
    +
    \Gamma_{\minus \plus \plus}
  \bigr) \Gamma_{\plus \minus \minus}
  + \bigl( \partial_\minus
    +
    \Gamma_{\plus \minus \minus}
  \bigr) \Gamma_{\minus \plus \plus}
  -
  \bigl(
    \Gamma_{\minus {\scriptstyle v} \plus}
    -
    \Gamma_{\plus {\scriptstyle v} \minus}
  \bigr) \Gamma_{\plus \minus {\scriptstyle u}}
  -
  \bigl(
    \Gamma_{\plus {\scriptstyle u} \minus}
    -
    \Gamma_{\minus {\scriptstyle u} \plus}
  \bigr) \Gamma_{\minus \plus {\scriptstyle v}}
\nonumber
\\
  &&
  - \,
  \Gamma_{\minus {\scriptstyle v} \plus} \,
  \Gamma_{\plus {\scriptstyle u} \minus}
  -
  \Gamma_{\plus {\scriptstyle v} \minus} \,
  \Gamma_{\minus {\scriptstyle u} \plus}
  +
  \Gamma_{\plus {\scriptstyle v} \plus} \,
  \Gamma_{\minus {\scriptstyle u} \minus}
  +
  \Gamma_{\minus {\scriptstyle v} \minus} \,
  \Gamma_{\plus {\scriptstyle u} \plus}
  \ ,
\label{npRb}
\\
  R_{v \plus u \minus}
  &=&
  - \, \bigl( \partial_{\scriptstyle v}
    - \Gamma_{{\scriptstyle u} {\scriptstyle v} {\scriptstyle v}}
    + \Gamma_{\minus {\scriptstyle v} \plus}
  \bigr) \Gamma_{\minus {\scriptstyle u} \plus}
  + \bigl( \partial_\plus
    + \Gamma_{\minus \plus \plus}
    + \Gamma_{\plus {\scriptstyle u} {\scriptstyle v}}
  \bigr) \Gamma_{\minus {\scriptstyle u} {\scriptstyle v}}
  +
  \Gamma_{\plus {\scriptstyle v} {\scriptstyle v}} \,
  \Gamma_{\minus {\scriptstyle u} {\scriptstyle u}}
  -
  \Gamma_{\plus {\scriptstyle v} \plus} \,
  \Gamma_{\minus {\scriptstyle u} \minus}
\nonumber
\\
  &=&
  - \, \bigl( \partial_{\scriptstyle u}
    - \Gamma_{{\scriptstyle v} {\scriptstyle u} {\scriptstyle u}}
    + \Gamma_{\plus {\scriptstyle u} \minus}
  \bigr) \Gamma_{\plus {\scriptstyle v} \minus}
  + \bigl( \partial_\minus
    + \Gamma_{\plus \minus \minus}
    + \Gamma_{\minus {\scriptstyle v} {\scriptstyle u}}
  \bigr) \Gamma_{\plus {\scriptstyle v} {\scriptstyle u}}
  +
  \Gamma_{\plus {\scriptstyle v} {\scriptstyle v}} \,
  \Gamma_{\minus {\scriptstyle u} {\scriptstyle u}}
  -
  \Gamma_{\plus {\scriptstyle v} \plus} \,
  \Gamma_{\minus {\scriptstyle u} \minus}
  \ ,
\label{npRc}
\\
  R_{v \minus u \plus}
  &=&
  - \, \bigl( \partial_{\scriptstyle v}
    - \Gamma_{{\scriptstyle u} {\scriptstyle v} {\scriptstyle v}}
    + \Gamma_{\plus {\scriptstyle v} \minus}
  \bigr) \Gamma_{\plus {\scriptstyle u} \minus}
  + \bigl( \partial_\minus
    + \Gamma_{\plus \minus \minus}
    + \Gamma_{\minus {\scriptstyle u} {\scriptstyle v}}
  \bigr) \Gamma_{\plus {\scriptstyle u} {\scriptstyle v}}
  +
  \Gamma_{\minus {\scriptstyle v} {\scriptstyle v}} \,
  \Gamma_{\plus {\scriptstyle u} {\scriptstyle u}}
  -
  \Gamma_{\minus {\scriptstyle v} \minus} \,
  \Gamma_{\plus {\scriptstyle u} \plus}
\nonumber
\\
  &=&
  - \, \bigl( \partial_{\scriptstyle u}
    - \Gamma_{{\scriptstyle v} {\scriptstyle u} {\scriptstyle u}}
    + \Gamma_{\minus {\scriptstyle u} \plus}
  \bigr) \Gamma_{\minus {\scriptstyle v} \plus}
  + \bigl( \partial_\plus
    + \Gamma_{\minus \plus \plus}
    + \Gamma_{\plus {\scriptstyle v} {\scriptstyle u}}
  \bigr) \Gamma_{\minus {\scriptstyle v} {\scriptstyle u}}
  +
  \Gamma_{\minus {\scriptstyle v} {\scriptstyle v}} \,
  \Gamma_{\plus {\scriptstyle u} {\scriptstyle u}}
  -
  \Gamma_{\minus {\scriptstyle v} \minus} \,
  \Gamma_{\plus {\scriptstyle u} \plus}
  \ ,
\label{npRd}
\\
  R_{v u \plus \minus}
  &=&
  \bigl( \partial_{\scriptstyle v}
    -
    \Gamma_{{\scriptstyle u} {\scriptstyle v} {\scriptstyle v}}
  \bigr)
  \Gamma_{\plus \minus {\scriptstyle u}}
  +
  \bigl( \partial_{\scriptstyle u}
    -
    \Gamma_{{\scriptstyle v} {\scriptstyle u} {\scriptstyle u}}
  \bigr)
  \Gamma_{\minus \plus {\scriptstyle v}}
  -
  \bigl(
    \Gamma_{\plus {\scriptstyle u} {\scriptstyle v}}
    -
    \Gamma_{\plus {\scriptstyle v} {\scriptstyle u}}
  \bigr)
  \Gamma_{\plus \minus \minus}
  -
  \bigl(
    \Gamma_{\minus {\scriptstyle v} {\scriptstyle u}}
    -
    \Gamma_{\minus {\scriptstyle u} {\scriptstyle v}}
  \bigr)
  \Gamma_{\minus \plus \plus}
\nonumber
\\
  &&
  - \,
  \Gamma_{\plus {\scriptstyle v} {\scriptstyle u}} \,
  \Gamma_{\minus {\scriptstyle u} {\scriptstyle v}}
  +
  \Gamma_{\plus {\scriptstyle u} {\scriptstyle v}} \,
  \Gamma_{\minus {\scriptstyle v} {\scriptstyle u}}
  +
  \Gamma_{\plus {\scriptstyle v} {\scriptstyle v}} \,
  \Gamma_{\minus {\scriptstyle u} {\scriptstyle u}}
  -
  \Gamma_{\minus {\scriptstyle v} {\scriptstyle v}} \,
  \Gamma_{\plus {\scriptstyle u} {\scriptstyle u}}
\nonumber
\\
  &=&
  \bigl( \partial_\plus 
    +
    \Gamma_{\minus \plus \plus}
  \bigr)
  \Gamma_{{\scriptstyle v} {\scriptstyle u} \minus}
  +
  \bigl( \partial_\minus 
    +
    \Gamma_{\plus \minus \minus}
  \bigr)
  \Gamma_{{\scriptstyle u} {\scriptstyle v} \plus}
  -
  \bigl(
    \Gamma_{\minus {\scriptstyle v} \plus}
    -
    \Gamma_{\plus {\scriptstyle v} \minus}
  \bigr)
  \Gamma_{{\scriptstyle v} {\scriptstyle u} {\scriptstyle u}}
  -
  \bigl(
    \Gamma_{\plus {\scriptstyle u} \minus}
    -
    \Gamma_{\minus {\scriptstyle u} \plus}
  \bigr)
  \Gamma_{{\scriptstyle u} {\scriptstyle v} {\scriptstyle v}}
\nonumber
\\
  &&
  + \,
  \Gamma_{\plus {\scriptstyle v} \minus} \,
  \Gamma_{\minus {\scriptstyle u} \plus}
  -
  \Gamma_{\minus {\scriptstyle v} \plus} \,
  \Gamma_{\plus {\scriptstyle u} \minus}
  -
  \Gamma_{\plus {\scriptstyle v} \plus} \,
  \Gamma_{\minus {\scriptstyle u} \minus}
  +
  \Gamma_{\minus {\scriptstyle v} \minus} \,
  \Gamma_{\plus {\scriptstyle u} \plus}
  \ ,
\label{npRe}
\\
  R_{\overset{{\scriptstyle v} \plus {\scriptstyle v} \minus}{{\scriptstyle u} \minus {\scriptstyle u} \plus}}
  &=&
  - \,
  \bigl( \partial_\vu
    + \Gamma_{\overset{{\scriptstyle u} {\scriptstyle v} {\scriptstyle v}}{{\scriptstyle v} {\scriptstyle u} {\scriptstyle u}}}
    + \Gamma_{\overset{\minus {\scriptstyle v} \plus}{\plus {\scriptstyle u} \minus}}
  \bigr) \Gamma_{\overset{\minus {\scriptstyle v} \plus}{\plus {\scriptstyle u} \minus}}
  + \bigl( \partial_\plusminus
    + \Gamma_{\overset{\minus \plus \plus}{\plus \minus \minus}}
    + 2 \, \Gamma_{\overset{{\scriptstyle u} {\scriptstyle v} \plus}{{\scriptstyle v} {\scriptstyle u} \minus}}
    + \Gamma_{\overset{\plus {\scriptstyle u} {\scriptstyle v}}{\minus {\scriptstyle v} {\scriptstyle u}}}
  \bigr) \Gamma_{\overset{\minus {\scriptstyle v} {\scriptstyle v}}{\plus {\scriptstyle u} {\scriptstyle u}}}
  +
  \Gamma_{\overset{\minus {\scriptstyle v} {\scriptstyle u}}{\plus {\scriptstyle u} {\scriptstyle v}}}
  \Gamma_{\overset{\plus {\scriptstyle v} {\scriptstyle v}}{\minus {\scriptstyle u} {\scriptstyle u}}}
  -
  \Gamma_{\overset{\plus {\scriptstyle v} \plus}{\minus {\scriptstyle u} \minus}}
  \Gamma_{\overset{\minus {\scriptstyle v} \minus}{\plus {\scriptstyle u} \plus}}
\nonumber
\\
  &=&
  - \,
  \bigl( \partial_\vu
    +
    \Gamma_{\overset{{\scriptstyle u} {\scriptstyle v} {\scriptstyle v}}{{\scriptstyle v} {\scriptstyle u} {\scriptstyle u}}}
    +
    \Gamma_{\overset{\plus {\scriptstyle v} \minus}{\minus {\scriptstyle u} \plus}}
  \bigr)
  \Gamma_{\overset{\plus {\scriptstyle v} \minus}{\minus {\scriptstyle u} \plus}}
  + \bigl( \partial_\minusplus
    +
    \Gamma_{\overset{\plus \minus \minus}{\minus \plus \plus}}
    +
    2 \, \Gamma_{\overset{{\scriptstyle u} {\scriptstyle v} \minus}{{\scriptstyle v} {\scriptstyle u} \plus}}
    +
    \Gamma_{\overset{\minus {\scriptstyle u} {\scriptstyle v}}{\plus {\scriptstyle v} {\scriptstyle u}}}
  \bigr)
  \Gamma_{\overset{\plus {\scriptstyle v} {\scriptstyle v}}{\minus {\scriptstyle u} {\scriptstyle u}}}
  +
  \Gamma_{\overset{\plus {\scriptstyle v} {\scriptstyle u}}{\minus {\scriptstyle u} {\scriptstyle v}}}
  \Gamma_{\overset{\minus {\scriptstyle v} {\scriptstyle v}}{\plus {\scriptstyle u} {\scriptstyle u}}}
  -
  \Gamma_{\overset{\minus {\scriptstyle v} \minus}{\plus {\scriptstyle u} \plus}}
  \Gamma_{\overset{\plus {\scriptstyle v} \plus}{\minus {\scriptstyle u} \minus}}
  \ ,
\label{npRf}
\\
  R_{\overset{{\scriptstyle{v u v} \plus}}{{\scriptstyle{u v u} \minus}}}
  &=&
  - \,
  \bigl( \partial_\vu
    -
    \Gamma_{\overset{\minus \plus {\scriptstyle v}}{\plus \minus {\scriptstyle u}}}
    +
    \Gamma_{\overset{\plus {\scriptstyle v} \minus}{\minus {\scriptstyle u} \plus}}
  \bigr)
  \Gamma_{\overset{\plus {\scriptstyle v} {\scriptstyle u}}{\minus {\scriptstyle u} {\scriptstyle v}}}
  +
  \bigl( \partial_\uv
    -
    2 \, \Gamma_{\overset{{\scriptstyle v} {\scriptstyle u} {\scriptstyle u}}{{\scriptstyle u} {\scriptstyle v} {\scriptstyle v}}}
    -
    \Gamma_{\overset{\minus \plus {\scriptstyle u}}{\plus \minus {\scriptstyle v}}}
  \bigr)
  \Gamma_{\overset{\plus {\scriptstyle v} {\scriptstyle v}}{\minus {\scriptstyle u} {\scriptstyle u}}}
  +
  \Gamma_{\overset{\plus {\scriptstyle v} \minus}{\minus {\scriptstyle u} \plus}} \,
  \Gamma_{\overset{\plus {\scriptstyle u} {\scriptstyle v}}{\minus {\scriptstyle v} {\scriptstyle u}}}
  +
  \bigl(
    \Gamma_{\overset{\minus {\scriptstyle u} {\scriptstyle v}}{\plus {\scriptstyle v} {\scriptstyle u}}}
    -
    \Gamma_{\overset{\minus {\scriptstyle v} {\scriptstyle u}}{\plus {\scriptstyle u} {\scriptstyle v}}}
  \bigr)
  \Gamma_{\overset{\plus {\scriptstyle v} \plus}{\minus {\scriptstyle u} \minus}}
\nonumber
\\
  &=&
  - \,
  \bigl( \partial_\vu
    -
    \Gamma_{\overset{\minus \plus {\scriptstyle v}}{\plus \minus {\scriptstyle u}}}
    +
    \Gamma_{\overset{\minus {\scriptstyle v} \plus}{\plus {\scriptstyle u} \minus}}
  \bigr)
  \Gamma_{\overset{{\scriptstyle u} {\scriptstyle v} \plus}{{\scriptstyle v} {\scriptstyle u} \minus}}
  +
  \bigl( \partial_\plusminus
    +
    \Gamma_{\overset{{\scriptstyle u} {\scriptstyle v} \plus}{{\scriptstyle v} {\scriptstyle u} \minus}}
    +
    \Gamma_{\overset{\plus {\scriptstyle u} {\scriptstyle v}}{\minus {\scriptstyle v} {\scriptstyle u}}}
  \bigr)
  \Gamma_{\overset{{\scriptstyle u} {\scriptstyle v} {\scriptstyle v}}{{\scriptstyle v} {\scriptstyle u} {\scriptstyle u}}}
\nonumber
\\
  &&
  + \,
  \Gamma_{\overset{\minus {\scriptstyle v} \plus}{\plus {\scriptstyle u} \minus}} \,
  \Gamma_{\overset{\plus {\scriptstyle u} {\scriptstyle v}}{\minus {\scriptstyle v} {\scriptstyle u}}}
  -
  \bigl(
    \Gamma_{\overset{{\scriptstyle v} {\scriptstyle u} {\scriptstyle u}}{{\scriptstyle u} {\scriptstyle v} {\scriptstyle v}}}
    +
    \Gamma_{\overset{\minus {\scriptstyle u} \plus}{\plus {\scriptstyle v} \minus}}
  \bigr)
  \Gamma_{\overset{\plus {\scriptstyle v} {\scriptstyle v}}{\minus {\scriptstyle u} {\scriptstyle u}}}
  +
  \bigl(
    \Gamma_{\overset{{\scriptstyle v} {\scriptstyle u} \minus}{{\scriptstyle u} {\scriptstyle v} \plus}}
    +
    \Gamma_{\overset{\minus {\scriptstyle u} {\scriptstyle v}}{\plus {\scriptstyle v} {\scriptstyle u}}}
  \bigr)
  \Gamma_{\overset{\plus {\scriptstyle v} \plus}{\minus {\scriptstyle u} \minus}}
  -
  \Gamma_{\overset{\minus {\scriptstyle v} {\scriptstyle v}}{\plus {\scriptstyle u} {\scriptstyle u}}} \,
  \Gamma_{\overset{\plus {\scriptstyle u} \plus}{\minus {\scriptstyle v} \minus}}
  \ ,
\label{npRg}
\\
  R_{\overset{{\scriptstyle{u v u} \plus}}{{\scriptstyle{v u v} \minus}}}
  &=&
  - \,
  \bigl( \partial_\uv
    -
    \Gamma_{\overset{\minus \plus {\scriptstyle u}}{\plus \minus {\scriptstyle v}}}
    +
    \Gamma_{\overset{\plus {\scriptstyle u} \minus}{\minus {\scriptstyle v} \plus}}
  \bigr)
  \Gamma_{\overset{\plus {\scriptstyle u} {\scriptstyle v}}{\minus {\scriptstyle v} {\scriptstyle u}}}
  +
  \bigl( \partial_\vu
    -
    2 \, \Gamma_{\overset{{\scriptstyle u} {\scriptstyle v} {\scriptstyle v}}{{\scriptstyle v} {\scriptstyle u} {\scriptstyle u}}}
    -
    \Gamma_{\overset{\minus \plus {\scriptstyle v}}{\plus \minus {\scriptstyle u}}}
  \bigr)
  \Gamma_{\overset{\plus {\scriptstyle u} {\scriptstyle u}}{\minus {\scriptstyle v} {\scriptstyle v}}}
  +
  \Gamma_{\overset{\plus {\scriptstyle u} \minus}{\minus {\scriptstyle v} \plus}} \,
  \Gamma_{\overset{\plus {\scriptstyle v} {\scriptstyle u}}{\minus {\scriptstyle u} {\scriptstyle v}}}
  +
  \bigl(
    \Gamma_{\overset{\minus {\scriptstyle v} {\scriptstyle u}}{\plus {\scriptstyle u} {\scriptstyle v}}}
    -
    \Gamma_{\overset{\minus {\scriptstyle u} {\scriptstyle v}}{\plus {\scriptstyle v} {\scriptstyle u}}}
  \bigr)
  \Gamma_{\overset{\plus {\scriptstyle u} \plus}{\minus {\scriptstyle v} \minus}}
\nonumber
\\
  &=&
  - \,
  \bigl( \partial_\uv
    -
    \Gamma_{\overset{\minus \plus {\scriptstyle u}}{\plus \minus {\scriptstyle v}}}
    +
    \Gamma_{\overset{\minus {\scriptstyle u} \plus}{\plus {\scriptstyle v} \minus}}
  \bigr)
  \Gamma_{\overset{{\scriptstyle v} {\scriptstyle u} \plus}{{\scriptstyle u} {\scriptstyle v} \minus}}
  +
  \bigl( \partial_\plusminus
    +
    \Gamma_{\overset{{\scriptstyle v} {\scriptstyle u} \plus}{{\scriptstyle u} {\scriptstyle v} \minus}}
    +
    \Gamma_{\overset{\plus {\scriptstyle v} {\scriptstyle u}}{\minus {\scriptstyle u} {\scriptstyle v}}}
  \bigr)
  \Gamma_{\overset{{\scriptstyle v} {\scriptstyle u} {\scriptstyle u}}{{\scriptstyle u} {\scriptstyle v} {\scriptstyle v}}}
\nonumber
\\
  &&
  + \,
  \Gamma_{\overset{\minus {\scriptstyle u} \plus}{\plus {\scriptstyle v} \minus}} \,
  \Gamma_{\overset{\plus {\scriptstyle v} {\scriptstyle u}}{\minus {\scriptstyle u} {\scriptstyle v}}}
  -
  \bigl(
    \Gamma_{\overset{{\scriptstyle u} {\scriptstyle v} {\scriptstyle v}}{{\scriptstyle v} {\scriptstyle u} {\scriptstyle u}}}
    +
    \Gamma_{\overset{\minus {\scriptstyle v} \plus}{\plus {\scriptstyle u} \minus}}
  \bigr)
  \Gamma_{\overset{\plus {\scriptstyle u} {\scriptstyle u}}{\minus {\scriptstyle v} {\scriptstyle v}}}
  +
  \bigl(
    \Gamma_{\overset{{\scriptstyle u} {\scriptstyle v} \minus}{{\scriptstyle v} {\scriptstyle u} \plus}}
    +
    \Gamma_{\overset{\minus {\scriptstyle v} {\scriptstyle u}}{\plus {\scriptstyle u} {\scriptstyle v}}}
  \bigr)
  \Gamma_{\overset{\plus {\scriptstyle u} \plus}{\minus {\scriptstyle v} \minus}}
  -
  \Gamma_{\overset{\minus {\scriptstyle u} {\scriptstyle u}}{\plus {\scriptstyle v} {\scriptstyle v}}} \,
  \Gamma_{\overset{\plus {\scriptstyle v} \plus}{\minus {\scriptstyle u} \minus}}
  \ ,
\label{npRh}
\\
  R_{\overset{\plus \minus {\scriptstyle v} \plus}{\minus \plus {\scriptstyle u} \minus}}
  &=&
  - \,
  \bigl( \partial_\plusminus
    +
    \Gamma_{\overset{{\scriptstyle u} {\scriptstyle v} \plus}{{\scriptstyle v} {\scriptstyle u} \minus}}
    +
    \Gamma_{\overset{\plus {\scriptstyle v} {\scriptstyle u}}{\minus {\scriptstyle u} {\scriptstyle v}}}
  \bigr)
  \Gamma_{\overset{\plus {\scriptstyle v} \minus}{\minus {\scriptstyle u} \plus}}
  +
  \bigl( \partial_\minusplus
    +
    2 \, \Gamma_{\overset{\plus \minus \minus}{\minus \plus \plus}}
    +
    \Gamma_{\overset{{\scriptstyle u} {\scriptstyle v} \minus}{{\scriptstyle v} {\scriptstyle u} \plus}}
  \bigr)
  \Gamma_{\overset{\plus {\scriptstyle v} \plus}{\minus {\scriptstyle u} \minus}}
  +
  \Gamma_{\overset{\minus {\scriptstyle v} \plus}{\plus {\scriptstyle u} \minus}} \,
  \Gamma_{\overset{\plus {\scriptstyle v} {\scriptstyle u}}{\minus {\scriptstyle u} {\scriptstyle v}}}
  +
  \bigl(
    \Gamma_{\overset{\minus {\scriptstyle u} \plus}{\plus {\scriptstyle v} \minus}}
    -
    \Gamma_{\overset{\plus {\scriptstyle u} \minus}{\minus {\scriptstyle v} \plus}}
  \bigr)
  \Gamma_{\overset{\plus {\scriptstyle v} {\scriptstyle v}}{\minus {\scriptstyle u} {\scriptstyle u}}}
\nonumber
\\
  &=&
  - \,
  \bigl( \partial_\vu
    -
    \Gamma_{\overset{\minus \plus {\scriptstyle v}}{\plus \minus {\scriptstyle u}}}
    +
    \Gamma_{\overset{\minus {\scriptstyle v} \plus}{\plus {\scriptstyle u} \minus}}
  \bigr)
  \Gamma_{\overset{\minus \plus \plus}{\plus \minus \minus}}
  +
  \bigl( \partial_\plusminus
    +
    \Gamma_{\overset{{\scriptstyle u} {\scriptstyle v} \plus}{{\scriptstyle v} {\scriptstyle u} \minus}}
    +
    \Gamma_{\overset{\plus {\scriptstyle u} {\scriptstyle v}}{\minus {\scriptstyle v} {\scriptstyle u}}}
  \bigr)
  \Gamma_{\overset{\minus \plus {\scriptstyle v}}{\plus \minus {\scriptstyle u}}}
\nonumber
\\
  &&
  + \,
  \Gamma_{\overset{\minus {\scriptstyle v} \plus}{\plus {\scriptstyle u} \minus}} \,
  \Gamma_{\overset{\plus {\scriptstyle u} {\scriptstyle v}}{\minus {\scriptstyle v} {\scriptstyle u}}}
  +
  \bigl(
    \Gamma_{\overset{\minus {\scriptstyle u} \plus}{\plus {\scriptstyle v} \minus}}
    +
    \Gamma_{\overset{\minus \plus {\scriptstyle u}}{\plus \minus {\scriptstyle v}}}
  \bigr)
  \Gamma_{\overset{\plus {\scriptstyle v} {\scriptstyle v}}{\minus {\scriptstyle u} {\scriptstyle u}}}
  +
  \bigl(
    \Gamma_{\overset{\plus \minus \minus}{\minus \plus \plus}}
    -
    \Gamma_{\overset{\minus {\scriptstyle u} {\scriptstyle v}}{\plus {\scriptstyle v} {\scriptstyle u}}}
  \bigr)
  \Gamma_{\overset{\plus {\scriptstyle v} \plus}{\minus {\scriptstyle u} \minus}}
  -
  \Gamma_{\overset{\minus {\scriptstyle v} {\scriptstyle v}}{\plus {\scriptstyle u} {\scriptstyle u}}} \,
  \Gamma_{\overset{\plus {\scriptstyle u} \plus}{\minus {\scriptstyle v} \minus}}
  \ ,
\label{npRi}
\\
  R_{\overset{\plus \minus {\scriptstyle u} \plus}{\minus \plus {\scriptstyle v} \minus}}
  &=&
  - \,
  \bigl( \partial_\plusminus
    +
    \Gamma_{\overset{{\scriptstyle v} {\scriptstyle u} \plus}{{\scriptstyle u} {\scriptstyle v} \minus}}
    +
    \Gamma_{\overset{\plus {\scriptstyle u} {\scriptstyle v}}{\minus {\scriptstyle v} {\scriptstyle u}}}
  \bigr)
  \Gamma_{\overset{\plus {\scriptstyle u} \minus}{\minus {\scriptstyle v} \plus}}
  +
  \bigl( \partial_\minusplus
    +
    2 \, \Gamma_{\overset{\plus \minus \minus}{\minus \plus \plus}}
    +
    \Gamma_{\overset{{\scriptstyle v} {\scriptstyle u} \minus}{{\scriptstyle u} {\scriptstyle v} \plus}}
  \bigr)
  \Gamma_{\overset{\plus {\scriptstyle u} \plus}{\minus {\scriptstyle v} \minus}}
  +
  \Gamma_{\overset{\minus {\scriptstyle u} \plus}{\plus {\scriptstyle v} \minus}} \,
  \Gamma_{\overset{\plus {\scriptstyle u} {\scriptstyle v}}{\minus {\scriptstyle v} {\scriptstyle u}}}
  +
  \bigl(
    \Gamma_{\overset{\minus {\scriptstyle v} \plus}{\plus {\scriptstyle u} \minus}}
    -
    \Gamma_{\overset{\plus {\scriptstyle v} \minus}{\minus {\scriptstyle u} \plus}}
  \bigr)
  \Gamma_{\overset{\plus {\scriptstyle u} {\scriptstyle u}}{\minus {\scriptstyle v} {\scriptstyle v}}}
\nonumber
\\
  &=&
  - \,
  \bigl( \partial_\uv
    -
    \Gamma_{\overset{\minus \plus {\scriptstyle u}}{\plus \minus {\scriptstyle v}}}
    +
    \Gamma_{\overset{\minus {\scriptstyle u} \plus}{\plus {\scriptstyle v} \minus}}
  \bigr)
  \Gamma_{\overset{\minus \plus \plus}{\plus \minus \minus}}
  +
  \bigl( \partial_\plusminus
    +
    \Gamma_{\overset{{\scriptstyle v} {\scriptstyle u} \plus}{{\scriptstyle u} {\scriptstyle v} \minus}}
    +
    \Gamma_{\overset{\plus {\scriptstyle v} {\scriptstyle u}}{\minus {\scriptstyle u} {\scriptstyle v}}}
  \bigr)
  \Gamma_{\overset{\minus \plus {\scriptstyle u}}{\plus \minus {\scriptstyle v}}}
\nonumber
\\
  &&
  + \,
  \Gamma_{\overset{\minus {\scriptstyle u} \plus}{\plus {\scriptstyle v} \minus}} \,
  \Gamma_{\overset{\plus {\scriptstyle v} {\scriptstyle u}}{\minus {\scriptstyle u} {\scriptstyle v}}}
  +
  \bigl(
    \Gamma_{\overset{\minus {\scriptstyle v} \plus}{\plus {\scriptstyle u} \minus}}
    +
    \Gamma_{\overset{\minus \plus {\scriptstyle v}}{\plus \minus {\scriptstyle u}}}
  \bigr)
  \Gamma_{\overset{\plus {\scriptstyle u} {\scriptstyle u}}{\minus {\scriptstyle v} {\scriptstyle v}}}
  +
  \bigl(
    \Gamma_{\overset{\plus \minus \minus}{\minus \plus \plus}}
    -
    \Gamma_{\overset{\minus {\scriptstyle v} {\scriptstyle u}}{\plus {\scriptstyle u} {\scriptstyle v}}}
  \bigr)
  \Gamma_{\overset{\plus {\scriptstyle u} \plus}{\minus {\scriptstyle v} \minus}}
  -
  \Gamma_{\overset{\minus {\scriptstyle u} {\scriptstyle u}}{\plus {\scriptstyle v} {\scriptstyle v}}} \,
  \Gamma_{\overset{\plus {\scriptstyle v} \plus}{\minus {\scriptstyle u} \minus}}
  \ ,
\label{npRj}
\\
  R_{\overset{{\scriptstyle v} \plus {\scriptstyle u} \plus}{{\scriptstyle u} \minus {\scriptstyle v} \minus}}
  &=&
  - \,
  \bigl( \partial_\vu
    -
    \Gamma_{\overset{{\scriptstyle u} {\scriptstyle v} {\scriptstyle v}}{{\scriptstyle v} {\scriptstyle u} {\scriptstyle u}}}
    -
    2 \, \Gamma_{\overset{\minus \plus {\scriptstyle v}}{\plus \minus {\scriptstyle u}}}
    +
    \Gamma_{\overset{\minus {\scriptstyle v} \plus}{\plus {\scriptstyle u} \minus}}
  \bigr)
  \Gamma_{\overset{\plus {\scriptstyle u} \plus}{\minus {\scriptstyle v} \minus}}
  + \bigl( \partial_\plusminus
    -
    \Gamma_{\overset{\minus \plus \plus}{\plus \minus \minus}}
    +
    \Gamma_{\overset{\plus {\scriptstyle u} {\scriptstyle v}}{\minus {\scriptstyle v} {\scriptstyle u}}}
  \bigr)
  \Gamma_{\overset{\plus {\scriptstyle u} {\scriptstyle v}}{\minus {\scriptstyle v} {\scriptstyle u}}}
  +
  \Gamma_{\overset{\plus {\scriptstyle u} {\scriptstyle u}}{\minus {\scriptstyle v} {\scriptstyle v}}} \,
  \Gamma_{\overset{\plus {\scriptstyle v} {\scriptstyle v}}{\minus {\scriptstyle u} {\scriptstyle u}}}
  -
  \Gamma_{\overset{\plus {\scriptstyle u} \minus}{\minus {\scriptstyle v} \plus}} \,
  \Gamma_{\overset{\plus {\scriptstyle v} \plus}{\minus {\scriptstyle u} \minus}}
\nonumber
\\
  &=&
  - \,
  \bigl( \partial_\uv
    -
    \Gamma_{\overset{{\scriptstyle v} {\scriptstyle u} {\scriptstyle u}}{{\scriptstyle u} {\scriptstyle v} {\scriptstyle v}}}
    -
    2 \, \Gamma_{\overset{\minus \plus {\scriptstyle u}}{\plus \minus {\scriptstyle v}}}
    +
    \Gamma_{\overset{\minus {\scriptstyle u} \plus}{\plus {\scriptstyle v} \minus}}
  \bigr)
  \Gamma_{\overset{\plus {\scriptstyle v} \plus}{\minus {\scriptstyle u} \minus}}
  + \bigl( \partial_\plusminus
    -
    \Gamma_{\overset{\minus \plus \plus}{\plus \minus \minus}}
    +
    \Gamma_{\overset{\plus {\scriptstyle v} {\scriptstyle u}}{\minus {\scriptstyle u} {\scriptstyle v}}}
  \bigr)
  \Gamma_{\overset{\plus {\scriptstyle v} {\scriptstyle u}}{\minus {\scriptstyle u} {\scriptstyle v}}}
  +
  \Gamma_{\overset{\plus {\scriptstyle v} {\scriptstyle v}}{\minus {\scriptstyle u} {\scriptstyle u}}} \,
  \Gamma_{\overset{\plus {\scriptstyle u} {\scriptstyle u}}{\minus {\scriptstyle v} {\scriptstyle v}}}
  -
  \Gamma_{\overset{\plus {\scriptstyle v} \minus}{\minus {\scriptstyle u} \plus}} \,
  \Gamma_{\overset{\plus {\scriptstyle u} \plus}{\minus {\scriptstyle v} \minus}}
  \ ,
\label{npRk}
\\
  R_{\overset{{\scriptstyle v} \plus {\scriptstyle v} \plus}{{\scriptstyle u} \minus {\scriptstyle u} \minus}}
  &=&
  - \,
  \Bigl(
  \partial_\vu
  +
  \Gamma_{\overset{{\scriptstyle u v v}}{{\scriptstyle v u u}}}
  -
  2 \, \Gamma_{\overset{\minus \plus {\scriptstyle v}}{\plus \minus {\scriptstyle u}}}
  +
  \Gamma_{\overset{\minus {\scriptstyle v} \plus}{\plus {\scriptstyle u} \minus}}
  +
  \Gamma_{\overset{\plus {\scriptstyle v} \minus}{\minus {\scriptstyle u} \plus}}
  \Bigr)
  \Gamma_{\overset{\plus {\scriptstyle v} \plus}{\minus {\scriptstyle u} \minus}}
  +
  \Bigl(
  \partial_\plusminus
  -
  \Gamma_{\overset{\minus \plus \plus}{\plus \minus \minus}}
  +
  2 \, \Gamma_{\overset{{\scriptstyle u v} \plus}{{\scriptstyle v u} \minus}}
  +
  \Gamma_{\overset{\plus {\scriptstyle u v}}{\minus {\scriptstyle v u}}}
  +
  \Gamma_{\overset{\plus {\scriptstyle v u}}{\minus {\scriptstyle u v}}}
  \Bigr)
  \Gamma_{\overset{\plus {\scriptstyle v} {\scriptstyle v}}{\minus {\scriptstyle u} {\scriptstyle u}}}
  \ ,
\label{npRl}
\\
  R_{\overset{{\scriptstyle u} \plus {\scriptstyle u} \plus}{{\scriptstyle v} \minus {\scriptstyle v} \minus}}
  &=&
  - \,
  \Bigl(
  \partial_\uv
  +
  \Gamma_{\overset{{\scriptstyle v u u}}{{\scriptstyle u v v}}}
  -
  2 \, \Gamma_{\overset{\minus \plus {\scriptstyle u}}{\plus \minus {\scriptstyle v}}}
  +
  \Gamma_{\overset{\minus {\scriptstyle u} \plus}{\plus {\scriptstyle v} \minus}}
  +
  \Gamma_{\overset{\plus {\scriptstyle u} \minus}{\minus {\scriptstyle v} \plus}}
  \Bigr)
  \Gamma_{\overset{\plus {\scriptstyle u} \plus}{\minus {\scriptstyle v} \minus}}
  +
  \Bigl(
  \partial_\plusminus
  -
  \Gamma_{\overset{\minus \plus \plus}{\plus \minus \minus}}
  +
  2 \, \Gamma_{\overset{{\scriptstyle v u} \plus}{{\scriptstyle u v} \minus}}
  +
  \Gamma_{\overset{\plus {\scriptstyle v u}}{\minus {\scriptstyle u v}}}
  +
  \Gamma_{\overset{\plus {\scriptstyle u v}}{\minus {\scriptstyle v u}}}
  \Bigr)
  \Gamma_{\overset{\plus {\scriptstyle u} {\scriptstyle u}}{\minus {\scriptstyle v} {\scriptstyle v}}}
  \ .
\label{npRm}
\end{eqnarray}
\end{subequations}

\subsection{Weyl currents}

The Newman-Penrose components of the Weyl currents $\Jz_{lmn}$
on the right hand sides of equations~(\ref{npWeyl})
for the evolution of the complexified Weyl tensor are
\begin{subequations}
\label{Jz}
\begin{eqnarray}
  \Jz_{\overset{{\scriptstyle v} {\scriptstyle v} {\scriptstyle u}}{{\scriptstyle u} {\scriptstyle u} {\scriptstyle v}}}
  \  = \ 
  \Jz_{\overset{{\scriptstyle v} \minus \plus}{{\scriptstyle u} \plus \minus}}
  &=&
  \frac{1}{4} \Bigl[
  \partial_\vu \bigl(
    {\textstyle \frac{1}{3}} G_{{\scriptstyle v} {\scriptstyle u}} + {\textstyle \frac{2}{3}} G_{\plus \minus}
  \bigr)
  - \bigl( \partial_\plusminus
    + \Gamma_{\overset{\minus \plus \plus}{\plus \minus \minus}}
    + \Gamma_{\overset{{\scriptstyle u} {\scriptstyle v} \plus}{{\scriptstyle v} {\scriptstyle u} \minus}}
    - 2 \, \Gamma_{\overset{\plus {\scriptstyle v} {\scriptstyle u}}{\minus {\scriptstyle u} {\scriptstyle v}}}
    + \Gamma_{\overset{\plus {\scriptstyle u} {\scriptstyle v}}{\minus {\scriptstyle v} {\scriptstyle u}}}
  \bigr) G_{\overset{{\scriptstyle v} \minus}{{\scriptstyle u} \plus}}
\nonumber
\\
  &&
  \quad
  - \, \bigl( \partial_\uv
    - 2 \, \Gamma_{\overset{{\scriptstyle v} {\scriptstyle u} {\scriptstyle u}}{{\scriptstyle u} {\scriptstyle v} {\scriptstyle v}}}
    + \Gamma_{\overset{\plus {\scriptstyle u} \minus}{\minus {\scriptstyle v} \plus}}
    - \Gamma_{\overset{\minus {\scriptstyle u} \plus}{\plus {\scriptstyle v} \minus}}
  \bigr) G_{\overset{{\scriptstyle v} {\scriptstyle v}}{{\scriptstyle u} {\scriptstyle u}}}
  + \bigl( \partial_\minusplus
    + \Gamma_{\overset{\plus \minus \minus}{\minus \plus \plus}}
    + \Gamma_{\overset{{\scriptstyle u} {\scriptstyle v} \minus}{{\scriptstyle v} {\scriptstyle u} \plus}}
    - \Gamma_{\overset{\minus {\scriptstyle u} {\scriptstyle v}}{\plus {\scriptstyle v} {\scriptstyle u}}}
    + 2 \, \Gamma_{\overset{\minus {\scriptstyle v} {\scriptstyle u}}{\plus {\scriptstyle u} {\scriptstyle v}}}
  \bigr) G_{\overset{{\scriptstyle v} \plus}{{\scriptstyle u} \minus}}
\nonumber
\\
  &&
  \quad
  + \, \bigl(
    \Gamma_{\overset{\minus {\scriptstyle v} \plus}{\plus {\scriptstyle u} \minus}}
    - \Gamma_{\overset{\plus {\scriptstyle v} \minus}{\minus {\scriptstyle u} \plus}}
  \bigr) \bigl(
    G_{{\scriptstyle v} {\scriptstyle u}} + G_{\plus \minus}
  \bigr)
  - \Gamma_{\overset{\minus {\scriptstyle v} {\scriptstyle v}}{\plus {\scriptstyle u} {\scriptstyle u}}}
  \, G_{\overset{{\scriptstyle u} \plus}{{\scriptstyle v} \minus}}
  - \Gamma_{\overset{\plus {\scriptstyle v} {\scriptstyle v}}{\minus {\scriptstyle u} {\scriptstyle u}}}
  \, G_{\overset{{\scriptstyle u} \minus}{{\scriptstyle v} \plus}}
  - \Gamma_{\overset{\minus {\scriptstyle v} \minus}{\plus {\scriptstyle u} \plus}}
  \, G_{\overset{\plus \plus}{\minus \minus}}
  + \Gamma_{\overset{\plus {\scriptstyle v} \plus}{\minus {\scriptstyle u} \minus}}
  \, G_{\overset{\minus \minus}{\plus \plus}}
  \Bigr]
\\
  = \ 
  \Jz_{\overset{\minus {\scriptstyle v} \plus}{\plus {\scriptstyle u} \minus}}
  &=&
  \frac{1}{2} \Bigl[
  \partial_\vu \bigl(
    {\textstyle \frac{2}{3}} G_{{\scriptstyle v} {\scriptstyle u}} + {\textstyle \frac{1}{3}} G_{\plus \minus}
  \bigr)
  - \bigl( \partial_\plusminus
    + \Gamma_{\overset{\minus \plus \plus}{\plus \minus \minus}}
    + \Gamma_{\overset{{\scriptstyle u} {\scriptstyle v} \plus}{{\scriptstyle v} {\scriptstyle u} \minus}}
    + \Gamma_{\overset{\plus {\scriptstyle u} {\scriptstyle v}}{\minus {\scriptstyle v} {\scriptstyle u}}}
  \bigr) G_{\overset{{\scriptstyle v} \minus}{{\scriptstyle u} \plus}}
\nonumber
\\
  &&
  \quad
  + \, \Gamma_{\overset{\minus {\scriptstyle v} \plus}{\plus {\scriptstyle u} \minus}}
  \bigl(
    G_{{\scriptstyle v} {\scriptstyle u}} + G_{\plus \minus}
  \bigr)
  + \Gamma_{\overset{\minus {\scriptstyle u} \plus}{\plus {\scriptstyle v} \minus}}
  \, G_{\overset{{\scriptstyle v} {\scriptstyle v}}{{\scriptstyle u} {\scriptstyle u}}}
  - \Gamma_{\overset{\minus {\scriptstyle u} {\scriptstyle v}}{\plus {\scriptstyle v} {\scriptstyle u}}}
  \, G_{\overset{{\scriptstyle v} \plus}{{\scriptstyle u} \minus}}
  - \Gamma_{\overset{\minus {\scriptstyle v} {\scriptstyle v}}{\plus {\scriptstyle u} {\scriptstyle u}}}
  \, G_{\overset{{\scriptstyle u} \plus}{{\scriptstyle v} \minus}}
  - \Gamma_{\overset{\plus {\scriptstyle v} {\scriptstyle v}}{\minus {\scriptstyle u} {\scriptstyle u}}}
  \, G_{\overset{{\scriptstyle u} \minus}{{\scriptstyle v} \plus}}
  + \Gamma_{\overset{\plus {\scriptstyle v} \plus}{\minus {\scriptstyle u} \minus}}
  \, G_{\overset{\minus \minus}{\plus \plus}}
  \Bigr]
  \ ,
  \qquad
\\
  \Jz_{\overset{\plus {\scriptstyle v} {\scriptstyle u}}{\minus {\scriptstyle u} {\scriptstyle v}}}
  \  = \ 
  \Jz_{\overset{\plus \minus \plus}{\minus \plus \minus}}
  &=&
  \frac{1}{4} \Bigl[
  - \,
  \partial_\plusminus \bigl(
    {\textstyle \frac{2}{3}} G_{{\scriptstyle v} {\scriptstyle u}} + {\textstyle \frac{1}{3}} G_{\plus \minus}
  \bigr)
  +
  \bigl( \partial_\vu
    - \Gamma_{\overset{{\scriptstyle u} {\scriptstyle v} {\scriptstyle v}}{{\scriptstyle v} {\scriptstyle u} {\scriptstyle u}}}
    - \Gamma_{\overset{\minus \plus {\scriptstyle v}}{\plus \minus {\scriptstyle u}}}
    - 2 \, \Gamma_{\overset{\plus {\scriptstyle v} \minus}{\minus {\scriptstyle u} \plus}}
    + \Gamma_{\overset{\minus {\scriptstyle v} \plus}{\plus {\scriptstyle u} \minus}}
  \bigr)
  G_{\overset{{\scriptstyle u} \plus}{{\scriptstyle v} \minus}}
\nonumber
\\
  &&
  \quad
  - \, \bigl( \partial_\uv
    - \Gamma_{\overset{{\scriptstyle v} {\scriptstyle u} {\scriptstyle u}}{{\scriptstyle u} {\scriptstyle v} {\scriptstyle v}}}
    + \Gamma_{\overset{\plus \minus {\scriptstyle u}}{\minus \plus {\scriptstyle v}}}
    - \Gamma_{\overset{\minus {\scriptstyle u} \plus}{\plus {\scriptstyle v} \minus}}
    + 2 \, \Gamma_{\overset{\plus {\scriptstyle u} \minus}{\minus {\scriptstyle v} \plus}}
  \bigr) G_{\overset{{\scriptstyle v} \plus}{{\scriptstyle u} \minus}}
  + \bigl( \partial_\minusplus
    + 2 \, \Gamma_{\overset{\plus \minus \minus}{\minus \plus \plus}}
    - \Gamma_{\overset{\minus {\scriptstyle u} {\scriptstyle v}}{\plus {\scriptstyle v} {\scriptstyle u}}}
    + \Gamma_{\overset{\minus {\scriptstyle v} {\scriptstyle u}}{\plus {\scriptstyle u} {\scriptstyle v}}}
  \bigr) G_{\overset{\plus \plus}{\minus \minus}}
\nonumber
\\
  &&
  \quad
  + \, \bigl( 
    \Gamma_{\overset{\plus {\scriptstyle v} {\scriptstyle u}}{\minus {\scriptstyle u} {\scriptstyle v}}}
    - \Gamma_{\overset{\plus {\scriptstyle u} {\scriptstyle v}}{\minus {\scriptstyle v} {\scriptstyle u}}}
  \bigr) \bigl(
    G_{{\scriptstyle v} {\scriptstyle u}} + G_{\plus \minus}
  \bigr)
  + \Gamma_{\overset{\plus {\scriptstyle u} {\scriptstyle u}}{\minus {\scriptstyle v} {\scriptstyle v}}}
  \, G_{\overset{{\scriptstyle v} {\scriptstyle v}}{{\scriptstyle u} {\scriptstyle u}}}
  - \Gamma_{\overset{\plus {\scriptstyle v} {\scriptstyle v}}{\minus {\scriptstyle u} {\scriptstyle u}}}
  \, G_{\overset{{\scriptstyle u} {\scriptstyle u}}{{\scriptstyle v} {\scriptstyle v}}}
  + \Gamma_{\overset{\plus {\scriptstyle v} \plus}{\minus {\scriptstyle u} \minus}}
  \, G_{\overset{{\scriptstyle u} \minus}{{\scriptstyle v} \plus}}
  + \Gamma_{\overset{\plus {\scriptstyle u} \plus}{\minus {\scriptstyle v} \minus}}
  \, G_{\overset{{\scriptstyle v} \minus}{{\scriptstyle u} \plus}}
  \Bigr]
\\
  = \ 
  \Jz_{\overset{{\scriptstyle u} {\scriptstyle v} \plus}{{\scriptstyle v} {\scriptstyle u} \minus}}
  &=&
  \frac{1}{2} \Bigl[
  - \,
  \partial_\plusminus \bigl(
    {\textstyle \frac{1}{3}} G_{{\scriptstyle v} {\scriptstyle u}} + {\textstyle \frac{2}{3}} G_{\plus \minus}
  \bigr)
  +
  \bigl( \partial_\vu
    - \Gamma_{\overset{{\scriptstyle u} {\scriptstyle v} {\scriptstyle v}}{{\scriptstyle v} {\scriptstyle u} {\scriptstyle u}}}
    + \Gamma_{\overset{\minus {\scriptstyle v} \plus}{\plus {\scriptstyle u} \minus}}
    - \Gamma_{\overset{\minus \plus {\scriptstyle v}}{\plus \minus {\scriptstyle u}}}
  \bigr) G_{\overset{{\scriptstyle u} \plus}{{\scriptstyle v} \minus}}
\nonumber
\\
  &&
  \quad
  - \, \Gamma_{\overset{\plus {\scriptstyle u} {\scriptstyle v}}{\minus {\scriptstyle v} {\scriptstyle u}}}
  \bigl(
    G_{{\scriptstyle v} {\scriptstyle u}} + G_{\plus \minus}
  \bigr)
  - \Gamma_{\overset{\plus {\scriptstyle v} {\scriptstyle v}}{\minus {\scriptstyle u} {\scriptstyle u}}}
  \, G_{\overset{{\scriptstyle u} {\scriptstyle u}}{{\scriptstyle v} {\scriptstyle v}}}
  + \Gamma_{\overset{\minus {\scriptstyle u} \plus}{\plus {\scriptstyle v} \minus}}
  \, G_{\overset{{\scriptstyle v} \plus}{{\scriptstyle u} \minus}}
  - \Gamma_{\overset{\minus {\scriptstyle u} {\scriptstyle v}}{\plus {\scriptstyle v} {\scriptstyle u}}}
  \, G_{\overset{\plus \plus}{\minus \minus}}
  + \Gamma_{\overset{\plus {\scriptstyle v} \plus}{\minus {\scriptstyle u} \minus}}
  \, G_{\overset{{\scriptstyle u} \minus}{{\scriptstyle v} \plus}}
  + \Gamma_{\overset{\plus {\scriptstyle u} \plus}{\minus {\scriptstyle v} \minus}}
  \, G_{\overset{{\scriptstyle v} \minus}{{\scriptstyle u} \plus}}
  \Bigr]
  \ ,
  \qquad
\\
  \Jz_{\overset{{\scriptstyle v} {\scriptstyle v} \plus}{{\scriptstyle u} {\scriptstyle u} \minus}}
  &=&
  \frac{1}{2} \Bigl[
  \bigl( \partial_\vu
    + \Gamma_{\overset{{\scriptstyle u} {\scriptstyle v} {\scriptstyle v}}{{\scriptstyle v} {\scriptstyle u} {\scriptstyle u}}}
    - \Gamma_{\overset{\minus \plus {\scriptstyle v}}{\plus \minus {\scriptstyle u}}}
    + 2 \, \Gamma_{\overset{\minus {\scriptstyle v} \plus}{\plus {\scriptstyle u} \minus}}
  \bigr) G_{\overset{{\scriptstyle v} \plus}{{\scriptstyle u} \minus}}
  - \bigl( \partial_\plusminus
    + 2 \, \Gamma_{\overset{{\scriptstyle u} {\scriptstyle v} \plus}{{\scriptstyle v} {\scriptstyle u} \minus}}
    + \Gamma_{\overset{\plus {\scriptstyle u} {\scriptstyle v}}{\minus {\scriptstyle v} {\scriptstyle u}}}
  \bigr) G_{\overset{{\scriptstyle v} {\scriptstyle v}}{{\scriptstyle u} {\scriptstyle u}}}
\nonumber
\\
  &&
  \quad
  - \, \Gamma_{\overset{\plus {\scriptstyle v} {\scriptstyle v}}{\minus {\scriptstyle u} {\scriptstyle u}}}
  \bigl(
    G_{{\scriptstyle v} {\scriptstyle u}} + G_{\plus \minus}
  \bigr)
  - \Gamma_{\overset{\minus {\scriptstyle v} {\scriptstyle v}}{\plus {\scriptstyle u} {\scriptstyle u}}}
  \, G_{\overset{\plus \plus}{\minus \minus}}
  + 2 \, \Gamma_{\overset{\plus {\scriptstyle v} \plus}{\minus {\scriptstyle u} \minus}}
  \, G_{\overset{{\scriptstyle v} \minus}{{\scriptstyle u} \plus}}
  \Bigr]
  \ ,
  \qquad
\\
  \Jz_{\overset{\plus {\scriptstyle v} \plus}{\minus {\scriptstyle u} \minus}}
  &=&
  \frac{1}{2} \Bigl[
  \bigl( \partial_\vu
    - 2 \, \Gamma_{\overset{\minus \plus {\scriptstyle v}}{\plus \minus {\scriptstyle u}}}
    + \Gamma_{\overset{\minus {\scriptstyle v} \plus}{\plus {\scriptstyle u} \minus}}
  \bigr) G_{\overset{\plus \plus}{\minus \minus}}
  - \bigl( \partial_\plusminus
    - \Gamma_{\overset{\minus \plus \plus}{\plus \minus \minus}}
    + \Gamma_{\overset{{\scriptstyle u} {\scriptstyle v} \plus}{{\scriptstyle v} {\scriptstyle u} \minus}}
    + 2 \, \Gamma_{\overset{\plus {\scriptstyle u} {\scriptstyle v}}{\minus {\scriptstyle v} {\scriptstyle u}}}
  \bigr) G_{\overset{{\scriptstyle v} \plus}{{\scriptstyle u} \minus}}
\nonumber
\\
  &&
  \quad
  + \, \Gamma_{\overset{\plus {\scriptstyle v} \plus}{\minus {\scriptstyle u} \minus}}
    \bigl( G_{{\scriptstyle v} {\scriptstyle u}} + G_{\plus \minus} \bigr)
  + \Gamma_{\overset{\plus {\scriptstyle u} \plus}{\minus {\scriptstyle v} \minus}}
    \, G_{\overset{{\scriptstyle v} {\scriptstyle v}}{{\scriptstyle u} {\scriptstyle u}}}
  - 2 \, \Gamma_{\overset{\plus {\scriptstyle v} {\scriptstyle v}}{\minus {\scriptstyle u} {\scriptstyle u}}}
    \, G_{\overset{{\scriptstyle u} \plus}{{\scriptstyle v} \minus}}
  \Bigr]
  \ .
  \qquad
\end{eqnarray}
\end{subequations}

\subsection{Electromagnetic energy-momentum tensor}

The Newman-Penrose components of the energy-momentum tensor
$T_{mn} = ( 1 / 4 \pi ) \bigl( F_{mk} {F_{n}}^k - \frac{1}{4} \gamma_{mn} F_{kl} F^{kl} \bigr)$
of the electromagnetic field
are given by:
\begin{subequations}
\label{npTem}
\begin{eqnarray}
  4 \pi T_{uv}
  =
  4 \pi T_{\plus \minus}
  &=&
  {\textstyle \frac{1}{2}}
  \bigl(
    F_{uv}^2 - F_{\plus\minus}^2
  \bigr)
  =
  2 \,
  \Fz^{}_\zero
  \Fz^\star_\zero
  \ ,
\\
  4 \pi T_{\overset{{\scriptstyle v v}}{{\scriptstyle u u}}}
  &=&
  2 \,
  F_{\overset{{\scriptstyle v} \plus}{{\scriptstyle u} \minus}}
  F_{\overset{{\scriptstyle v} \minus}{{\scriptstyle u} \plus}}
  =
  2 \,
  \Fz^{}_{\plusminus \one}
  \Fz^\star_{\plusminus \one}
  \ ,
\\
  4 \pi T_{\overset{{\scriptstyle v} \plus}{{\scriptstyle u} \minus}}
  &=&
  \bigl(
    F_{\overset{{\scriptstyle v u}}{{\scriptstyle u v}}} + F_{\overset{\plus\minus}{\minus\plus}}
  \bigr)
  F_{\overset{{\scriptstyle v} \plus}{{\scriptstyle u} \minus}}
  =
  \mp \, 2 \,
  \Fz^\star_{\zero}
  \Fz^{}_{\plusminus \one}
  \ ,
\\
  4 \pi T_{\overset{{\scriptstyle v} \minus}{{\scriptstyle u} \plus}}
  &=&
  \bigl(
    F_{\overset{{\scriptstyle v u}}{{\scriptstyle u v}}} + F_{\overset{\minus\plus}{\plus\minus}}
  \bigr)
  F_{\overset{{\scriptstyle v} \minus}{{\scriptstyle u} \plus}}
  =
  \mp \, 2 \,
  \Fz^{}_{\zero}
  \Fz^\star_{\plusminus \one}
  \ ,
\\
  4 \pi T_{\overset{\plus \plus}{\minus \minus}}
  &=&
  2 \,
  F_{\overset{{\scriptstyle v} \plus}{{\scriptstyle u} \minus}}
  F_{\overset{{\scriptstyle u} \plus}{{\scriptstyle v} \minus}}
  =
  2 \,
  \Fz^{}_{\plusminus \one}
  \Fz^\star_{\minusplus \one}
  \ .
\end{eqnarray}
\end{subequations}

\subsection{Scalar energy-momentum tensor}

The Newman-Penrose components of the energy-momentum tensor
$T_{mn} = ( 1 / 4 \pi ) \bigl( \pi_m \pi_n - \frac{1}{2} \gamma_{mn} \pi_k \pi^k \bigr)$
of a massless scalar field $\psi$,
with gradient (momentum)
$\pi_m \equiv \partial_m \psi$,
are given by:
\begin{subequations}
\label{npTscalar}
\begin{eqnarray}
  4 \pi T_{uv}
  &=&
  \pi_\plus \pi_\minus
  \ ,
\\
  4 \pi T_{\plus \minus}
  &=&
  \pi_v \pi_u
  \ ,
\\
  4 \pi T_{\overset{{\scriptstyle v v}}{{\scriptstyle u u}}}
  &=&
  \pi_\vu \pi_\vu
  \ ,
\\
  4 \pi T_{\overset{{\scriptstyle v} \plus}{{\scriptstyle u} \minus}}
  &=&
  \pi_\vu \pi_\plusminus
  \ ,
\\
  4 \pi T_{\overset{{\scriptstyle v} \minus}{{\scriptstyle u} \plus}}
  &=&
  \pi_\vu \pi_\minusplus
  \ ,
\\
  4 \pi T_{\overset{\plus \plus}{\minus \minus}}
  &=&
  \pi_\plusminus \pi_\plusminus
  \ .
\end{eqnarray}
\end{subequations}

\subsection{Some operator commutators}

The following equation is referred to in \S\ref{evolutionweyl},
equation~(\ref{linDeltaC}):
\begin{eqnarray}
\label{DeltaC}
\lefteqn{
  \Bigl(
  \nDelta{\zero}^\prime_\plusminus \, \nDelta{\zero}_\vu
  -
  \nDelta{\zero}^\prime_\vu \, \nDelta{\zero}_\plusminus
  \Bigr)
  A^{-1} C
}
  &&
\nonumber
\\
  &=&
  3 A C
  \Bigl[
  2 \, \Cz_{\plusminus \one}
  -
  \Bigl(
  \partial_\uv
  -
  2 \,
  \Gamma_{\overset{{\scriptstyle v u u}}{{\scriptstyle u v v}}}
  +
  \Gamma_{\overset{\plus \minus {\scriptstyle u}}{\minus \plus {\scriptstyle v}}}
  -
  \Gamma_{\overset{\minus {\scriptstyle u} \plus}{\plus {\scriptstyle v} \minus}}
  \Bigr)
  \Gamma_{\overset{\plus {\scriptstyle v} {\scriptstyle v}}{\minus {\scriptstyle u} {\scriptstyle u}}}
  +
  \Bigl(
  \partial_\minusplus
  +
  2 \,
  \Gamma_{\overset{\plus \minus \minus}{\minus \plus \plus}}
  -
  \Gamma_{\overset{{\scriptstyle v u} \minus}{{\scriptstyle u v} \plus}}
  -
  \Gamma_{\overset{\minus {\scriptstyle u} {\scriptstyle v}}{\plus {\scriptstyle v} {\scriptstyle u}}}
  \Bigr)
  \Gamma_{\overset{\plus {\scriptstyle v} \plus}{\minus {\scriptstyle u} \minus}}
  \Bigr]
\nonumber
\\
  &&
  + \,
  A^3
  \Bigl[
  - \,
  \Gamma_{\overset{\plus {\scriptstyle v} {\scriptstyle v}}{\minus {\scriptstyle u} {\scriptstyle u}}}
  \Bigl(
  \partial_\uv
  +
  3 \,
  \Gamma_{\overset{\plus {\scriptstyle u} \minus}{\minus {\scriptstyle v} \plus}}
  \Bigr)
  +
  \Gamma_{\overset{\plus {\scriptstyle v} \plus}{\minus {\scriptstyle u} \minus}}
  \Bigl(
  \partial_\minusplus
  +
  3 \,
  \Gamma_{\overset{\minus {\scriptstyle v u}}{\plus {\scriptstyle u v}}}
  \Bigr)
  \Bigr]
  A^{-2} C
  \ .
\end{eqnarray}
The following equation is referred to in \S\ref{evolutionweyl},
equation~(\ref{linDeltaCa}):
\begin{eqnarray}
\label{DeltaCa}
\lefteqn{
  \Bigl(
  \nDelta{\plus \one}^\prime_u \, \nDelta{\zero}_v
  -
  \nDelta{\plus \one}^\prime_\minus \, \nDelta{\zero}_\plus
  -
  \nDelta{\plus \one}^{\prime \star}_u \, \nDelta{\zero}^\star_v
  +
  \nDelta{\plus \one}^{\prime \star}_\minus \, \nDelta{\zero}^\star_\plus
  \Bigr)
  A^{-1} C
}
  &&
\nonumber
\\
  &=&
  3 A C
  \Bigl[
  2 \, \Cz^{(a)}_\zero
  +
  \Gamma_{\plus v v} \,
  \Gamma_{\minus u u}
  -
  \Gamma_{\plus v \plus} \,
  \Gamma_{\minus u \minus}
  -
  \Gamma_{\minus v v} \,
  \Gamma_{\plus u u}
  +
  \Gamma_{\minus v \minus} \,
  \Gamma_{\plus u \plus}
  \Bigr]
\nonumber
\\
  &&
  + \,
  2 A^3
  \Bigl[
  - \,
  (
    \Gamma_{\minus v \plus}
    -
    \Gamma_{\plus v \minus}
  )
  \bigl(
    \partial_u
    + {\textstyle \frac{3}{2}} \, \Gamma_{\plus u \minus}
    + {\textstyle \frac{3}{2}} \, \Gamma_{\minus u \plus}
  \bigl)
  \, + \,
  (
    \Gamma_{\plus u v}
    -
    \Gamma_{\plus v u}
  )
  \bigl(
    \partial_\minus
    + {\textstyle \frac{3}{2}} \, \Gamma_{\minus v u}
    + {\textstyle \frac{3}{2}} \, \Gamma_{\minus u v}
  \bigl)
\nonumber
\\
  &&
  \quad\quad\,
  - \,
  (
    \Gamma_{\plus u \minus}
    -
    \Gamma_{\minus u \plus}
  )
  \bigl(
    \partial_v
    + {\textstyle \frac{3}{2}} \, \Gamma_{\minus v \plus}
    + {\textstyle \frac{3}{2}} \, \Gamma_{\plus v \minus}
  \bigl)
  \, + \,
  (
    \Gamma_{\minus v u}
    -
    \Gamma_{\minus u v}
  )
  \bigl(
    \partial_\plus
    + {\textstyle \frac{3}{2}} \, \Gamma_{\plus u v}
    + {\textstyle \frac{3}{2}} \, \Gamma_{\plus v u}
  \bigl)
  \Bigr]
  A^{-2} C
  \ .
\end{eqnarray}
The following equation is referred to in \S\ref{maxwellsequations},
equation~(\ref{linfrakDQ})
[compare equation~(\ref{DeltaC}), which is quite similar]:
\begin{eqnarray}
\label{frakDQ}
\lefteqn{
  \Bigl(
  \nfrakD{\zero}^\prime_\plusminus \, \nfrakD{\zero}_\vu
  -
  \nfrakD{\zero}^\prime_\vu \, \nfrakD{\zero}_\plusminus
  \Bigr)
  A^{-1} Q
}
  &&
\nonumber
\\
  &=&
  2 A Q
  \Bigl[
  2 \, \Cz_{\plusminus \one}
  - \,
  \Bigl(
  \partial_\uv
  -
  2 \,
  \Gamma_{\overset{{\scriptstyle v u u}}{{\scriptstyle u v v}}}
  +
  \Gamma_{\overset{\plus \minus {\scriptstyle u}}{\minus \plus {\scriptstyle v}}}
  -
  \Gamma_{\overset{\minus {\scriptstyle u} \plus}{\plus {\scriptstyle v} \minus}}
  \Bigr)
  \Gamma_{\overset{\plus {\scriptstyle v} {\scriptstyle v}}{\minus {\scriptstyle u} {\scriptstyle u}}}
  +
  \Bigl(
  \partial_\minusplus
  +
  2 \,
  \Gamma_{\overset{\plus \minus \minus}{\minus \plus \plus}}
  -
  \Gamma_{\overset{{\scriptstyle v u} \minus}{{\scriptstyle u v} \plus}}
  -
  \Gamma_{\overset{\minus {\scriptstyle u} {\scriptstyle v}}{\plus {\scriptstyle v} {\scriptstyle u}}}
  \Bigr)
  \Gamma_{\overset{\plus {\scriptstyle v} \plus}{\minus {\scriptstyle u} \minus}}
  \Bigr]
\nonumber
\\
  &&
  + \,
  A^2
  \Bigl[
  - \,
  \Gamma_{\overset{\plus {\scriptstyle v} {\scriptstyle v}}{\minus {\scriptstyle u} {\scriptstyle u}}}
  \Bigl(
  \partial_\uv
  +
  2 \,
  \Gamma_{\overset{\plus {\scriptstyle u} \minus}{\minus {\scriptstyle v} \plus}}
  \Bigr)
  +
  \Gamma_{\overset{\plus {\scriptstyle v} \plus}{\minus {\scriptstyle u} \minus}}
  \Bigl(
  \partial_\minusplus
  +
  2 \,
  \Gamma_{\overset{\minus {\scriptstyle v u}}{\plus {\scriptstyle u v}}}
  \Bigr)
  \Bigr]
  A^{-1} Q
  \ .
\end{eqnarray}
The following equation is referred to in \S\ref{maxwellsequations},
equation~(\ref{linfrakDQa})
[compare equation~(\ref{DeltaCa}), which is quite similar]:
\begin{eqnarray}
\label{frakDQa}
\lefteqn{
  \Bigl(
  \nfrakD{\plus \one}^\prime_u \, \nfrakD{\zero}_v
  -
  \nfrakD{\plus \one}^\prime_\minus \, \nfrakD{\zero}_\plus
  -
  \nfrakD{\plus \one}^{\prime \star}_u \, \nfrakD{\zero}^\star_v
  +
  \nfrakD{\plus \one}^{\prime \star}_\minus \, \nfrakD{\zero}^\star_\plus
  \Bigr)
  A^{-1} Q
}
  &&
\nonumber
\\
  &=&
  2 A C
  \Bigl[
  2 \, \Cz^{(a)}_\zero
  +
  \Gamma_{\plus v v} \,
  \Gamma_{\minus u u}
  -
  \Gamma_{\plus v \plus} \,
  \Gamma_{\minus u \minus}
  -
  \Gamma_{\minus v v} \,
  \Gamma_{\plus u u}
  +
  \Gamma_{\minus v \minus} \,
  \Gamma_{\plus u \plus}
  \Bigr]
\nonumber
\\
  &&
  + \,
  A^2
  \Bigl[
  - \,
  (
    \Gamma_{\minus v \plus}
    -
    \Gamma_{\plus v \minus}
  )
  \bigl(
    \partial_u
    + \Gamma_{\plus u \minus}
    + \Gamma_{\minus u \plus}
  \bigl)
  \, + \,
  (
    \Gamma_{\plus u v}
    -
    \Gamma_{\plus v u}
  )
  \bigl(
    \partial_\minus
    + \Gamma_{\minus v u}
    + \Gamma_{\minus u v}
  \bigl)
\nonumber
\\
  &&
  \quad\quad\,
  - \,
  (
    \Gamma_{\plus u \minus}
    -
    \Gamma_{\minus u \plus}
  )
  \bigl(
    \partial_v
    + \Gamma_{\minus v \plus}
    + \Gamma_{\plus v \minus}
  \bigl)
  \, + \,
  (
    \Gamma_{\minus v u}
    -
    \Gamma_{\minus u v}
  )
  \bigl(
    \partial_\plus
    + \Gamma_{\plus u v}
    + \Gamma_{\plus v u}
  \bigl)
  \Bigr]
  A^{-1} Q
  \ .
\end{eqnarray}


\end{document}